\documentclass[a4paper,fleqn,usenatbib]{mnras}

\usepackage{newtxtext,newtxmath}
\usepackage{graphicx}
\usepackage[T1]{fontenc}

\usepackage{subfig}
\usepackage[flushleft]{threeparttable}

\DeclareRobustCommand{\VAN}[3]{#2}
\let\VANthebibliography\thebibliography
\def\thebibliography{\DeclareRobustCommand{\VAN}[3]{##3}\VANthebibliography}

\title[Relics of buckling events]{Testing for relics of past strong buckling events in edge-on galaxies: Simulation predictions and data from S$^4$G} 

\author[Cuomo et al.]{
Virginia Cuomo,$^{1}$\thanks{E-mail: virginia.cuomo@uda.cl}
Victor P. Debattista,$^{2}$
Sarah Racz,$^{3}$
Stuart Robert Anderson,$^{2}$
\newauthor Peter Erwin,$^{4,5}$
Oscar A. Gonzalez,$^{6}$
J. W. Powell,$^{7}$
Enrico Maria Corsini,$^{8,9}$
\newauthor Lorenzo Morelli,$^{1}$
and Mark A. Norris$^{2}$
\\
$^{1}$Instituto de Astronom\'ia y Ciencias Planetarias, Universidad de Atacama, Avenida Copayapu 485, 1350000 Copiap\'o, Chile\\
$^{2}$Jeremiah Horrocks Institute, University of Central Lancashire, Preston, PR1 2HE, UK\\
$^{3}$Department of Physics, The University of Texas at Austin, Austin, TX, 78712, USA\\
$^{4}$Max-Planck-Insitut f\"{u}r extraterrestrische Physik, Giessenbachstra{\ss}e, D-85748 Garching, Germany\\
$^{5}$Universit\"{a}ts-Sternwarte M\"{u}nchen, Scheinerstra{\ss}e 1, D-81679 M\"{u}nchen, Germany\\
$^{6}$UK Astronomy Technology Centre, Royal Observatory, Blackford Hill, Edinburgh EH9 3HJ, UK\\
$^{7}$Physics Department, Reed College, Portland, OR, 97202, USA\\
$^{8}$Dipartimento di Fisica e Astronomia ``G. Galilei'', Università di Padova, vicolo dell'Osservatorio 3, 35122 Padova, Italy\\
$^{9}$INAF – Osservatorio Astronomico di Padova, vicolo dell'Osservatorio 2, 35122 Padova, Italy
}

\pubyear{2022}

\hypersetup{draft}
\begin{document}
\label{firstpage}
\pagerange{\pageref{firstpage}--\pageref{lastpage}}
\maketitle

\begin{abstract}
The short-lived buckling instability is responsible for the formation of at least some box/peanut (B/P) shaped bulges, which are observed in most massive, $z=0$, barred galaxies. Nevertheless, it has also been suggested that B/P bulges form via the slow trapping of stars onto vertically extended resonant orbits. The key difference between these two scenarios is that when the bar buckles, symmetry about the mid-plane is broken for a period of time. We use a suite of simulations (with and without gas) to show that when the buckling is sufficiently strong, a residual mid-plane asymmetry persists for several Gyrs after the end of the buckling phase, and is visible in simulation images. On the other hand, images of B/P bulges formed through resonant trapping and/or weak buckling remain symmetric about the mid-plane. We develop two related diagnostics to identify and quantify mid-plane asymmetry in simulation images of galaxies that are within 3\degr{} of edge-on orientation, allowing us to test whether the presence of a B/P-shaped bulge can be explained by a past buckling event.
We apply our diagnostics to two nearly edge-on galaxies with B/P bulges from the {\it Spitzer} Survey of Stellar Structure in Galaxies, finding no mid-plane asymmetry, implying these galaxies formed their bulges either by resonant trapping or by buckling more than $\sim 5$ Gyr ago. We conclude that the formation of B/P bulges through strong buckling may be a rare event in the past $\sim 5$ Gyr.

\end{abstract}



\section{Introduction}

Most barred galaxies with stellar mass $M_{*} \gtrsim 2.5\times10^{10}~{\rm M}_{\sun}$ host box/peanut-shaped bulges (hereafter B/P bulges, \citealt{Erwin2013,Laurikainen2014,Erwin2017,Kruk2019}). B/P bulges can easily be identified in edge-on galaxies from the shape that gives them their name \citep{Bureau1999,Lutticke2000,Chung2004,Bureau2006,Yoshino2015}. The Milky Way itself hosts a B/P bulge, which is sometimes referred to as the X-shaped bulge because our unique perspective allows us to distinguish the two arms of the density peaks along the line-of-sight \citep{Nataf2010,McWilliam2010,Saito2011,Ness2016}. 

The first 3-D simulations of barred galaxies revealed that bars readily form a B/P bulge \citep{Combes1981}, which can be supported by several resonant and non-resonant orbit families \citep{Pfenniger1984,Combes1990,Pfenniger1991,Patsis2002,Athanassoula2005,Portail2015,Abbott2017}. Two main mechanisms have been proposed to explain how barred galaxies acquire a B/P bulge: the buckling (bending) instability of the bar \citep{Toomre1966,Raha1991, Merritt1994, Debattista2006, Saha2013, Lokas2020, Collier2020} and the secular trapping of stars onto resonant orbits as the bar evolves \citep{Combes1981, Combes1990, Quillen2002, Quillen2014}. Recently, \cite{Sellwood2020} unambiguously demonstrated the formation of B/P bulges via secular trapping of stars at the 2:1 vertical resonance, without any buckling. 

The buckling instability represents a short-lived phase during which a bar bends and develops vertical asymmetry, typically with the inner portion moving upwards (downwards) while the outer portion moves downwards (upwards) in an $m=2$ pattern as seen from above \citep{Raha1991, Merritt1994}. This instability arises because the formation of the bar increases the radial random motion but does not appreciably increase the vertical motions. The resulting anisotropic distribution is unstable and drives the buckling, which in pure $N$-body simulations (i.e. without gas) often occurs shortly after the bar forms \citep{Raha1991, Debattista2006, Saha2013}. Buckling leads to significant vertical heating, and thus thickening, while the bar itself becomes more centrally concentrated. 
The buckling instability can occur more than once during the secular growth of a bar. \cite{Martinez-Valpuesta2006} showed that continued growth of the bar can trigger a second buckling event, leading to the formation of a new B/P bulge. During the second buckling, the bar remains vertically asymmetric for a longer time compared with the first one. %
The buckling instability has been suggested to explain the B/P bulge of our Galaxy \citep[e.g.,][]{Li2015,Khoperskov2019}. \cite{Debattista2017} showed that even weak recurrent buckling is able to produce many of the trends observed in the Milky Way's bulge, such as an X shape traced only by the metal-rich stars, a vertical metallicity gradient, and a weaker bar in the oldest stars \citep[e.g.,][]{Zoccali2008,Johnson2013,Ness2013,Gonzalez2015,Rojasarriagada2017,Zoccali2017}.

B/P bulges may also form via the slow trapping of stars onto vertically extended resonant orbits during the secular growth of the bar \citep{Combes1990, Quillen2002, Quillen2014, Sellwood2020}. No deviation from symmetry about the mid-plane is induced by this mode of B/P formation. While the expected trends from the trapping scenario have not been explored in detail \citep[but see][]{Chiba2021}, it is likely to lead to results similar to a mild buckling case. The key difference between these two B/P formation scenarios is that in the strong buckling case, mid-plane symmetry is broken for a period of time and a mid-plane asymmetry is produced, which is not present in the resonant trapping case. 

Because the buckling event is relatively short-lived, observationally testing the importance of buckling is difficult. Less than ten barred galaxies are known to be currently buckling \citep{Erwin2016, Li2017, Xiang2021}, which is insufficient to infer which is the dominant mechanism for forming B/P bulges. \cite{Erwin2016} presented a photometric and kinematic analysis of two buckling galaxies, NGC~4569 and NGC~3227. Based on their initial sample, they estimated that the fraction of barred galaxies with B/P bulges is consistent with all of them having formed via buckling. Observational studies showed that the fraction of B/P bulges depends strongly on a galaxy's stellar mass, $M_*$, (\citealt{Erwin2017}; Erwin et al. 2022, {\it in preparation}), with a sharp rise at $M_{*}\simeq  2.5\times10^{10}~{\rm M}_{\sun}$ \citep[see also][]{Li2017}, and no apparent dependence on the gas fraction, despite theoretical studies showing that gas can suppress buckling \citep{Berentzen1998, Debattista2006, Berentzen2007, Wozniak2009}. This critical mass seems to be unchanged out to a redshift $z\sim1$ \citep{Kruk2019}. Whereas a mass dependence for instabilities is quite plausible, why galaxies would avoid secular trapping below a particular mass may seem harder to understand. Nevertheless, the Illustris TNG50 cosmological simulation suggests that the fraction of B/P bulges decreases at lower mass because the corresponding bars form later and may not yet have had time to buckle (Anderson et al. 2022, {\it in preparation}). 

To gain further insight into the formation of a B/P bulge, it would be useful to be able to distinguish whether this central structure formed via strong buckling or via secular trapping. Since strong buckling breaks the mid-plane symmetry  \citep{Raha1991, oneill2003, Debattista2006}, the rate of vertical scattering of stars does not have to be equal on the two sides of the mid-plane. As a result, large scale breaking of symmetry during strong buckling may leave a long lasting density mid-plane asymmetry. Such mid-plane asymmetries have been already pointed out in many simulations, including the best fit Milky Way model of \cite{Gardner2014}. Likewise, the Milky Way model of \cite{Shen2010} is visibly asymmetric, as can be seen in the visualisation of \cite{Li2015}. Further examples come from the models of \cite{Saha2013} and \cite{Smirnov2018}, which are clearly asymmetric, as well as from cosmological simulations \citep{Fragkoudi2020}.
In this study we use a suite of simulations (with and without gas) to investigate the connection between the mode of B/P formation and the resulting mid-plane asymmetry/symmetry. Therefore, we present mid-plane asymmetry diagnostics to identify the formation mechanism of B/P bulges, suitable for galaxies that are within $3\degr$ of the edge-on orientation.

We then consider a sample of nearly edge-on galaxies from the {\it Spitzer} Survey of Stellar Structure in Galaxies \citep[S$^4$G,][]{Sheth2010, MunozMateos2013, Querejeta2015} hosting a B/P bulge to conduct a pilot study. We apply the mid-plane asymmetry diagnostics to discuss the formation scenario of the observed B/P bulges and compare to predictions from the simulations.

This paper is organised as follows: in Sec.~\ref{sec:models} we present the set of simulations and the diagnostics developed to identify mid-plane asymmetries, and we discuss the prospects for detecting such asymmetries in real galaxies in Sec.~\ref{sec:obs_compl}. In Sec.~\ref{sec:s4g_galaxies} we present the sample of galaxies used to hunt for mid-plane asymmetries, while in Sec.~\ref{sec:evidences_real} we present the results. We discuss our results in Sec.~\ref{sec:discussion} and summarise and conclude in Sec.~\ref{sec:conclusion}. Finally, we include in a series of appendices available online, \ref{app:sim}, \ref{appendix:b}, and \ref{appendix:c}, complementary material related to our analysis. 
Where necessary, we adopt the cosmological parameters $\Omega_{m} = 0.308, \Omega_{\Lambda} = 0.692$, and $H_0 = 67.8$ km s$^{-1}$ Mpc$^{-1}$ \citep{Planck2016}. 

\section{Mid-plane asymmetry formation and evolution in models}
\label{sec:models}

A buckling event is characterised by a period when the morphology of the bar deviates from mid-plane symmetry. For a strong buckling event the asymmetry is visible by eye and may affect a large region of the galaxy, whereas gentle buckling causes only a mild deviation from symmetry, and requires a careful analysis to identify. 

We explore asymmetries using two high resolution simulations from \cite{Debattista2017}. First, we consider the pure $N$-body simulation (no gas or star formation) referred to as model D5 in \cite{Debattista2017}, which suffered a modest buckling event during its evolution. Buckling in model D5 occurs at $t=4$ Gyr: the global amplitude $A_{\rm buck}$, defined as the $m = 2$ bending amplitude \citep{Raha1991, Debattista2006, Debattista2020}, reaches a value of $\sim0.04$ kpc at $\sim4$ Gyr. The bar abruptly weakens as the buckling occurs, with the normalized amplitude of the $m=2$ density distribution, $A_{\rm bar}$, dropping from $\sim0.2$ to $\sim0.1$. After the instability, the bar slowly recovers strength, reaching $A_{\rm bar} \sim0.15$ by $t=10$~Gyr \citep[see Figure 2 in][where model D5 is denoted as model 1]{Debattista2020}.

We also consider the star-forming simulation from \cite{Debattista2017} which, following \citet{Gardner2014}, we refer to as model HG1. This model undergoes no strong buckling event but appears to suffer from weak, but recurrent, small-scale buckling activity: $A_{\rm buck}$ remains below $\sim0.0008$ kpc, while $A_{\rm bar}$ oscillates between $0.1-0.15$ during the 10 Gyr of its evolution \citep[see][]{Anderson2022}. Nevertheless, a B/P bulge develops by $t=5$ Gyr. This model includes gas, from which all stars form. 

\begin{figure*}
    \centering
    \includegraphics[scale=0.5]{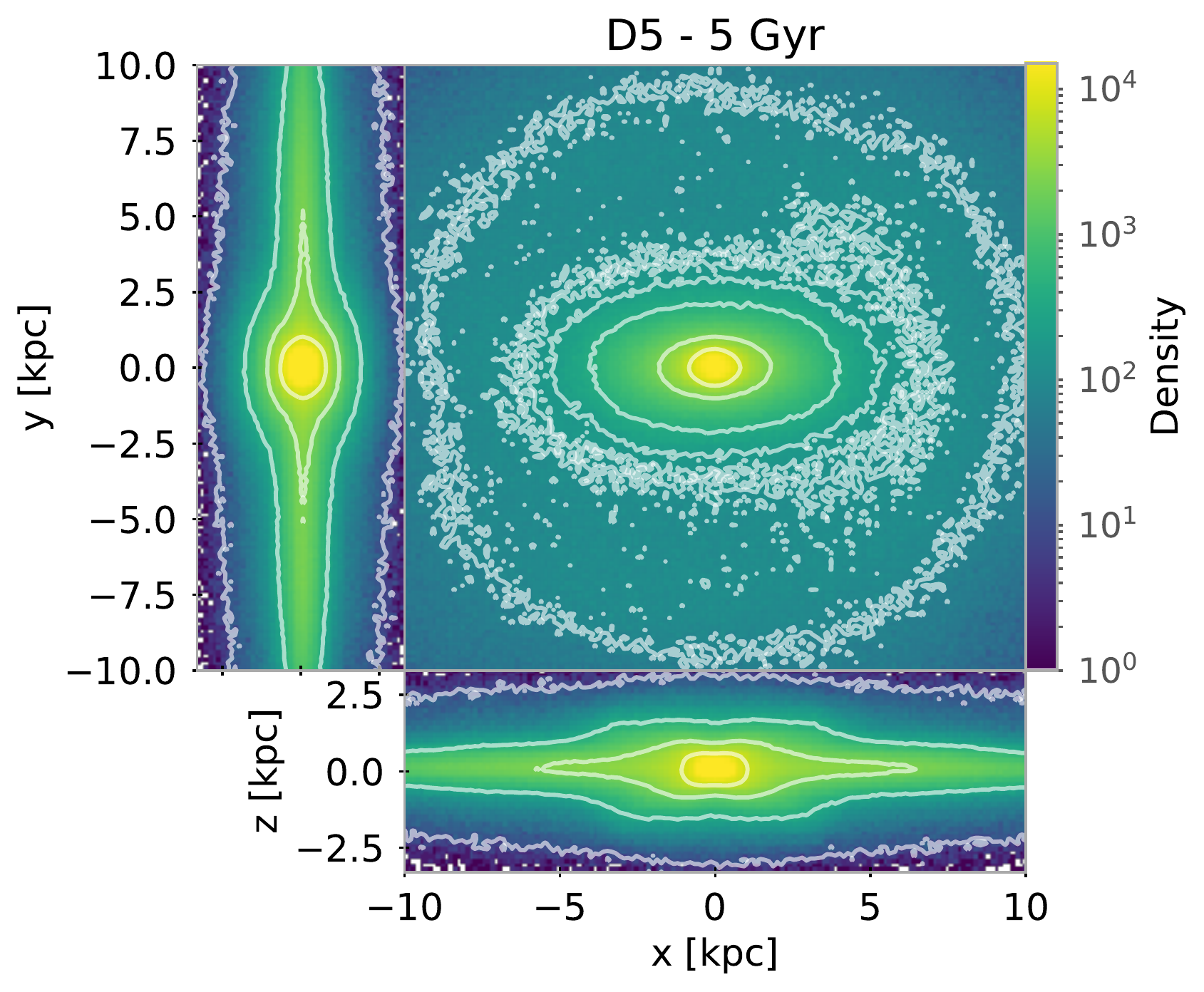}
    \includegraphics[scale=0.5]{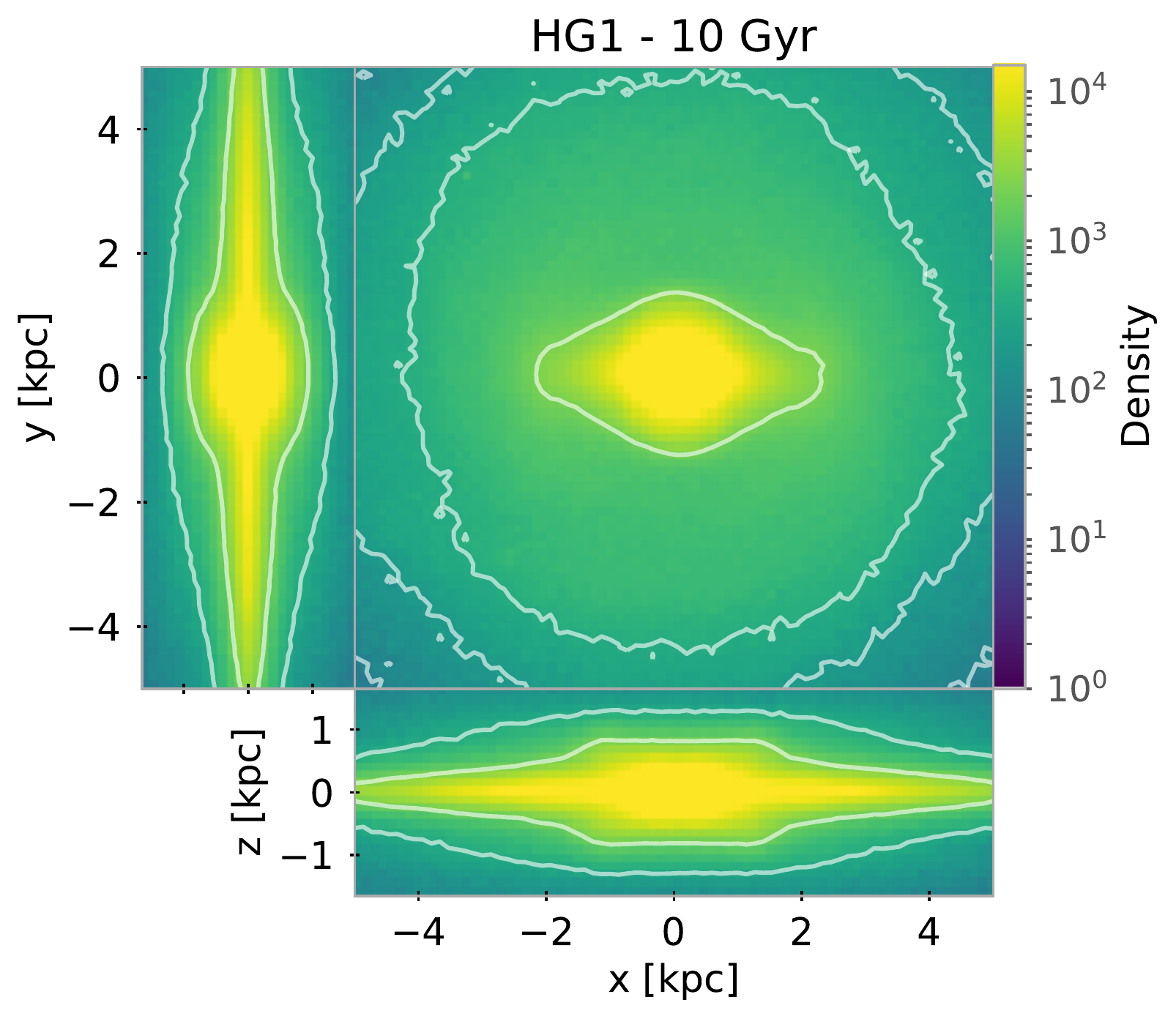}
    \caption{Maps of the surface number density of models D5 (at 5 Gyr, left-hand column) and HG1 (at 10 Gyr, right-hand column) seen face-on (upper right-hand panel), side-on (lower right-hand panel), and end-on (left-hand panel). Some isocontours of the surface number density (white lines) are shown for each viewing geometry. In all the plots the bar major axis coincides with the $x$ axis.} 
    \label{fig:sim_3d_view}
\end{figure*}

Besides these two models, which are typical of the type of behaviours expected, we have repeated our analysis using several other barred simulations. In particular, we have analysed the pure $N$-body models D8, T1, and the SD1 which are studied in detail in \cite{Anderson2022}, where they are denoted as model 4, T1, and SD1,  respectively. We have verified that the results are qualitatively consistent with those presented here. In Appendix~\ref{app:sim} we show the results for all the remaining models. We base our asymmetry analysis on model D5 because its asymmetries are rather mild and should be more difficult to detect them observationally; this reference model allows us to explore the detectability of the mid-plane asymmetry and to test the limits of our diagnostics.  

Fig.~\ref{fig:sim_3d_view} shows the face-on, side-on and end-on views of the projected logarithmically-scaled surface number density of the particles of the two models, centred on the origin of the $(x,y)$ plane, with the bar aligned with the $x$ axis, where the galaxy centre is located at $(x,y,z)=(0,0,0)$. We consider the snapshot at 5 Gyr (1 Gyr after the buckling event) for model D5 and the snapshot at 10 Gyr (the end of the simulation) for model HG1. Surface density contours are shown in white: they appear boxy in the central regions of both models in the $(x,z)$ plane (the side-on view, where the disc position angle, PA, is aligned with the $x$ axis and $i=90\degr$, while the bar is aligned with the $x$ axis). A slight asymmetry with respect to the mid-plane is barely visible by eye in model D5, which we explore below. 

\subsection{Density mid-plane asymmetry diagnostics}
\label{sec:diagnostics}

We start by applying the median filtering unsharp mask technique to the logarithmically-scaled projected number density of the particles for the edge-on view of the models, with the bar seen edge-on. This technique has been used on both observational and simulated data \citep[e.g.,][]{Athanassoula2005,Bureau2006} because it highlights the X-shaped structure in the central regions of the galaxies, which is associated with a B/P bulge \citep{Bureau2006}. We convolve the image of the projected number density of the particles of the models with a circular Gaussian and manually vary the corresponding standard deviation, to highlight the X shape. We use the \textsc{python-scipy} function \textsc{gaussian\_filter} and subtract the convolved image from the original one. 

To better visualise the mid-plane asymmetry observed in Fig.~\ref{fig:sim_3d_view} for model D5 we develop two related diagnostics. We first map the mid-plane asymmetry in the surface density, $A_\Sigma \left( x,z \right)$ as:
\begin{equation}
\label{eqn:asym}
A_\Sigma \left(x,z\right) = \frac{\Sigma \left( x,z \right)-\Sigma \left( x,-z \right)}{\Sigma \left( x,z \right)+\Sigma \left( x,-z \right)}
\end{equation}
where $\Sigma \left( x,z \right)$ is the projected surface number density of the particles at each position of the image of the edge-on view of the models. We refer to the resulting figure as the `mid-plane asymmetry map'. 

\begin{figure*}
    \centering
    \includegraphics[scale=0.53]{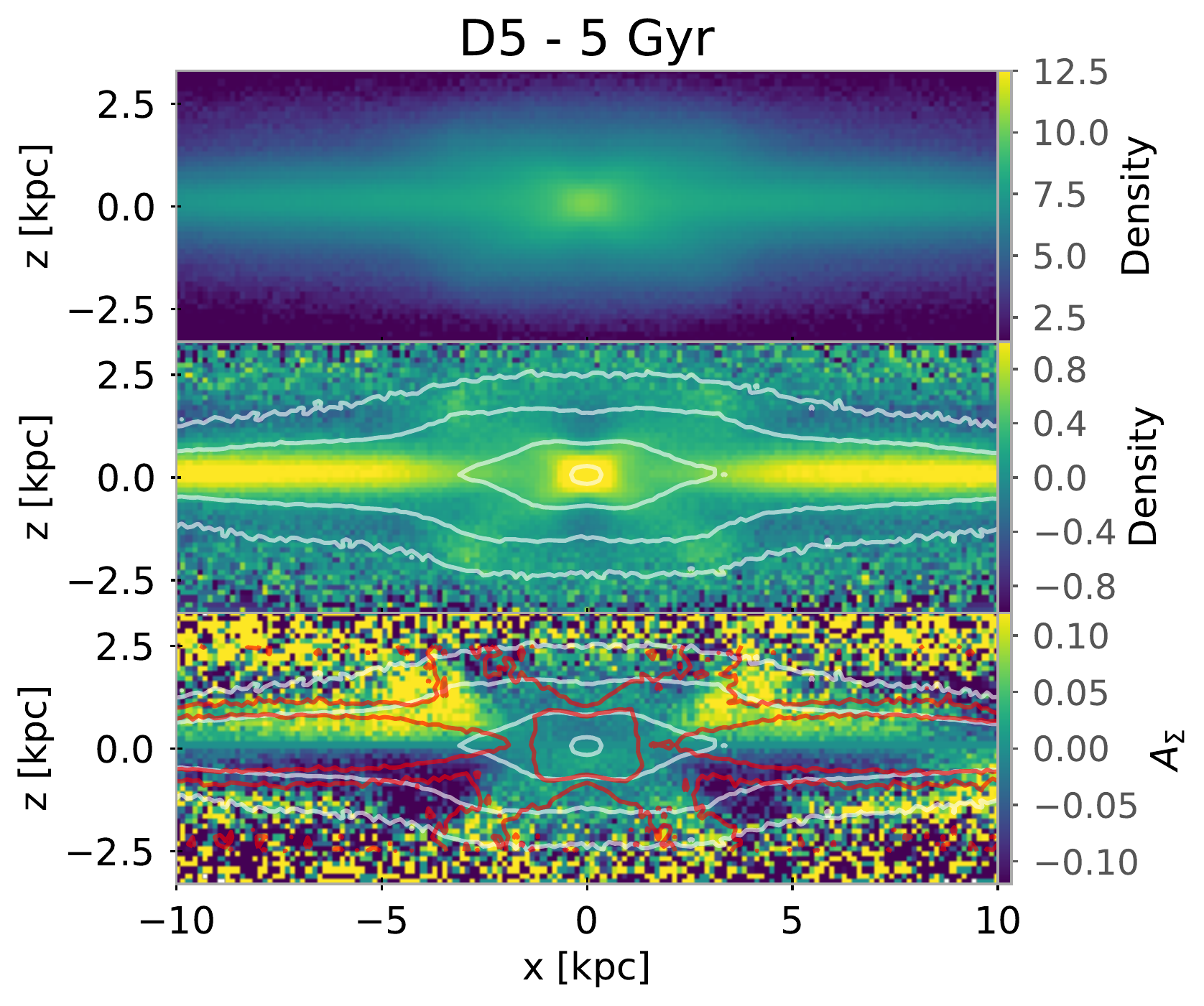}
    \includegraphics[scale=0.53]{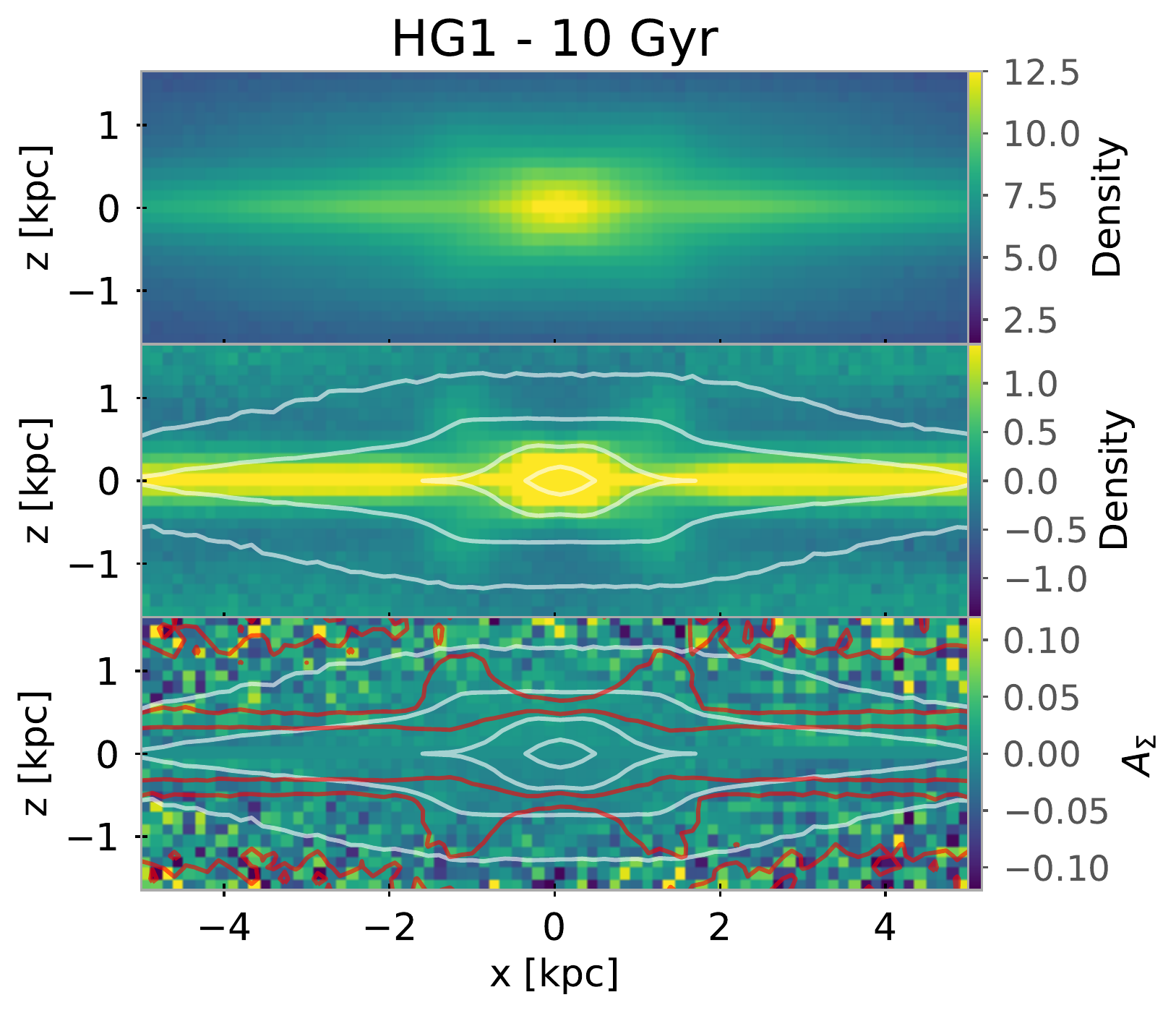}
    \caption{Map of the surface number density (top panel), its unsharp masked version (middle panel), and mid-plane asymmetry map (bottom panel) of models D5 (at 5 Gyr, left-hand column) and HG1 (at 10 Gyr, right-hand column) seen side-on. Some isocontours of the surface number density (white lines) and unsharp mask (red lines) maps are shown for both models. In all the plots the bar major axis coincides with the $x$ axis.}
    \label{fig:dia}
\end{figure*}

Fig.~\ref{fig:dia} shows the side-on views (top row) of models D5 (left column) and HG1 (right column), and the corresponding unsharp masked images (middle row), which reveal the characteristic X-shaped structure in both models. The bottom row shows the mid-plane asymmetry maps. Some surface density contours of the side-on views are shown in white, while some contours of the unsharp mask are in red. Inspection of the unsharp mask of model D5 confirms that the X shape is asymmetric with respect to the mid-plane. In particular, the extent of the mid-plane asymmetry covers the entire region of the B/P bulge, starting just within the arms of the X shape to roughly twice the extent of the X shape, $2< |x|/\mathrm{kpc} < 5$. The mid-plane asymmetry maps thus clearly identify the extension and strength of the asymmetric regions, if present. The mid-plane asymmetry map of model D5 shows asymmetries reaching values of $| A_{\Sigma}(x,z) | \sim 0.15$. The maxima in $|A_{\Sigma}(x,z)|$ are located just within the arms of the X shape.

To quantify the mid-plane asymmetry, we define a related diagnostic describing the `mid-plane asymmetry profile', $\mathcal{A}_\Sigma (x)$, by collapsing the vertical distribution within the region of the X shape:
%
\begin{equation}
\mathcal{A}_\Sigma (x) = \sum_{z_{\min}}^{z_{\max}} A_\Sigma(x,z)
\end{equation}
where the sum is calculated along the $x$ axis within the vertical interval $z_{\rm min}<z<z_{\rm max}$. The region near the mid-plane needs to be excluded (i.e. $z_{\rm min} > 0$) because in real galaxies it often hosts dust and/or star forming features which produce spurious signatures not associated with mid-plane asymmetry produced by buckling. The upper limit, $z_{\rm max}$, covers the vertical extent of the X shape (along the $z$ axis), i.e. its semi-minor axis. It is measured by analysing various profiles along the $x$ axis extracted at different vertical heights from the surface number density and unsharp images, along the region where the X shape traces a trapezoidal shape. We estimate the uncertainties on the mid-plane asymmetry profile with a Monte Carlo simulation by generating 100 noisy mock side-on maps of the surface number density $\Sigma \left( x,z \right)$, assuming the galaxy centre is at $(x,y,z)=(0,0,0)$. 
We calculate the Poisson noise for each location $(x,z)$ of the density map and build the mock maps by assuming the value of the density has a Gaussian distribution centred on the original value of $\Sigma \left( x,z \right)$ and with standard deviation equal to the corresponding Poisson noise. We then generate 100 mid-plane asymmetry profiles from the mock maps and adopt the root mean square of the distribution of measured values at each location along the $x$ axis as the error for the profile. To estimate the semi-major axis of the X shape (along the $x$ axis), we extract various profiles of the surface number density and unsharp images along the $x$ axis at given values of $z$ in the region of the B/P bulge. These profiles clearly look double-peaked. The semi-major axis of the X shape corresponds to the distance between the two peaks. The B/P bulge in model D5 has a semi-major axis (along the $x$ axis) of 3.1 kpc and a semi-minor one (along the $z$ axis) of 2.4 kpc. The B/P bulge in model HG1 has a semi-major axis of 1.3 kpc and a semi-minor one of 1.2 kpc. 

\begin{figure*}
    \centering
    \includegraphics[scale=0.43]{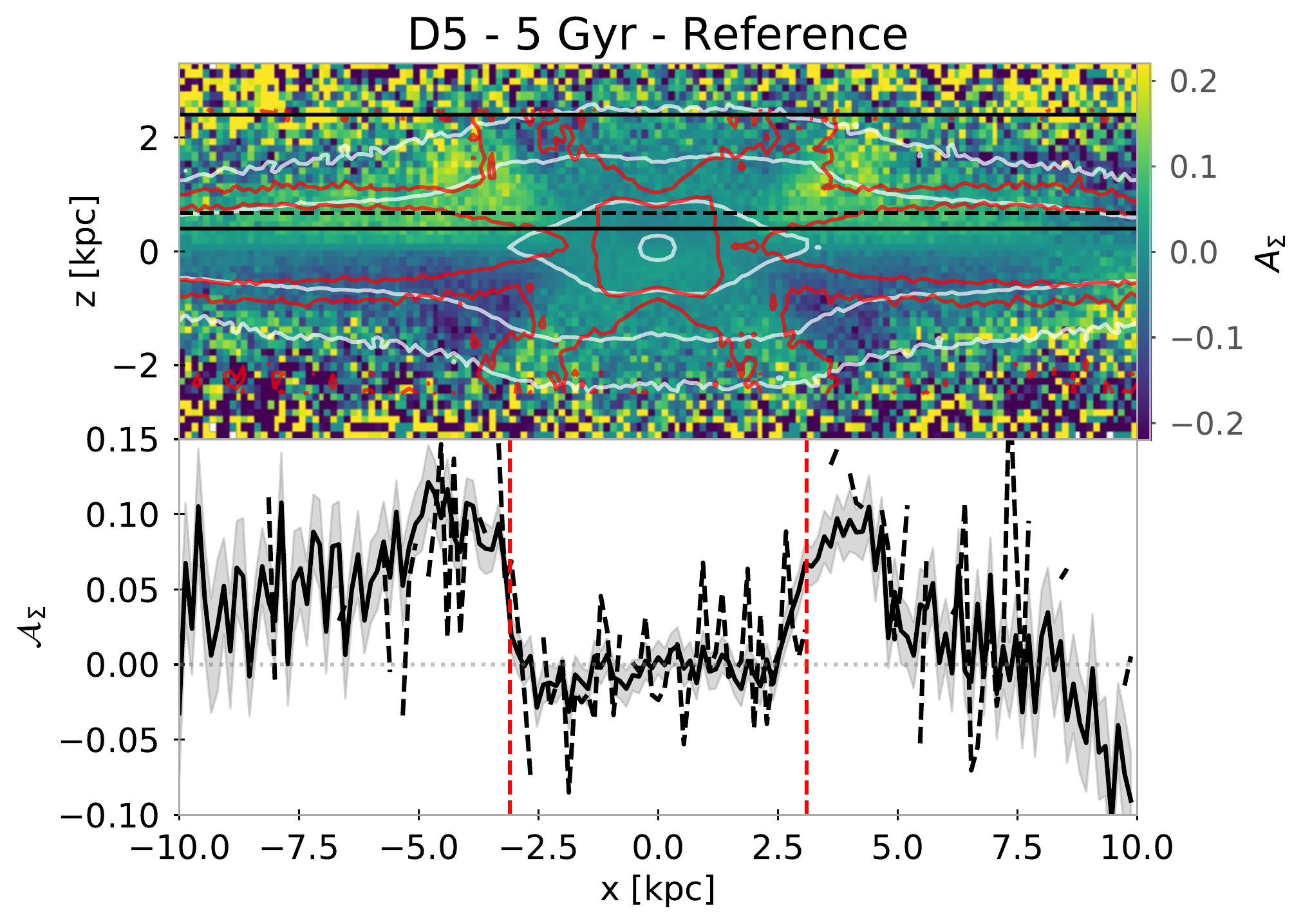}
    \includegraphics[scale=0.43]{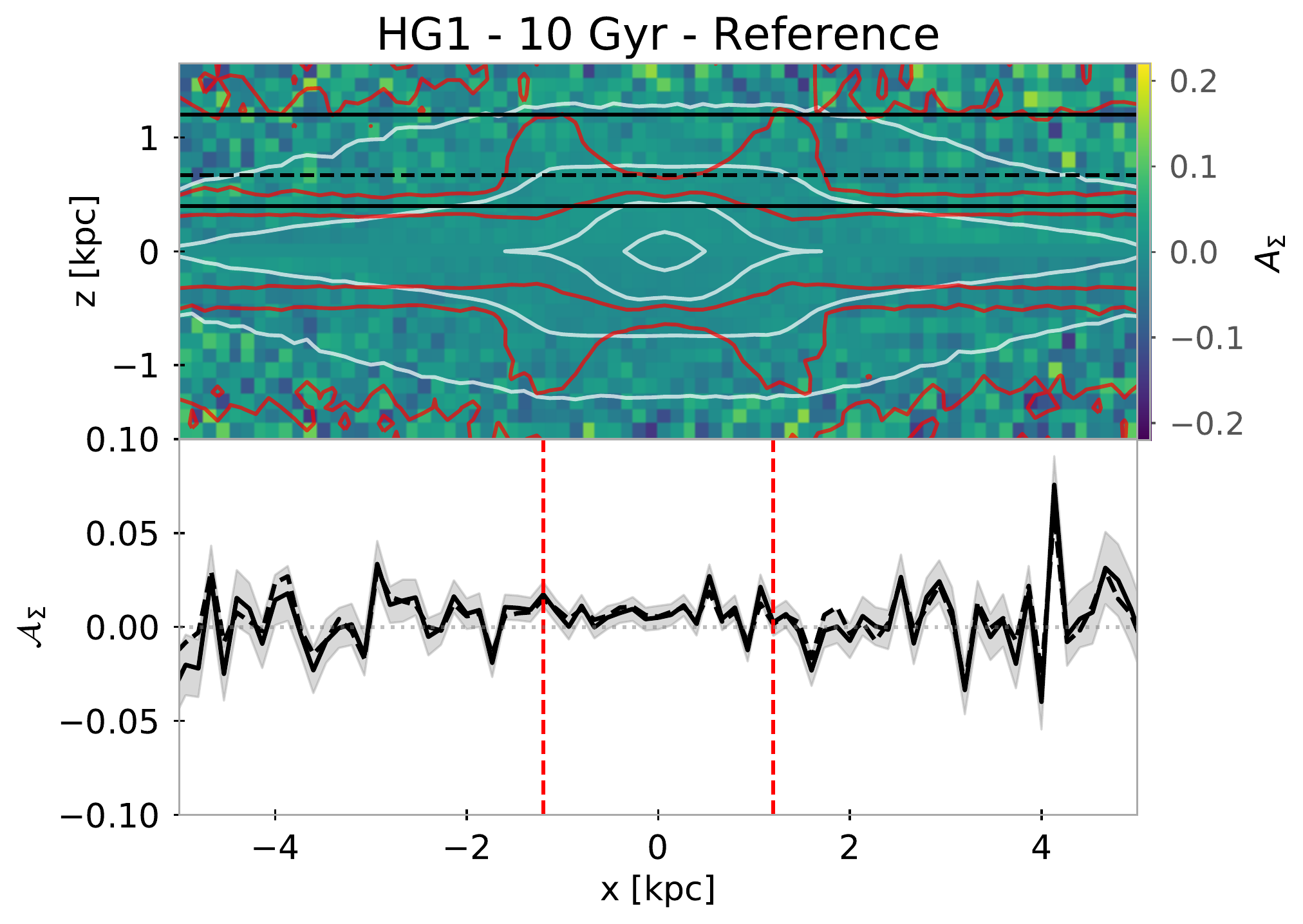}
    \caption{Left-hand column: mid-plane asymmetry map (upper panel) of model D5 (at 5 Gyr) with some isocontours of the surface number density (white lines) and its unsharp mask (red lines) maps. The horizontal lines mark the vertical extension of the regions adopted to extract the mid-plane asymmetry radial profiles shown in the lower panel.
    Mid-plane asymmetry radial profiles (lower panel) derived in the vertical ranges between $0.4<z<2.4$ kpc (solid line with grey error bars) and $0.7<z<2.4$ kpc (dashed line). The vertical red dashed lines mark the extension of the X shape along the $x$ axis.
    Right-hand column: Same as in the left-hand column, but for model HG1 (at 10 Gyr) with $0.4<z<1.1$ kpc (solid line, lower panel) and $0.6<z<1.1$ kpc (dashed line, lower panel).}
    \label{fig:sim_dia}
\end{figure*}

Fig.~\ref{fig:sim_dia} shows the mid-plane asymmetry diagnostics for models D5 (left column) and HG1 (right column). The two profiles in each model (solid and dashed black lines) are calculated in slightly different vertical regions, both including the vertical extension of the B/P bulge (these regions are highlighted by the horizontal lines in the top panels), but varying the extension of the excluded region near the mid-plane.
In particular, we adopt the vertical ranges between
$0.4 < z < 2.4$~kpc and $0.7 < z < 2.4$~kpc for model D5 and  $0.4 < z < 1.1$~kpc and $0.6 < z < 1.1$~kpc for model HG1, to calculate the mid-plane asymmetry radial profiles.
The results of the different profiles are broadly in agreement: the extension of the excluded region near the mid-plane does not affect significantly the mid-plane asymmetry profile. We will refer to the profile traced with a solid black line as the reference one for subsequent comparisons.
The mid-plane asymmetry profiles in model D5 exhibit two clear peaks, reaching $\mathcal{A}_\Sigma (x) \sim 0.10-0.12$. The profiles are flat with $\mathcal{A}_\Sigma (x) \sim 0.0$ for $|x| < 2.5$ kpc, increase steeply to peaks at $|x|\sim 5.0$ kpc and then decrease gently farther out, reaching $\mathcal{A}_\Sigma (x) \sim 0.00-0.05$ at $|x| \sim 7.5$ kpc. At $|x| > 7.5$ kpc, further out in the disc, the profile remains slightly asymmetric when comparing the positive and negative $x$ range.
The mid-plane asymmetry profiles for model HG1 reveal no coherent mid-plane asymmetry. The mid-plane asymmetry profile remains flat, with $\mathcal{A}_\Sigma (x) \lesssim 0.02$. 

\subsection{Time evolution of the mid-plane asymmetry}

\begin{figure*}
    \centering
    \includegraphics[scale=0.5]{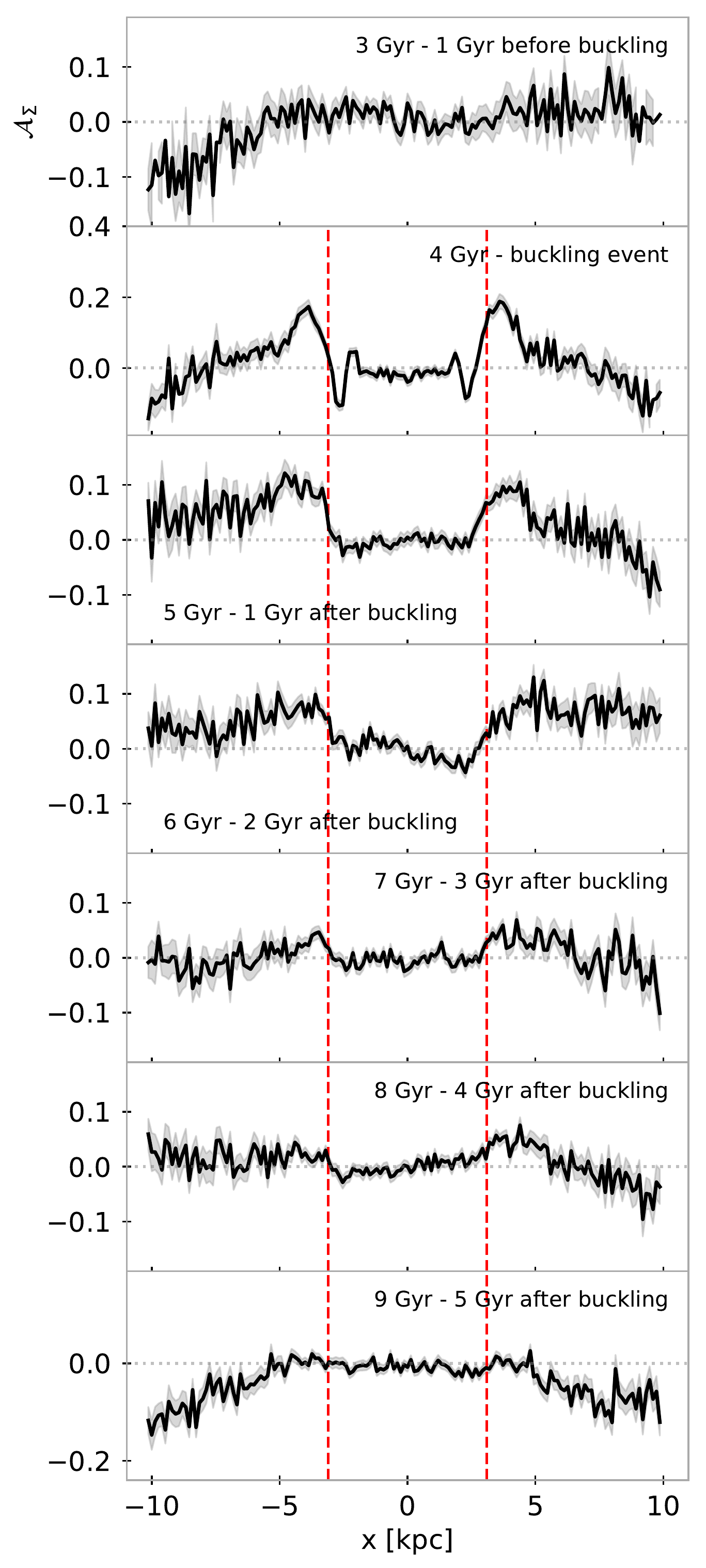}
    \includegraphics[scale=0.5]{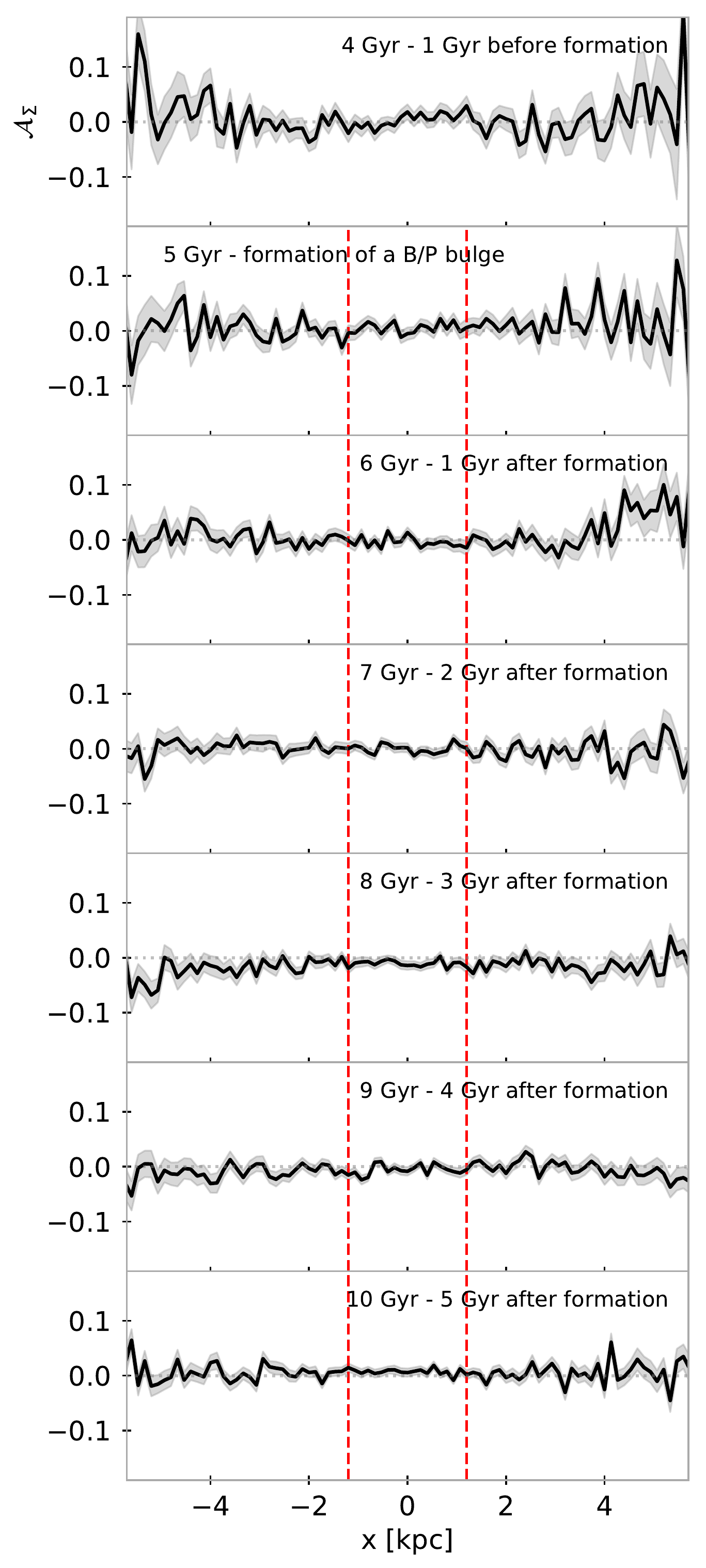}    
    \caption{Time evolution of the mid-plane asymmetry radial profile of models D5 (left-hand column) and HG1 (right-hand column) over six Gyrs bracketing the formation of the B/P bulge. The vertical red dashed lines mark the extension of the X shape along the $x$ axis.}
    \label{fig:evol}
\end{figure*}

In order to understand how long-lasting the mid-plane asymmetries  are, we compute the asymmetry diagnostics at intervals of 1 Gyr to track their evolution in both models. 
Fig.~\ref{fig:evol} shows the evolution of the mid-plane asymmetry profiles for models D5 (left column) and HG1 (right column). The onset of a B/P bulge via buckling in model D5 is clearly visible in the strongly asymmetric shape of the profile, where two peaks symmetric with respect to the $x$ axis appear, at $\sim4$ Gyr. The mid-plane asymmetry persists for $\sim4$ Gyr after the buckling event, but decreases in strength and disappears at 9 Gyr: it reaches values as large as $\mathcal{A}_\Sigma (x) \sim 0.2$ at $\sim4$ Gyr, decreases rapidly during the first 1 Gyr after the buckling event, reaching $\mathcal{A}_\Sigma (x) \sim 0.10-0.15$, and decreases further but slowly in the following 3 Gyr, reaching $\mathcal{A}_\Sigma (x) \sim 0.05$ at 8 Gyr. Some asymmetry is visible in the outer part of the disc at 9 Gyr, not associated to the B/P bulge and possibly due to a weak warp in the disc. In Fig.~\ref{fig:evol_detailed} we show the time evolution of the mid-plane asymmetry with higher time resolution around the buckling event. In particular, we analyse the interval between 3.5 Gyr and 4.5 Gyr with an interval of 0.1 Gyr. The buckling lasts $\sim 0.5$~Gyr and mid-plane asymmetry is visible within the region of the B/P bulge, and during the early phase of the instability. During the buckling event the inner part of the bar bends, which scatters stars, breaking the symmetric distribution of stars with respect to the mid-plane, and causing a visible asymmetry to appear at the corner of the B/P region. This asymmetry slowly declines with time. The timescale for this decline is a conservative lower limit due to the fact that the evolution of $N$-body simulations, used here, is known to be fast, because of the absence of gas \citep[e.g.,][]{Erwin2019}. The rate at which resonance crossing within the bar occurs may be responsible for the lifetime of the asymmetry, but this requires a detailed orbital study to quantify (Beraldo e Silva et al. 2022, {\it in preparation)}. On the other hand, no significant asymmetries are visible during the entire evolution of star-forming model HG1 (see Fig.~\ref{fig:evol}, right panel). 

In Appendix~\ref{app:sim} we also present the time evolution of the pure $N$-body models D8 and T1 from \citealt{Anderson2022}. Model D8 suffers a weak buckling event, with $A_{\rm buck}\sim0.04$, similar to model D5. The mid-plane asymmetry visible 1~Gyr after buckling is similar to that in model D5; it decreases with time but remains visible for at least 5 Gyr after buckling (the simulation ends at 9 Gyr). This result means the mid-plane asymmetry remains visible for at least 1~Gyr longer than in model D5. On the other hand, model T1, which suffers the weakest buckling ($A_{\rm buck}\sim0.02$), has a weaker mid-plane asymmetry which disappears already 1~Gyr after the instability. A weaker buckling event in model T1 thus produces a smaller mid-plane asymmetry, which does not persist as long.

\section{Observational considerations}
\label{sec:obs_compl}

Having demonstrated long-lasting mid-plane asymmetries  produced by buckling in some simulations, we now explore the prospects for detecting such asymmetries in real galaxies. In this section we consider some of the observational effects that may complicate measurements of mid-plane asymmetries. We will conclude that these difficulties can be surmounted, but the selection of nearly edge-on galaxies is crucial.

\subsection{Effect of galaxy orientation and centre}
\label{sec:prove}

We start by varying the parameters used to build the mid-plane asymmetry maps in the simulated data to test how they influence the mid-plane asymmetry diagnostics. In particular, we want to identify the observational conditions under which the mid-plane asymmetry due to a past buckling event is clearly identifiable. These tests are needed since real galaxies are unlikely to be observed perfectly edge-on ($i=90\degr$) and with the bar oriented side-on (i.e., the bar aligned with the $x$ axis). 

\subsubsection{Effect of the galaxy inclination}

A galaxy is often considered edge-on when the disc inclination $i>85\degr$, while a nearly edge-on view refers to $80\degr<i<85\degr$ \citep[e.g.][]{Makarov2021}. The probability of observing a galaxy within $1\degr$ of exactly edge-on is less than $2\%$. Disc inclination cannot be measured exactly from its photometry, unless its intrinsic thickness is well known, while any dust extinction further complicates the determination of $i$ \citep{Padilla2008,Unterborn2008}. 
We therefore explore what happens to the diagnostics when the model is not seen perfectly edge-on, assuming the inclination of the galaxy to vary between $81\degr <i<99\degr$. 

\begin{figure*}
    \centering
    \includegraphics[scale=0.43]{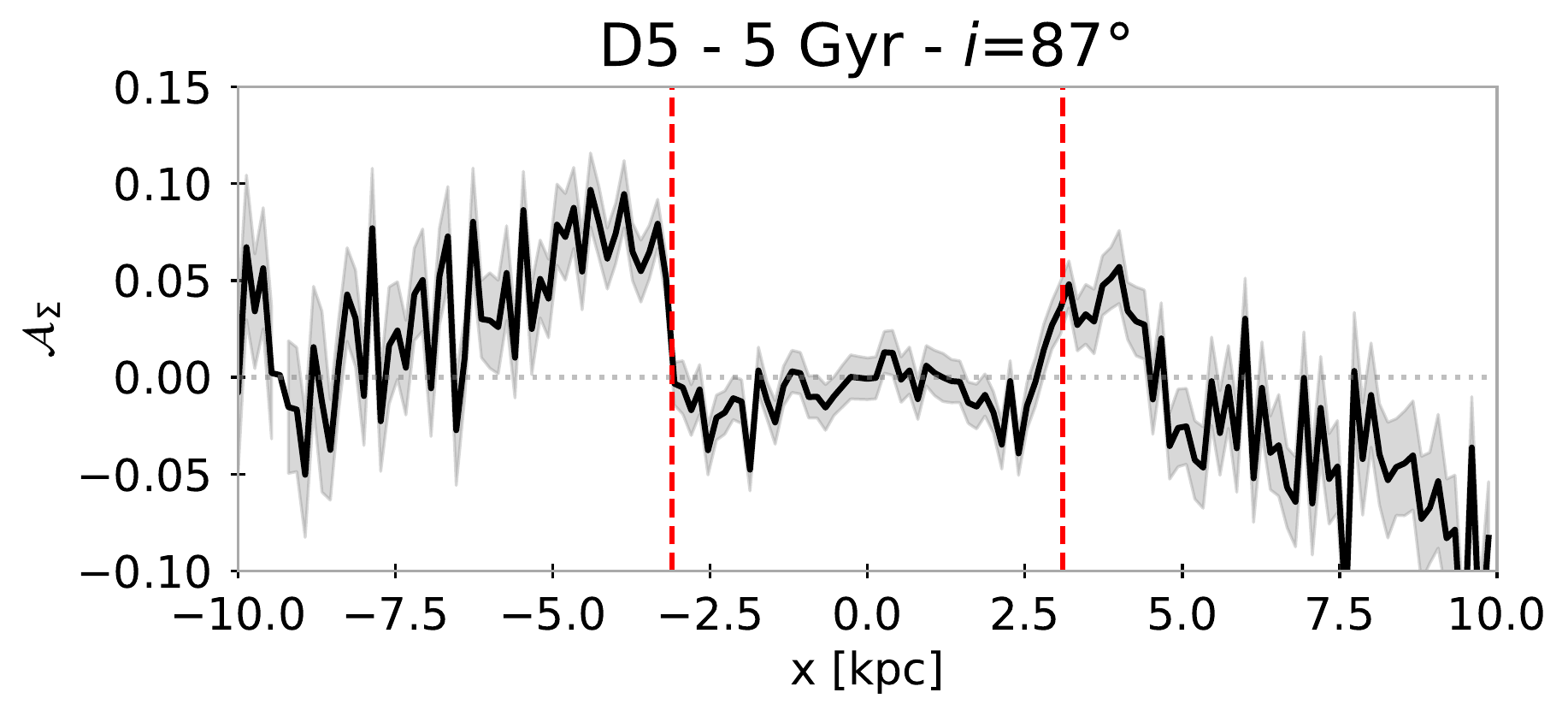}
    \includegraphics[scale=0.43]{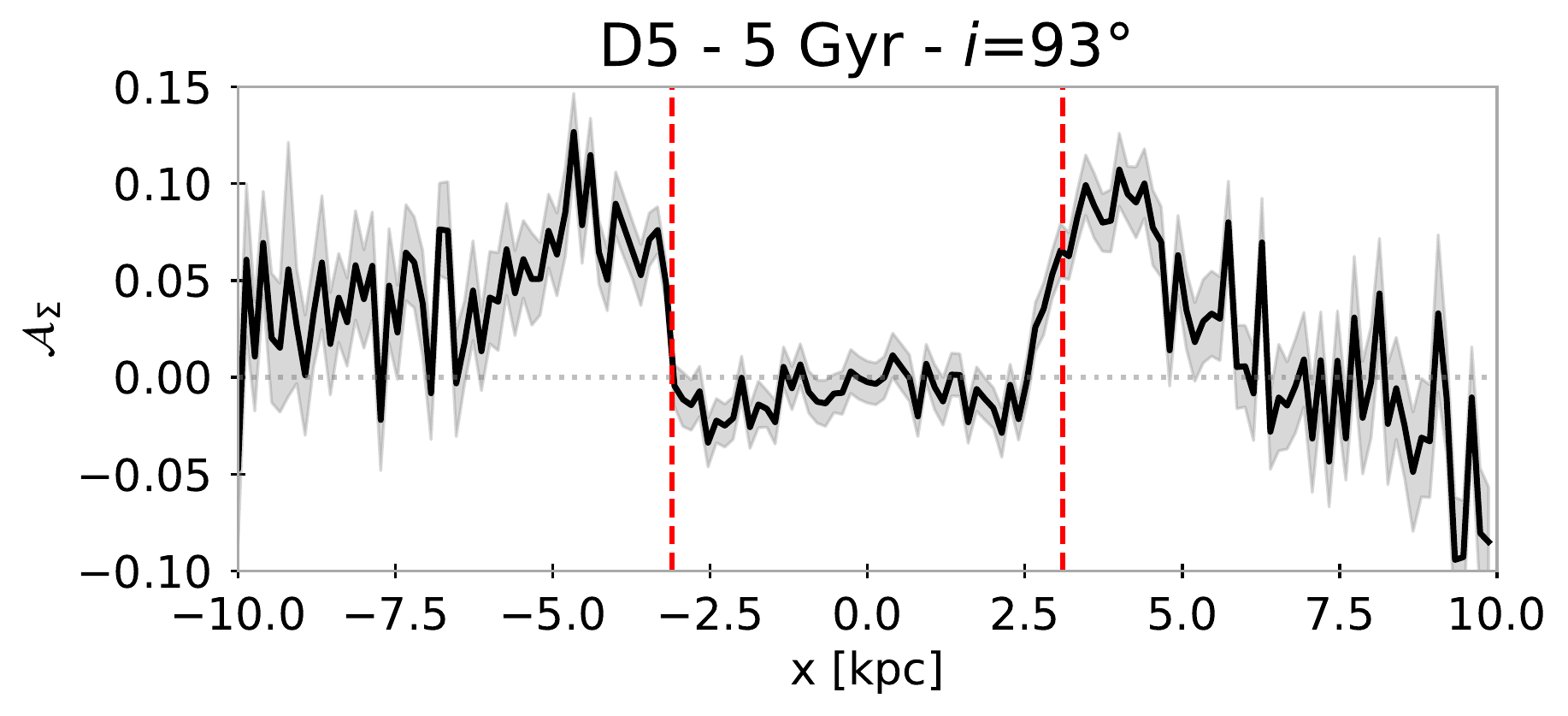}
    \includegraphics[scale=0.43]{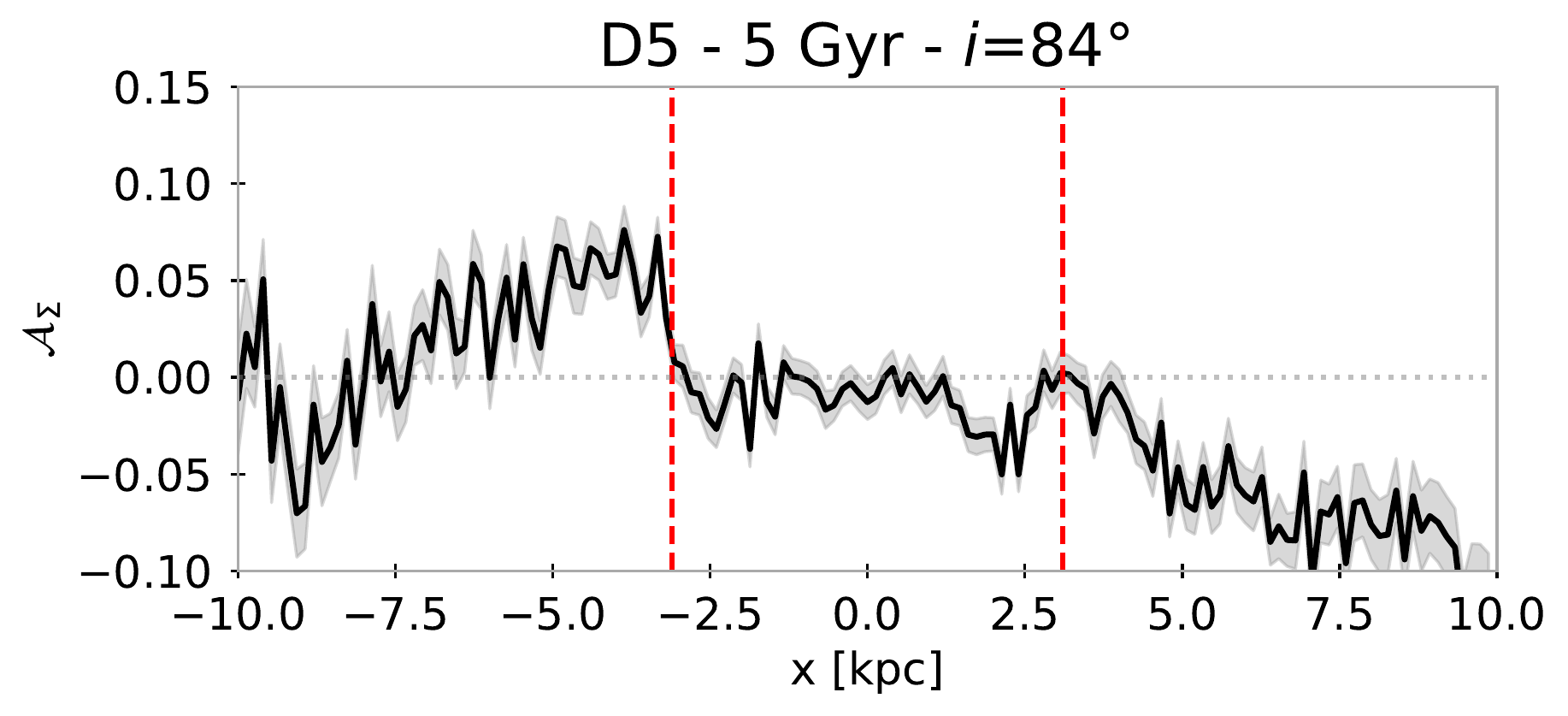}
    \includegraphics[scale=0.43]{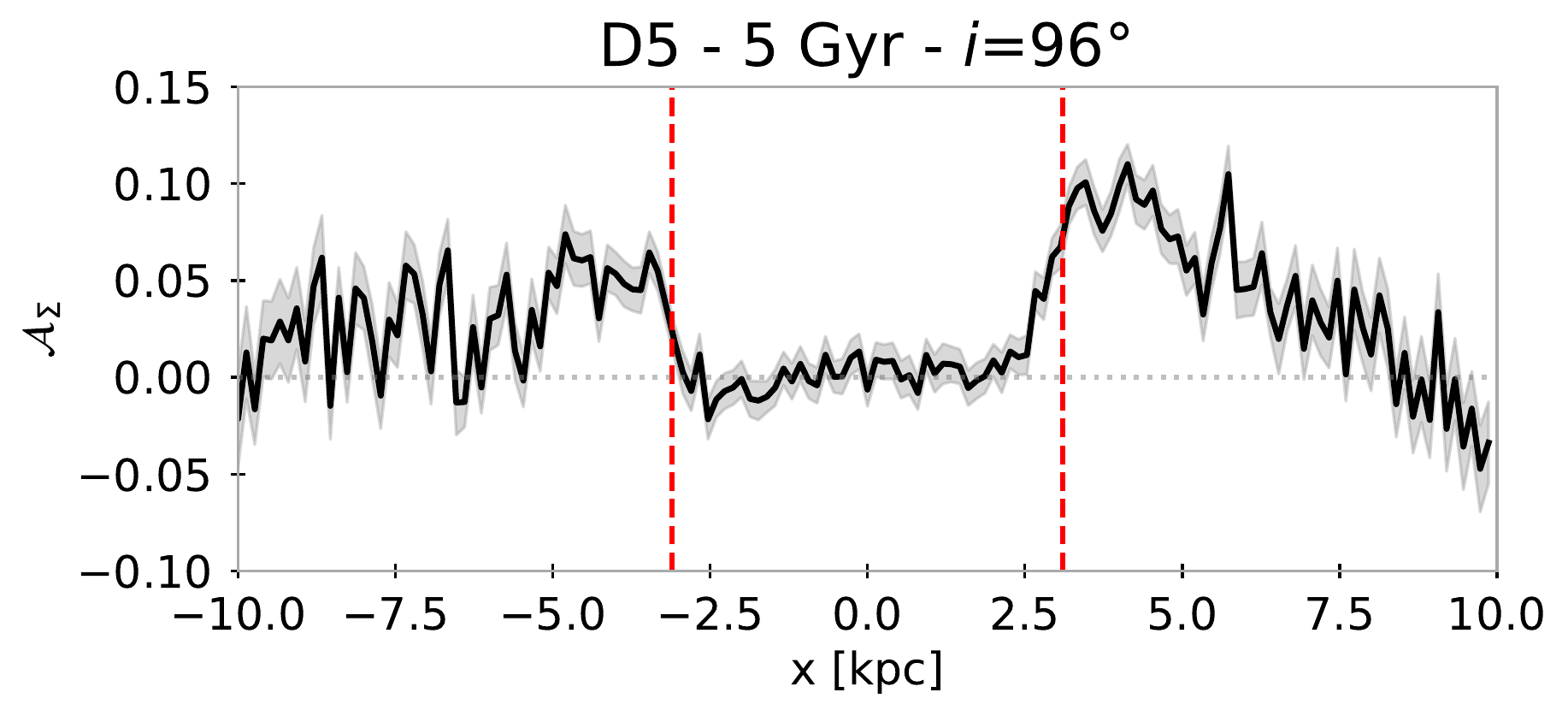}
    \includegraphics[scale=0.43]{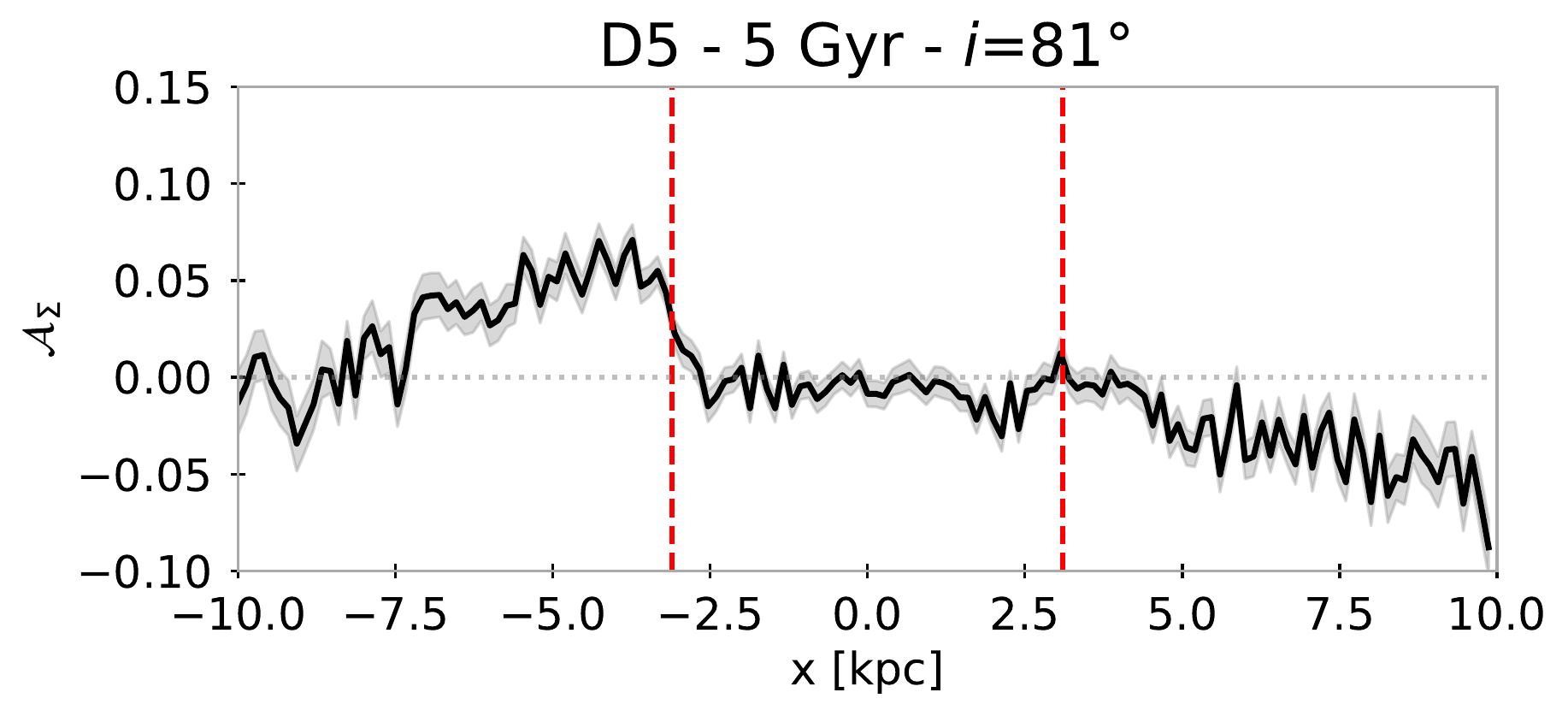}
    \includegraphics[scale=0.43]{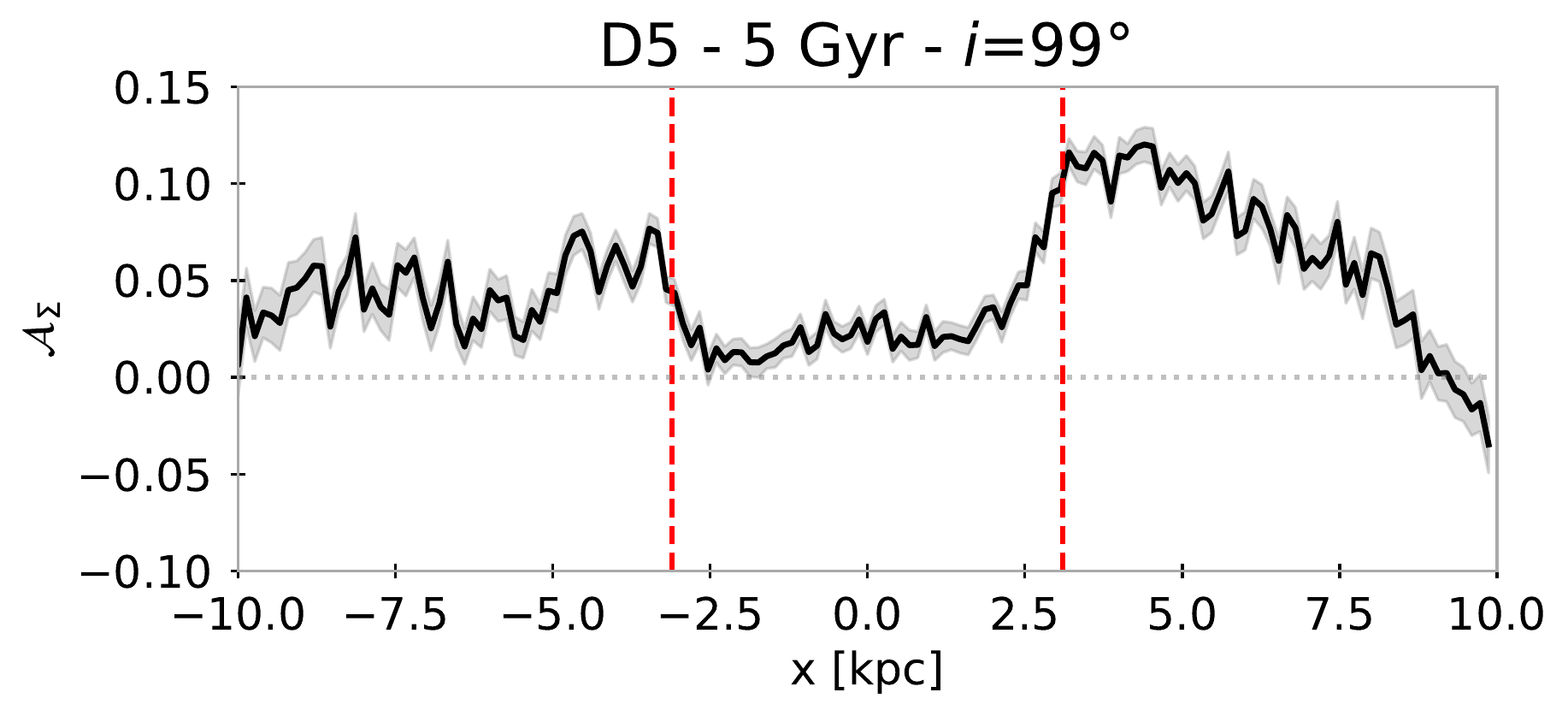}
    \caption{Mid-plane asymmetry profiles as in Fig.~\ref{fig:sim_dia} lower panel, but for model the D5 (at 5 Gyr) with different inclinations with respect to the side-on view ($\Delta i\pm3\degr$, top row; $\Delta i\pm6\degr$, middle row; $\Delta i\pm9\degr$, bottom row).}
    \label{fig:sim_d5_inc}
\end{figure*}

In Fig.~\ref{fig:sim_d5_inc} we present the resulting diagnostics for model D5, assuming $i=90\degr\pm3\degr,~\pm6\degr$, and $\pm9\degr$. The B/P bulge's extent is $3.1\times2.4$ kpc. While a mid-plane asymmetry is still visible, it decreases in strength as the inclination deviates from $90\degr$. Nevertheless, the characteristic double-peaked shape of the mid-plane asymmetry profiles remains easily identifiable. Deviation from a perfectly edge-on view produces a non-zero contribution to the flat part of the mid-plane asymmetry profile, i.e. within the region of the X shape, which starts to be clearly visible for $ |90\degr-i|>6\degr$. There is also a slight asymmetry with respect to the $z$ axis: for $i>90\degr$ the stronger peak is on the right side, while for $i<90\degr$ it is on the left. 

\begin{figure*}
    \centering
    \includegraphics[scale=0.43]{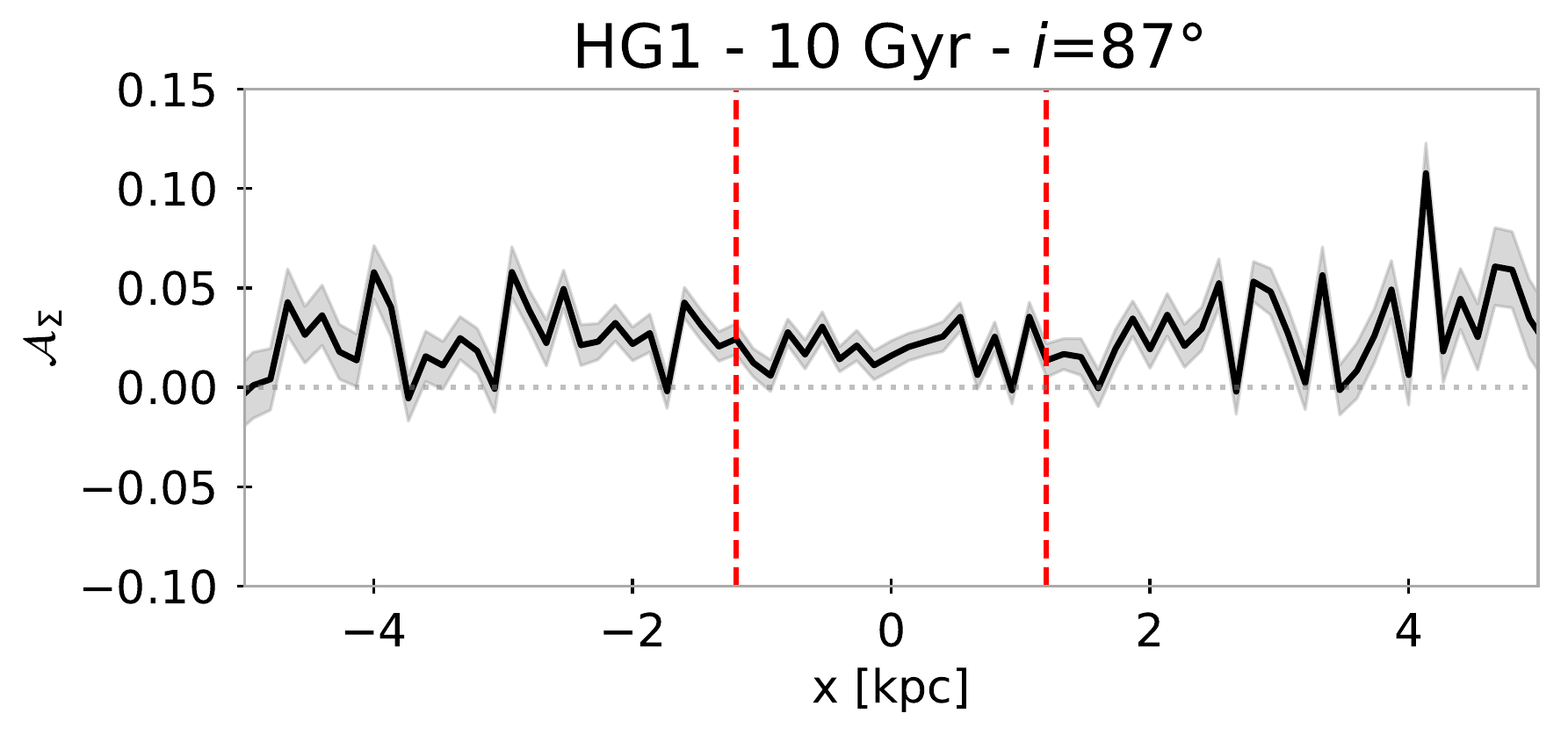}
    \includegraphics[scale=0.43]{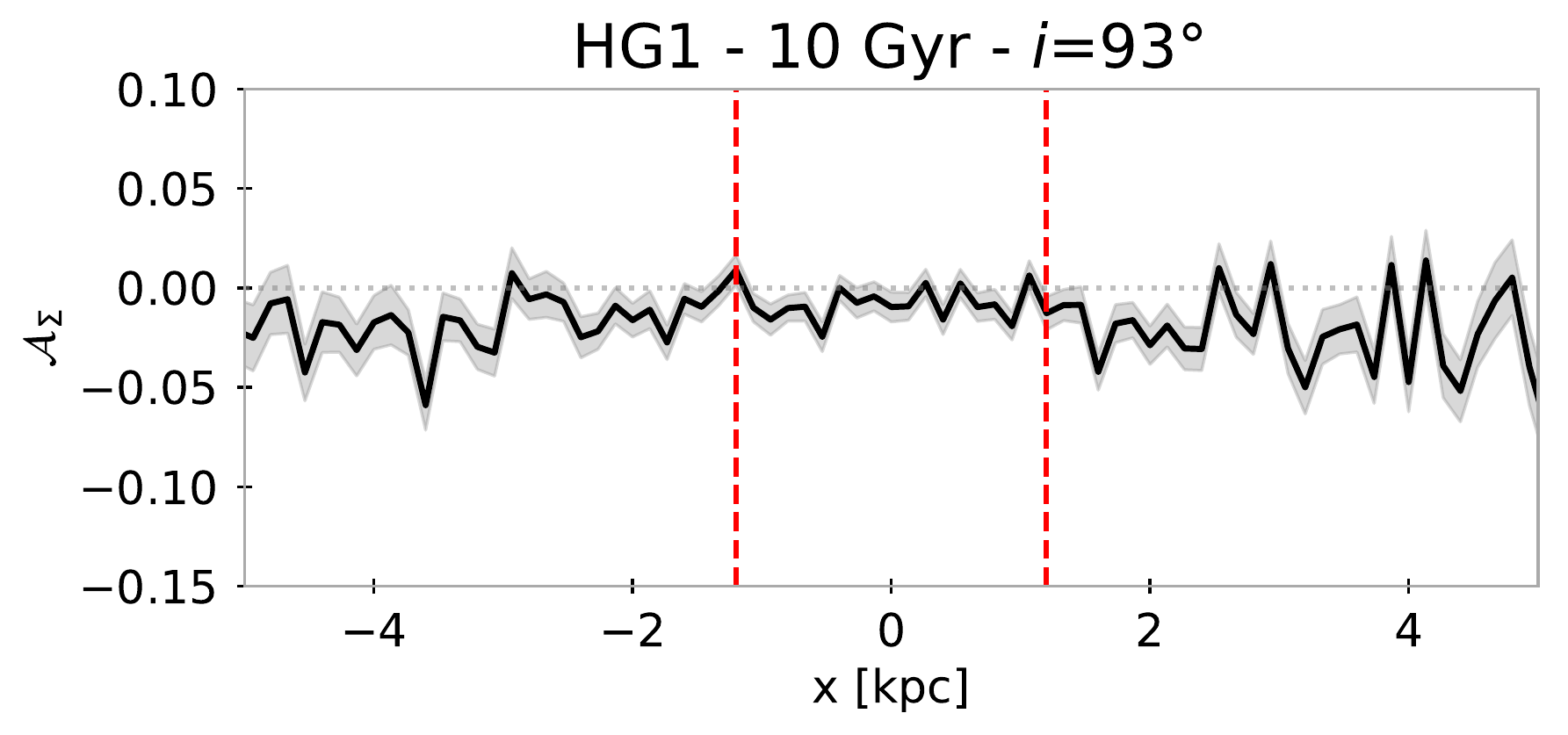}
    \includegraphics[scale=0.43]{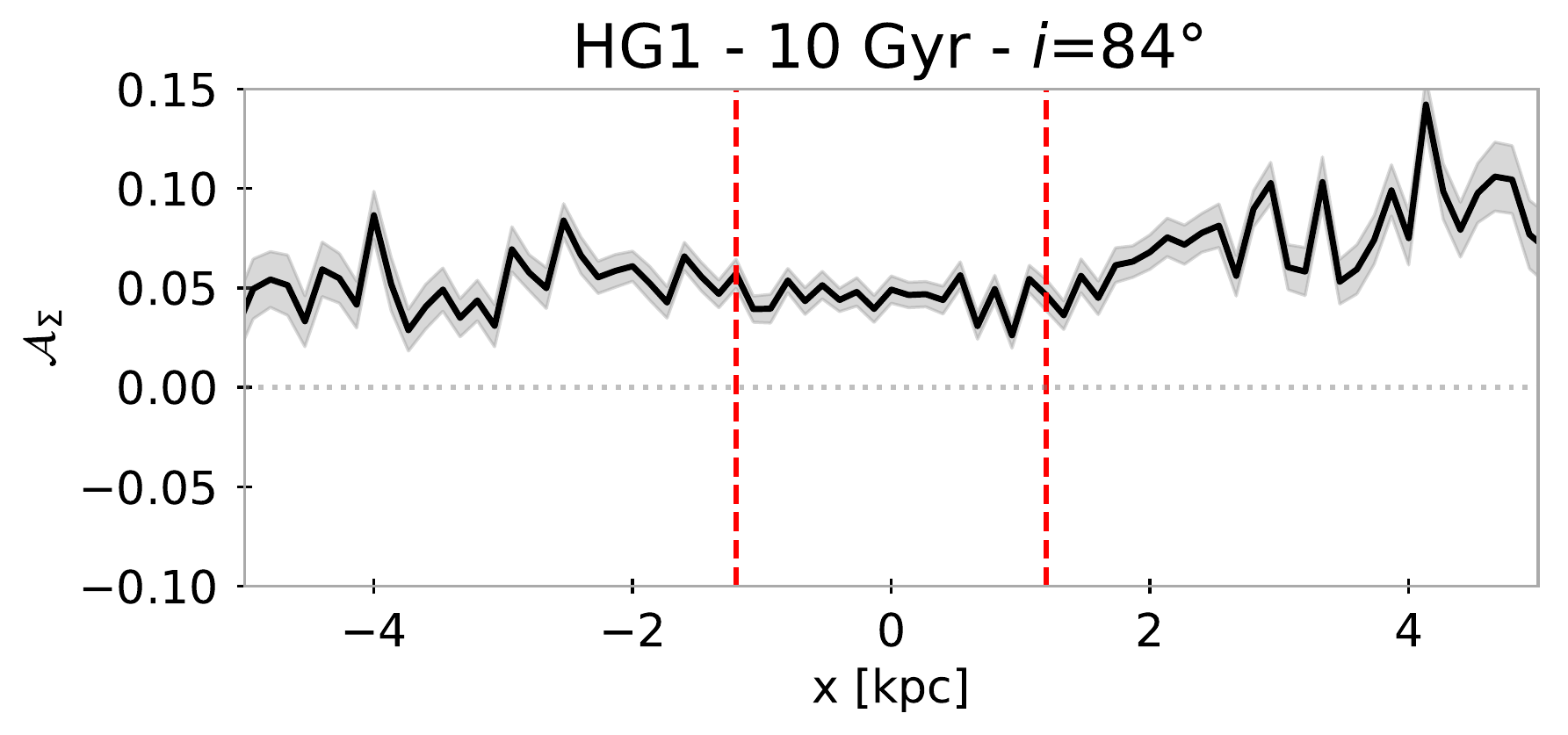}
    \includegraphics[scale=0.43]{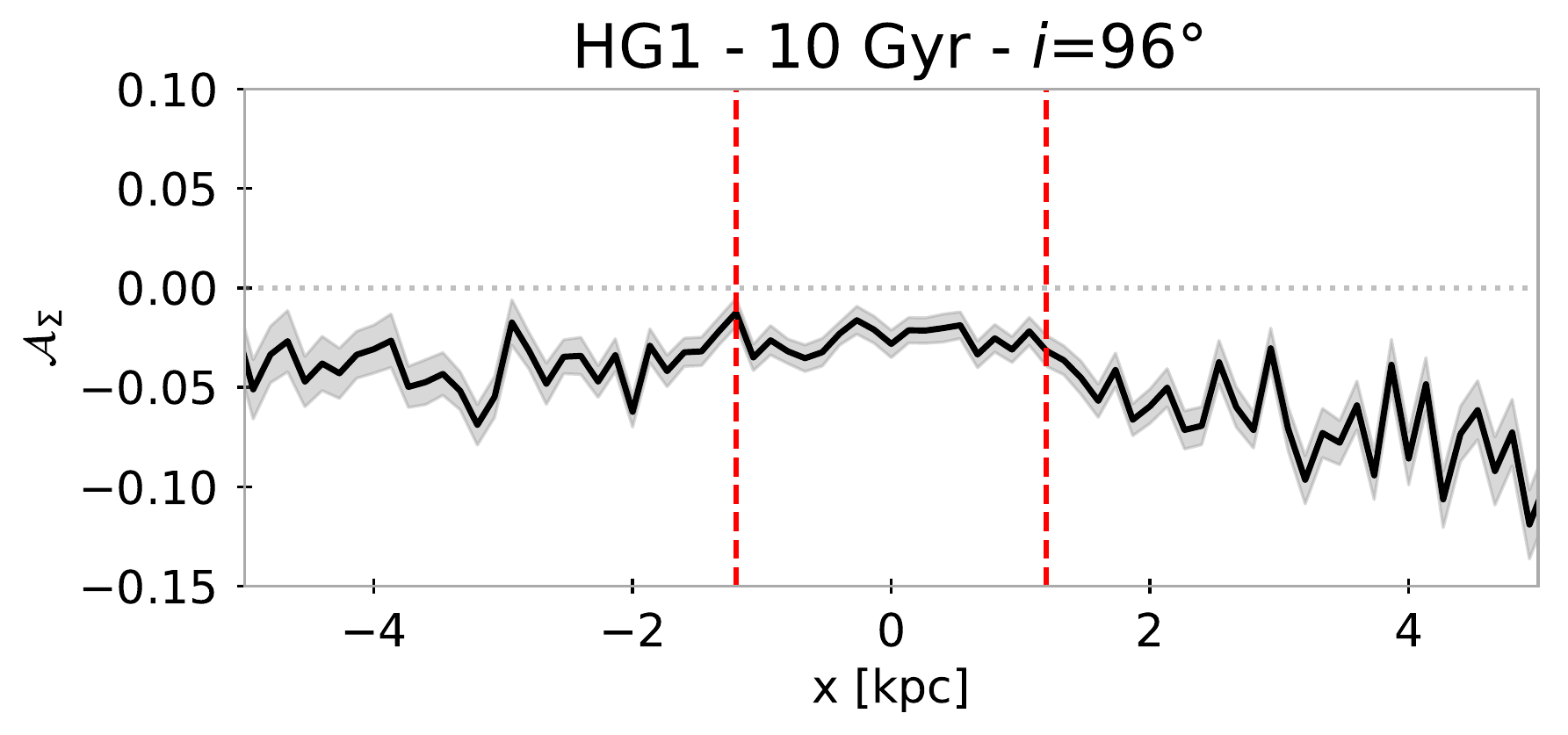}
    \includegraphics[scale=0.43]{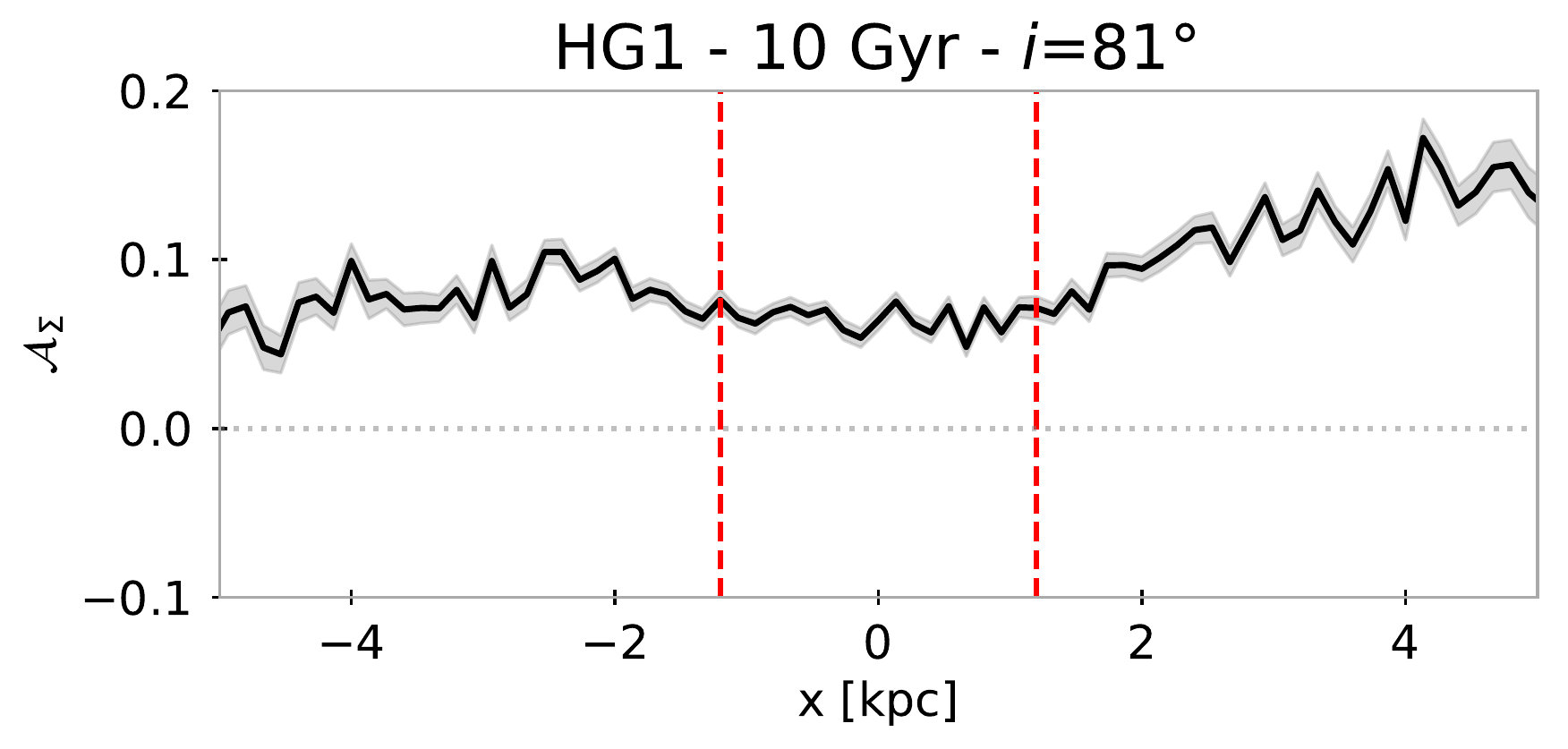}
    \includegraphics[scale=0.43]{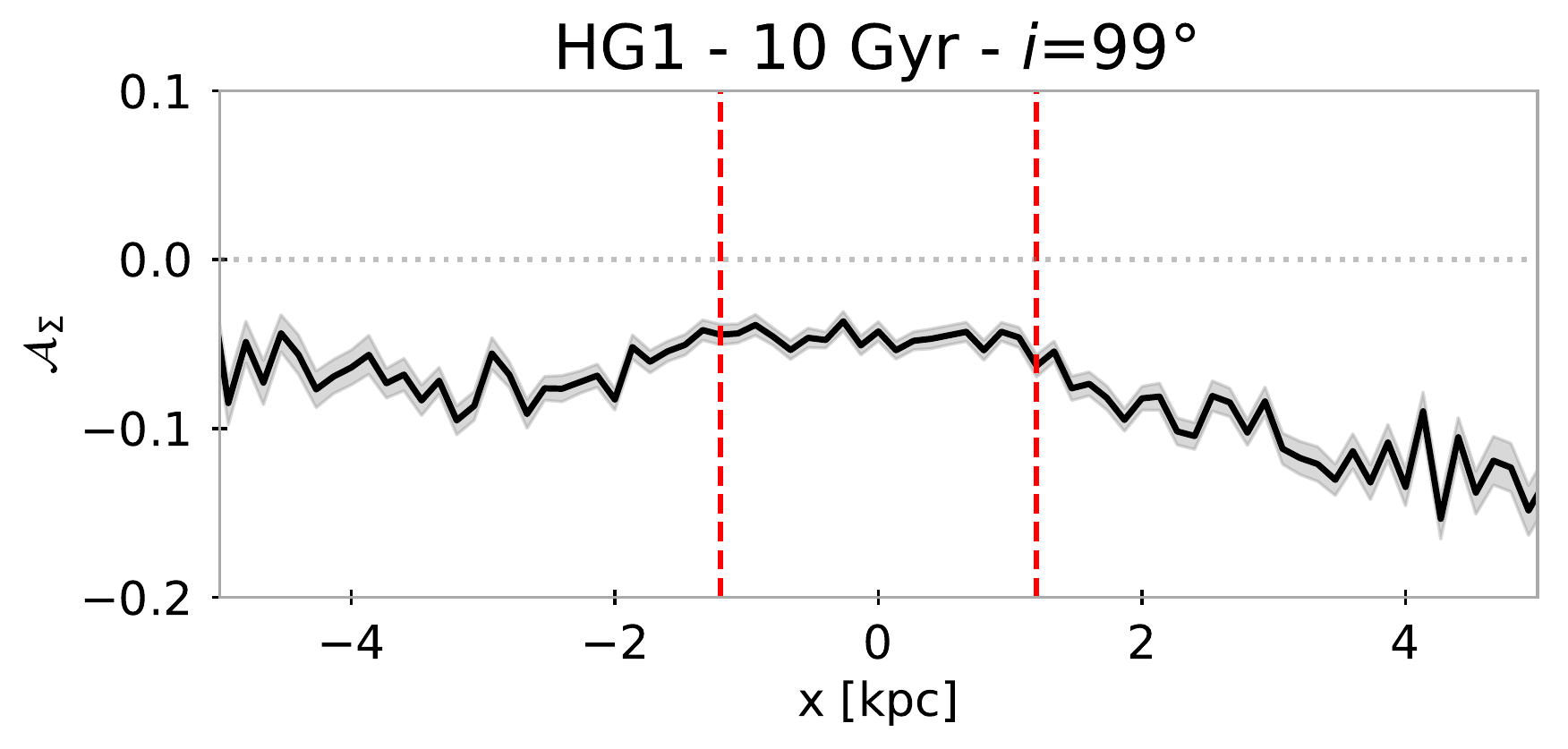}
    \caption{Mid-plane asymmetry profiles as in Fig.~\ref{fig:sim_dia} lower panel, but for model HG1 (at 10 Gyr) with different inclinations with respect to the side-on view ($\Delta i\pm3\degr$, top row; $\Delta i\pm6\degr$, middle row; $\Delta i\pm9\degr$, bottom row).}
    \label{fig:sim_hg1_inc}
\end{figure*}

In Fig.~\ref{fig:sim_hg1_inc} we present a similar analysis for model HG1. The B/P bulge's extent is now $1.3\times1.2$ kpc. A very weak mid-plane asymmetry appears in model HG1 for $i=90\degr\pm3\degr$, reaching $|\mathcal{A}_\Sigma (x)|\sim 0.02-0.04$. This is $10-20\%$ of the corresponding values found for model D5. Moreover, a stronger deviation from $\mathcal{A}_\Sigma (x) = 0$ in the region of the X shape is visible at all the inclinations compared with model D5. For $i=90\degr\pm6\degr$ and $90\degr\pm9\degr$, an asymmetry persists in the region of the disc at $x>2$ kpc as well: this asymmetry does not present the double-peaked shape expected for the mid-plane asymmetry produced by the buckling event and extends further out in the disc than the X shape (which extends only up to $x<1.2$ kpc), so it is reasonable to assume that it is not associated with the X shape. Larger departures from $i=90\degr$ produce larger mid-plane asymmetries  all along the disc. The disc region outside the bar in model HG1 is not axisymmetric but contains spirals, as seen in the face-on view of Fig.~\ref{fig:sim_3d_view}. These spirals produce the observed features in this region of the mid-plane asymmetry map when the model is seen at small deviations from perfectly edge-on.

Although a mid-plane asymmetry produced by a past buckling event is easily revealed by our diagnostics in perfectly edge-on galaxies, it is still recognisable up to at least $9\degr$ away from a perfectly edge-on view, based on the double-peaked shape of the mid-plane asymmetry profile. Nevertheless, asymmetries not associated with buckling may also arise from deviations from edge-on views larger than $\pm3\degr$. These results imply that mid-plane asymmetries  can be safety identified for images that are within $3\degr$ of the edge-on orientation for realistic galaxies.

\subsubsection{Effect of the bar orientation}

\begin{figure*}
    \centering
    \includegraphics[scale=0.43]{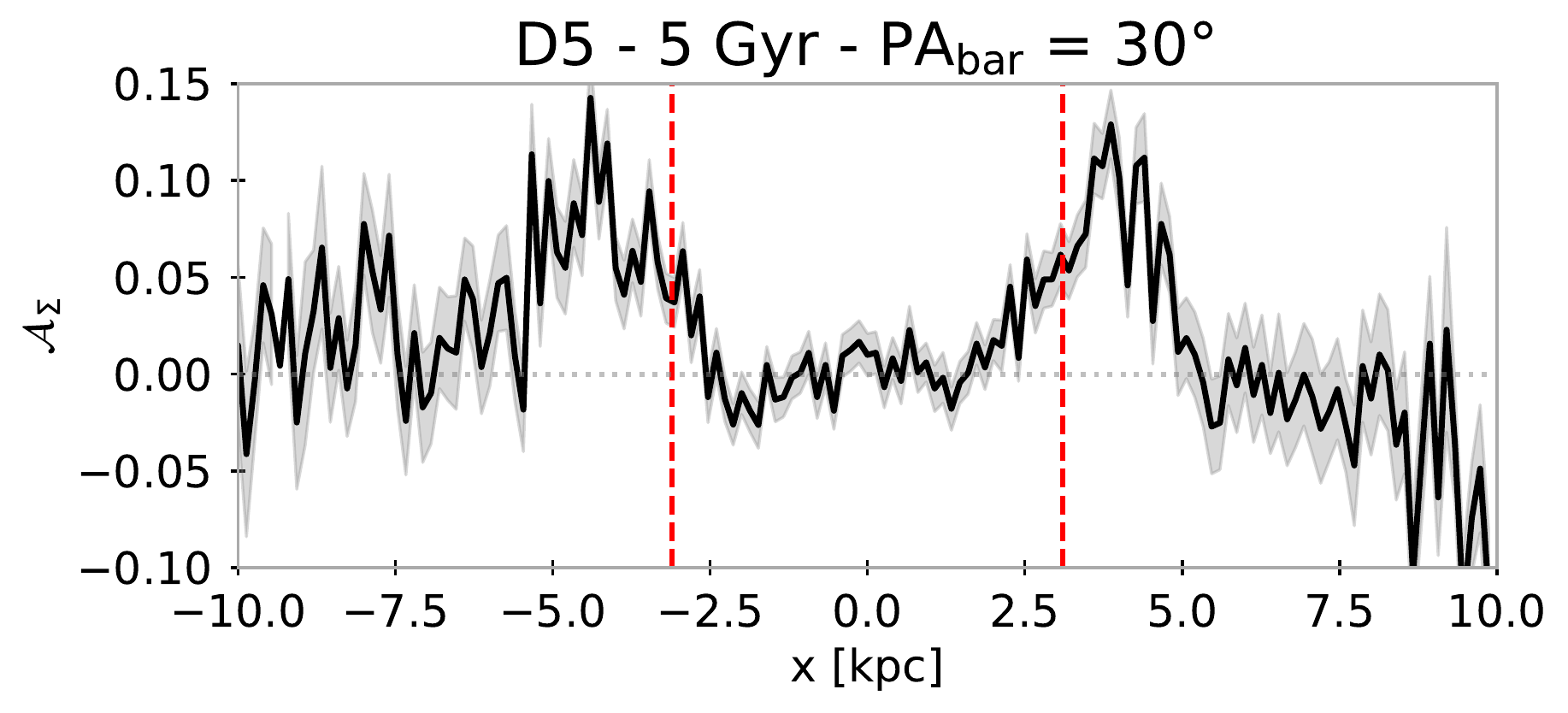}
    \includegraphics[scale=0.43]{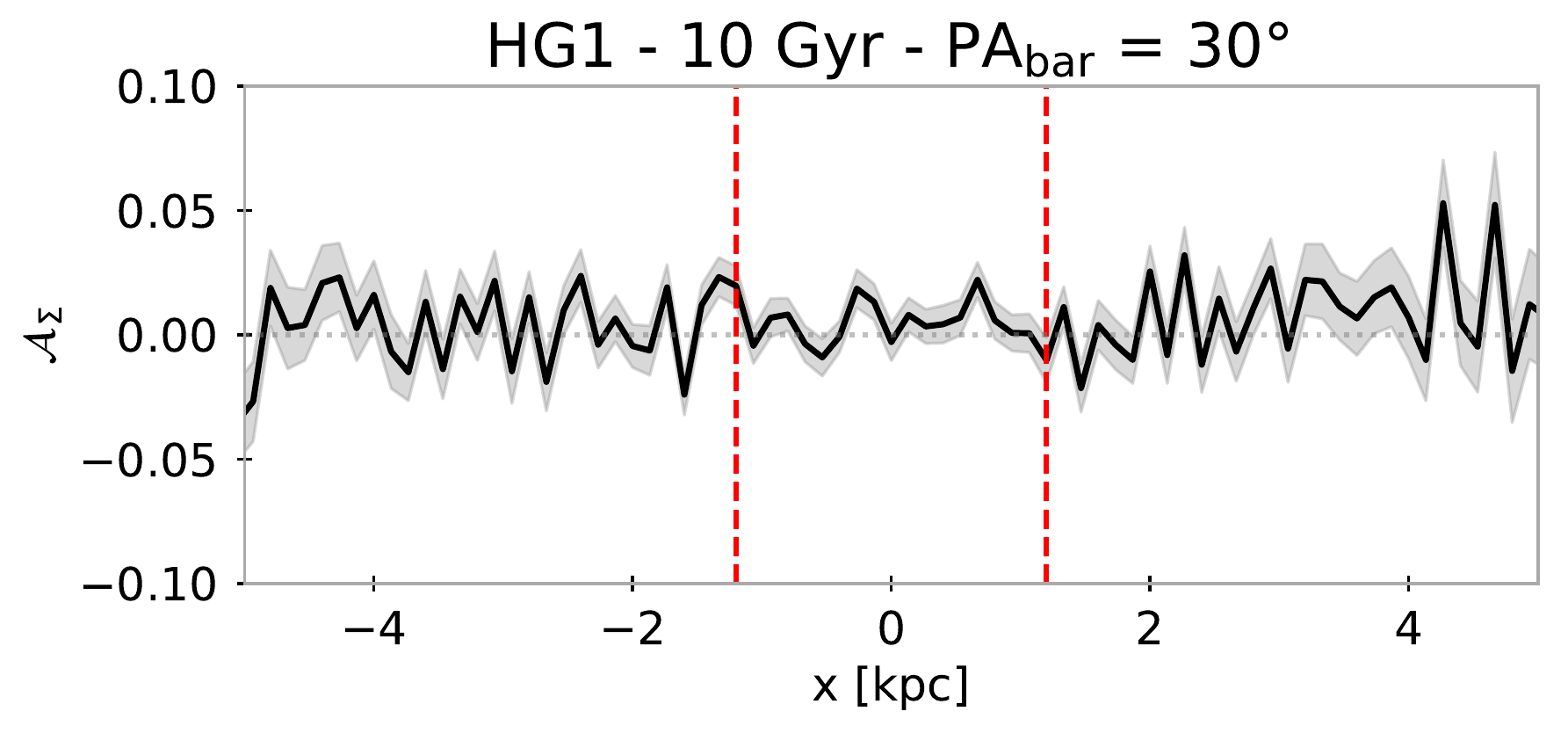}
    \includegraphics[scale=0.43]{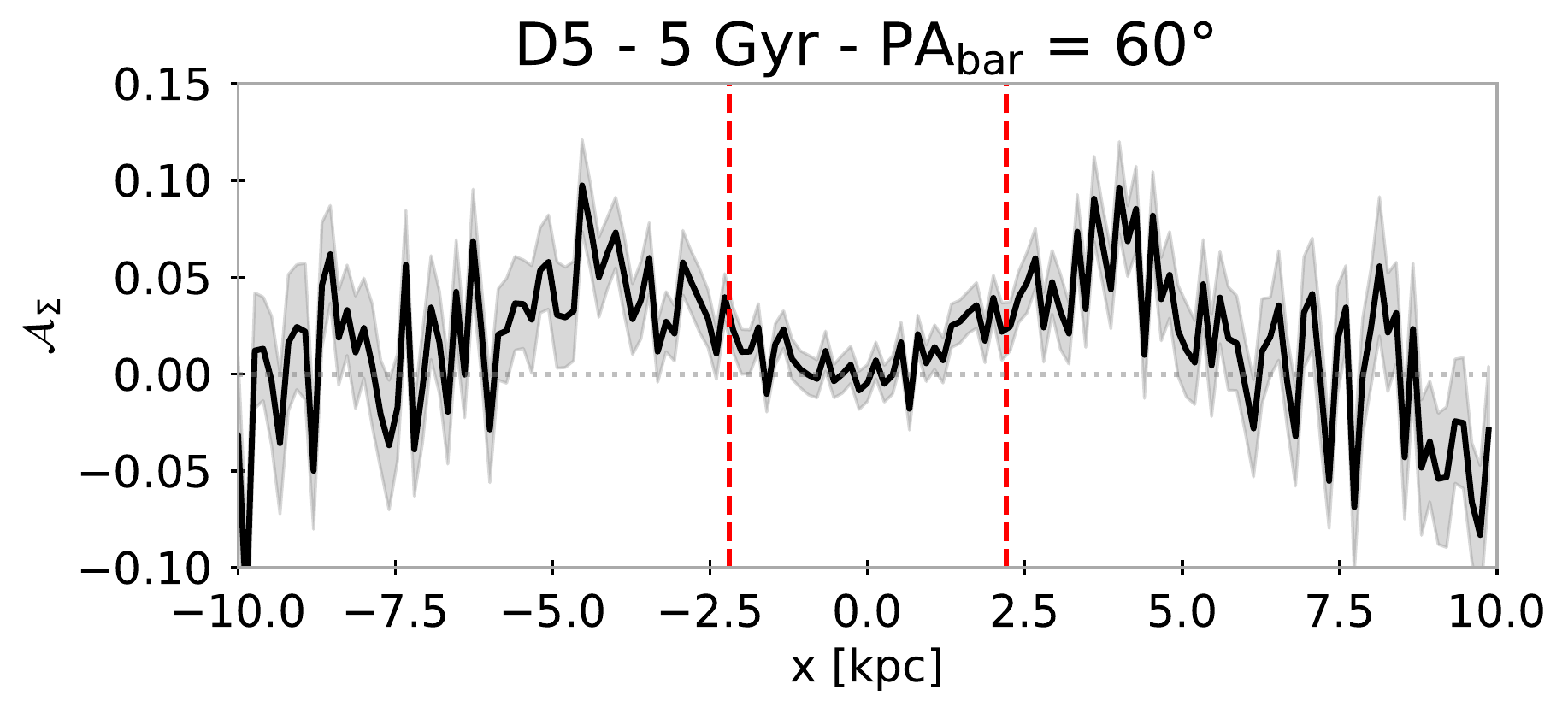}
    \includegraphics[scale=0.43]{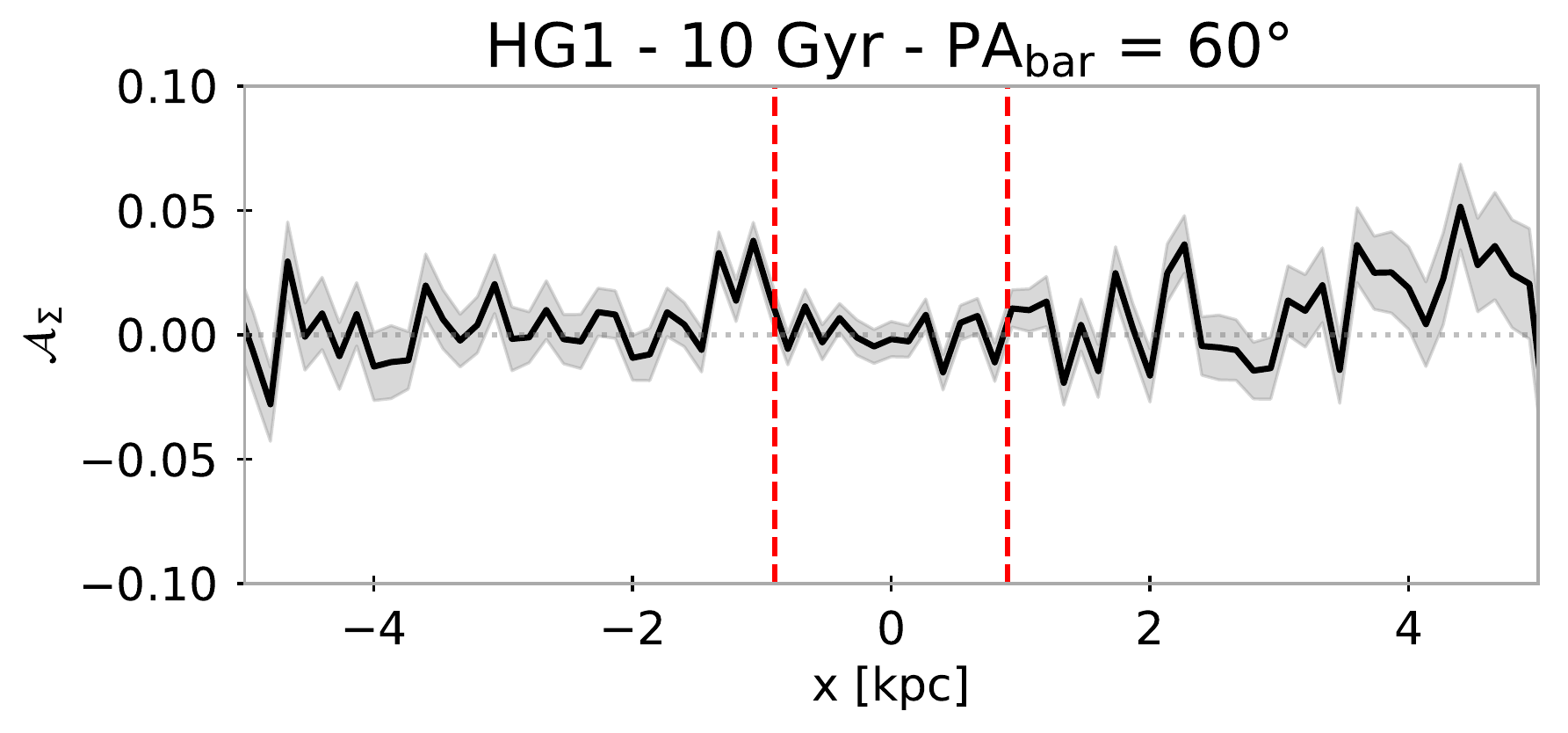}
    \includegraphics[scale=0.43]{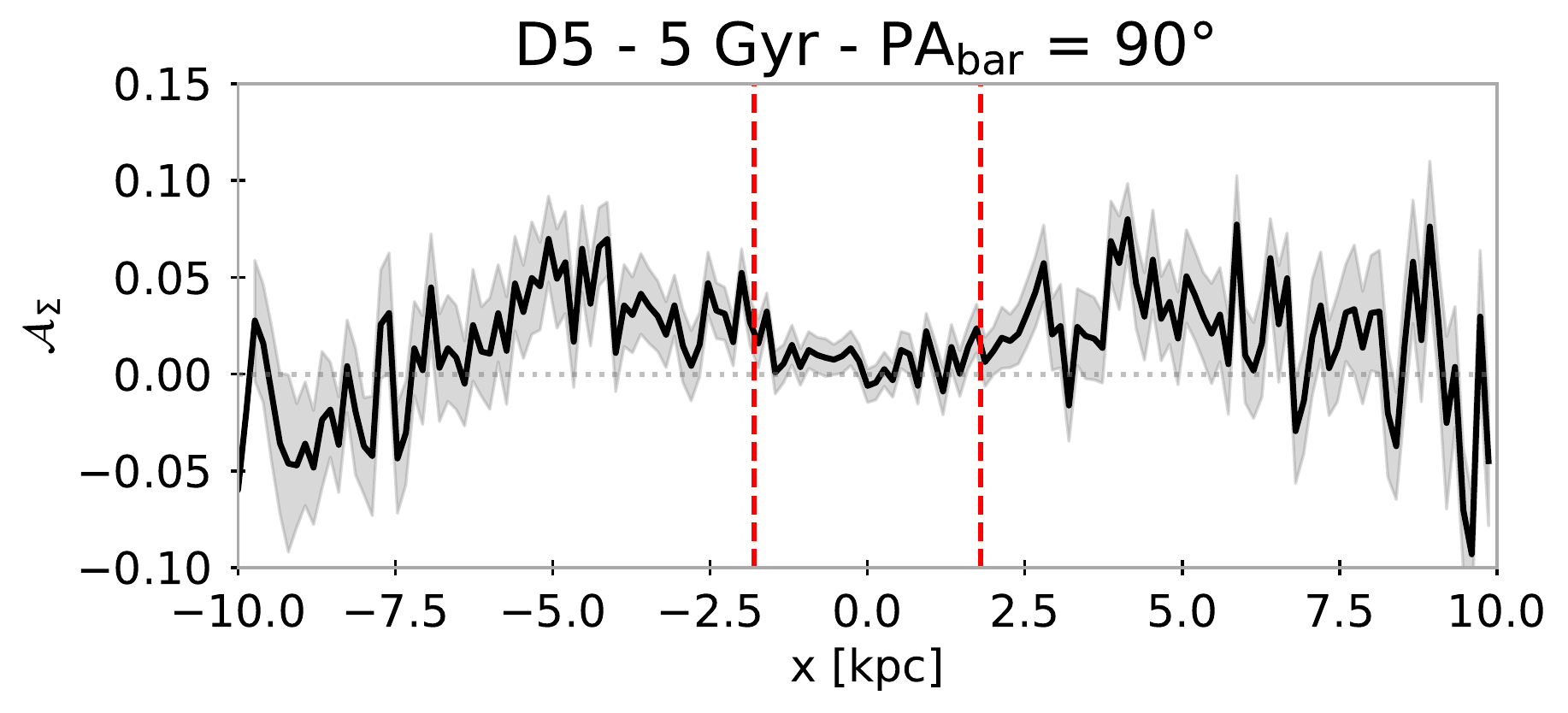}
    \includegraphics[scale=0.43]{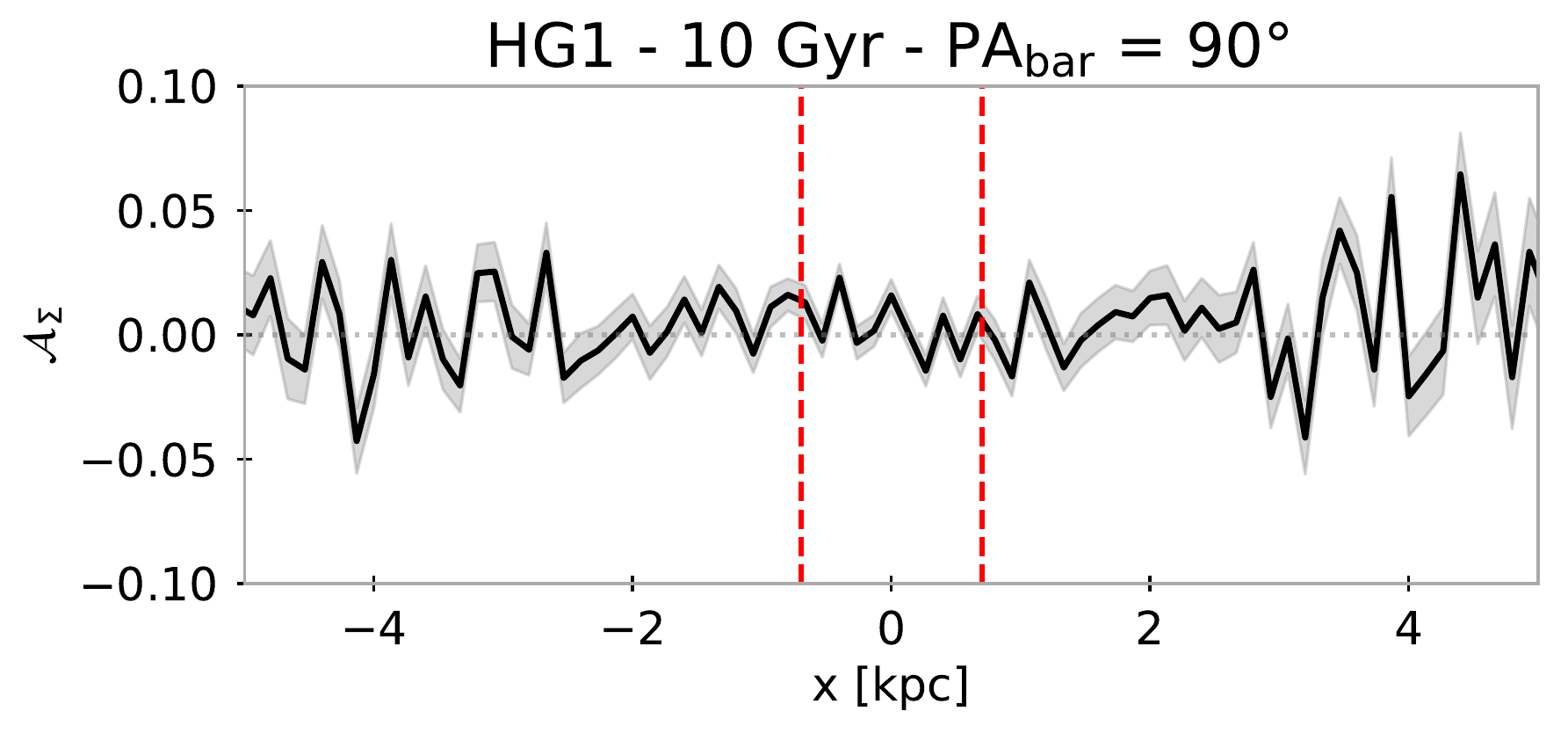}
    \caption{Mid-plane asymmetry profiles as in Fig.~\ref{fig:sim_dia} lower panel, but for models D5 (at 5 Gyr, left-hand column) and HG1 (at 10 Gyr, right-hand column) with different bar orientations in the edge-on view (${\rm PA}_{\rm bar}=30\degr$, top row; ${\rm PA}_{\rm bar}=60\degr$, middle row; ${\rm PA}_{\rm bar}=90\degr$, bottom row).}
    \label{fig:sim_bar}
\end{figure*}

The orientation of the bar in real galaxies will also be randomly distributed between side-on (i.e., the bar is aligned with the $x$ axis) and end-on (i.e., the bar is perpendicular to the $x$ axis). In the first case, the associated B/P bulge, if present, has the expected X shape, while in the second case, it usually appears boxy \citep[e.g.,][]{Combes1981,Bureau1999,Lutticke2000,Chung2004,Laurikainen2016}. To explore the effects of the bar's orientation on the mid-plane asymmetry, we now explore the effect of ${\rm PA}_{\rm bar}=30\degr,~60\degr,$ and $90\degr$, where ${\rm PA}_{\rm bar}$ is measured rotating the bar in the ($x,y$) plane and around the $z$ axis, where ${\rm PA}_{\rm bar}=0\degr$ means the bar is aligned to the $x$ axis in the edge-on view. The resulting asymmetry diagnostics for models D5 (left column) and HG1 (right column) are presented in Fig.~\ref{fig:sim_bar}. The apparent dimension of the B/P bulge along the $x$ axis progressively decreases for ${\rm PA}_{\rm bar}=30\degr,~60\degr,$ and $90\degr$: it becomes 3.1, 2.2, and 1.8 kpc for model D5 and 1.2, 0.9, and 0.7 kpc for model HG1.

In model D5, when ${\rm PA}_{\rm bar}=30\degr$ the mid-plane asymmetry looks similar to the reference case from Fig.~\ref{fig:sim_dia}. When ${\rm PA}_{\rm bar}=60\degr$ both the X shape traced by the red contours and the asymmetries appear weaker. The flat part of the mid-plane asymmetry profile around $x=0$ disappears, while the profile increases less steeply to the two peaks, when compared to the side-on case of Fig.~\ref{fig:sim_dia}. Moreover, the extension along the $x$ axis of the X shape decreases and the same behaviour is adopted by the asymmetric regions. At ${\rm PA}_{\rm bar}=90\degr$ the X shape is rather weak but still distinguishable as a boxy structure. The asymmetries are weak (reaching $\mathcal{A}_\Sigma (x) \sim 0.05-0.07$, half the values in Fig.~\ref{fig:sim_dia}), but the profile continues to show the typical double-peaked shape. 

No asymmetries appear for model HG1 at any ${\rm PA}_{\rm bar}$ (Fig.~\ref{fig:sim_bar}, right).  An X-shaped structure is clearly visible when ${\rm PA}_{\rm bar}=30\degr$ and $60\degr$, and it becomes boxy at ${\rm PA}_{\rm bar}=90\degr$.

\subsubsection{Effect of the disc position angle}
\label{sec:discPA}

The identification of the correct disc PA is crucial for detecting a mid-plane asymmetry because it defines the mid-plane relative to which the diagnostics are measured. The disc PA in real galaxies is usually determined by fitting ellipses to the isophotes of the outer disc \citep[see e.g.][]{Aguerri2015}. The typical resulting uncertainties on the disc PA for nearly edge-on galaxies are lower than $1\degr$ \citep{Salo2015}. Nevertheless, outer disc distortions such as warps, thick discs, or asymmetric features in the disc (such as dust along the mid-plane and bright foreground stars) may increase the uncertainty on the PA of the disc.
Therefore next we test the effect on the mid-plane asymmetry diagnostics of errors on the disc PA, by assuming that the disc is not perfectly aligned with the $x$ axis.

\begin{figure*}
    \centering
    \includegraphics[scale=0.43]{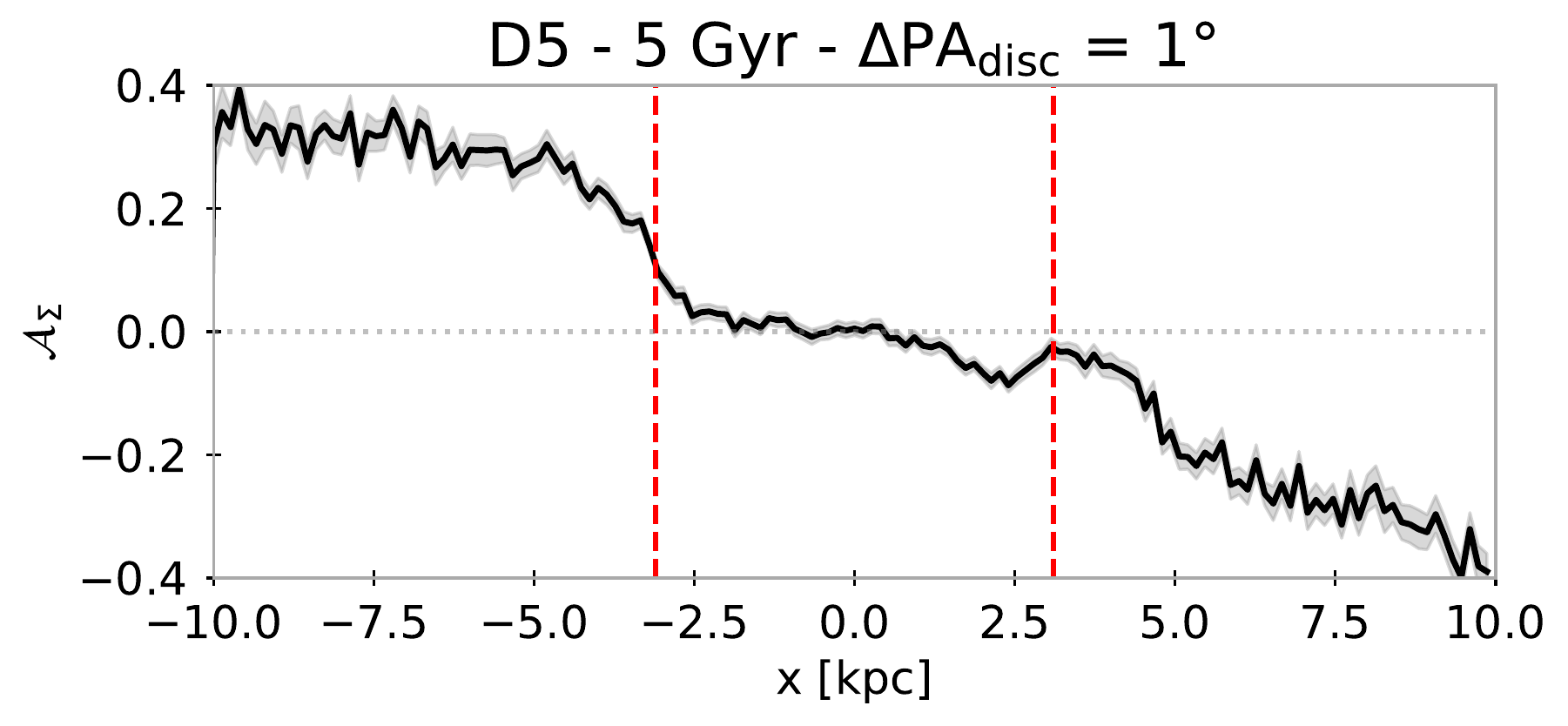}
    \includegraphics[scale=0.43]{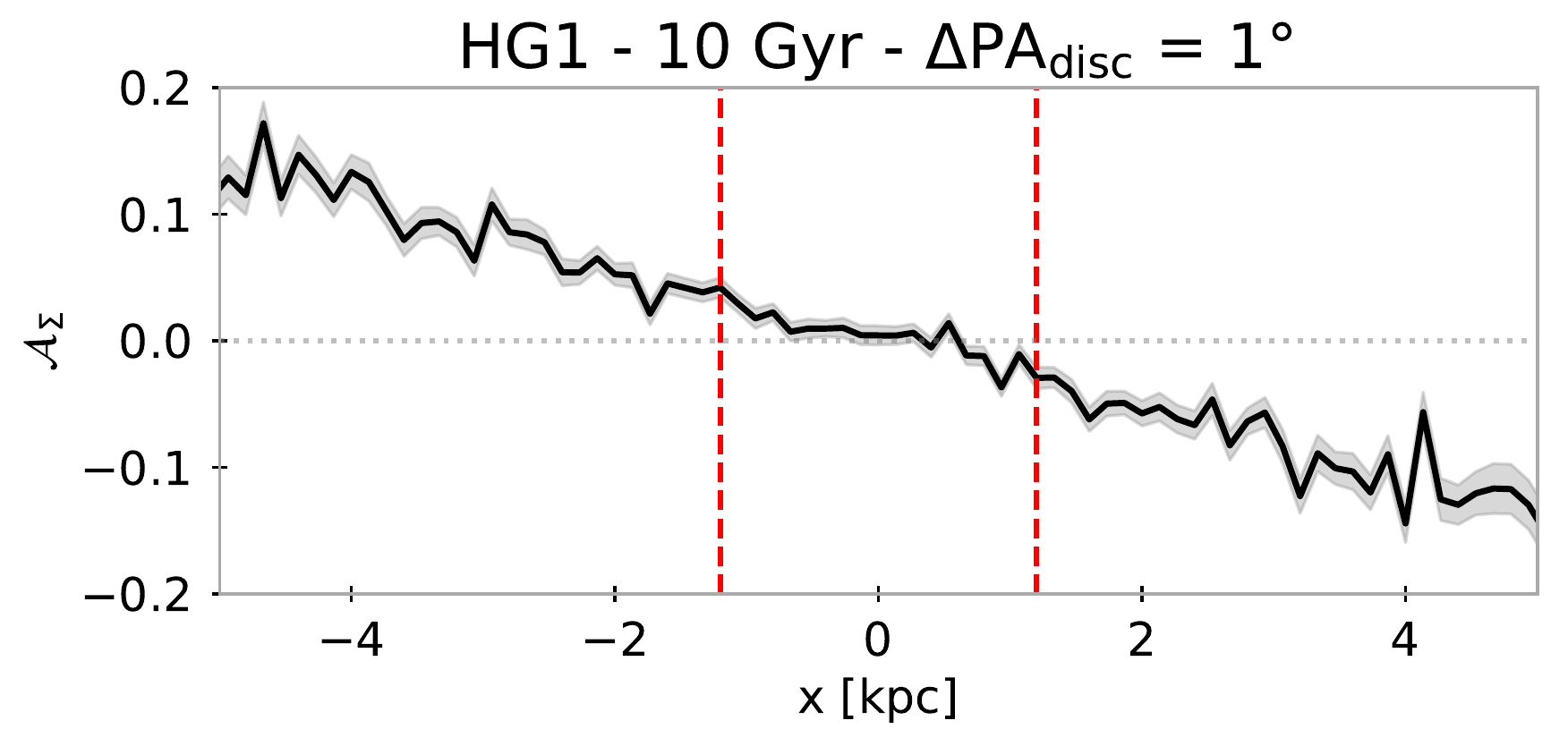}
    \includegraphics[scale=0.43]{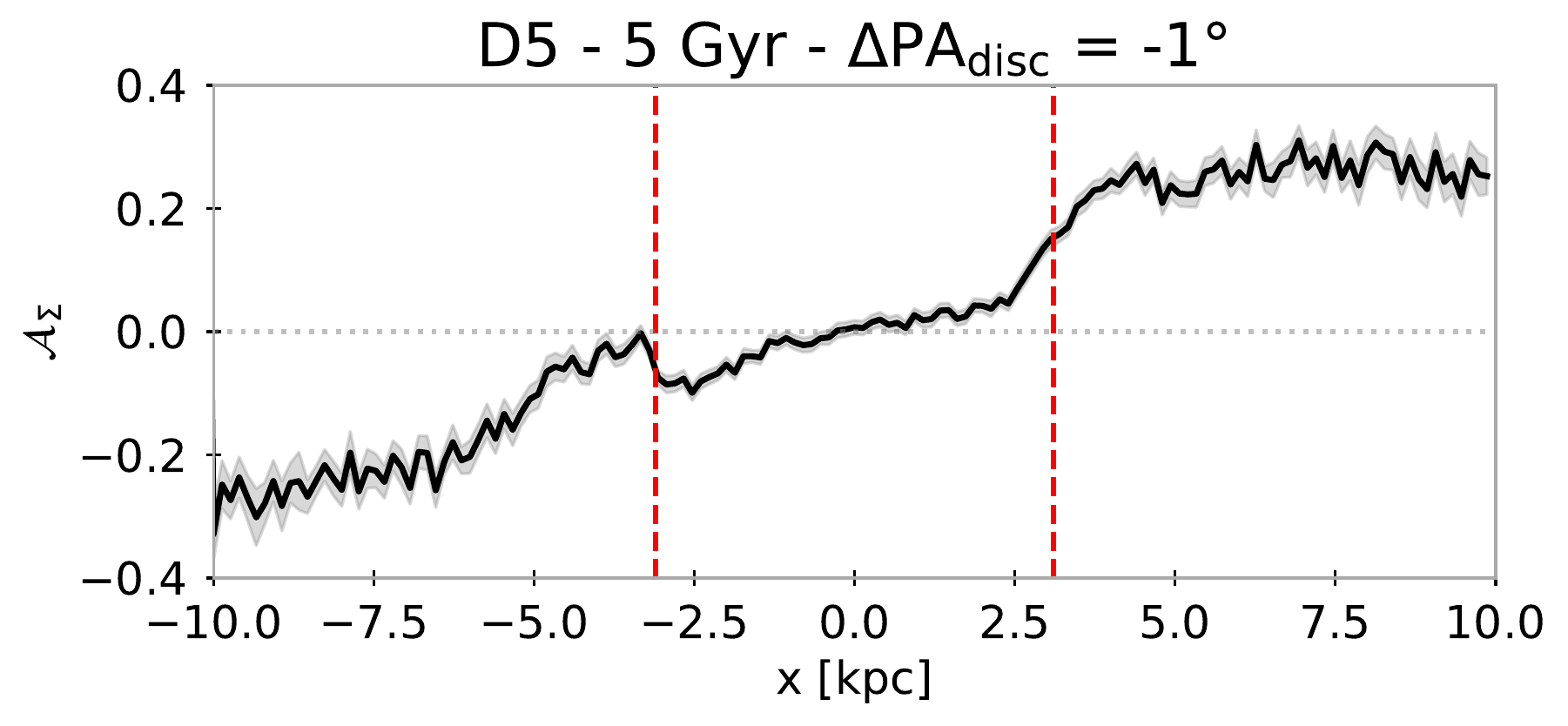}
    \includegraphics[scale=0.43]{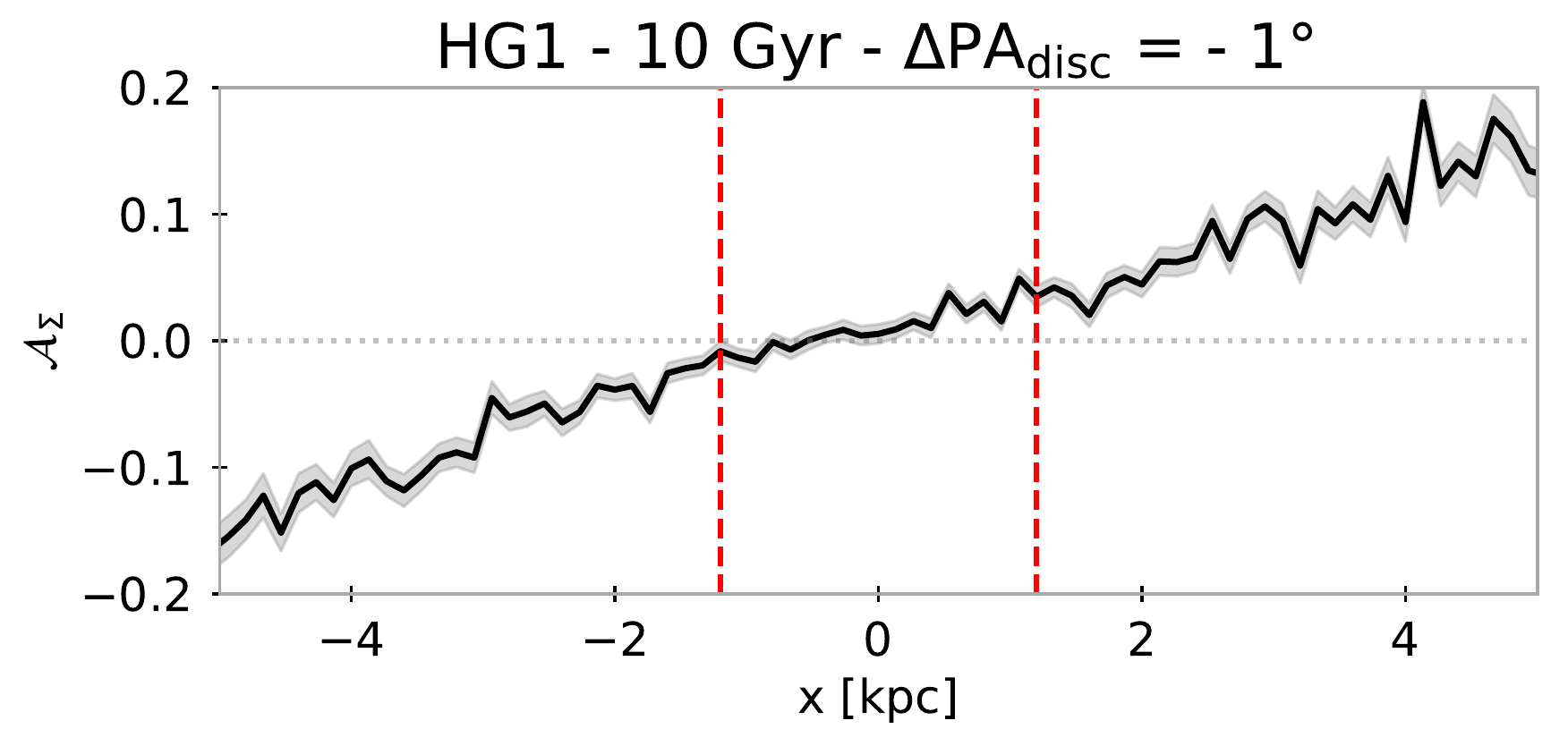}
    \caption{Mid-plane asymmetry profiles as in Fig.~\ref{fig:sim_dia} lower panel, but for models D5 (at 5 Gyr, left-hand column) and HG1 (at 10 Gyr, right-hand column) with different misalignments of the disc major axis with respect to $x$ axis ($\Delta{\rm PA}_{\rm disc}=+1\degr$, top row; $\Delta{\rm PA}_{\rm disc}=-1\degr$, bottom row).}
    \label{fig:sim_pa}
\end{figure*}

Fig.~\ref{fig:sim_pa} presents the mid-plane asymmetry diagnostics assuming a ${\rm PA}_{\rm disc}=+1\degr$ (upper panels) and $-1\degr$ (bottom panels) for models D5 (left column) and HG1 (right column). As might be expected, the effect is dramatic for both models: a strong quadrupolar signal appears in the disc region in the mid-plane asymmetry maps while the mid-plane asymmetry profiles are sloped along the entire radial range, especially in the region of the disc. The slope of the mid-plane asymmetry profiles changes sign when varying the disc PA in opposite directions ($\pm1\degr$), while the quadrupolar signal changes its orientation with respect to the mid-plane. 

The peaks in the mid-plane asymmetry profile of model D5 are altered by the incorrect disc PA, with both of them flattened and broadened to different extents. The mid-plane asymmetry profile of model HG1 behaves similarly but lacks the bumps corresponding to the two peaks. Despite both the mid-plane asymmetry map and mid-plane asymmetry profile being strongly affected by the wrong PA, the presence of the bumps/peaks in model D5 helps distinguish between mid-plane asymmetry and mid-plane symmetry. We conclude that, in spite of the large artificial asymmetries produced by a misaligned disc, the shape of the mid-plane asymmetry profile can still be used to identify mid-plane asymmetry produced by buckling.

We also test the effect of a wrong disc orientation (disc ${\rm PA}\pm1\degr$), together with an imperfectly edge-on view ($i=\pm3\degr$). For each model the dominant effect is that produced by the wrong disc PA. A strong quadrupolar signal appears in the disc region, and the mid-plane asymmetry profile is sloped along its entire extent. Asymmetries associated with the B/P bulge are now harder to distinguish in model D5, since only a weak bump appears on the steep mid-plane asymmetry profile.

\subsubsection{Identification of the disc position angle}

Therefore correcting for small tilts is preferable to merely using the mid-plane asymmetry profile with a tilt in place. To identify the disc PA, we develop a simple algorithm which we test on simulations. We consider the edge-on view of the surface density of the models. We select two areas in the $(x,z)$ plane, covering the extension of the X shape along the $x$ axis and symmetric with respect to the galaxy centre and integrate the flux along the $x$ axis to obtain the collapsed profiles of the density. These profiles are peaked around the galaxy centre and decrease along the $z$ axis. We then vary the disc PA and run this test until the two profiles have minimal difference along their extent. We consider as the best value of the disc PA that for which the two profiles are identical. The two profiles do not coincide, and the peaks produced by the central region of the galaxy do not line up, if the disc is not aligned with the $x$ axis. We vary the assumed disc PA by $0.2\degr$ within $\pm2\degr$ from the reference value to identify the disc PA. Once we identify the best disc PA, we vary the extension of the two symmetric portions of the surface density plot, to test that the profiles remain superimposed.

\begin{figure}
    \centering
    \includegraphics[scale=0.4]{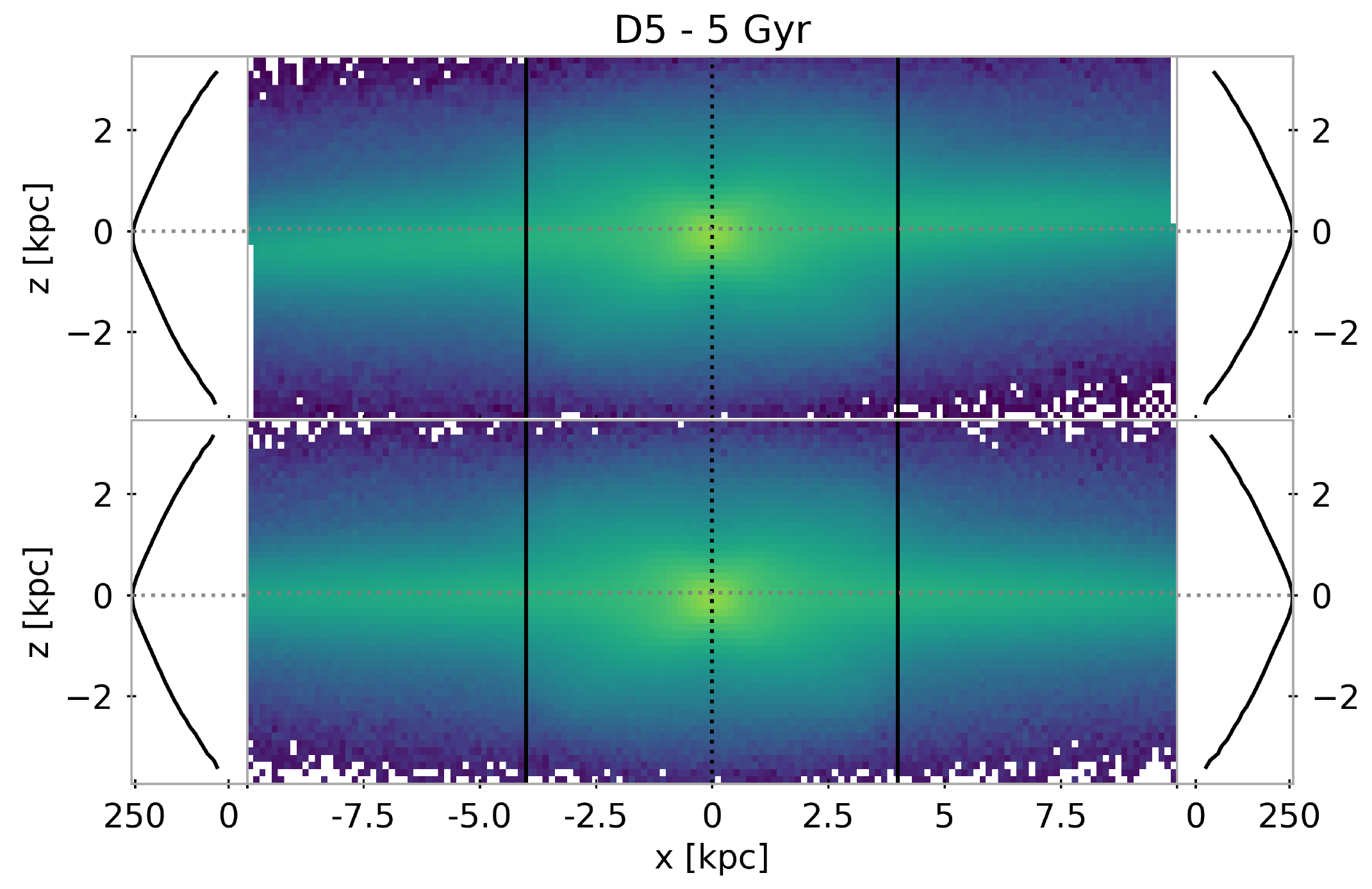}
    \caption{Maps of the surface number density of model D5 (at 5 Gyr) seen side-on with the disc major axis misaligned by $\Delta{\rm PA}=-2\degr$ (upper panel) and aligned (lower panel) with respect to the $x$ axis .
    The vertical lines define the two symmetric radial regions ($|x|>4$ kpc) where we obtained the vertical profiles of mean surface number density plotted on both sides of the panels.}
    \label{fig:pa}
\end{figure}

Fig.~\ref{fig:pa} shows an example of the application of this method to identify the disc PA of model D5, with a wrong (upper panel), and the well-identified disc PA (lower panel). The profiles plotted on the sides of the image are obtained by collapsing the two portions of the image within the three vertical black lines. The portions shown here are for $|x|< 4$ kpc but we have varied these regions. The method correctly determines the disc orientation; in subsequent sections of this paper we adopt this method for our observational sample of galaxies, and conservatively reject galaxies for which the method does not work, i.e. when the PA is not constant when varying the regions along the $x$ axis. 

\subsubsection{Effect of a mis-identified centre}
\label{sec:centre}

The correct determination of the centre of the galaxy is also crucial for defining the mid-plane. We therefore test the effect of a mis-centering on the mid-plane asymmetry diagnostics by varying the centre used to build the mid-plane asymmetry map along the $x$ and $y$ axes by 1 pixel (corresponding to 0.13 kpc in both models) simultaneously, $(\Delta x,\Delta y) = (1,1)$ pixel. 

\begin{figure*}
    \centering
    \includegraphics[scale=0.43]{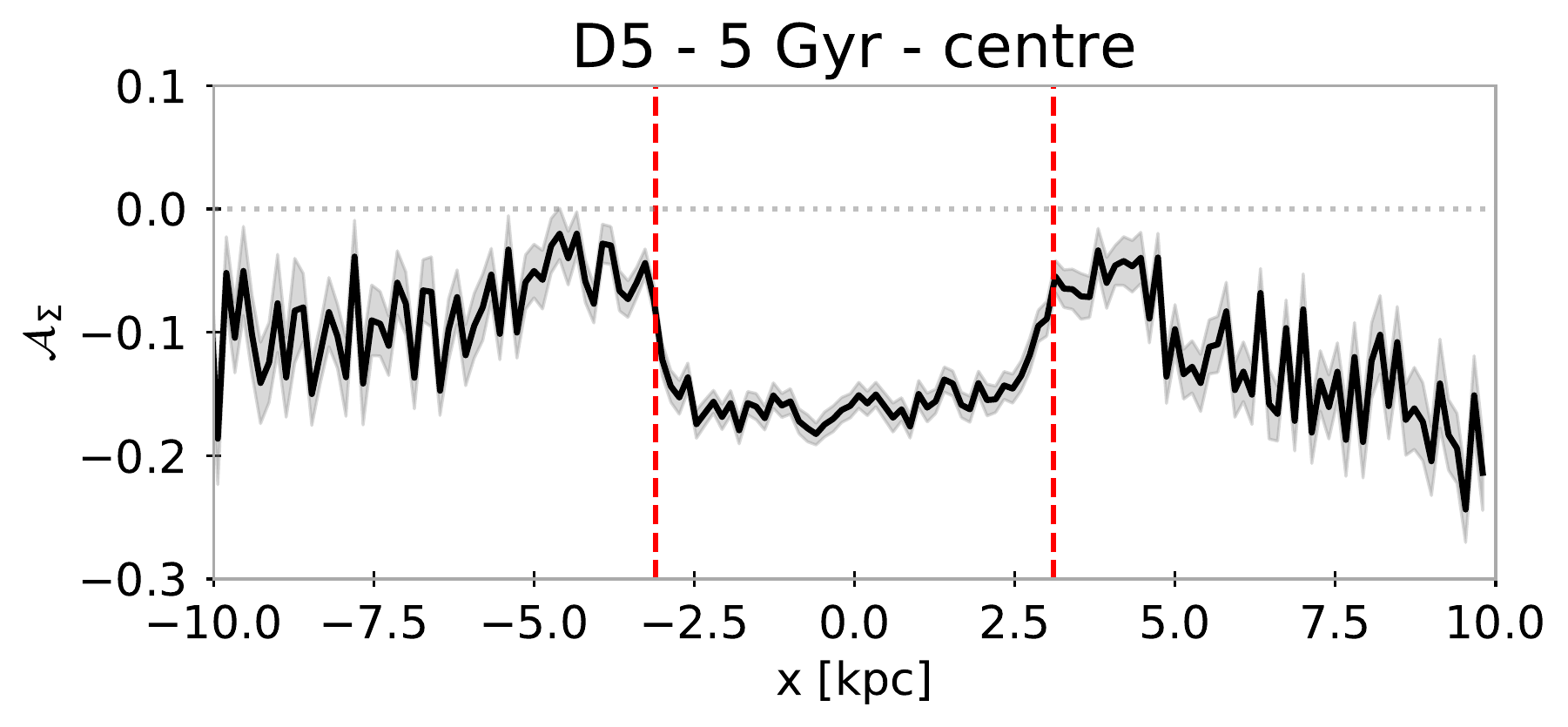}
    \includegraphics[scale=0.43]{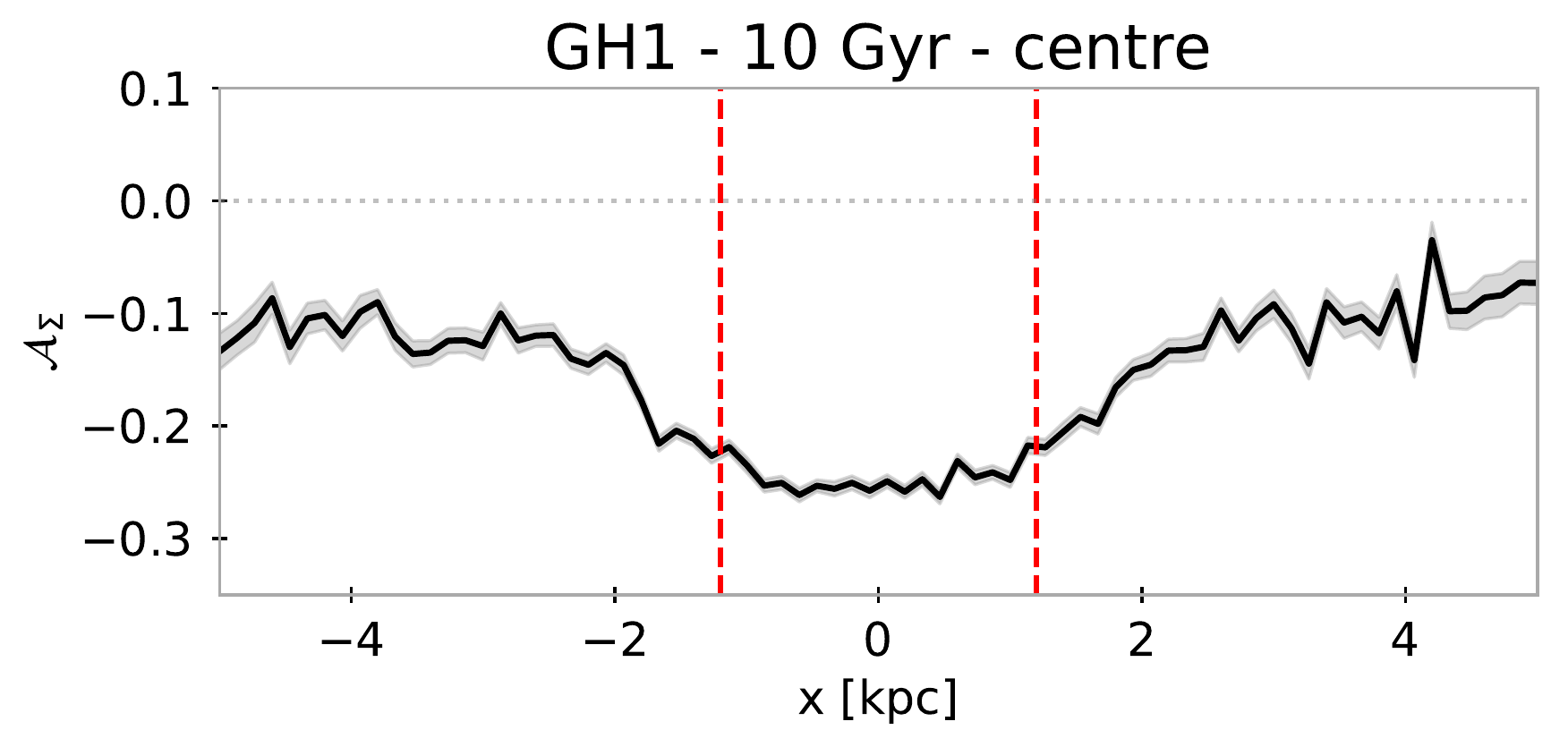}
    \caption{Mid-plane asymmetry profiles as in Fig.~\ref{fig:sim_dia} lower panel, but for models D5 (at 5 Gyr, left-hand column) and HG1 (at 10 Gyr, right-hand column) with a miscentring of $(\Delta x,\Delta y) = (1,1)$ pixel, (where 1 pixel corresponds to 0.13 kpc in both models.}
    \label{fig:centre}
\end{figure*}

Fig.~\ref{fig:centre} shows the mid-plane asymmetry maps and mid-plane asymmetry profiles assuming a mis-centering by $(\Delta x,\Delta y) = (1,1)$ pixel for models D5 (left column) and HG1 (right column). In both models, a strong but artificial signal appears in the central region of the mid-plane asymmetry map with a characteristic dipolar shape extending to $\sim10$ pixels ($\sim1.5$ kpc). Strong artificial asymmetries are visible also in the disc region, while the mid-plane asymmetry profiles are non-zero everywhere, with a strong peak near the centre. 
The same result is found when the mid-plane asymmetry map is mis-centered along just the $y$ axis, assuming $(\Delta x,\Delta y) = (0,1)$ pixel, while no dipolar signal is present when the mid-plane asymmetry map is mis-centered along the $x$ axis, assuming $(\Delta x,\Delta y) = (1,0)$ pixel.

In order to identify the best centre for observed galaxies, we develop another simple algorithm. We perform a Fourier analysis on the central regions of the mid-plane asymmetry map, by expanding its azimuthal radial profile, $I(r<R_{\rm max},\phi)$, as:
\begin{equation}
I(r<R_{\rm max},\phi)=\sum_{m=-\infty}^{\infty}[c_m e^{({im\phi})}].
\end{equation}
The corresponding Fourier components are given by:
\begin{equation}
c_m = \frac{1}{2\pi}\int_{-\pi}^{\pi}I(r<R_{\rm max},\phi)e^{-im\phi}{\rm d}\phi
\end{equation}
%
with the Fourier amplitude of the $m$-th component given by $I_m= |c_m|$.


%
A large value of the $m=1$ component represents a feature with dipolar asymmetry, which corresponds to the artificial signal due to mis-centering that we want to minimise. We proceed as follows: we build a set of mid-plane asymmetry maps by varying the centre by 0.5 pixels around the correct value by 5 pixels along both $x$ and $y$ axes. We measure the $m=1$ component within the central part of the image, out to $R_{\rm max}\sim20$~pixels, covering the extension of the dipolar structure. The $m=1$ component is minimised at the true centre. We repeat this analysis varying $R_{\rm max}$ between 50\% to 125\% of the extension of the dipolar structure to check that the resulting centre truly minimises the $m=1$ component. We apply this method to find the centres of all the galaxies in our observational sample.

\subsection{Further observational effects}
\label{sec:other_effects}

\subsubsection{Effect of image rotation}

In order to build the mid-plane asymmetry map we need to align the edge-on galaxy with the $x$ axis. While it is trivial in simulations to rotate the system to the $x$ axis before constructing the image, images of real galaxies are obtained with the galaxies at random orientations on a CCD, and need to be rotated in order to align the disc PA to the $x$ axis. To perform the rotation, we use the \textsc{idl} procedure \textsc{rot}, which involves an interpolation method based on a cubic convolution \citep{Park1983}. 

To test whether this rotation procedure produces any artifacts in the mid-plane asymmetry maps, we build the edge-on view for our models with the disc PA randomly oriented. We then rotate this image using the same \textsc{rot} procedure to the required orientation and build the corresponding asymmetry diagnostics. After comparing these with the diagnostics presented in Fig.~\ref{fig:sim_dia} we have verified that they are the same (see Fig.~\ref{fig:rot}). We also calculate the normalised difference between the original edge-on view of the models and those produced through a rotation of a misaligned image, and find that the differences due to image rotation are less than 1\%. 

\subsubsection{Effect of the spatial sampling}

Secondly, we explore the effect of the spatial sampling on the asymmetry diagnostics. We both halve and double the spatial sampling, assuming a pixel scale of 0.07 and 0.25 kpc, respectively. We find no noteworthy effects when varying the spatial sampling within reasonable values: in all cases model D5 presents strong asymmetries similar to those observed in Fig.~\ref{fig:sim_dia}, while no asymmetries appear for model HG1 (see Fig.~\ref{fig:hr}). 

\subsubsection{Effect of seeing}

When comparing the results of simulations to observational studies, it is necessary to take into account the effects of seeing (instrumental and/or atmospheric). To mimic the observational conditions presented in the following sections, we rescale the images of the projected number density of the particles of the models to a distance up to $\sim50$~Mpc. Then, we convolve the images with a Gaussian filter using the \textsc{python-scipy} function \textsc{gaussian\_filter} with a varying full width at half maximum (FWHM). At a FWHM of 0.5 kpc a slight smoothing of the mid-plane asymmetry profile is detectable (upper panel of Fig.~\ref{fig:seeing}). With increasing FWHM the width of the asymmetry rises while the peak declines, as expected. At FWHM $\sim3.0$ kpc, the double-peaked shape of the mid-plane asymmetry profile has been smoothed to such an extent that it becomes barely distinguishable and appears consistent with a flat profile (see bottom row of Fig.~\ref{fig:seeing}). This limiting value corresponds to almost one third of the radial separation between the peaks observed along the mid-plane asymmetry profile just outside the region of the B/P bulge in model D5 (compare e.g., Fig.~\ref{fig:dia} with \ref{fig:seeing}, bottom row). 
In turn, this limiting seeing corresponds to a FWHM of $12.2$~arcsec, for galaxies at a distance of $\sim50$~Mpc. This radial distance corresponds to half of the extension of the smallest observed B/P bulge analysed in this work.

\subsubsection{Effect of dust}

The mid-plane of star-forming galaxies is generally dusty. In barred galaxies, within the bar dust gets swept into dust lanes along the leading edges of the bar \citep[see e.g.,][]{athanassoula92,Smith2016}. Dust attenuates the stellar light along the line-of-sight (LOS), particularly at short wavelengths. We explore the effect of dust on the mid-plane asymmetry diagnostics by modelling an extended dust disc together with dust lanes along the bar in our models. We compute the resulting extinction along the LOS and the consequent mid-plane asymmetry diagnostics.

We assume the dust disc has a double-exponential profile in cylindrical coordinates \citep{Wainscoat1989}:
\begin{equation}
    D_{\rm disc}(R, z) = D_{\rm 0,disc} ~ 
    e^{-{ R/h_{\rm R,disc}}} ~e^{-|z|/h_{\rm z,disc}},
\label{eq:dust}
\end{equation}
where $h_{\rm R,disc}$ and $h_{\rm z,disc}$ are the radial and the vertical scalelengths of the dust disc, respectively, while $D_{\rm 0,disc}$ is the central dust density.

To mimic the bar dust lanes, we define rectangular areas on the bar's leading edges, mimicking the straight dust lanes of \citet{athanassoula92}. We assume a given width, $w_{\rm dl}$, an angle between the dust lane and the bar major axis, $s_{\rm dl}$, and the minimum and maximum extent of the dust lane, $x_{\rm dl,min}$ and $x_{\rm dl,max}$, respectively, along the $x$ axis, where the bar is located. In particular, we assume $x_{\rm dl,max}$ to reach the extension of the bar, while $x_{\rm dl,min}$ is negative, given the observation that dust lanes can extend past the centre of the bar and wind around the bulge. Within the dust lanes, we assume that the dust distribution is described by the double-exponential model from Eq.~\ref{eq:dust}, and zero outside, as in \cite{Gerssen2007}. In Fig.~\ref{fig:dust_lanes} we show the maps of the stellar surface number density of models D5 and HG1, superimposing the contours of the dust lanes.

To build the mid-plane asymmetry map taking into account dust extinction, we start from the projected surface number density of the particles at each position of the image of the (nearly) edge-on view of the models. When a particle is obscured by dust, it contributes with a lower weight to the projected surface number density, which depends on the amount of intervening dust located along the LOS between the observer and the given particle. For each particle, we integrate the projected dust distribution along the LOS between the observer and the particle. Assuming a unit mass absorption coefficient, this corresponds to the optical depth ${\tau_i}$. The corresponding particle weight is then $w_i=e^{-\tau_i}$.

We adopt the highest typical observed value for the central extinction $A_V\sim5.0$~mag \citep{Holwerda2005}, which corresponds to the face-on optical depth for a particle infinitely far behind the disc, $\tau_0=0.921*A_V$. We assume a thin dust disc which radially extends beyond the bar ($h_{\rm R,disc}=1$~kpc and $h_{\rm z,disc} = 0.1$~kpc), after checking a more extended disc (up to $1.5\times$ the initial galaxy disc scalelength) does not change substantially our conclusions.

\begin{figure*}
    \centering
    \includegraphics[scale=0.43]{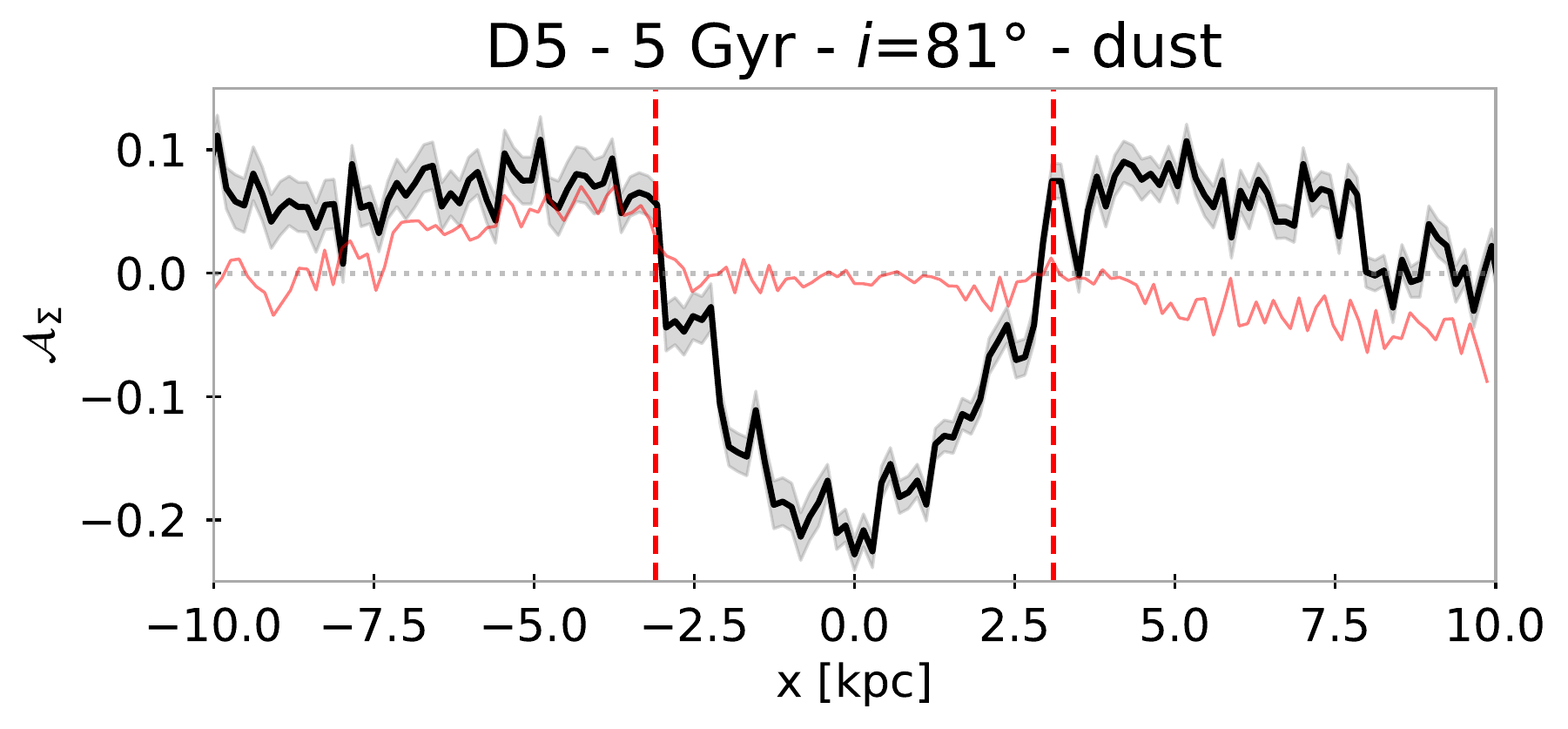}
    \includegraphics[scale=0.43]{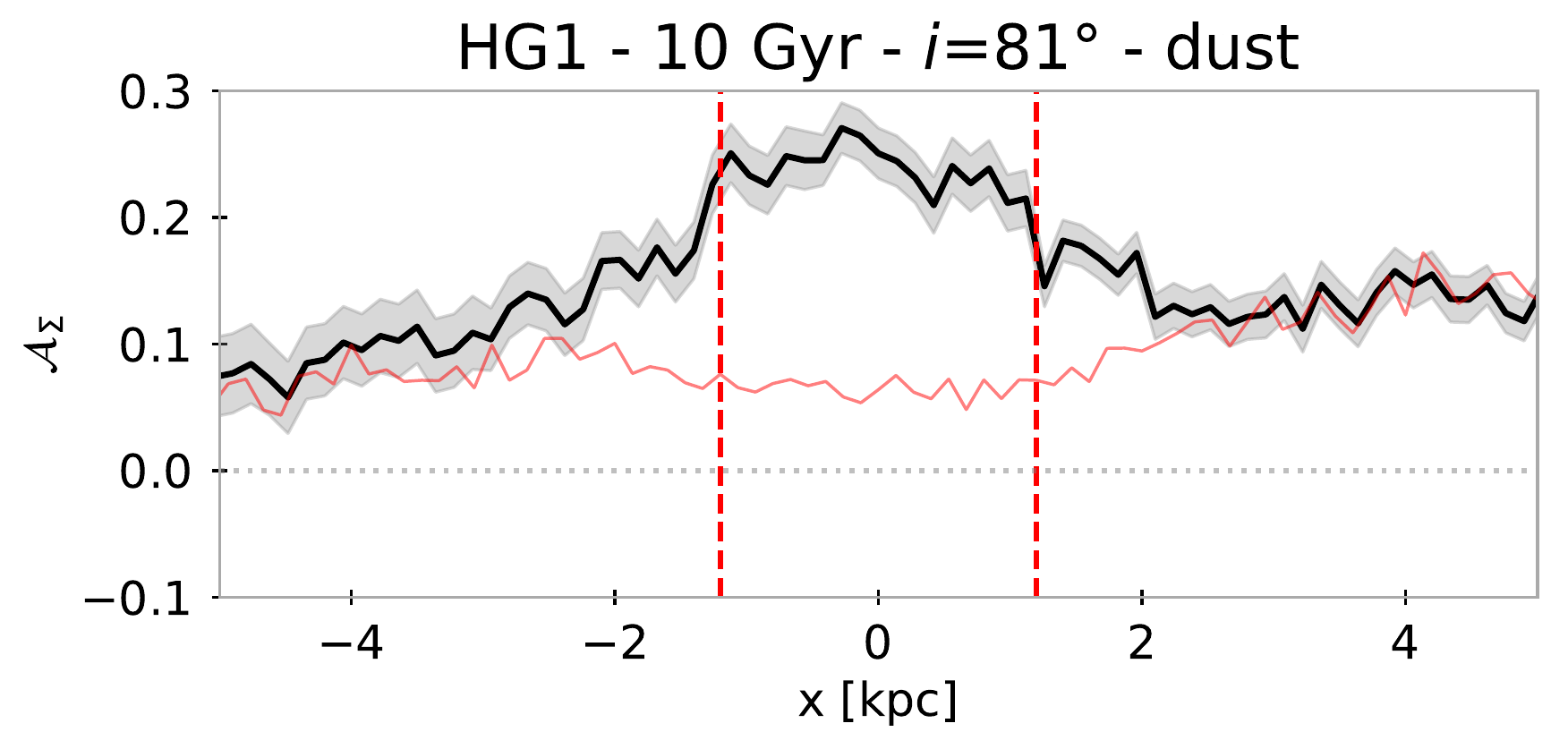}
    \includegraphics[scale=0.43]{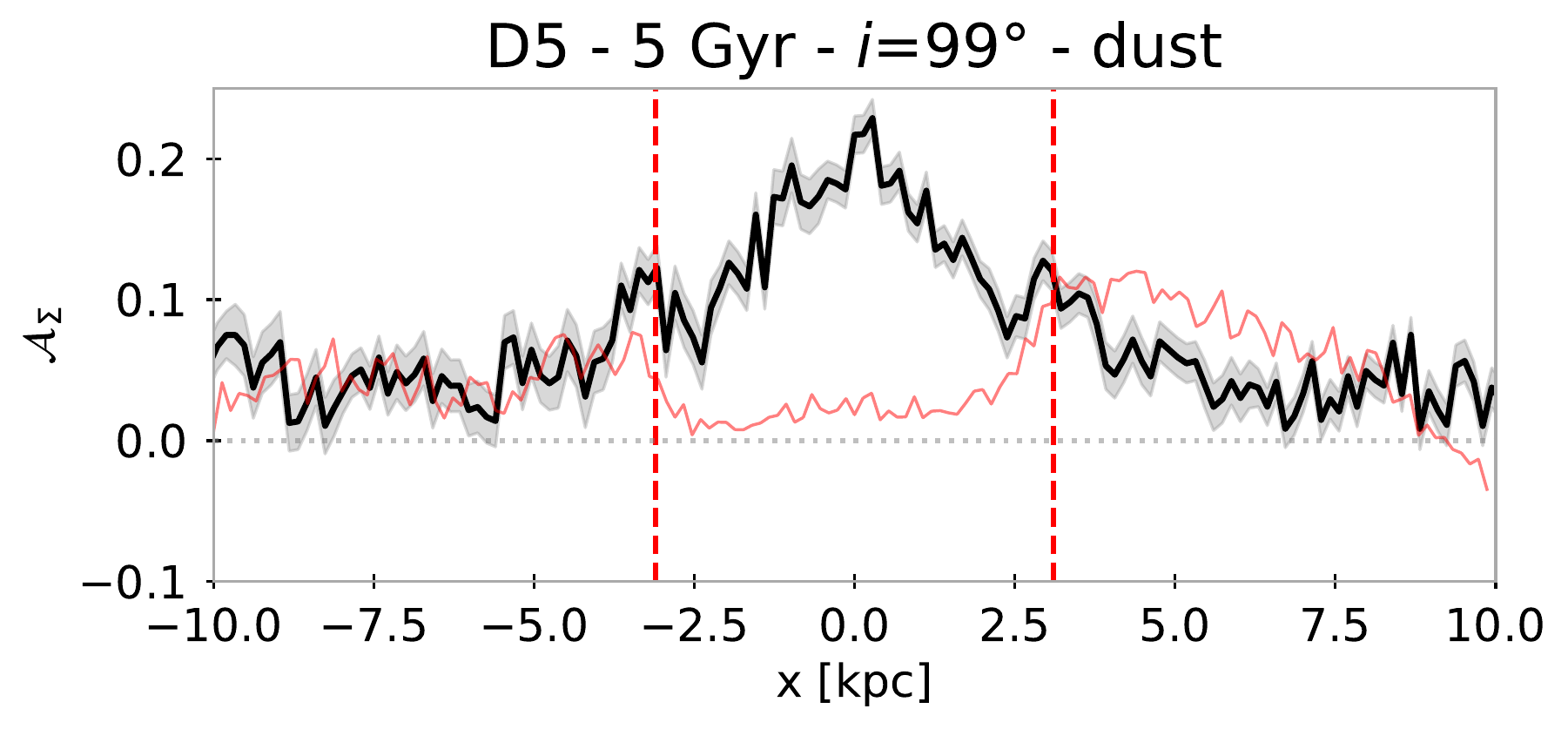}
    \includegraphics[scale=0.43]{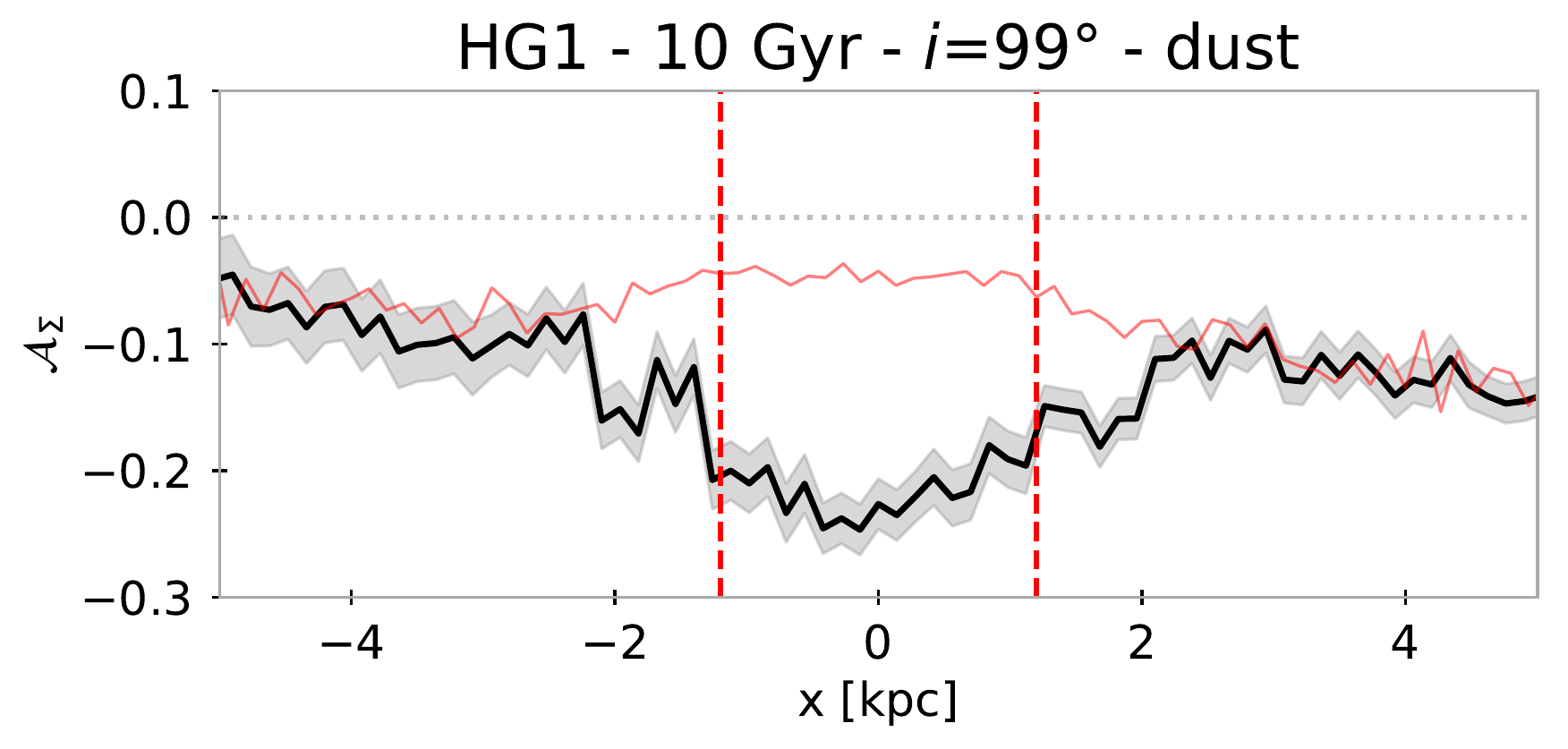}
    \caption{Mid-plane asymmetry profiles as in Fig.~\ref{fig:sim_dia} lower panel, with a different inclination of $\Delta i=\pm~9~\degr$ with respect to the side-on view and taking into account the effect of dust. Superimposed in red are shown the corresponding profiles obtained without taking into account the effect of the dust.}
    \label{fig:dust_pa}
\end{figure*}

We build the mid-plane asymmetry diagnostics using the extincted map of the surface number density (after assuming a mass-to-light ratio equal to one). The dust produces an asymmetry with respect to the mid-plane in both models when they are observed not perfectly edge-on. In Fig.~\ref{fig:dust_pa} we show the mid-plane asymmetry profiles for models D5 and HG1 with $i = 90\degr \pm 9\degr$; a strong asymmetry, with a single peak centred at $x = 0$, is evident, while the double-peaked asymmetry profile of D5 is lost.

In Figs.~\ref{fig:dust_pa_d5} and \ref{fig:dust_pa_hg1} (top panels) we show the effect of the dust for varying inclination in the range $84\degr < i < 96\degr$. The shape of the asymmetry produced by the dust is now clearly distinguishable from the asymmetry due to buckling since it peaks on the minor axis. The double peaks in the asymmetry profile due to buckling is preserved in model D5, but asymmetry appears near the minor axis as well for $\Delta i \geq 3\degr$. We also note that, in model HG1, the dust does not produce the characteristic double-peaked mid-plane asymmetry profile arising from an intrinsic buckling asymmetry. In Figs.~\ref{fig:dust_pa_d5} and \ref{fig:dust_pa_hg1} (bottom rows) we show as well the effect of a radially more extended dust disc (with $h_{\rm R,disc}=1.5\times$ the initial galaxy disc scalelength) for $i=90\degr$. No appreciable effects of the dust are observable. Of course, the vertical (horizontal) extension of the central asymmetry arising from the dust is related to the scaleheight (scalelength) of the dust disc, $h_{\rm z,disc}$ ($h_{\rm R,disc}$), with larger $h_{\rm z,disc}$ ($h_{\rm R,disc}$) producing an asymmetry extending to larger distances along the $z$ ($R$) axis in particular for deviation from the edge-on view. Since the central dust asymmetry decreases away from the mid-plane, it is possible to adjust the vertical extent within which the mid-plane asymmetry profile is computed to exclude the region most affected by the dust. Since the asymmetry that appears in the intrinsically symmetric model HG1 in the presence of dust does not present the double-peaked shape associated with a recent buckling event, we conclude that the effect of a smooth distribution of dust is distinct from that of a buckling-induced asymmetry.

\subsection{Summary of the simulation results}

Using simulations, we have shown that buckling produces long lasting mid-plane asymmetries. Inasmuch as bars in pure $N$-body simulations such as D5 generally evolve rapidly, for instance becoming too large \citep{Erwin2019}, it is likely that we underestimate the duration of the asymmetry. We have shown that the mid-plane asymmetry is not very sensitive to small deviations ($\sim 3\degr$) from a perfectly edge-on orientation, but it becomes difficult to interpret at larger deviations. Mid-plane asymmetries  are present for all bar position angles from side-on to end-on, even though the characteristic X shape is not clearly visible if the bar is end-on. Small errors in identifying the disc position angle produce very large asymmetries but these have a different distribution than those due to strong buckling and can be recognised readily by the strong quadrupolar signal in the mid-plane asymmetry map, and by the steep mid-plane asymmetry along the disc extent. Likewise, an error in identifying the correct centre is apparent in the large dipole in the mid-plane asymmetry map produced by vertical offsets, whereas small offsets in the horizontal direction do not change the mid-plane asymmetry diagnostics much. The rotation of an image using the {\sc idl} procedure {\sc rot} does not introduce significant variation in the measured mid-plane asymmetry. Moreover, reasonable variations in the image resolution also do not significantly alter the measured asymmetry. Lastly, the effect of a diffuse dust disc and dust lanes associated with the bar can produce an asymmetry with respect to the mid-plane with a single peak centred near the galaxy centre, which is very different from the asymmetry due to the buckling and which can be easily avoided by adjusting the region along the $z$ axis within which the mid-plane asymmetry profile is computed. We conclude that the mid-plane asymmetry map and profile are viable ways to test for past buckling events in real B/P galaxies, but it is crucial to select edge-on galaxies.

\section{A sample of edge-on B/P galaxies in S$^4$G}
\label{sec:s4g_galaxies}

Having demonstrated that asymmetry about the mid-plane is produced by strong buckling but not by weak buckling or resonant trapping, and that this asymmetry is long-lasting and can be detected, we present a pilot project to explore whether in practice the detection of mid-plane asymmetry is feasible in real galaxies. 
We apply our diagnostics to a sample of nearby, nearly edge-on galaxies from S$^4$G \citep{Sheth2010,MunozMateos2013,Querejeta2015}.

S$^4$G is a volume-, magnitude-, and size-limited survey of $\sim2800$ galaxies spanning a wide range in Hubble type, mass, color, size and disc inclination performed using the Infrared Array Camera (IRAC) at 3.6 and 4.5 $\micron$ mounted on the {\it Spitzer} space telescope. The galaxies have distances $D < 40$ Mpc (for $H_0 = 75$ km s$^{-1}$ Mpc$^{-1}$), blue light isophotal diameters $D_{25} > 1.0$ arcmin, blue photographic magnitudes $m_B < 15.5$ mag (corrected for internal extinction), and Galactic latitudes $b > 30\degr$. The main goal of the survey is to unveil the assembly history and evolution of galaxies \citep{Sheth2010}. The wavelengths used to record the images are reasonably free of dust emission; this allows clear identification of morphological features of the old stellar populations. 

S$^4$G images have a pixel scale of 0.75 arcsec, and angular resolution with a ${\rm FWHM}=2.1$ arcsec. They are processed through a uniform pipeline and a variety of analyses are already publicly available \citep[e.g.,][]{Querejeta2015,Salo2015}. We download the images from Pipeline 4 \citep{Salo2015}, which includes a careful sky subtraction, from the NASA/IPAC Infrared Science Archive\footnote{The S$^4$G images are available at \url{https://irsa.ipac.caltech.edu/data/SPITZER/S4G/index.html}}. 

\subsection{Sample selection}

\begin{figure*}
    \centering
    \includegraphics[scale=0.31]{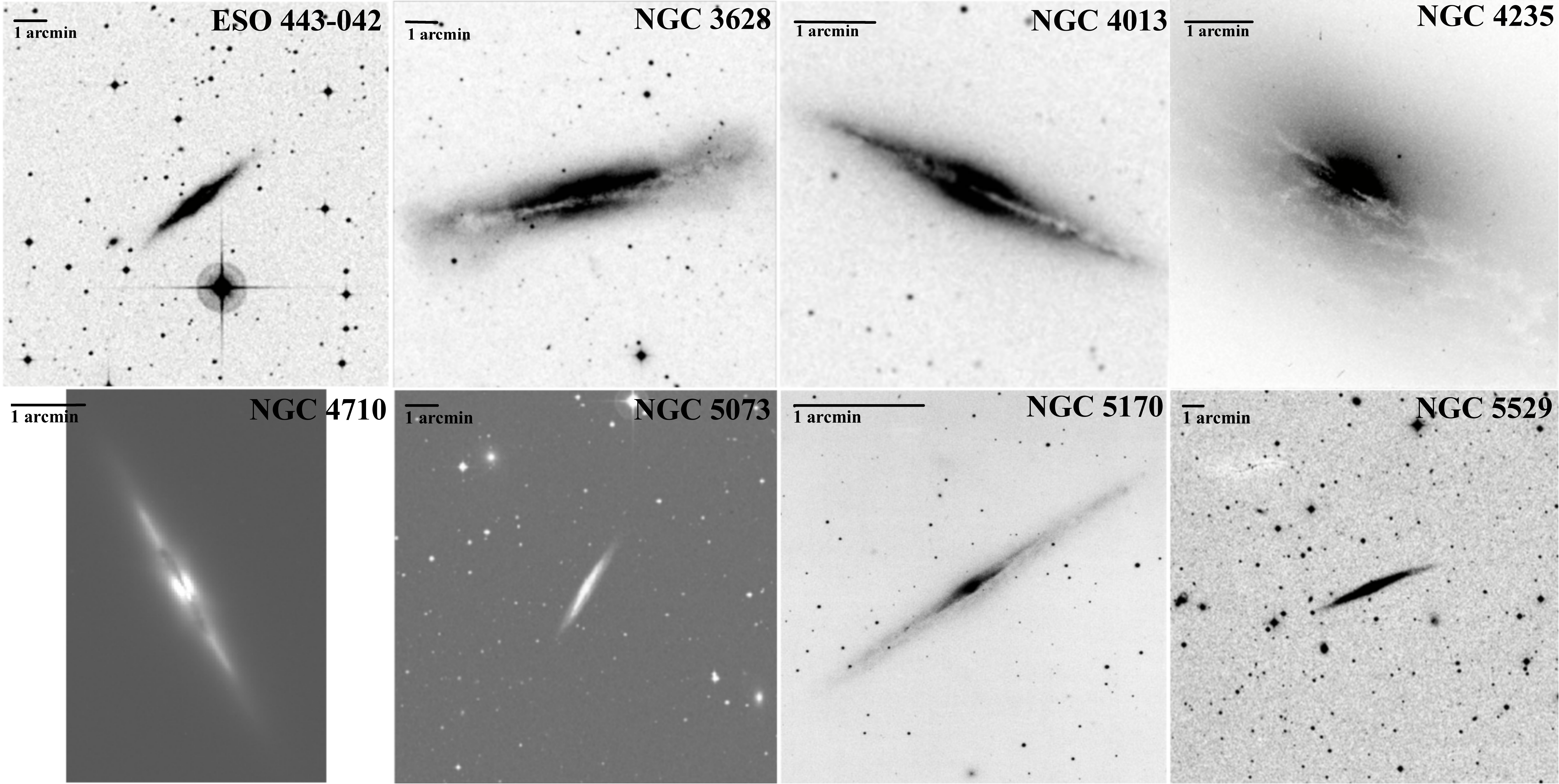}
    \caption{Optical images of the parent sample of galaxies from S$^4$G. All the images are oriented with the North at the top and the East at the left.
    Upper row (from left to right): ESO~443-042 (DSS), NGC~3628 (DSS), NGC~4013 (DSS), and NGC~4235 ({\it HST}).
    Lower row (from left to right): NGC~4710 (WIYN), NGC~5073 (DSS), NGC~5170 (DSS), and NGC~5529 (DSS).}
    \label{fig:optic}
\end{figure*}

\cite{Buta2015} performed a morphological analysis of the S$^4$G galaxies, adopting a modified version of the \cite{deVaucouleurs1959} revised Hubble-Sandage system \citep{Buta2007}, to identify and describe features such as lenses, inner and outer rings, as well as pseudorings, X-shaped structures and B/P bulges. 
They identified 60 nearly edge-on galaxies hosting an X shape. Among these, we select galaxies which appear almost edge-on based on visual inspection, where dust does not obscure large portions of the X shape and which do not present extreme asymmetric features in the disc. This results in a parent sample of eight objects, listed in Table~\ref{tab:galaxies_poperties}, for which optical images, obtained with different instruments/surveys (Digitized Sky Survey - DSS, {\it Hubble Space Telescope - HST}, WIYN telescope), are available from NASA/IPAC Extragalactic Database (NED\footnote{Available at \url{https://ned.ipac.caltech.edu/}}) and are shown in Fig.~\ref{fig:optic}. A visual inspection of the optical images of the sample reveals that the selected galaxies are almost edge-on (the disc inclination is difficult to derive from the available images). They often host dust bands along the disc which may help to visually identify the disc inclination and position angle. Extended dust lanes are visible along the disc of NGC~3628, NGC~4013, NGC~4235, and NGC~5170, while dust lanes are visible in the central part of the disc of NGC~4710. The dust lanes are well aligned with the disc major axis in the galaxy NGC~4013, whereas for the remaining objects they are not aligned, or are warped, which may indicate that the host galaxy is not seen perfectly edge-on. 
Moreover, the discs are not perfectly symmetric, but the B/P bulge (and/or a boxy structure) in the central portion of the galaxies is clearly recognisable. In particular, the outer discs of ESO~443-042, NGC~5073, and NGC~5529 are slightly warped, while the disc of NGC~3628 remains thick in its outer part.

We start the mid-plane asymmetry analysis of the S$^4$G images using the calibrated and sky-subtracted 3.6 $\micron$ images. \cite{Querejeta2015} obtained stellar mass maps for all our galaxies except NGC~5170 by separating the dominant light produced by the old stars and the dust emission using 3.6 $\micron$ images, and measuring [3.6]-[4.5] $\micron$ colours. We also download these stellar mass maps from the NASA/IPAC Infrared Science Archive.

We perform an unsharp mask analysis on the images by convolving them with a circular Gaussian varying the FWHM manually, to highlight their X shape. We construct the unsharp masked images by subtracting the convolved images from the originals, as we did for the simulations. We quantify the extension of the X-shaped structure in each case, i.e. its semi-major and semi-minor axes, analysing both the rotated images and unsharp masked ones, as done in Sec~\ref{sec:diagnostics}. We apply our mid-plane asymmetry diagnostics to the rotated images.

In Fig.~\ref{fig:sample} we present our analysis of the sample: for each galaxy we show the original S$^4$G calibrated and sky-subtracted 3.6 $\micron$ image (top row), the unsharp mask map (middle row) and the mid-plane asymmetry map (bottom row). The images are rotated to have the disc PA aligned with the $x$ axis and cut, in order to be centred on the galaxy centres. A similar unsharp mask analysis on our sample has already been performed by \cite{Laurikainen2017}, who provided an estimate of the extension of the X-shaped structure in each case, i.e. its semi-major and semi-minor axes. Our results from the unsharp mask analysis are qualitatively in agreement with \cite{Laurikainen2017}, when comparing the extension of the B/P bulges in Fig.~\ref{fig:sample} and reported in Table~\ref{tab:galaxies_poperties} and the semi-major and semi-minor axes of the X-shaped features from \cite{Laurikainen2017}.

\renewcommand{\tabcolsep}{0.18cm}
\begin{table*}
\caption{\label{tab:galaxies_poperties} Properties of the sample of galaxies from S$^4$G.}
\begin{tabular}{ccccccccc}
\hline
Galaxy & Morph. type & Distance & Disc PA & Disc $\epsilon$ & X shape & Disc PA from profile & Feature & Final sample\\
 &  & [Mpc] & [degree] & & [kpc] & [degree] & \\
(1) & (2) & (3) & (4) & (5) & (6) & (7) & (8) & (9)\\
\hline
ESO~443-042 & S$_{\bf x}$0/\underline{a} spw/E(d)8 & 47.4 & 127.7 & 0.859 & $3.5\times3.3$ & 127.7 & warped disc & yes\\
NGC~3628 & SB$_{\bf x}$(nd)bc sp/E(b)8 pec & 17.6 & 102.6 & 0.822 & $3.0\times 3.0$ & 104.0 & warped disc & no\\
NGC~4013 & SA\underline{B}$_x$a spw/E(d)7 & 15.6 & 65.1 & 0.818 & $1.0\times1.0$ & 65.1 & bright central star & yes\\
NGC~4235 & SAB:$_{\bf x}$0$^+$ sp & 38.5 & 48.5 & 0.691 & $4.5\times 4.2$ & 49.3 & warped disc & yes\\
NGC~4710 & SB$_{\bf xa}$(nd)0$^+$ sp/E(d)7 & 20.9  & 27.4 & 0.785  & $2.5\times1.7$ & 27.8 & dust + damaged image & no \\
NGC~5073 & SA\underline{B}$_{\bf xa}$0/a sp & 45.2 & 149.8 & 0.814  & $3.0\times 2.8$ & 148.8 & lopsided + warped disc & no\\
NGC~5170 & (R’)SAB$_{\bf x}$ (r\underline{l})0/a sp & 26.6 & 126.3 & 0.853 & $2.5\times1.7$ & 126.3 & -- & yes\\
NGC~5529 & SB$_{\bf x}$ab spw & 45.1 & 115.0 & 0.850  & $2.5\times3.3$ & 115.0 & warped disc & no\\
\hline
\end{tabular}
\begin{tablenotes}
      \item \textbf{Notes.} (1) Galaxy name. (2) Morphological type from \cite{Buta2015}. (3) Distance from NED, as obtained from the radial velocity with respect to the cosmic microwave background reference frame. (4) Disc position angle from \cite{Salo2015}. (5) Disc ellipticity from \cite{Salo2015}. (6) Semi-major (along $z$) and semi-minor (along $x$) axes of the X-shaped structure. (7) Disc position angle from the comparison of the mean vertical profiles of the surface brightness. (8) Features identified by inspecting the galaxy image. (9) Inclusion in the final sample of galaxies. 
\end{tablenotes}
\end{table*}

\begin{figure*}
    \centering
    \includegraphics[scale=0.6]{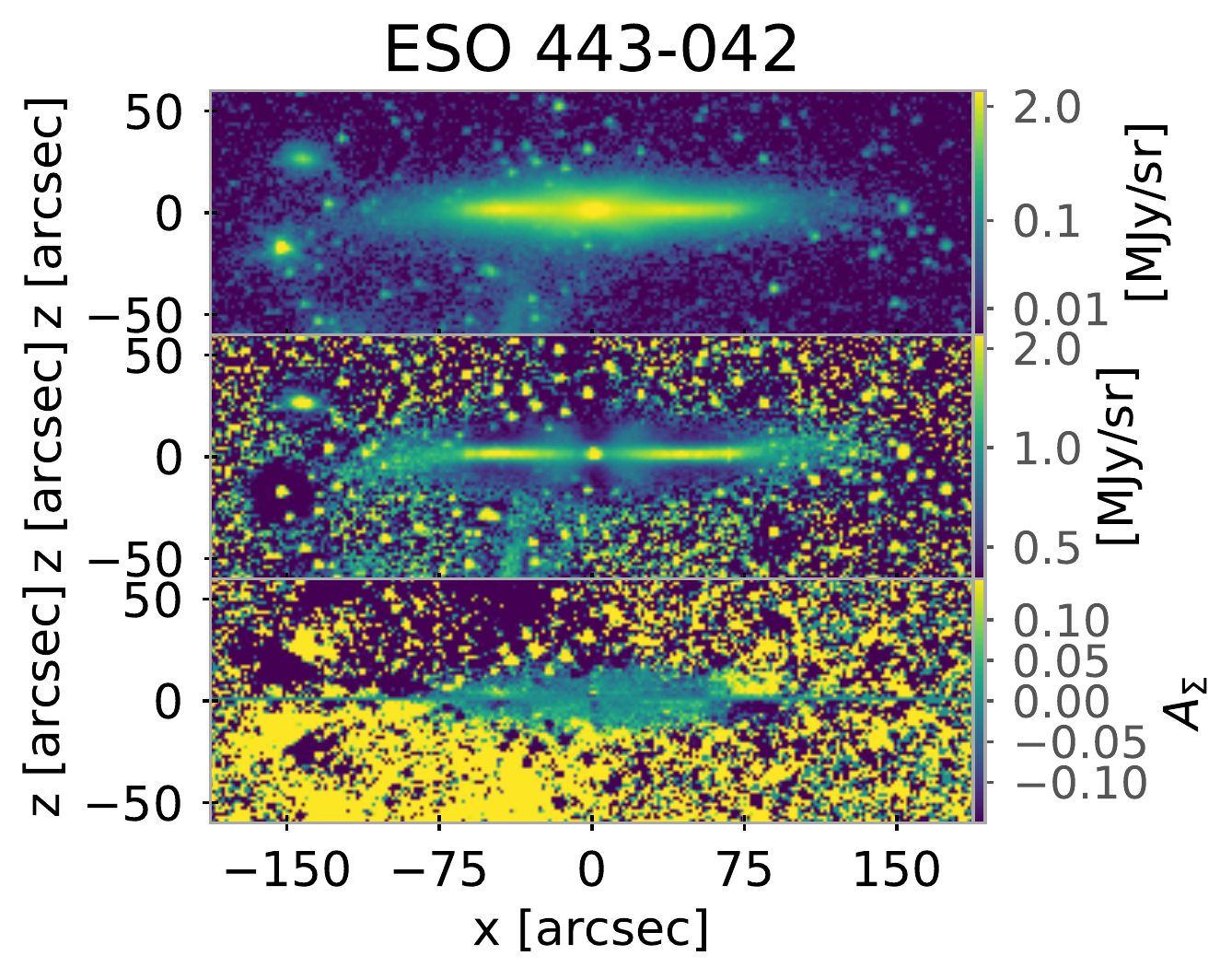}
    \includegraphics[scale=0.6]{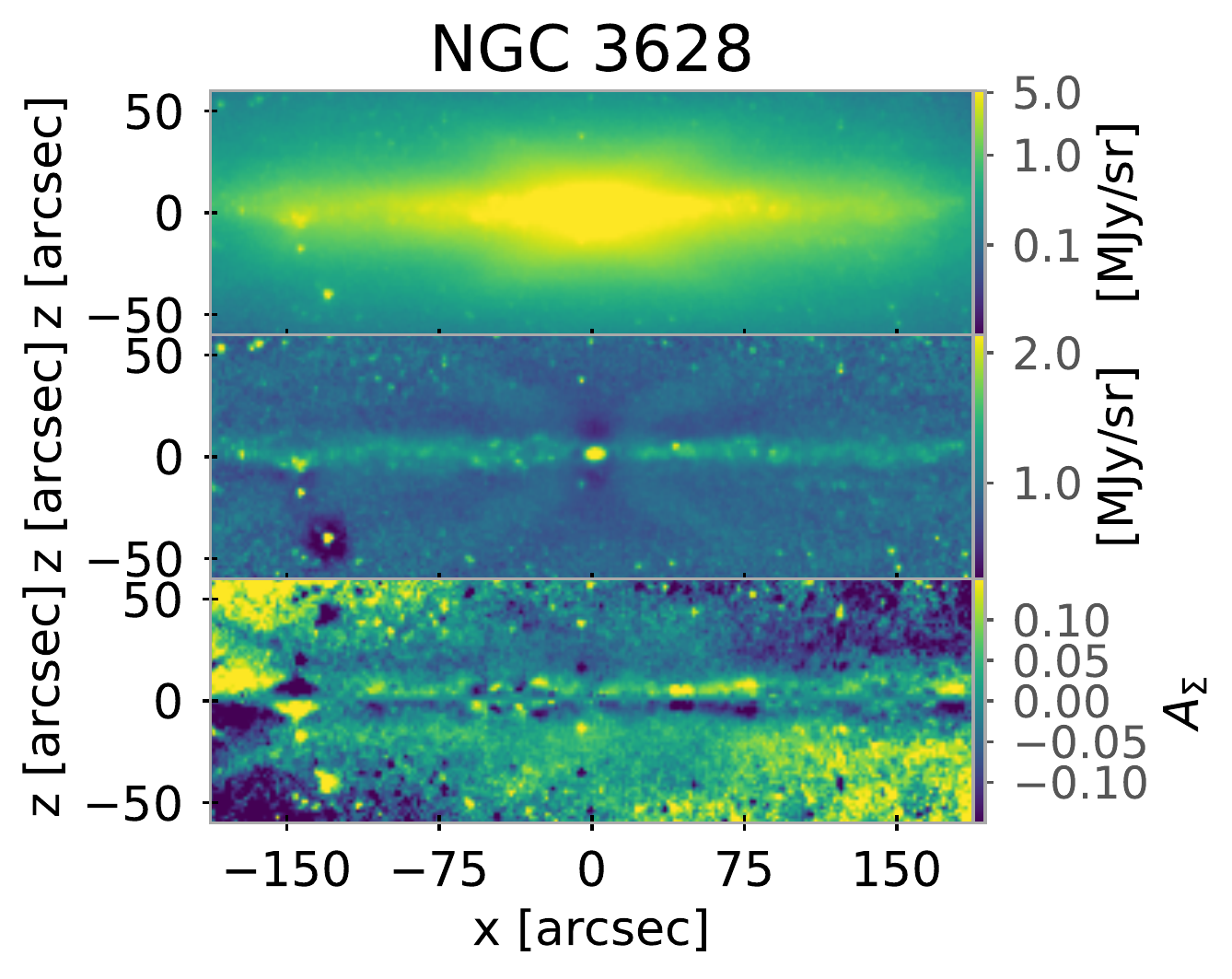}
    \includegraphics[scale=0.6]{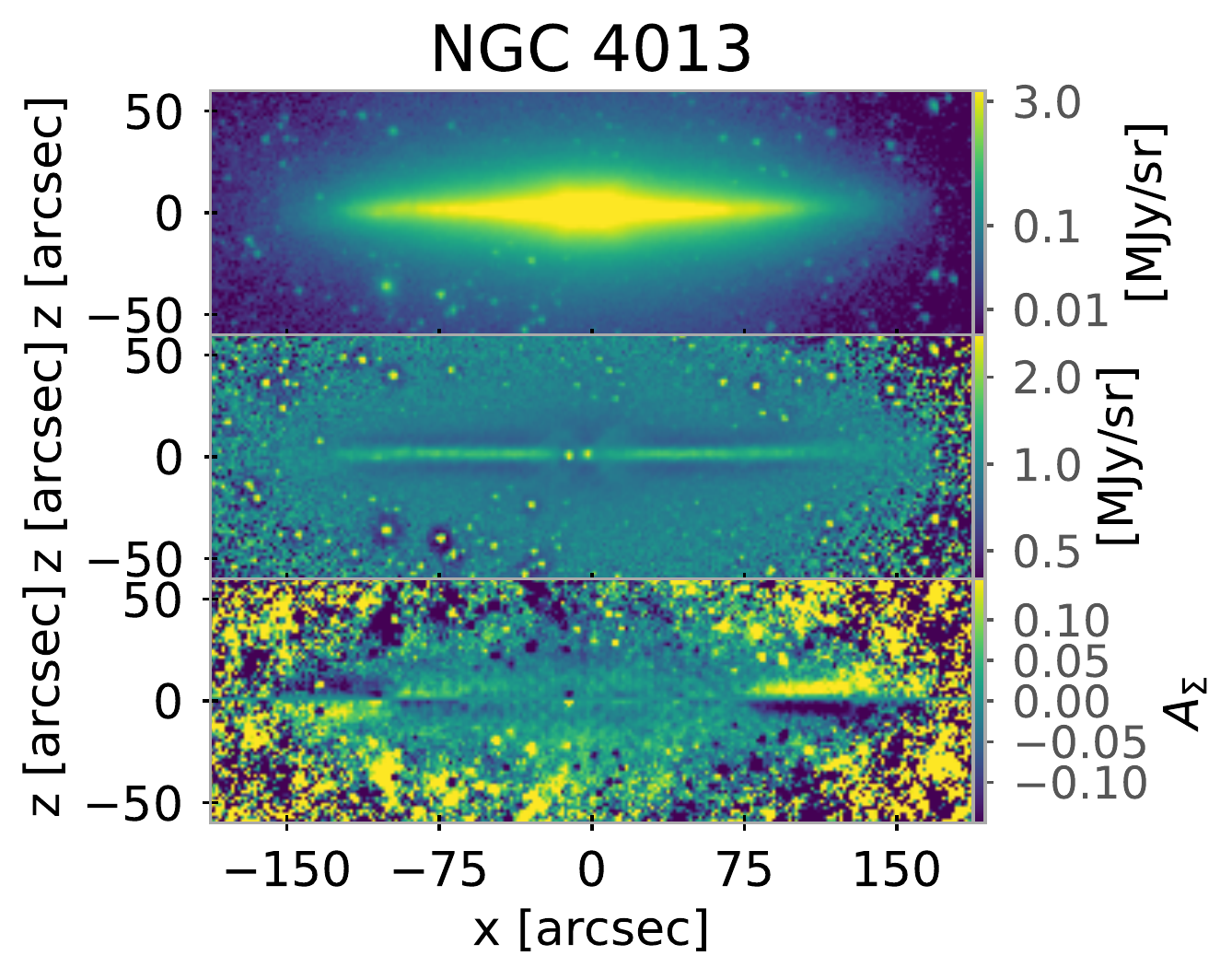}
    \includegraphics[scale=0.6]{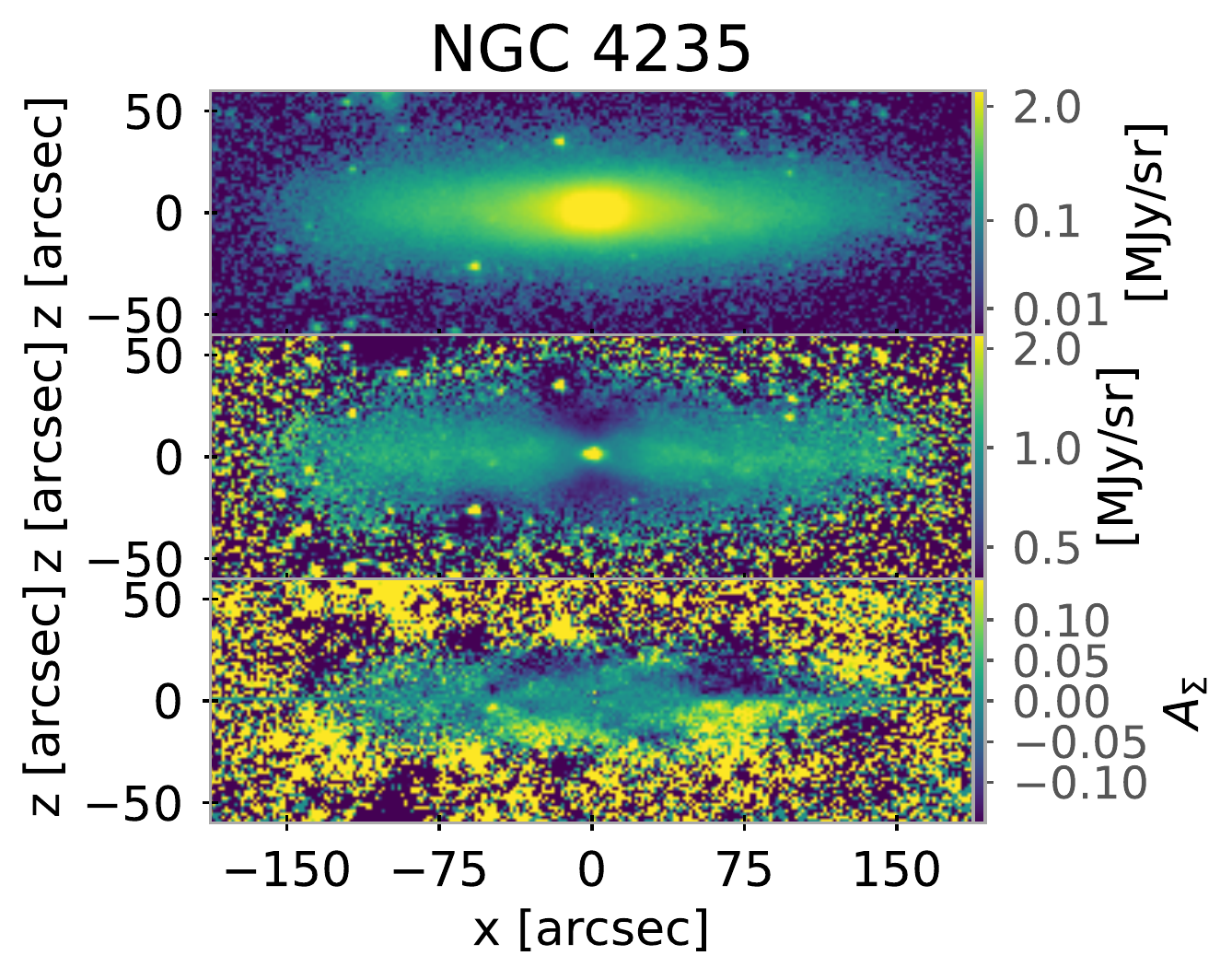}
\caption{Map of the surface brightness of the 3.6-$\micron$ image (top panel), its unsharp masked version (middle panel), and mid-plane asymmetry map (bottom panel) of the parent sample of galaxies from S$^4$G. Each image is rotated and centred to have the disc major axis aligned with the $x$ axis.}
    \label{fig:sample}
\end{figure*}

\begin{figure*}
	\ContinuedFloat
    \includegraphics[scale=0.6]{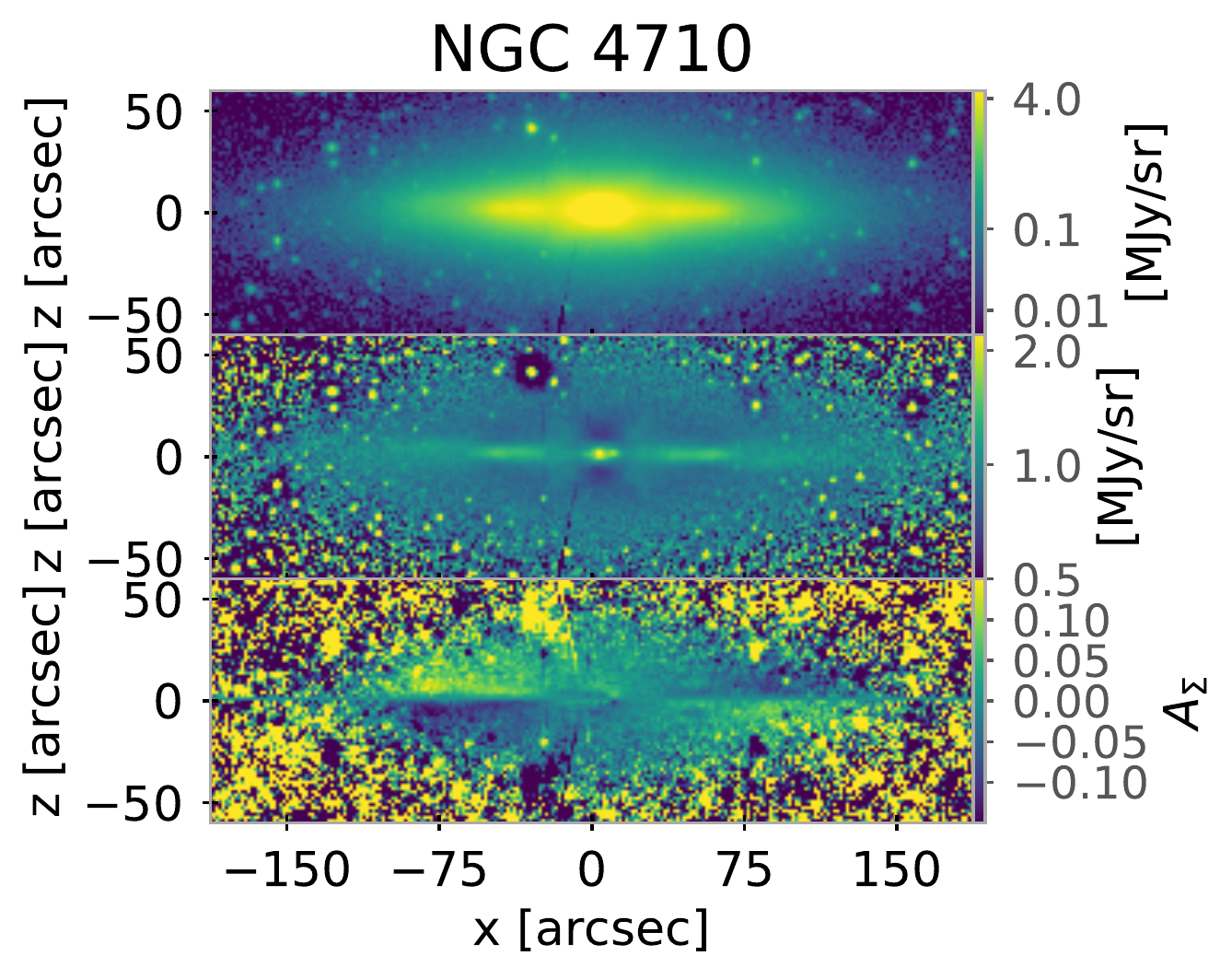}
    \includegraphics[scale=0.6]{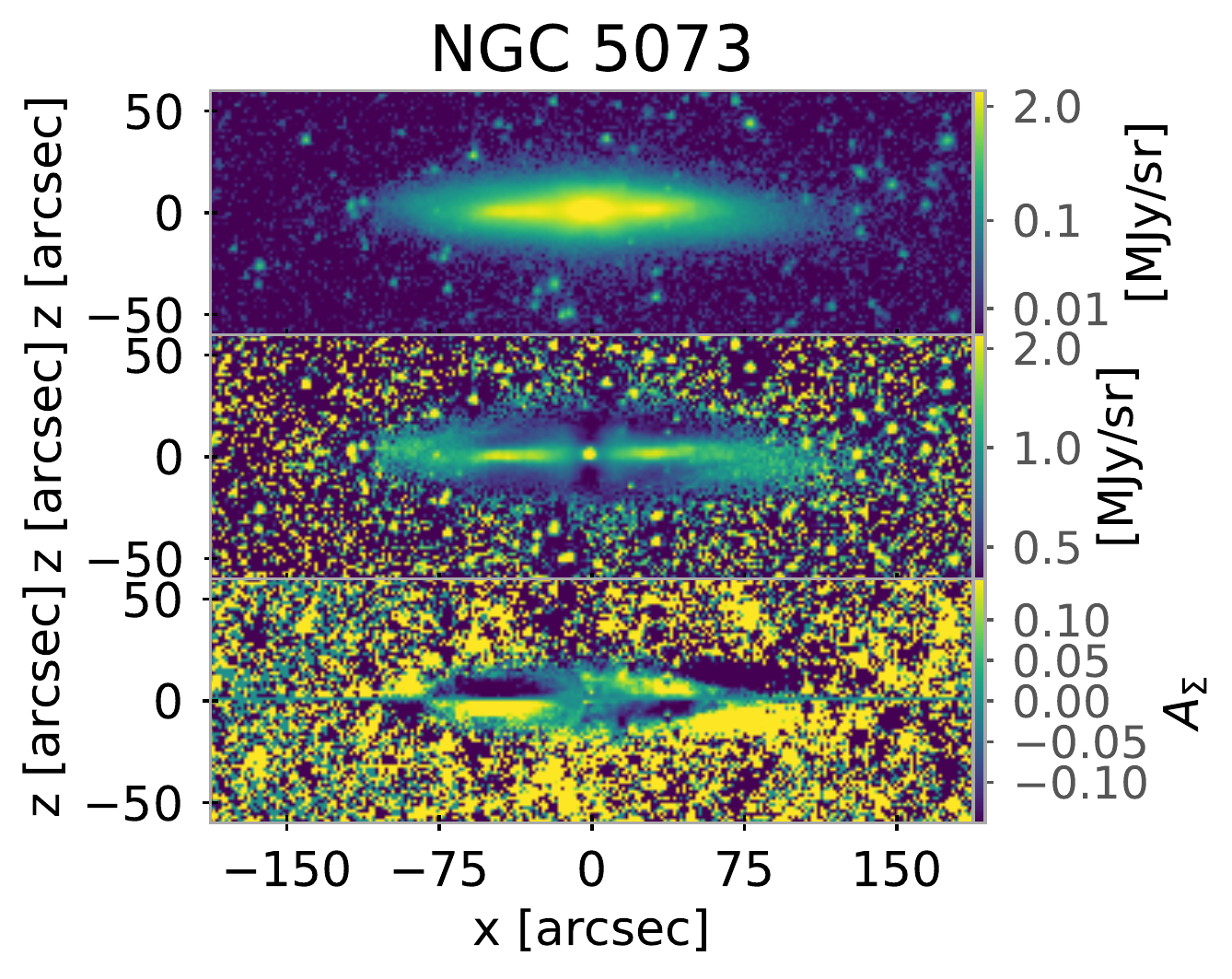}
    \includegraphics[scale=0.6]{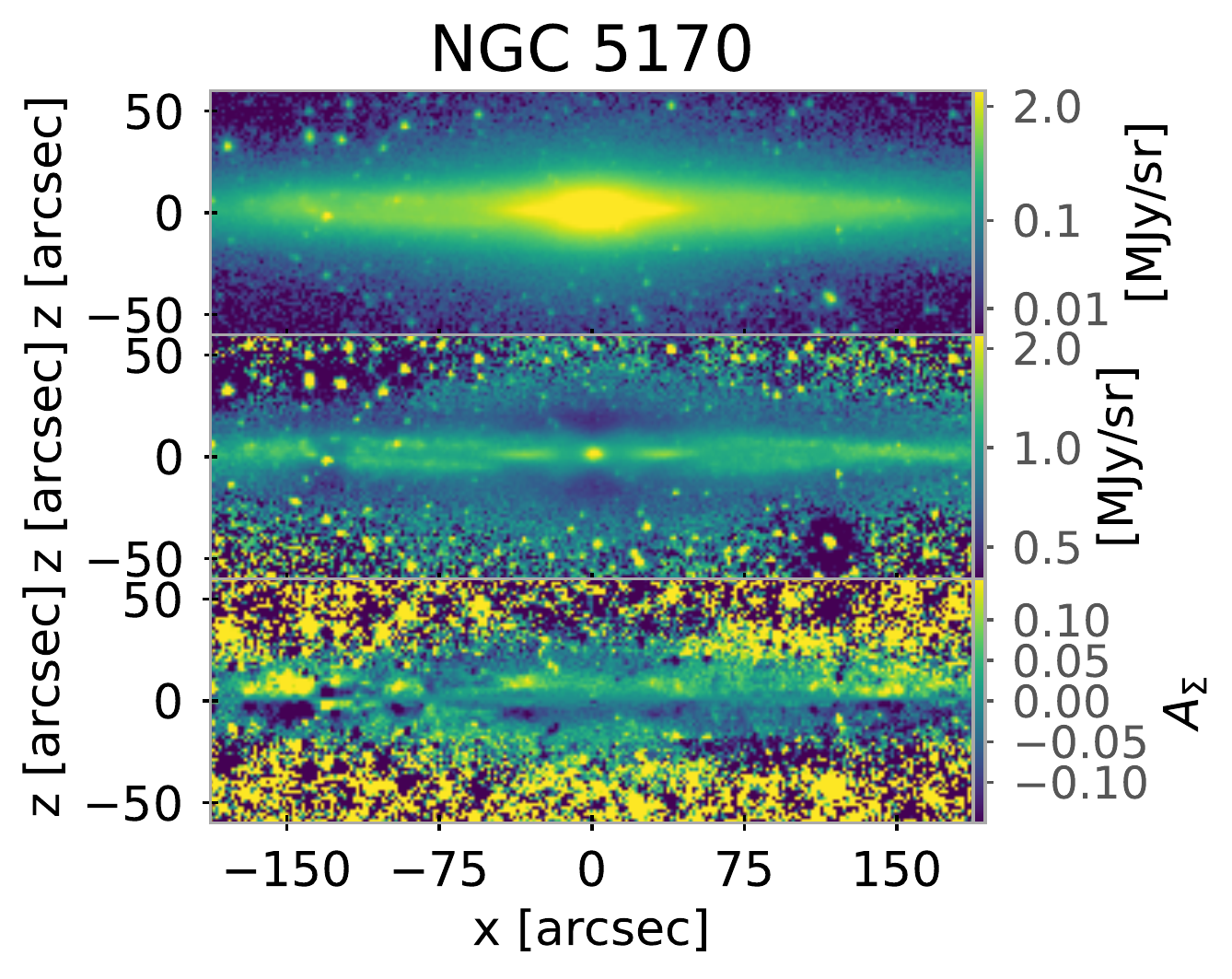}
    \includegraphics[scale=0.6]{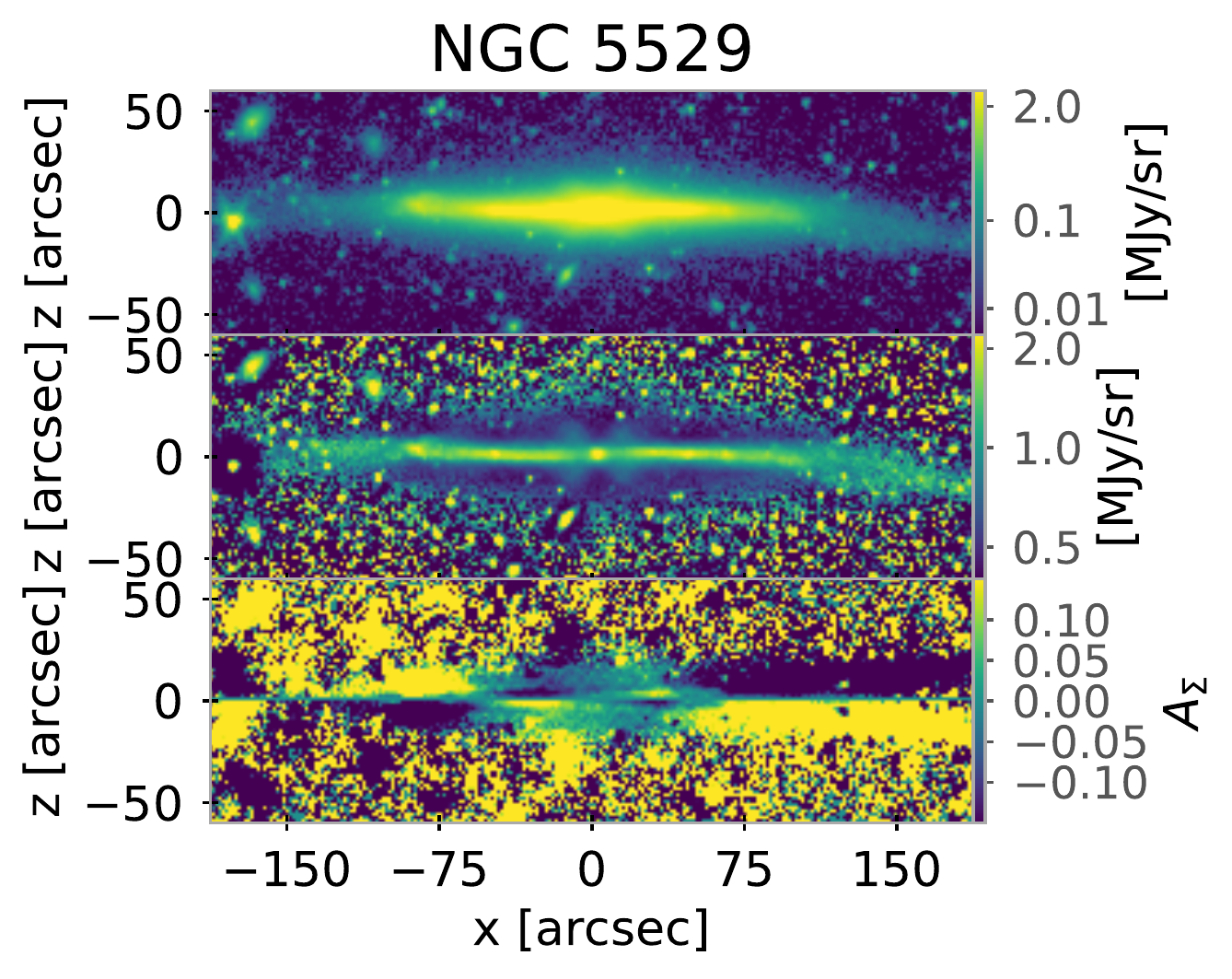}
    \caption{(continued).}
\end{figure*}    
    

\subsection{Galaxy parameters}
\label{sec:gala_param}

As discussed in Sect.~\ref{sec:prove}, the right identification of the galaxy centre and disc PA is necessary to build an effective mid-plane asymmetry map.
\cite{Salo2015} provided the galaxy parameters for our sample of galaxies: the centres of the galaxies were calculated as the position where the surface brightness gradient is zero, and the disc PAs were found from the orientation of the outer isophotes fitted with ellipses. The galaxy parameters from \cite{Salo2015} are reported in Table~\ref{tab:galaxies_poperties}. 

In building the mid-plane asymmetry maps for our sample of galaxies using these parameters, it becomes clear that they are not sufficiently accurate for our purposes. Indeed, a dipolar signal (with typical size of 4-6 pixels) appears in the central part of many of the mid-plane asymmetry maps, which can be explained by a small error in the centre, as shown in Sec.~\ref{sec:centre}. Likewise, using the \cite{Salo2015} parameters leads to a mis-identification of the disc PA, resulting in a strong quadrupolar signal in the mid-plane asymmetry maps, similar to those discussed in Sect.~\ref{sec:discPA}. A more detailed analysis to identify the correct galaxy centre and disc PA is therefore necessary.  

\subsubsection{Galaxy centre}

To recover the correct galaxy centre, we perform the Fourier analysis described in Sec.~\ref{sec:centre}. Specifically, the centre is allowed to vary by 0.5 pixel by up to 10 pixels (along both the $x$ and $y$ axes) and the disc PA by $0.2\degr$ within $\pm2\degr$ of the values of \cite{Salo2015}. We vary the extension of the central region $R_{\rm max}$ to perform the Fourier analysis (with $R_{\rm max}$ extending from $\sim50\%$ to $\sim125\%$ of the extension of the dipolar structure) and test that the identified centre is stable. 

As a further check that our measurements of the galaxy centres are reliable, we calculate the asymmetry of the central part of the images \citep{Conselice2000}
\begin{equation}
 A=\frac{|I-R|}{I}
\end{equation}
where $I$ is the portion of the original image considered for the calculation and $R$ is $I$ rotated  $180\degr$. We vary the centre and the radial extension along which to perform the analysis and identify the correct centre as the one which minimises $A$. We successfully confirm the centres identified with the Fourier analysis for most of the galaxy sample using this test. This confirmation of centres is not the case for NGC~4710, NGC~5073, NGC~5529 (which are discarded from the final analysis in the following) and NGC~4013, hosting a bright star nearby the centre, which makes it more difficult to apply our methods.

\subsubsection{Disc position angle}

When the correct disc PA is used to build the mid-plane asymmetry map, no quadrupolar signal is present in the disc region and the corresponding mid-plane asymmetry profile is not sloped, as shown in Sec.~\ref{sec:prove}. 
The initial guess of disc PA used in this work is based on an isophotal analysis of the outer region of the galaxies: a constant value of PA measured in the external radial range is assumed to be characteristic of the disc \citep{Salo2015}. Real galaxies often have warped discs and are also sometimes lopsided. This warping complicates the identification of the correct disc PA and the measurement of the mid-plane asymmetry diagnostic.

To refine the measurement of the disc PA, we use the approach described in Sec.~\ref{sec:discPA}. In particular, we allow the disc PA to vary by $0.2\degr$ within $2\degr$ starting from the \citet{Salo2015} values, and compare the resulting profiles, after excluding the portion of the image hosting the galaxy halo. We also vary the extension of the two symmetric portions of the image, to test that the profiles remain superimposed when the disc PA is correctly identified.

To double check our estimate of the disc PA, we perform a Fourier analysis on the mid-plane asymmetry map using the radial extension of the disc and assuming the galaxy centre previously identified. We allow the disc PA to vary. We consider the right disc PA as the one that minimises the $m=2$ component. We test if the identified disc PA remains constant when varying the radial extension of the portion of the image used for the Fourier analysis by 10-20 pixels and if it corresponds to the disc PA identified with the first approach. The disc PAs recovered from these two methods do not always coincide since sometimes the Fourier analysis is unable to identify a constant PA or it does not correspond to the previous estimate. This discrepancy may be due to the fact that we use a large portion of the image to perform the Fourier analysis here, since we need to include the extension of the disc, so we are inevitably including a portion of sky in the same analysis, where spurious features are located which can influence the value of the $m=2$ component. Unfortunately this is unavoidable, but we check that in the cases where the disc PA from the Fourier analysis remains constant, that it corresponds to the one obtained with the first approach. In Table~\ref{tab:galaxies_poperties} we report the value of the disc PA from \cite{Salo2015} used as a first guess and the values obtained from our analysis.

\section{Measurements of B/P asymmetries in the S$^4$G sample}
\label{sec:evidences_real}

Here we present the results of the B/P mid-plane asymmetry measurements for our parent sample of eight galaxies. 
In Fig.~\ref{fig:sample} (bottom panels) we show the resulting mid-plane asymmetry maps. 

For NGC~3628, NGC~4710, NGC~5073, and NGC~5529 the mid-plane asymmetry maps are strongly affected by a quadrupolar signal (which affects at least half the extent of the galaxy itself). This signal implies that we were unable to correctly identify the disc PA, despite our careful analysis. 
Visual inspection of the S$^4$G images of these galaxies and corresponding unsharp masked images (Fig.~\ref{fig:sample}, middle panels) identifies strong and intrinsically asymmetric features in the discs (such as warps and/or lopsided discs) which may have affected the measurement of the disc PA. 

In particular, NGC~3628 hosts a warped disc. The distortion is visible in the image at $|x|\gtrsim100$ arcsec, while a strong quadrupolar signal appears at $|x|\gtrsim75$ arcsec. NGC~4710 hosts a double peak near the centre, which may produce the small dipolar signal in the central region of the mid-plane asymmetry map. This feature is associated with an asymmetric dust lane, on each side of the centre, as suggested by \cite{Gonzalez2016}. The outer regions of the disc appear more regular, despite a weak quadrupolar signal visible at $|x|\sim75$ arcsec. Moreover, the image of NGC~4710 suffers from columns of pixels with damaged/missing signal, affecting its entire left side. Masking this region results in not being able to build effective mid-plane asymmetry diagnostics. NGC~5073 is warped at $|x|\gtrsim75$ arcsec, and lopsided, being more extended on the left side. A resulting strong quadrupolar signal appears both in the central and in the outer regions of the disc, with an opposite sign, in the mid-plane asymmetry map. Finally, NGC~5529 also hosts a clearly warped disc, which produces a strong quadrupolar signal along both the central and the outer regions of the disc, with an opposite sign in the mid-plane asymmetry map.
In Appendix~\ref{appendix:b} we present the mid-plane asymmetry maps and corresponding mid-plane asymmetry profiles for these galaxies. 

For the remaining four galaxies, which build our final sample, ESO~443-042, NGC~4013, NGC~4235, and NGC~5170, we are able to identify the correct mid-plane. Their mid-plane asymmetry maps are unaffected by strong dipoles or quadrupoles. Nonetheless, weak artefacts can be seen: for example NGC~4013 has bright foreground stars near the centre which produce a bipolar signal in the mid-plane asymmetry map, while small outer portions of the discs in ESO~443-042, NGC~4013, and NGC~4235 are affected by a weak quadrupolar signal (i.e., with a limited extension with respect to the galaxy and/or affecting only the very outer part), which can be due to local asymmetries in the disc. In particular, the discs of ESO~443-042 and NGC~4235 are weakly warped. Nonetheless, these features are very localised and far out in the disc, so the inner disc PA can be identified, while their effects do not compromise our subsequent analysis.  
Table~\ref{tab:galaxies_poperties} lists the features identified from visual inspection of the images and whether the galaxies are included or not in our final sample. 

\subsection{Galaxies with a mid-plane asymmetry}
\label{sec:asy}

We identify a mild mid-plane asymmetry in NGC~4235 and NGC~5170 similar to what we found in model D5. 

\noindent {\bf NGC~4235} is a highly-inclined S0 galaxy, hosting a prominent X-shaped structure (visible in the unsharp mask in Fig.~\ref{fig:sample}) associated with a B/P bulge \citep{Buta2015}. The X-shaped structure has a semi-major axis of 4.5 kpc and a semi-minor axis of 4.2 kpc.  The disc is weakly warped in the very outer region resulting in a weak quadrupolar signal at $|x|\sim 130$ arcsec.

\begin{figure*}
    \centering
    \includegraphics[scale=0.43]{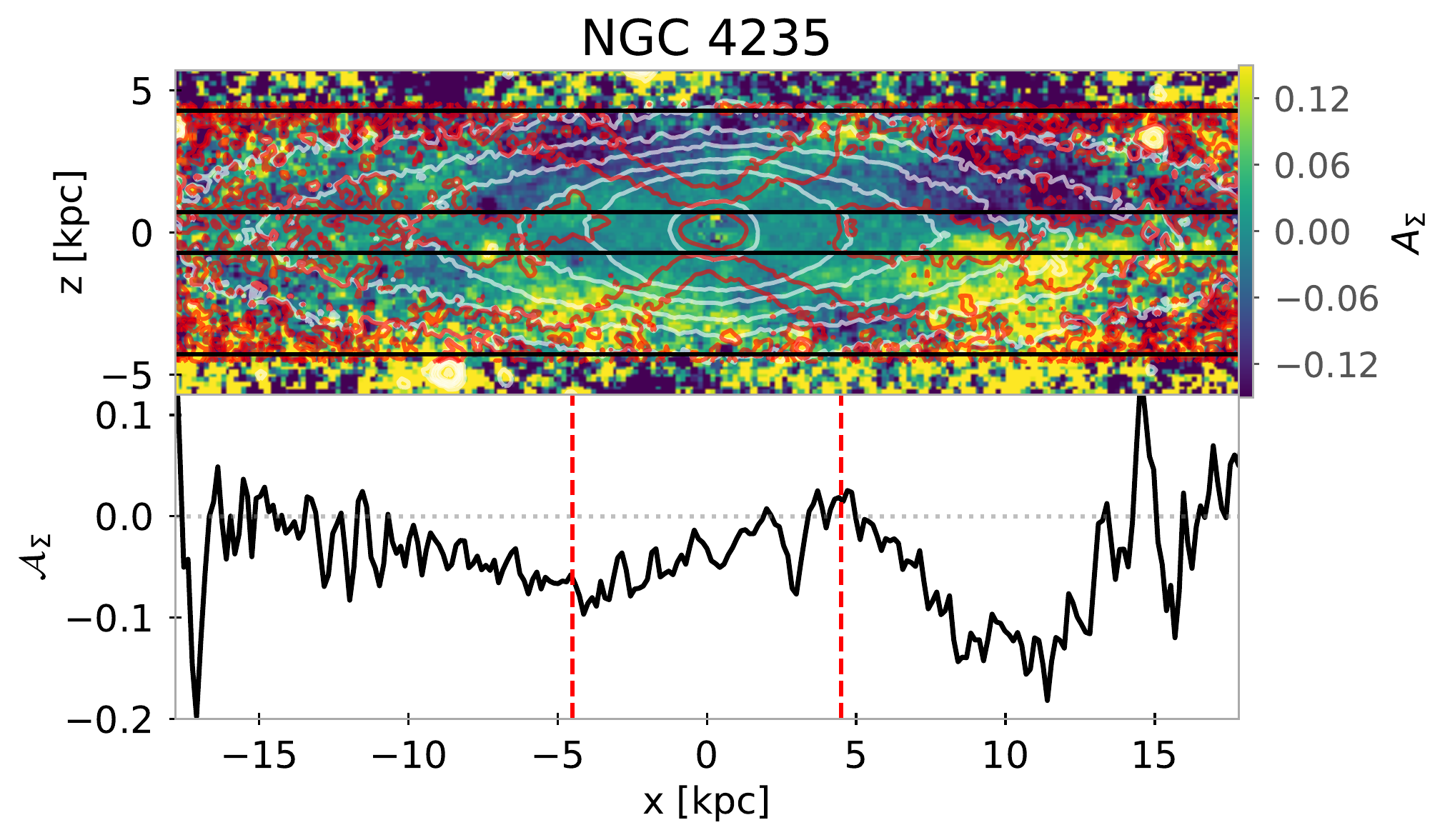}
    \includegraphics[scale=0.43]{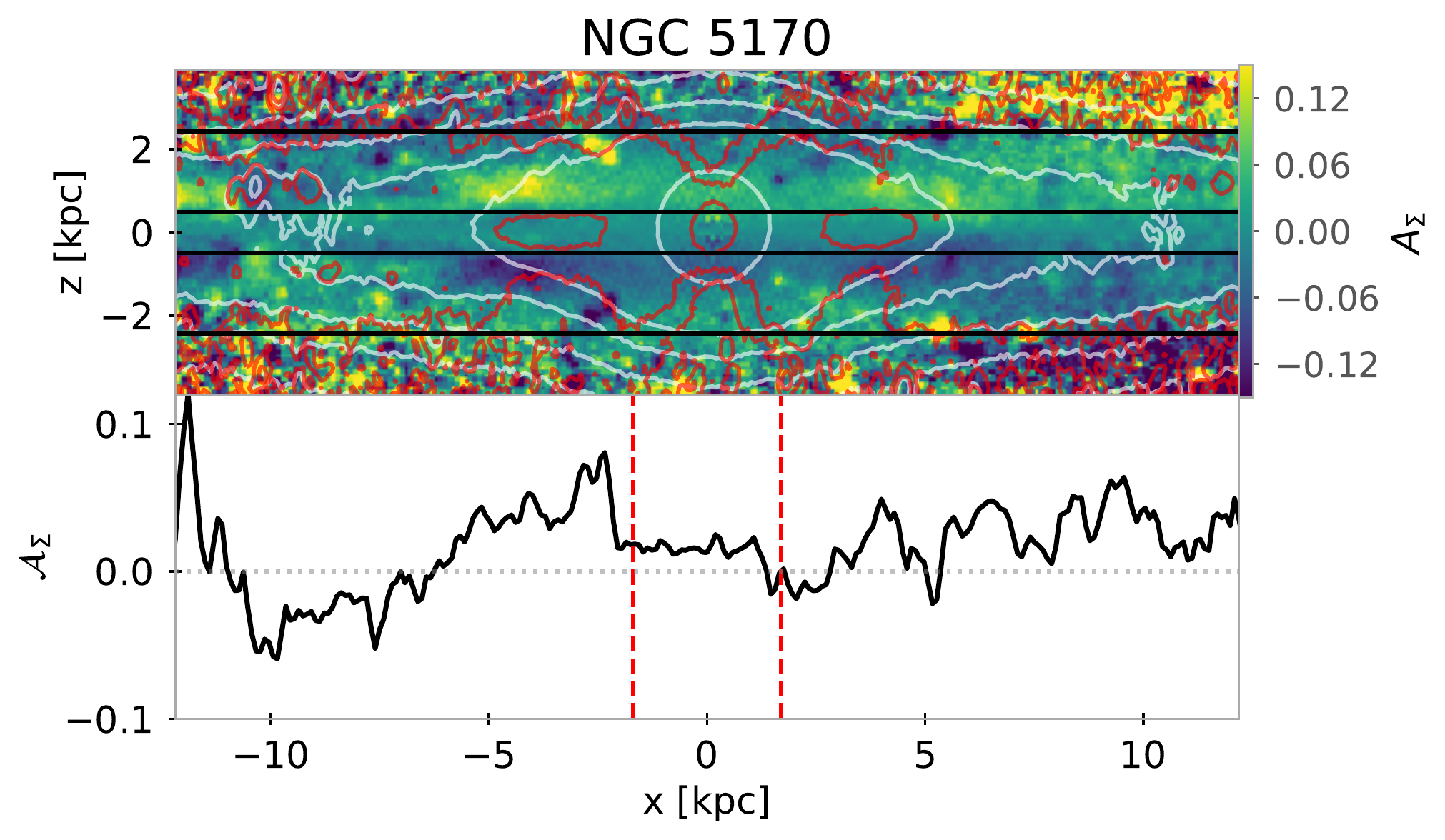}
    \includegraphics[scale=0.43]{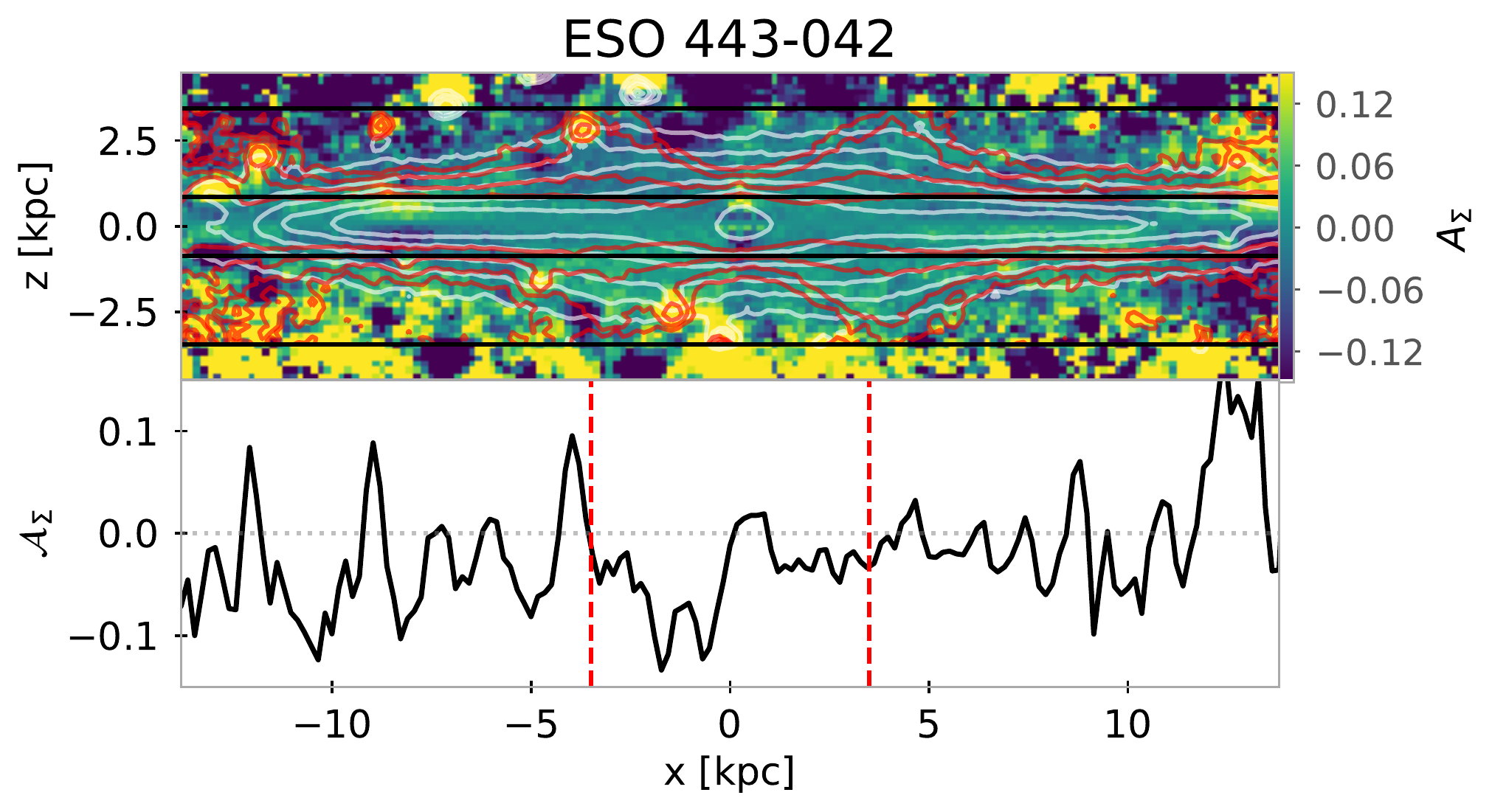}
    \includegraphics[scale=0.43]{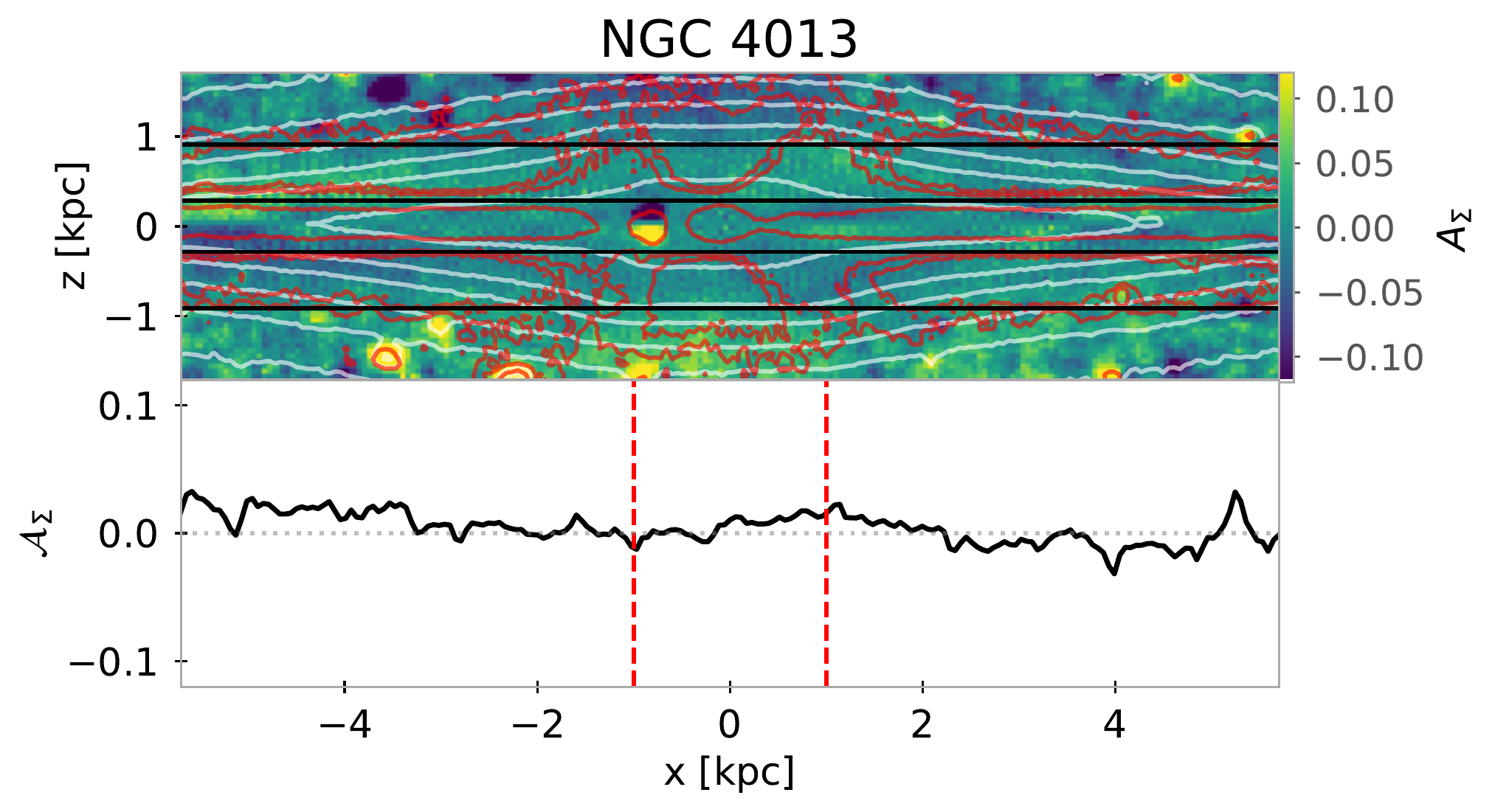}
    \caption{Mid-plane asymmetry diagnostics for our final sample. Top-left panel: mid-plane asymmetry map (upper panel) of the 3.6-$\micron$ image of NGC 4235 with some isocontours of the surface brightness (white lines) and its unsharp mask (red lines) maps. The horizontal solid black lines mark the vertical extension of the regions (at $0.7<z<4.2$ kpc) adopted to extract the mid-plane asymmetry radial profile (lower panel). Top-left panel: same as in the top-left panel, but for NGC~5170 with $0.5<z<2.5$ kpc. Bottom-left panel: same as in the top-left panel, but for ESO~443-042 with $0.9< z<3.3$ kpc. Bottom-right panel: same as in the top-left panel, but for NGC~4013 with $0.3<z<1.0$ kpc.}
    \label{fig:AMP_real_galaxies}
\end{figure*}

The mid-plane asymmetry diagnostics are shown in Fig.~\ref{fig:sample} and Fig.~\ref{fig:AMP_real_galaxies}, top-left panel. We superimpose on the mid-plane asymmetry map the contours of the surface brightness of the galaxy (in white) and the contours of the unsharp mask (in red), which highlight the X shape. The red contours themselves appear asymmetric with respect to the mid-plane. The mid-plane asymmetry map reveals asymmetries at the edges of the X shape and further out, which extend till $25 \lesssim \lvert x\rvert \lesssim 75$ arcsec before they disappear further out. However, these asymmetries are not symmetric with respect to the $z$ axis, being stronger and extending further on the right side compared to the left (the peaks are located at $x\sim-5$ kpc and at $x\sim10$ kpc). We calculate the mid-plane asymmetry profile adopting $z_{\rm min}=0.7$ kpc and  $z_{\rm max}=4.2$ kpc: two peaks are present in the mid-plane asymmetry profile, reaching values of $\mathcal{A}_\Sigma (x)\sim0.1$ on the left side and $\sim0.15$ on the right side, before decreasing to $\mathcal{A}_\Sigma (x)\sim0$ further out in the disc. The mid-plane asymmetry profile resembles those observed at $\sim3-4$ Gyr after the buckling event for models D5 and D8 (see Figs.~\ref{fig:evol} and \ref{fig:d8_evol}, respectively), or the one when the bar is not seen perfectly side-on for model D8 (Fig.~\ref{fig:d8_bar}). 

\noindent {\bf NGC~5170} is a S0/a galaxy, hosting an inner ring-lens and a near-outer ring made of spiral arms. A prominent X shape, associated with its B/P bulge, is clearly visible from the unsharp mask in Fig.~\ref{fig:sample}, together with evidence of ansae just beyond it \citep{Buta2015}. The B/P bulge has a small extension compared to the host galaxy. 

The mid-plane asymmetry diagnostics are shown in Fig.~\ref{fig:sample} and Fig.~\ref{fig:AMP_real_galaxies}, top-right panel. The red contours appear asymmetric with respect to the mid-plane, extending to $\lvert x\rvert \lesssim 3.3$ kpc. Further out, a mild mid-plane asymmetry in the surface brightness is visible, extending to $|x|\lesssim 6.5$ kpc. This asymmetry disappears in the disc region on the left side at $x<-6.5$ kpc, while it remains weak and extends further out on the left side at $x>6.5$ kpc. The corresponding mid-plane asymmetry profile is calculated adopting $z_{\rm min}=0.5$ kpc and $z_{\rm max}=2.5$ kpc and it shows a clear peak reaching $\mathcal{A}_\Sigma (x)\sim0.04$, at $x\sim-4$ kpc, which however does not present a complementary one at $x\sim4$ kpc. 

\subsection{Mid-plane symmetric galaxies}
\label{sec:no_asy}

We do not find any evidence of mid-plane asymmetry in ESO~443-042 and NGC~4013, similar to the results for model HG1. 

\noindent {\bf ESO~443-042} is a highly-inclined S0/a galaxy, hosting an X-shaped structure (visible in the unsharp mask in Fig.~\ref{fig:sample}) associated with a B/P bulge \citep{Buta2015}. The disc shows evidence of warping. The corresponding distortion is located in the very outer region and is responsible for the weak quadrupolar signal visible at $|x|\gtrsim 17.3$ kpc in the mid-plane asymmetry map. 

The mid-plane asymmetry diagnostics are shown in Fig.~\ref{fig:sample} and Fig.~\ref{fig:AMP_real_galaxies}, bottom-left panel. The red contours in the mid-plane asymmetry map highlight the X-shaped structure, extending to $\lvert x\rvert \lesssim 7$ kpc. No asymmetries in the surface brightness are evident. The corresponding mid-plane asymmetry profile is calculated adopting $z_{\rm min}=0.9$ kpc and $z_{\rm max}=3.3$ kpc; it remains flat over the entire range (i.e., $\lvert x\rvert < 13.8$ kpc). 

\noindent {\bf NGC~4013} is a highly-inclined Sa galaxy, hosting an X-shaped structure (visible in the unsharp mask in Fig.~\ref{fig:sample}) associated with a B/P bulge, of small size compared to the galaxy \citep{Buta2015}. The disc shows evidence of weak warping, at $|x|\gtrsim 6.0$ kpc, which produces a weak quadrupolar signal visible in the outer part of the mid-plane asymmetry map. Moreover, there is a bright star near the centre of the galaxy, which causes a clear dipolar signal in the central region of the asymmetry image. 

The mid-plane asymmetry diagnostics are shown in Fig.~\ref{fig:sample} and Fig.~\ref{fig:AMP_real_galaxies}, bottom-right panel. The X-shaped feature extends to $\lvert x\rvert \lesssim 1.5$~kpc. A clear dipolar signal is visible at $x\sim -0.8$ kpc, due to presence of the bright star near the galaxy's centre. No asymmetries in the surface brightness are visible. We adopt $z_{\rm min}=0.3$ kpc and $z_{\rm max}=1.0$ kpc to calculate the mid-plane asymmetry profile, which remains flat in the entire range (i.e. $\lvert x\rvert < 5.7$~kpc). 

\subsection{Mid-plane asymmetries from the stellar mass maps}

We repeat our analysis and build the mid-plane asymmetry diagnostics using the old stellar population mass maps from \cite{Querejeta2015}. The authors provide the processed 3.6 $\micron$ images from S$^4$G, after identifying and subtracting the dust emission, which reliably trace the old stellar flux. The processed images are available for our final sample of galaxies, except for NGC~5170: we find comparable results to the ones obtained using the 3.6 $\micron$ images for NGC~4013 and NGC~4235, for which we present the resulting diagnostics in Appendix~\ref{appendix:c}. An exception is ESO~442-043, where the corresponding stellar mass map is distorted, and the X shape is barely distinguishable. The dust mainly affects the region along the disc and the mid-plane asymmetry maps built using the stellar maps are strongly affected by dust near the mid-plane. 

Since we obtain compatible results when using 3.6 $\micron$ images and the stellar mass maps except for the portion of the image very close to the mid-plane, and given these regions are excluded by definition from the calculation of our diagnostic (see Eq.~\ref{eqn:asym}), we conclude that the dust effect is reasonably well controlled using this method.

\section{Discussion}
\label{sec:discussion}

\subsection{The time interval to detect evidence of a buckling event}

The fraction of barred galaxies with $M_* > 10^{10}M_\odot$ hosting B/P bulges increases from $\sim10\%$ at $z\sim1$ to $\sim70\%$ at $z=0$ \citep{Kruk2019}. This means most B/P bulges formed during the last $\sim7$ Gyr. We have explored the time evolution of the mid-plane asymmetry produced by buckling in simulations, finding that it can be detected up to 4-5 Gyr after the event, which is a sufficiently long time to observe lingering asymmetries if a significant fraction of B/P bulges formed by buckling \citep[e.g.][]{Anderson2022}. 
On the other hand, most of the models discussed here (including model D5), are pure $N$-body simulations, which evolve rapidly, because of the absence of gas. Therefore, the duration of the detectability of mid-plane asymmetries we find in these models is probably a lower limit for real galaxies. 

We conclude that asymmetries produced by buckling should survive long enough to be measured for a sufficiently long time interval. It would be fruitful to explore asymmetries at $z\sim0.5-0.7$, when the occurrence of buckling events is expected to peak \citep{Erwin2016,Xiang2021}. 

\subsection{Different mechanisms forming a B/P bulge}

We have presented models which develop a B/P bulge via strong buckling, resulting in a mid-plane asymmetry. Instead, models which form B/P bulges either via weak (but recurrent) buckling or by resonant capture \citep{Quillen2002, Sellwood2020} (such as models HG1 or SD1 in Appendix~\ref{app:sim}) give rise to mid-plane symmetry. At the same time, the mid-plane asymmetries  produced by strong buckling weaken over time intervals varying from 1.5 Gyr to 5 Gyr.
As a bar grows after the buckling event, it transfers angular momentum from the disc to the halo, while stars are trapped into resonance  \citep{Athanassoula2003, Ceverino2007}. 
Therefore it is not possible to conclude whether the observed mid-plane symmetry is the result of the B/P bulge forming via weak buckling, or via resonant capture, or is merely the decayed remnant of a mid-plane asymmetry from strong buckling.

Simulations find that the buckling instability may happen more than once in the life of a galaxy \citep{Martinez-Valpuesta2006, Lokas2019}. Due to the secular transport of angular momentum between the bar and the halo, after buckling the bar can grow and become unstable again to further buckling. The secondary buckling event lasts longer, so the bar bending could be easier to detect, and involves the outer part of the bar \citep{Martinez-Valpuesta2006}. If second bucklings are common, then this may increase the fraction of galaxies with significant mid-plane asymmetries  at the present time. 

\subsection{Caveat for the mid-plane asymmetry diagnostics}

The diagnostics we have developed require the correct identification of a galaxy centre and disc PA, which is not always trivial for observed galaxies, since many host warps and/or asymmetric discs, as well as dust, star forming complexes and foreground stars. These can introduce artefacts in a mid-plane asymmetry map. In Sec.~\ref{sec:obs_compl} we have carefully explored the effects of an erroneous identification of the mentioned galaxy parameters, concluding that they can be easily recognised and corrected, assuming the analysed galaxy is neither distorted nor hosts a strongly warped disc, as is the case in the models analysed here. 

The effect of a non-perfectly edge-on view may be dramatic for the mid-plane asymmetry diagnostics. Indeed, asymmetries may appear in the mid-plane asymmetry map due to the presence of disc features, such as spiral arms or weak lopsidedness, when they are not observed edge-on, as in the case of the symmetric model HG1. On the other hand, mid-plane asymmetries  due to a past strong buckling event weaken if the model is not observed perfectly edge-on, but remain distinguishable. This result implies that a spurious mid-plane asymmetry due to inclination effects is not easily distinguishable from the genuine one produced by the buckling instability. Our diagnostics are thus not effective when applied to a galaxy which is not very close to edge-on. Determining whether a galaxy is indeed edge-on can be facilitated by the visual inspection of the orientation of dust lanes, which are often aligned along the disc major axis. 

We emphasise that the effect of a wrong identification of the disc inclination $i$ is the biggest weakness of the diagnostics developed here: it is not possible to distinguish between a mid-plane asymmetry produced by a genuine past buckling event and inclination effect already for a deviation of $\pm3\degr$ from an edge-on view. Such as view happens for only $5\%$ of a random distribution of galaxy orientations. 

\subsection{Different shape and strength of the mid-plane asymmetries }

We have selected galaxies from S$^4$G, which has the largest available database of deep, homogeneous mid-infrared images of $\sim2800$ galaxies of all types. \cite{Buta2015} identified more than 60 nearly edge-on galaxies with visually recognised X-shaped structures. Amongst these, we select the most edge-on galaxies, ending up with eight galaxies suitable for our mid-plane asymmetry analysis. However, we are able to build reliable mid-plane asymmetry maps for just four galaxies, since the other four are either strongly warped or cannot be centred properly (because they are intrinsically lopsided or distorted). Indeed, we found a residual quadrupole in half of the sample, suggesting that the mid-plane was not properly determined. For this reason we conservatively decided to exclude these objects from our analysis. This highlights a possible difficulty in applying our mid-plane asymmetry diagnostics to larger samples. 

Our final sample consists of four galaxies. Two of them, NGC~4235 and NGC~5170 have significant mid-plane asymmetries  in the distribution of the surface brightness, which may indicate that a strong buckling event occurred sometime in their past.

Both recovered mid-plane asymmetries  are peculiar in different ways. The mid-plane asymmetry in NGC~5170 is relatively weak, with the mid-plane asymmetry profile presenting just a single clear peak. NGC~5170 also exhibits a weak mid-plane asymmetry on the right side of the disc, which continues out the region of the B/P bulge. A similar feature can be identified in model T1 (see Appendix~\ref{app:sim}), where a double-peaked mid-plane asymmetry associated with the buckling event is visible, alongside a strong asymmetry further out in the disc only on its left side. This one-side asymmetry is due to two weak spiral arms, as evident from inspecting the face-on view of the model. These produce a mid-plane asymmetry on one side of the disc region. Conversely, the asymmetries in NGC~4235 are stronger (reaching twice the values in NGC~5170), two peaks are clearly present along the mid-plane asymmetry profile, but they are asymmetric with respect to the $z$ axis. Mid-plane asymmetry profiles which are not symmetric with respect to the $z$ axis also occur during the evolution of models D5, D8, and T1. This asymmetry between the two peaks in the mid-plane asymmetry profile occurs only over a short time interval when the galaxy is perfectly edge-on, and always just after the buckling event. A small deviation from an edge-on view may also produce this left-right asymmetry, as found for model D5 (Sec.~\ref{sec:prove}, at $i=87\degr$).

The strength of the mid-plane asymmetry varies with time. In model D5 the asymmetries weaken rapidly during the first 1 Gyr after the buckling event and then more slowly during the following 4 Gyr, completely disappearing 5 Gyr after the buckling event. Moreover, weaker buckling generally produces weaker asymmetries (see e.g., the case of model T1 in Appendix~\ref{app:sim}, whose mid-plane asymmetry disappears after just 1.5 Gyr). We therefore cannot exclude the possibility that NGC~5170 suffered a weak but recent buckling event. Indeed, the strength of the mid-plane asymmetry does not distinguish between a strong buckling long ago and a more recent, weaker buckling. Furthermore, we can not exclude the possibility that buckling is also responsible for the formation of the B/P bulge in the other two galaxies, ESO~443-042 and NGC~4013, since the mid-plane asymmetries  may have decayed by the current epoch.

\subsection{Evidence of strong buckling or inclination effects?}

Despite a clear mid-plane asymmetry in NGC~4235 and NGC~5170, we cannot exclude the possibility that they are caused by a non-perfectly edge-on view of the galaxies. Inspecting the optical images presented in Fig.~\ref{fig:optic}, both galaxies present dust lanes along the disc and asymmetries with respect to the disc mid-plane. The optical dust lane itself cannot be used to unambiguously measure the disc inclination. However, the observed asymmetric dust lanes in both galaxies suggest that they are not perfectly edge-on. As a consequence, we conclude that we cannot distinguish if the observed mid-plane asymmetries  are produced by a past buckling event, rather than the effect of inclination. Thus the asymmetric galaxies NGC~4235 and NGC~5170 cannot be used to infer whether a strong buckling event occurred in them. The diagnostics presented here are efficient in identifying mid-plane asymmetry due to a past buckling event when the galaxy is observed almost edge-on. Since galaxies are randomly oriented in the sky, just a small fraction will be suitable for this analysis.

We find two clear examples, ESO~443-042 and NGC~4013, of galaxies with clear mid-plane symmetry. Finding two symmetric cases suggests that strong buckling instabilities may be rare events in the past $\sim 5$~Gyr. 

Although we have used a small sample, our results suggest that B/P bulges either formed via weak buckling or resonant capture, or that very strong buckling events may take place preferentially a long time ago, longer than $\sim5$~Gyr, the expected time during which mid-plane asymmetries  are expected to be visible (Fig.~\ref{fig:evol}).

\subsection{Future prospects} 

While a handful of galaxies have now been found that are currently undergoing buckling \citep{Erwin2016, Li2017, Xiang2021}, buckling is so short-lived that we are unlikely to find many bars during this point in their evolution \citep{Erwin2016}. We have shown that buckling produces small but long-lasting asymmetries in the density distribution about the mid-plane. They allow B/P bulges formed by buckling to be identified in edge-on galaxies long after the buckling. To compare the relative efficiency of the two methods, the two buckling bar galaxies of \citet{Erwin2016} are drawn from a sample of 84 galaxies with orientations suitable for detecting B/P bulges (buckling or not). Instead the four galaxies with successful application of our diagnostics are drawn from a parent sample of eight galaxies suitable for this analysis, closer to edge-on. Thus the presence of mid-plane asymmetries  may provide a powerful probe to test the origin of B/P bulges. 


However, nearly edge-on galaxies are rare, but remain interesting laboratories for studying the physical processes involved in the formation and evolution of galaxies. Attention has been devoted to identifying and cataloguing these systems \citep[e.g.][]{Mitronova2004, Bizyaev2014, Makarov2021,Marchuk2022}. High-resolution infrared imaging will soon be available with the {\it James Webb Space Telescope} which will help explore current edge-on B/P samples across cosmic time. Meanwhile, future large space-based surveys (such as the {\it Euclid} mission operating in optical and near-infrared bands) will observe a larger number of galaxies. This large data-set will enable more statistically meaningful studies of the occurrence of strong buckling in B/P bulges,  but they require a careful selection of galaxies, including galaxies with a precise estimate of disc inclination, since the diagnostics presented here are efficient only for $|i-90\degr|<3\degr$.

\section{Conclusions}
\label{sec:conclusion}

The main goal of this paper is to develop diagnostics for identifying B/P bulges formed by buckling long after the event. We use simulations of barred galaxies which form a B/P bulge due both to strong buckling and recurrent weak buckling/resonant trapping. Strong buckling leaves an observable asymmetric distribution in the stellar density about the mid-plane. This asymmetry persists for between $\sim 1.5$ and $5$ Gyr. No similar asymmetry results when the B/P bulge is produced by either weak buckling or resonant capture. 
Based on this result, we develop two diagnostics to identify and quantify the asymmetries: the mid-plane asymmetry map and the mid-plane asymmetry profile, which we present in Sec.~\ref{sec:diagnostics}. 

We explore the effects of varying galaxy orientation and other observational effects on the mid-plane asymmetry diagnostics, to identify the best conditions under which to apply them to real galaxies. Despite large artefacts resulting from an incorrect identification of the mid-plane and of the centre of the galaxy used to build the mid-plane asymmetry map, we demonstrate how these can be recognised in real galaxies. However, spurious mid-plane asymmetries  may appear when the galaxy is not observed perfectly edge-on, for deviations larger than $\pm3\degr$. This issue is the strongest caveat when applying our method to observed galaxies, which are unlikely to be observed at $i=90\degr$, or for which it is difficult to derive the disc inclination. 

We construct a sample of nearly edge-on galaxies in the $S^4$G catalog, based on previous identifications of X-shaped structures. We identify a parent sample of eight galaxies which are sufficiently edge-on for our analysis.
Nonetheless, we are only able to obtain reliable mid-plane asymmetry diagnostics in our final sample of four of these galaxies. Of these, two galaxies, NGC~4235 and NGC~5170, exhibit asymmetries resembling those found in the strongly buckling models. We cannot exclude that the observed mid-plane asymmetries  result from a non-perfectly edge-on orientation of the galaxies since asymmetric dust lanes are visible in optical images of discs, hinting that the observed mid-plane asymmetries  are probably due to inclination effects. Thus we cannot obtain strong conclusions about the origin of the mid-plane asymmetries  in these two objects. In the other two galaxies, ESO~443-042 and NGC~4013, we do not find any significant mid-plane asymmetries. We cannot exclude the possibility that these two symmetric galaxies suffered a strong buckling a long time ago, and that the asymmetries have since disappeared. This possibility suggests that strong buckling is relatively rare in the past $5$~Gyr and that B/P bulges are more likely to have formed before then, or formed via weak buckling and/or resonant trapping. 


This paper is a pilot study to test the applicability of our diagnostics; stronger conclusions about the role of the buckling instability in the formation of B/P bulges requires a larger sample of very nearly edge-on galaxies. 

\section*{Acknowledgements}

We thank the anonymous referee for valuable comments, which improved the paper. We are grateful to L. Beraldo e Silva for fruitful discussion.
VC acknowledges support from the Chilean FONDECYT postdoctoral programme 2022 and from ESO-Government of Chile Joint Committee programme ORP060/19. 
VPD is supported by STFC Consolidated grant \# ST/R000786/1. The simulations in this paper were run at the DiRAC Shared Memory Processing system at the University of Cambridge, operated by the COSMOS Project at the Department of Applied Mathematics and Theoretical Physics on behalf of the STFC DiRAC HPC Facility (www.dirac.ac.uk). This equipment was funded by BIS National E-infrastructure capital grant ST/J005673/1, STFC capital grant ST/H008586/1, and STFC DiRAC Operations grant ST/K00333X/1. DiRAC is part of the National E-Infrastructure. Model HG1 was run at the High Performance Computing Facility of the University of Central Lancashire. %
SAR acknowledges support by the National Science Foundation Grant Number PHY-1914679 and was supported for part of this work by Reed College.
EMC acknowledges support by Padua University grants DOR1935272/19, DOR2013080/20, and DOR2021 and by MIUR grant PRIN 2017 20173ML3WW-001.

\section*{Data availability}

The simulation data underlying this article will be shared on reasonable request to VPD (\thanks{vpdebattista@gmail.com}). The data from S$^4$G are available at \url{https://irsa.ipac.caltech.edu/data/SPITZER/S4G/index.html}, while the derived data will be shared on reasonable request to VC.

\bibliographystyle{mnras}

\clearpage

\clearpage
\begin{appendix}

ONLINE MATERIAL

\section{Asymmetries diagnostics for simulated data}
\label{app:sim}
Here we present the asymmetry diagnostics, and the same tests as in Sec.~\ref{sec:prove}, for other models.

\subsection*{Time evolution around the buckling event}

\begin{figure}
    \centering
    \includegraphics[scale=0.5]{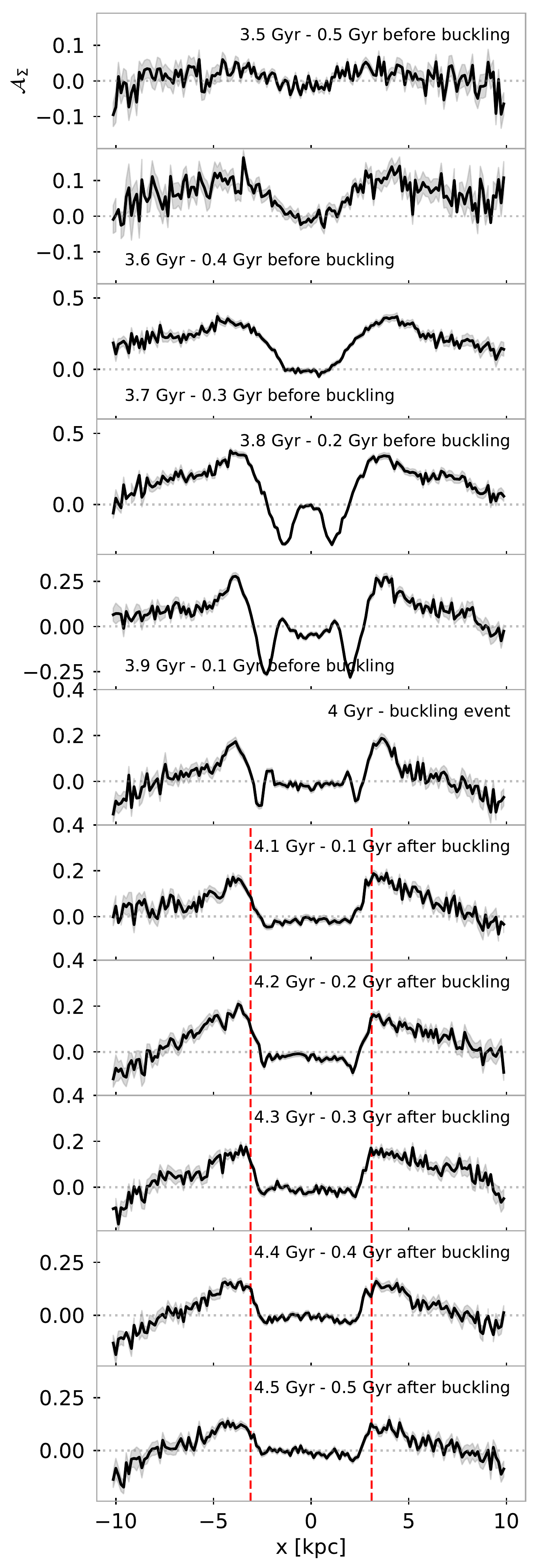}
    \caption{Time evolution of the mid-plane asymmetry radial profile of models D5 over 1 Gyr bracketing the formation of the B/P bulge.}
    \label{fig:evol_detailed}
\end{figure}

The buckling instability produces a peak in the time evolution of the buckling amplitude $A_{\rm buck}$ \citep{Debattista2020,Anderson2022} and consequently the occurrence of the B/P bulge, which marks the time of the buckling event. In Fig.~\ref{fig:evol_detailed} we present the mid-plane asymmetry profiles traced at high time resolution around the buckling. In particular, we analyse, every 0.1~Gyr, the interval between 3.5~Gyr and 4.5~Gyr. A clear and strong bending of the bar is already visible at 3.5~Gyr, while mid-plane asymmetry starts to be visible at 3.6~Gyr, associated with the typical double-peaked shape of the mid-plane asymmetry profile. Mid-plane asymmetry strongly increases between 3.8-4.0~Gyr, reaching a maximum value of  $\mathcal{A}_\Sigma (x) \sim0.45$ at 3.8~Gyr. During this phase, the central portion of the mid-plane asymmetry map, where the B/P bulge appears, is asymmetric as well as producing the four-peaked shape observed in the mid-plane asymmetry profile. At 4~Gyr the B/P bulge is clearly distinguishable from the unsharp mask. After that, the mid-plane asymmetry profiles again show two peaks and they start to slowly decline.

\subsection*{Effect of rotation}

\begin{figure}
    \centering
    \includegraphics[scale=0.41]{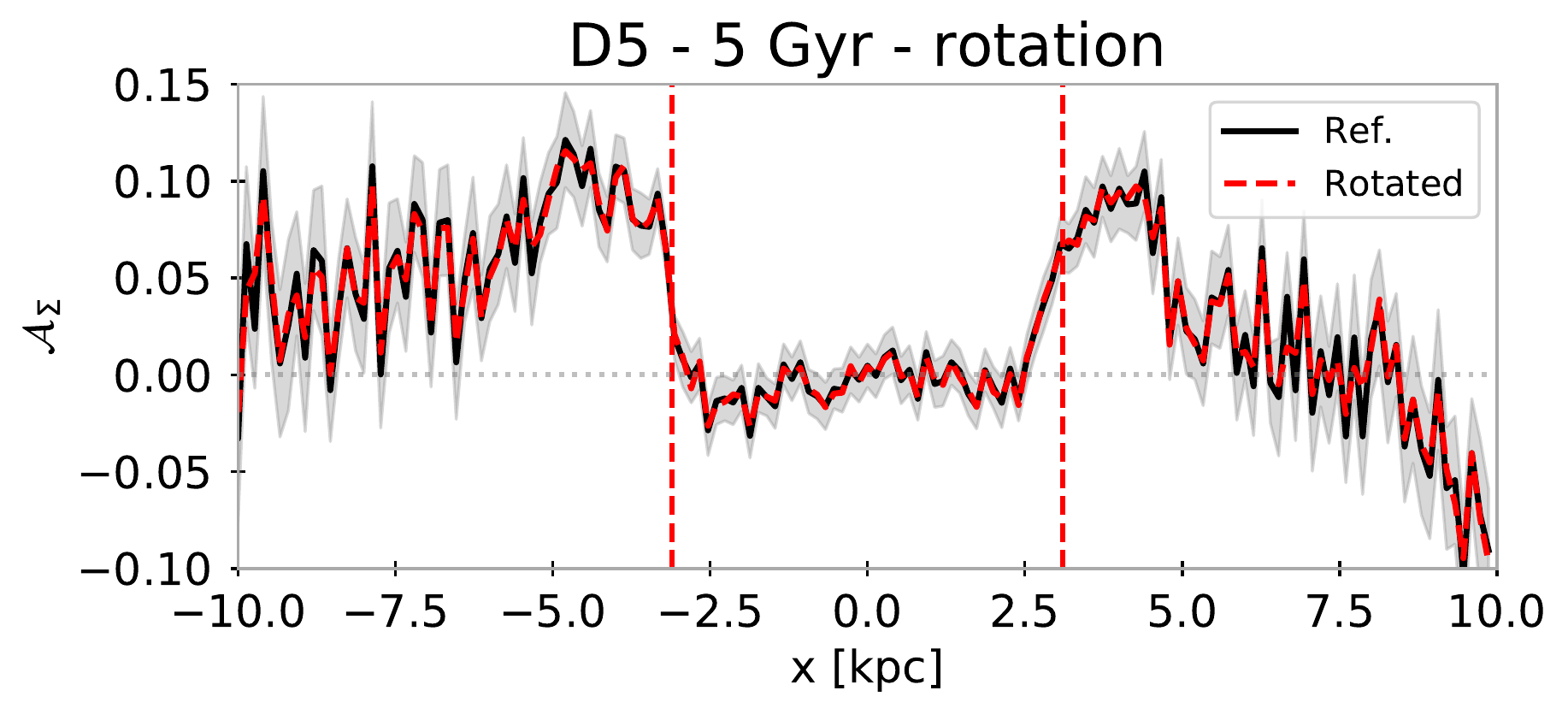}
    \caption{Mid-plane asymmetry profiles as in Fig.~\ref{fig:sim_dia} lower panel, but for testing the effect of rotation on model D5. The mid-plane asymmetry radial profiles extracted from the original unrotated image of model D5 (at 5 Gyr) seen side-on (black solid line) is compared to that obtained from the rotated image to align the disc position angle to the $x$ axis (red dashed line).}
    \label{fig:rot}
\end{figure}

In Fig.~\ref{fig:rot} we present the effect of the rotation procedure \textsc{rot} on the mid-plane asymmetry diagnostics. The procedure, written in {\sc idl}, uses an interpolation method based on a cubic convolution \citep{Park1983}.

\subsection*{Effect of the spatial sampling}

\begin{figure}
    \centering
    \includegraphics[scale=0.41]{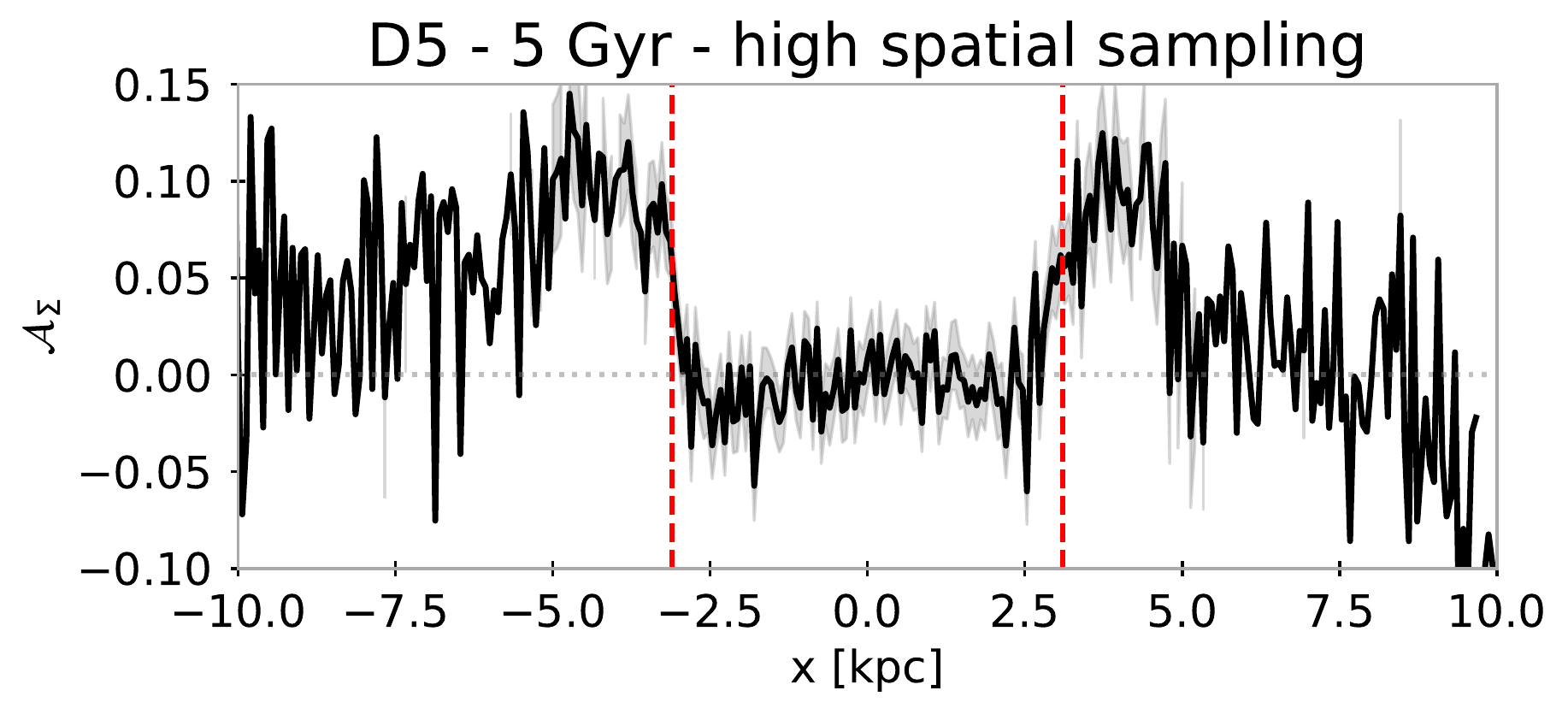}
    \includegraphics[scale=0.41]{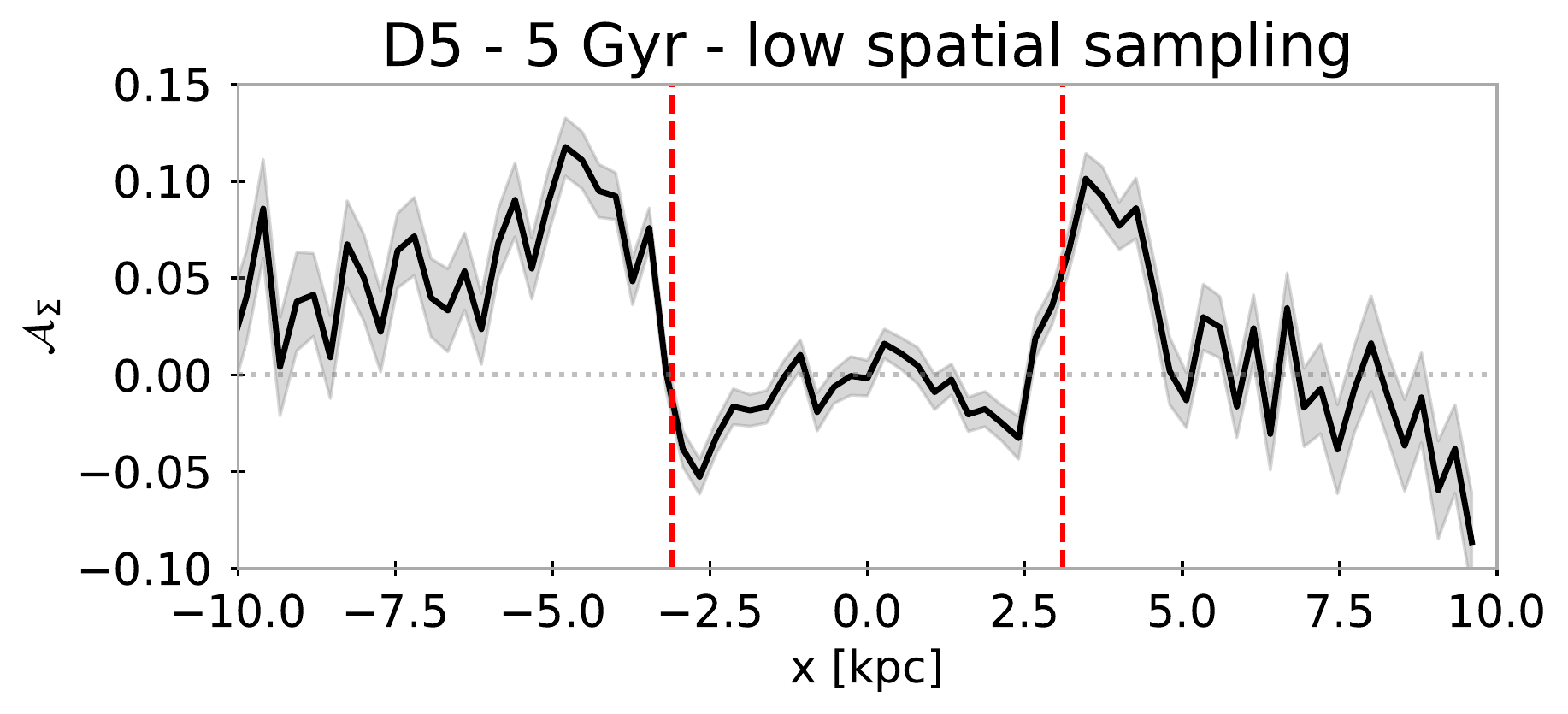}
    \caption{Mid-plane asymmetry profiles as in Fig.~\ref{fig:sim_dia} lower panel, but for model D5 (at 5 Gyr) mapped with a 2$\times$ better spatial sampling (0.065 kpc pixel$^{-1}$, upper panel) and with a 2$\times$ worse spatial sampling (0.26 kpc pixel$^{-1}$, lower panel).}
    \label{fig:hr}
\end{figure}

In Fig.~\ref{fig:hr} we show the effect of varying the spatial sampling on the mid-plane asymmetry diagnostics.

\subsection*{Effect of the seeing}

\begin{figure}
    \centering
    \includegraphics[scale=0.41]{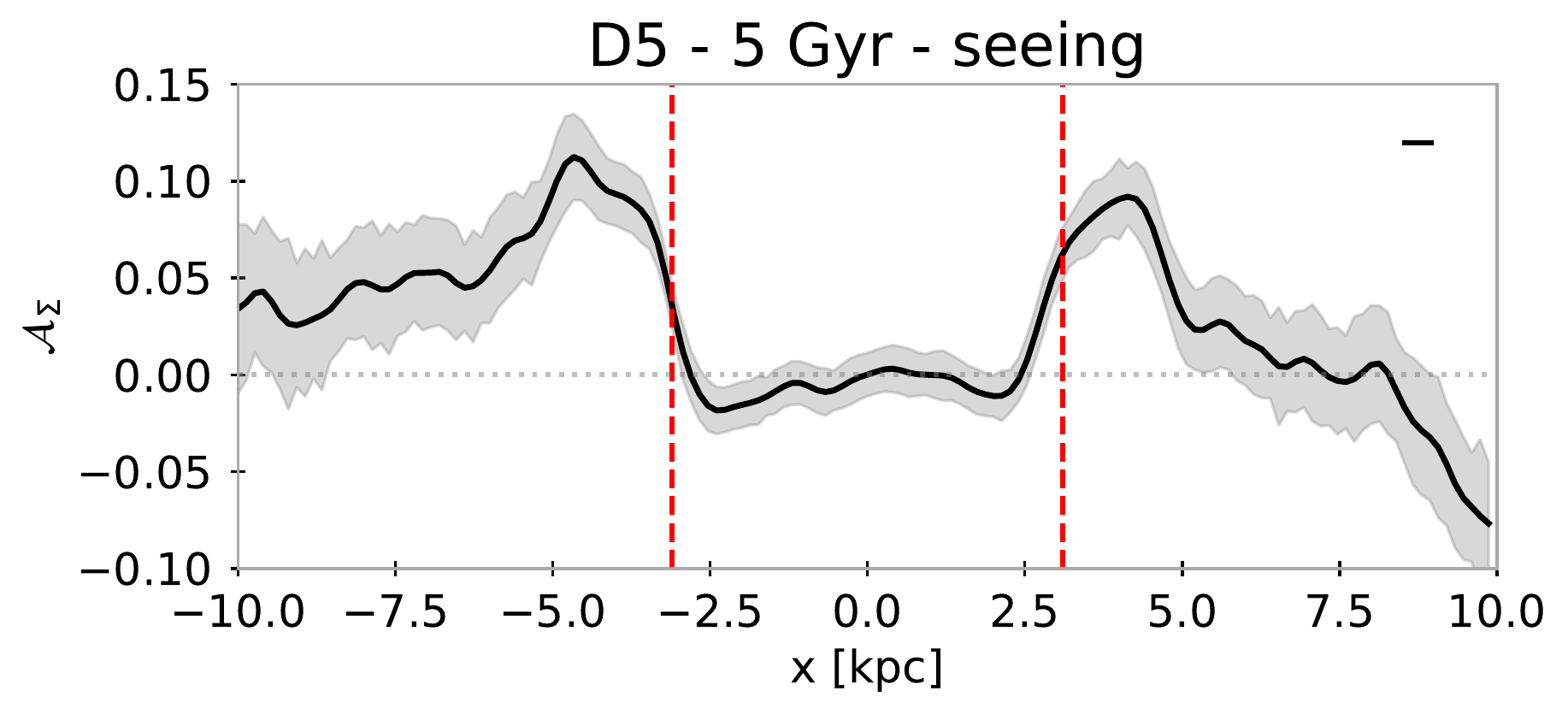}
    \includegraphics[scale=0.41]{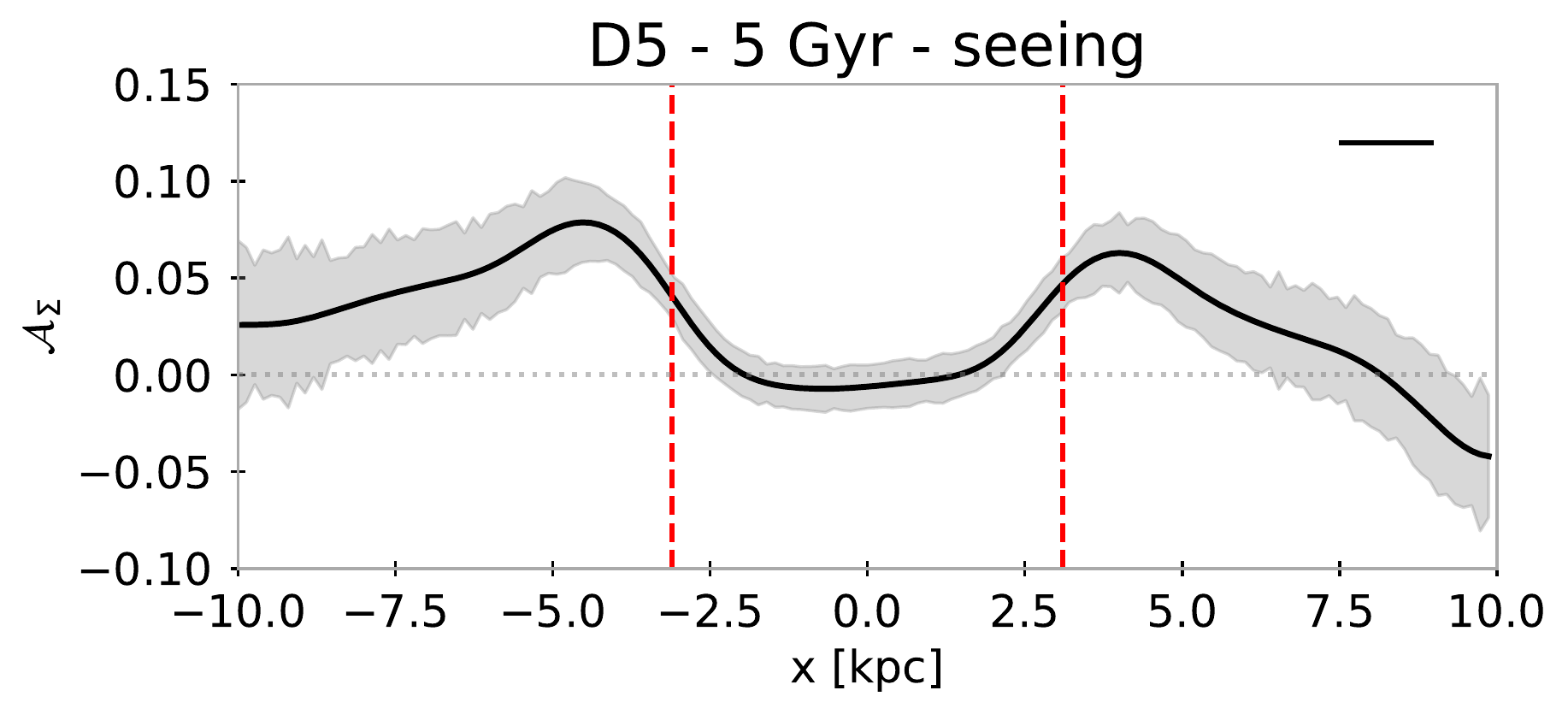}
    \includegraphics[scale=0.41]{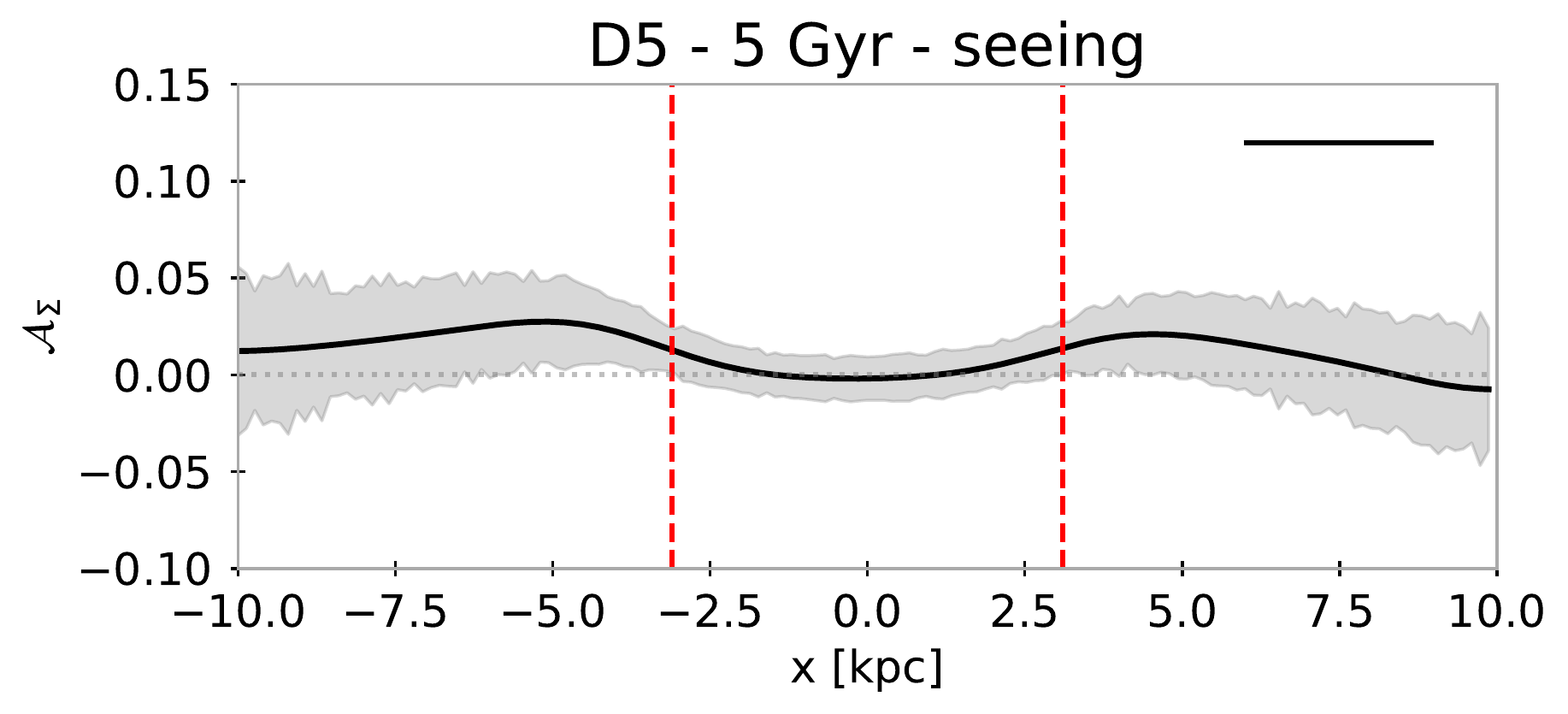}
    \caption{Same as Fig.~\ref{fig:sim_dia} lower panel, but for model D5 (at 5 Gyr) after convolving the image of the projected number density of the particles rescaled at a distance of $50$ Mpc with a Gaussian filter with a FWHM=0.5 kpc (corresponding to 2.1 arcsec, upper panel), 1.5 kpc (6.3 arcsec, central panel), and 3.1 kpc (12.6 arcsec, lower panel). The horizontal black segments at top right in each panel indicate the adopted FWHM.}
    \label{fig:seeing}
\end{figure}

In Fig.~\ref{fig:seeing} we show the effect of the seeing on the mid-plane asymmetry diagnostics.

\subsection*{Effect of dust}

\begin{figure*}
    \centering
    \includegraphics[scale=0.5]{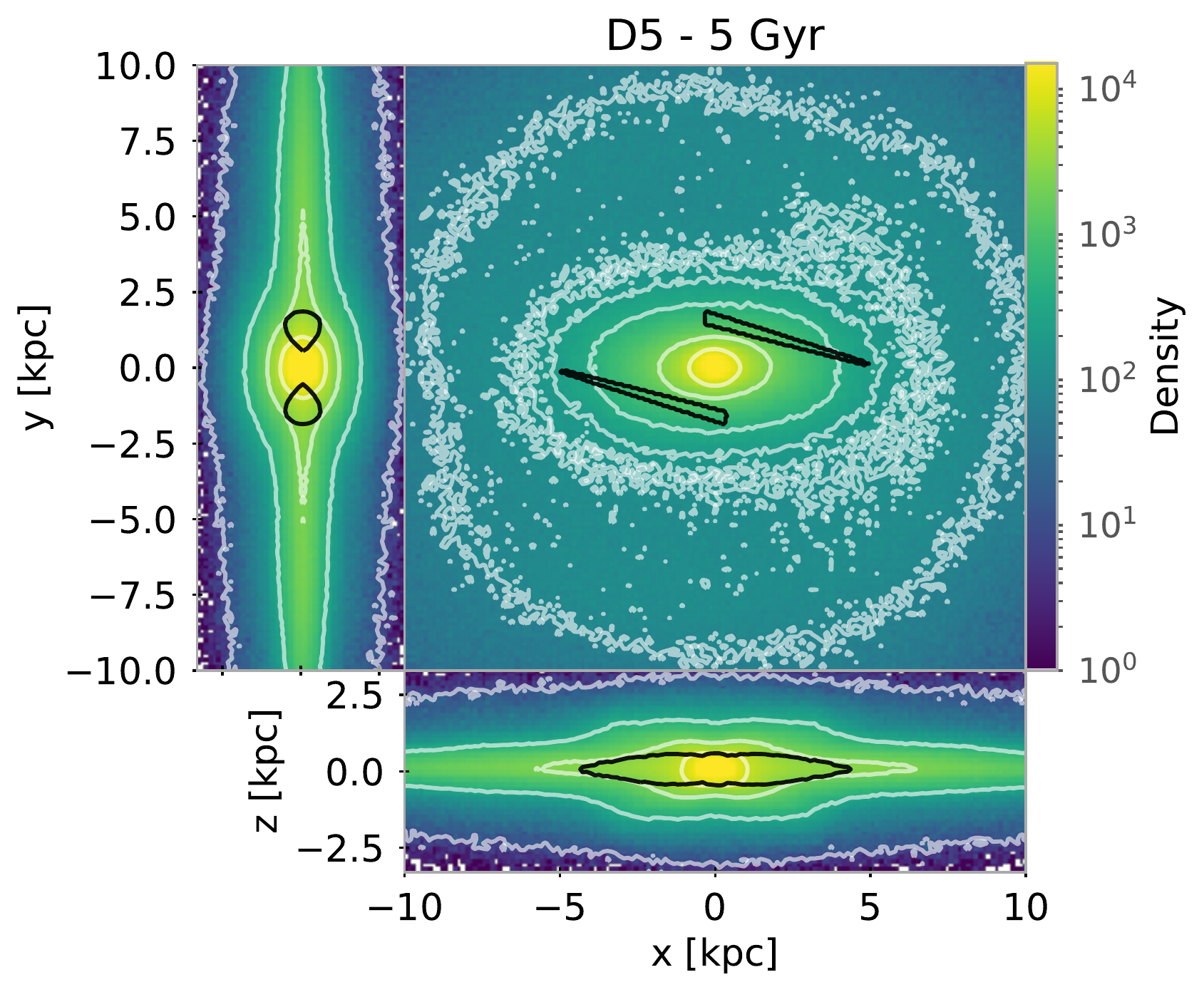}
    \includegraphics[scale=0.5]{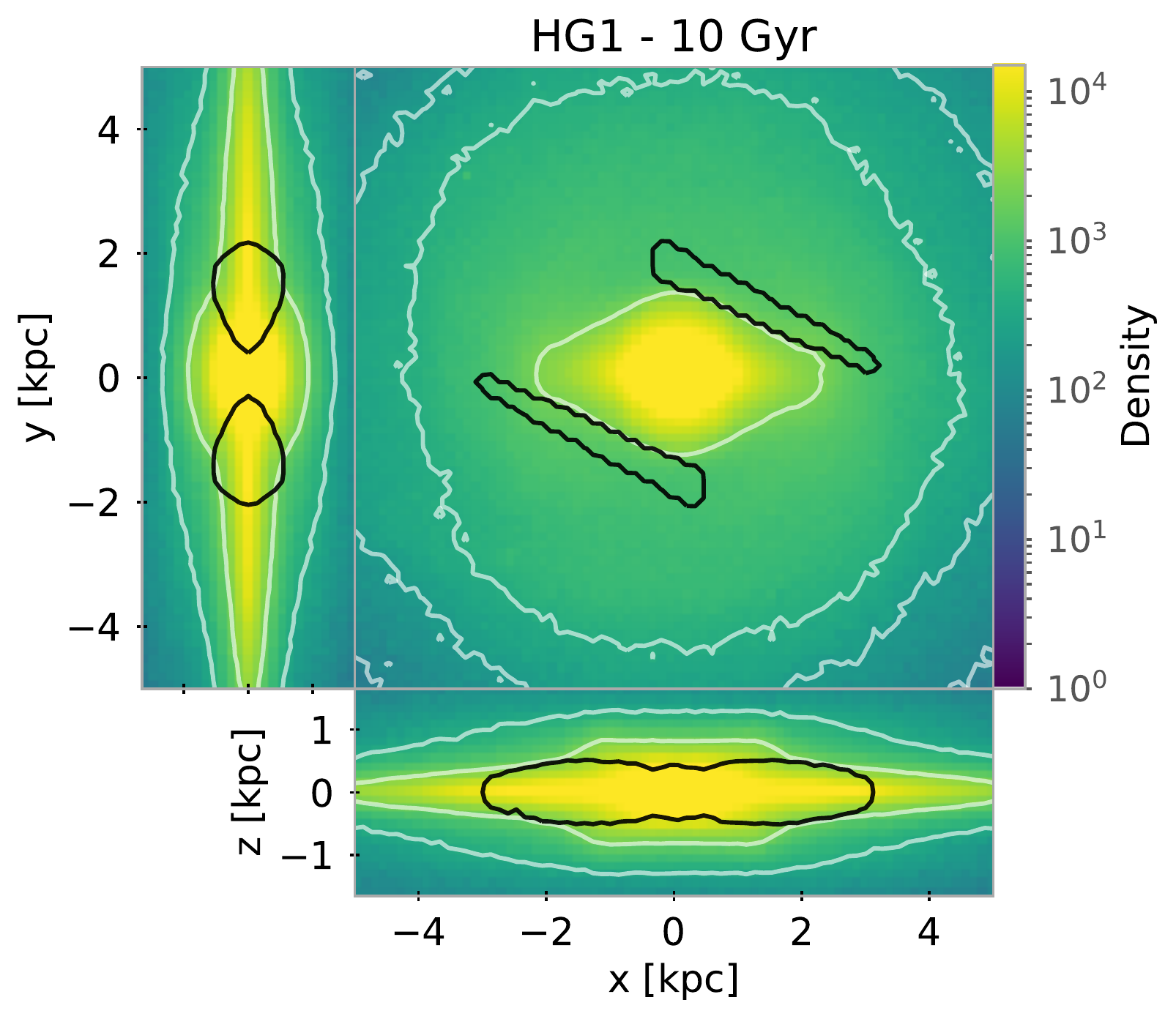}
    \caption{Same as Fig.~\ref{fig:sim_3d_view} but with some isocontours of the surface number density (black lines) of the dust lanes for each viewing geometry.}
    \label{fig:dust_lanes}
\end{figure*}

In Fig.~\ref{fig:dust_lanes} we show the face-on, side-on and end-on views of the projected
logarithmically-scaled surface number density of the particles of the two models, centred on the origin of the $(x, y)$ plane, with the bar aligned with the $x$ axis, and superimposing some of the isocontours of the dust lanes. We assume for both the dust disc and the dust lanes a double-exponential profile described by Eq.~\ref{eq:dust}. To describe the geometry dust lanes we assume for model D5 $w_{\rm dl} = 1.7 $~kpc, $s_{\rm dl} = 15\degr$, $x_{\rm dl,min} = -0.3$ kpc, and $x_{\rm dl,max} = 5.0$ kpc, while for model HG1 $w_{\rm dl} = 1.2$ kpc, $s_{\rm dl} = 25 \degr$, $x_{\rm dl,min} = -0.3$ kpc, and $x_{\rm dl,max} = 3.0$ kpc.

\begin{figure*}
    \centering
    \includegraphics[scale=0.43]{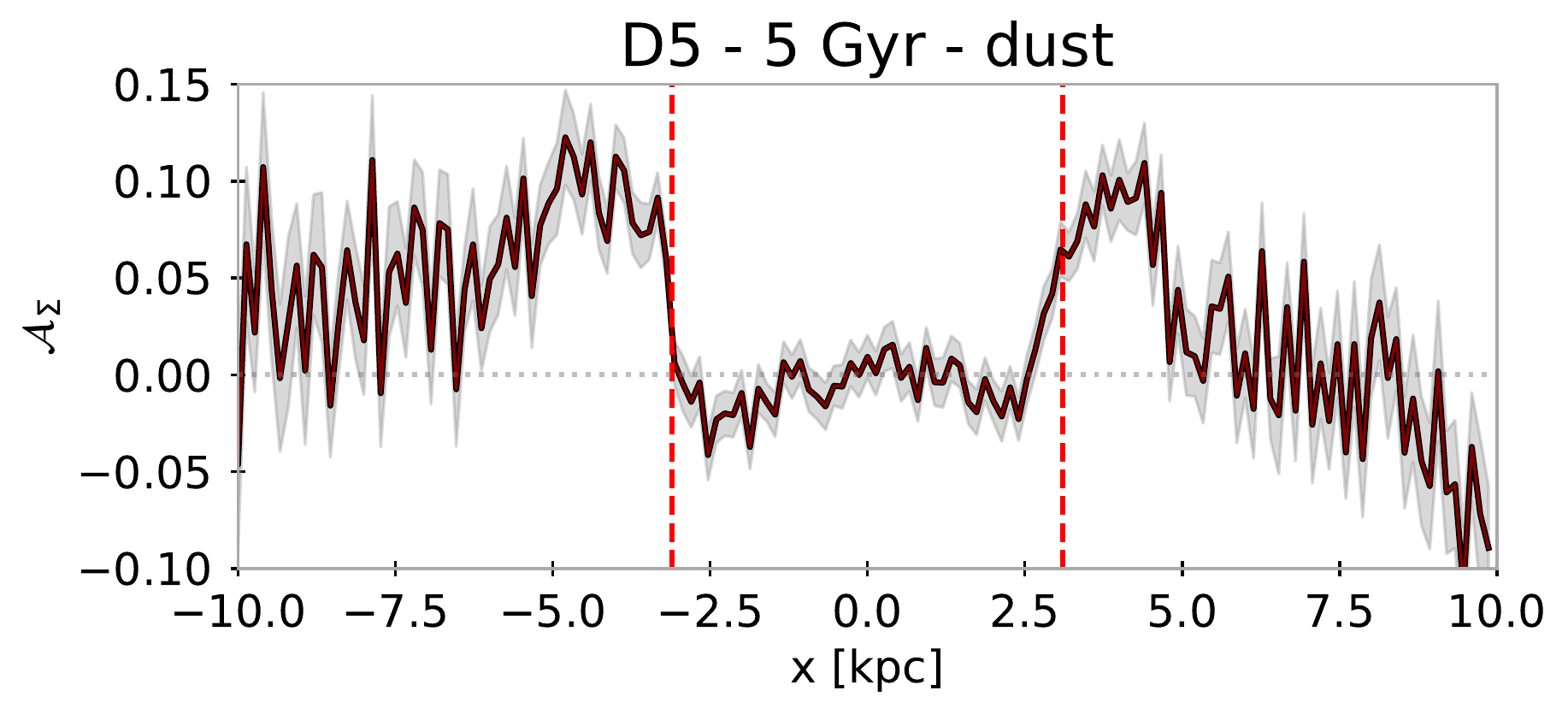} \\
    \includegraphics[scale=0.43]{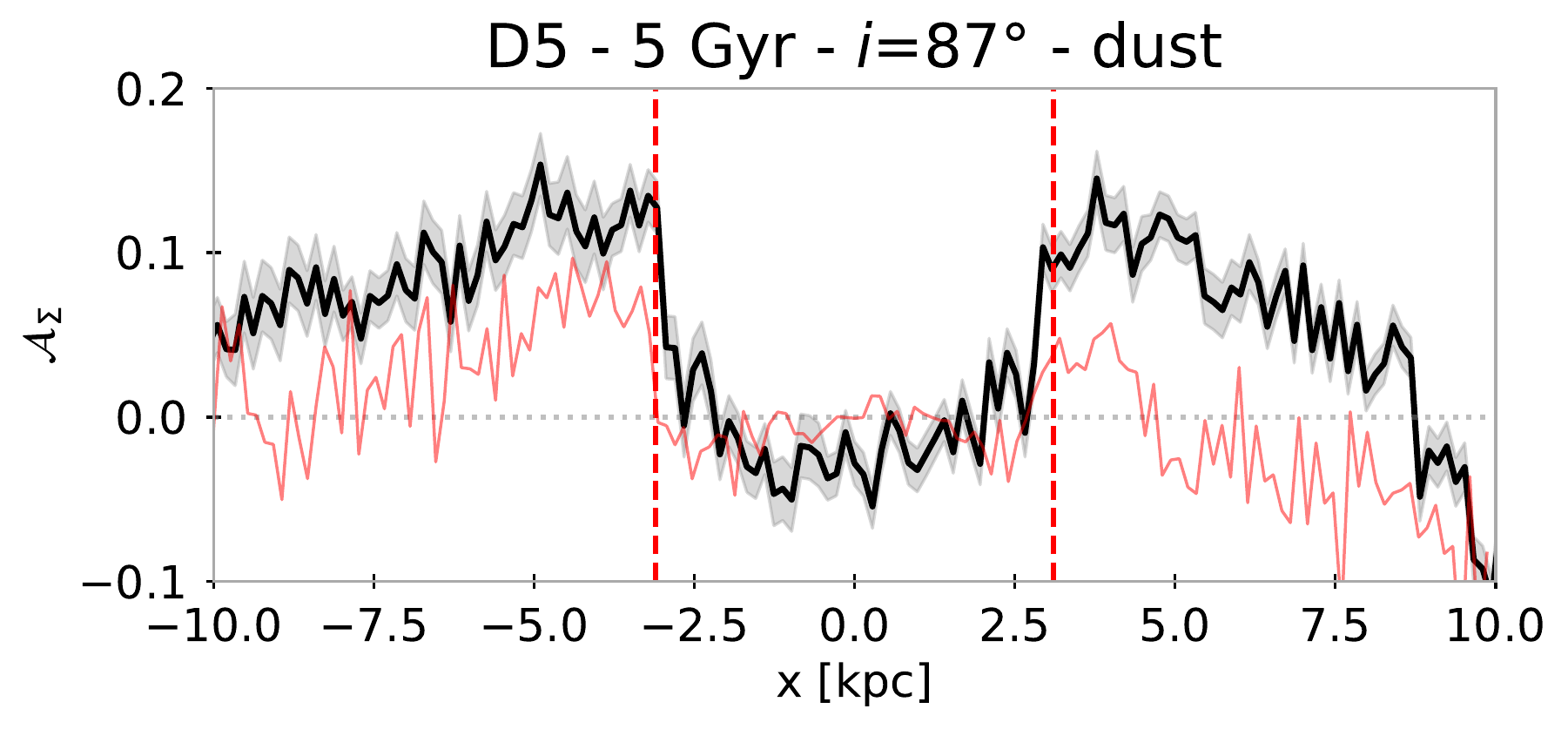}
    \includegraphics[scale=0.43]{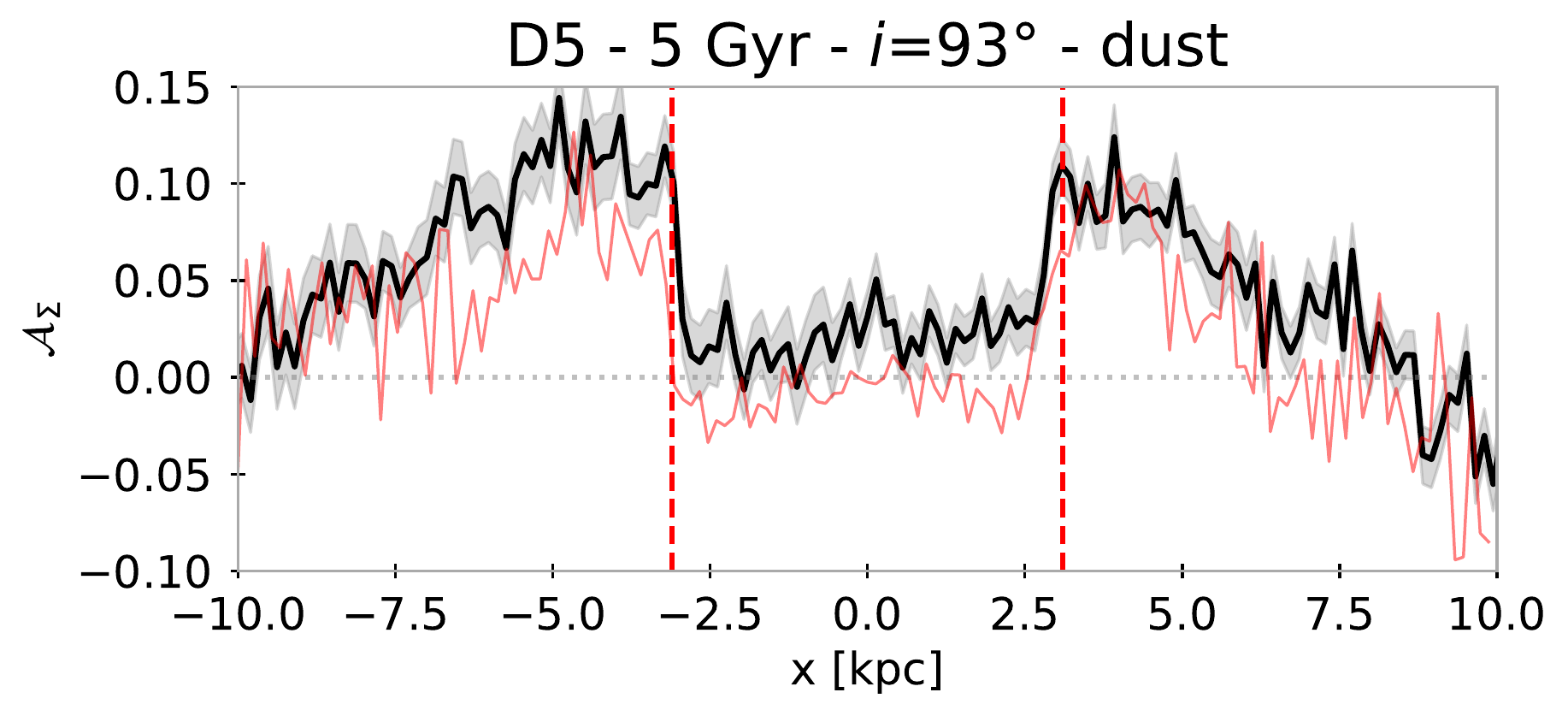}
    \includegraphics[scale=0.43]{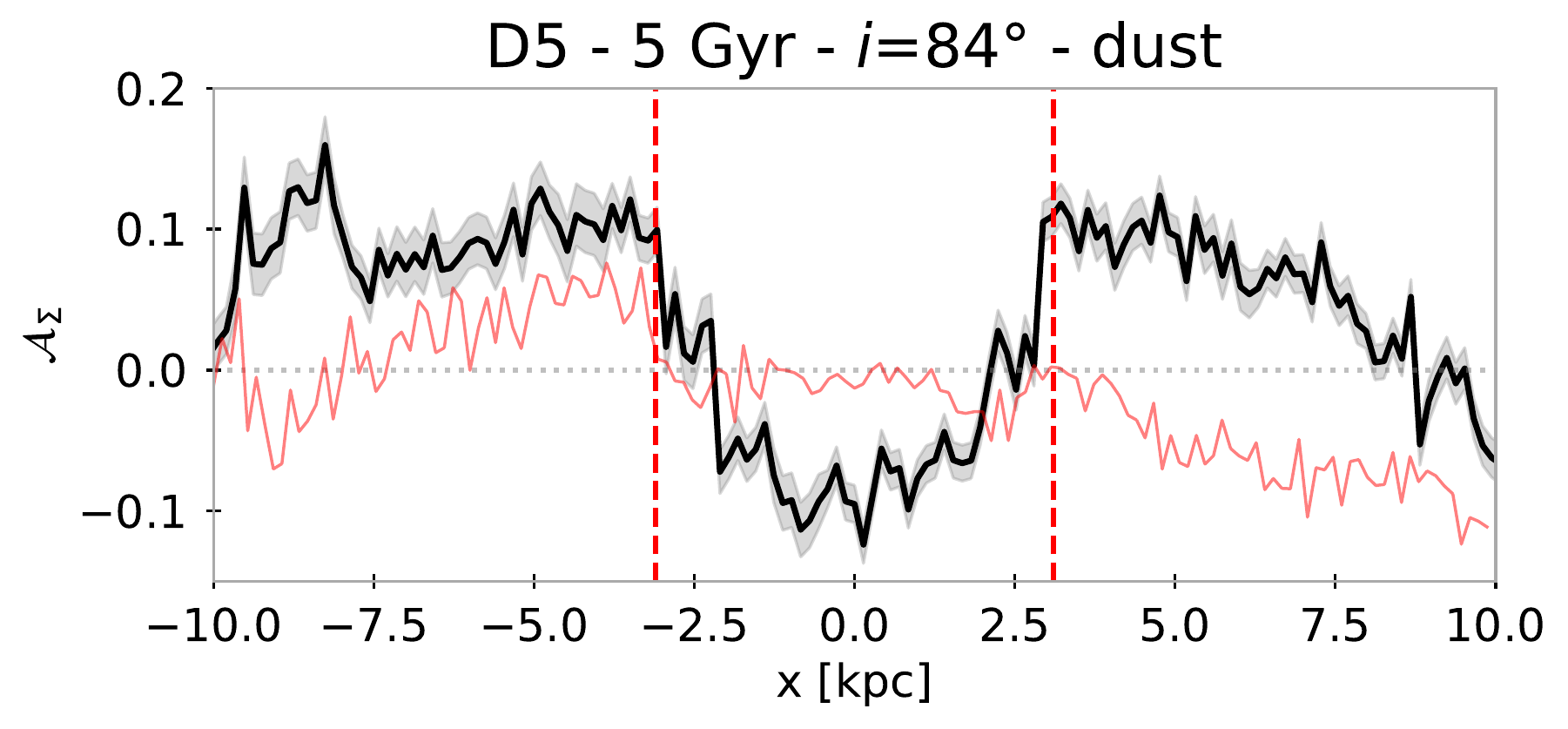}
    \includegraphics[scale=0.43]{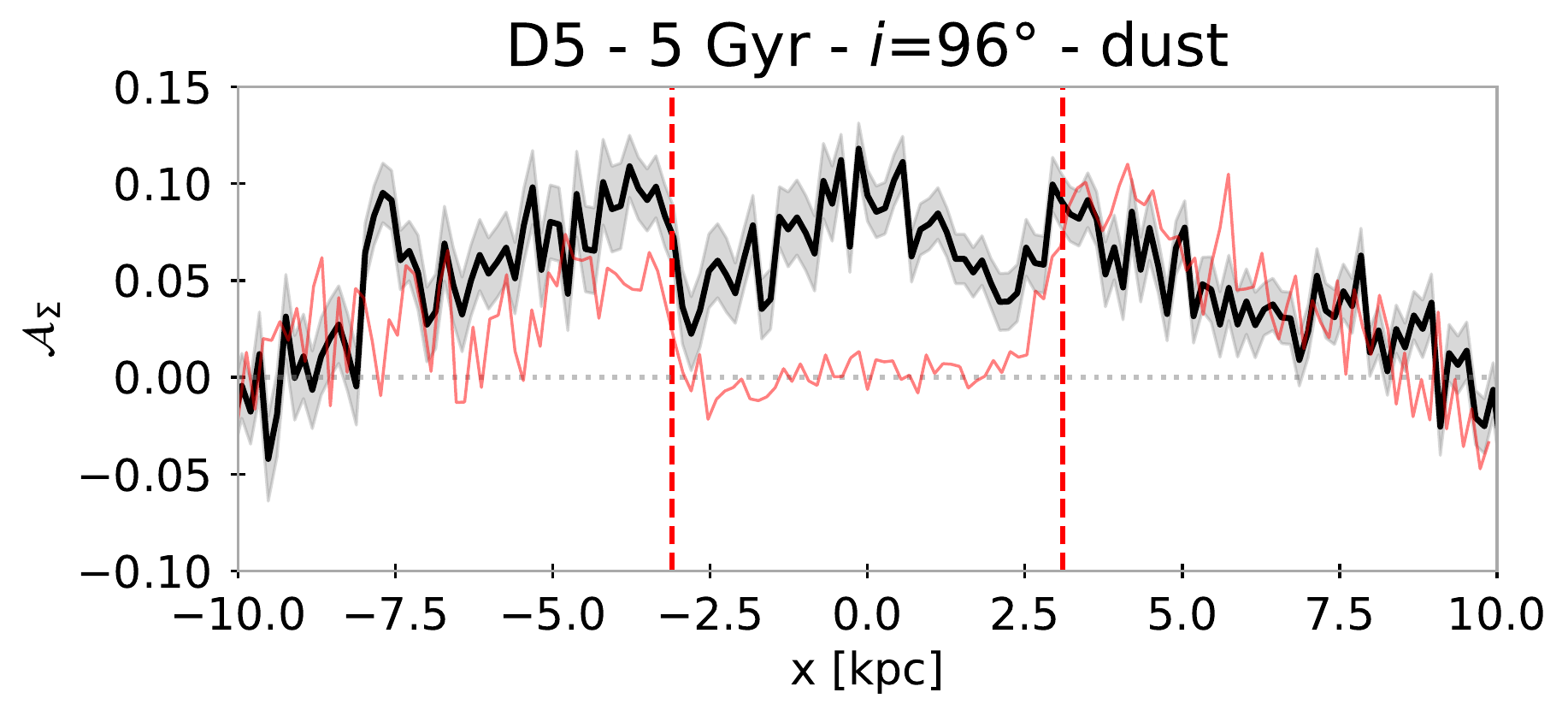}\\
    \includegraphics[scale=0.43]{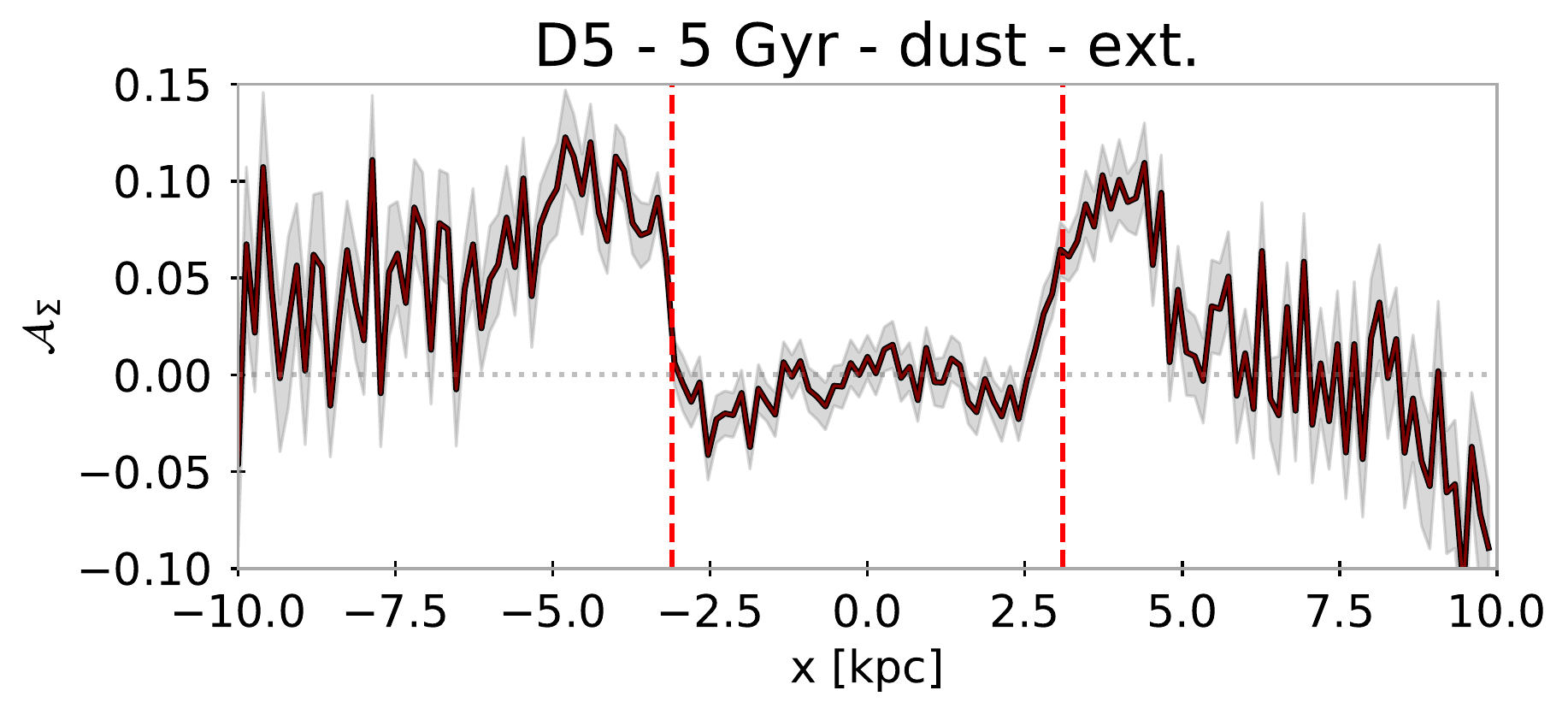}
    \caption{Mid-plane asymmetry profiles as in Fig.~\ref{fig:sim_dia} lower panel, but for model the D5 (at 5 Gyr) after taking into account the effect of the dust (disc with $h_{\rm R,disc}=1$~kpc and $h_{\rm z,disc} = 0.1$~kpc + dust lanes), with different inclinations with respect to the side-on view ($i=90\degr$, first row; $\Delta i\pm3\degr$, second row row; $\Delta i\pm6\degr$, third row) and with a radially very extended dust disc with $h_{\rm R,disc}=1.5\times$ initial galaxy disc scalelength ($i=90\degr$, fourth row). Superimposed in red are shown the corresponding profiles obtained without taking into account the effect of the dust.}
    \label{fig:dust_pa_d5}
\end{figure*}

\begin{figure*}
    \centering
    \includegraphics[scale=0.43]{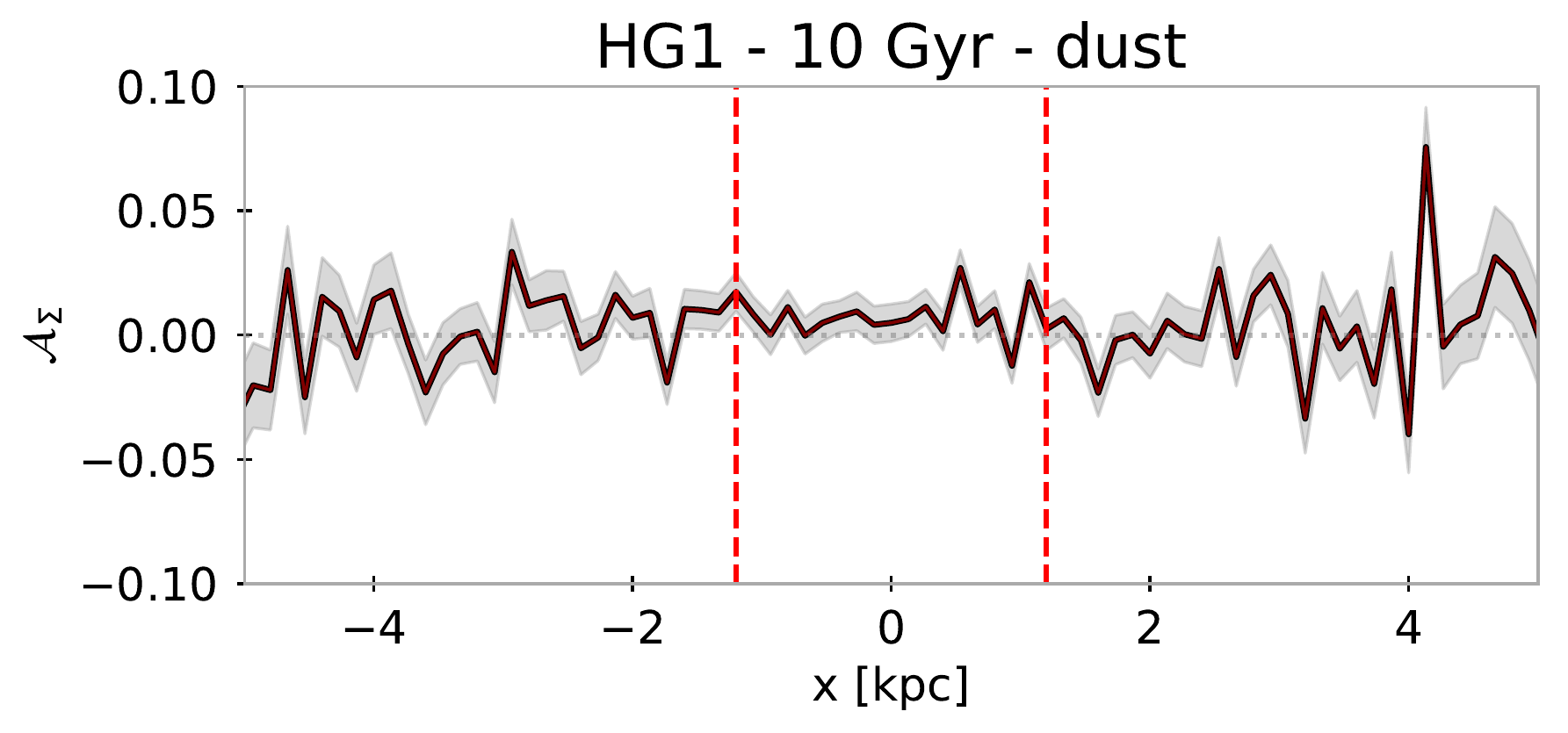} \\
    \includegraphics[scale=0.43]{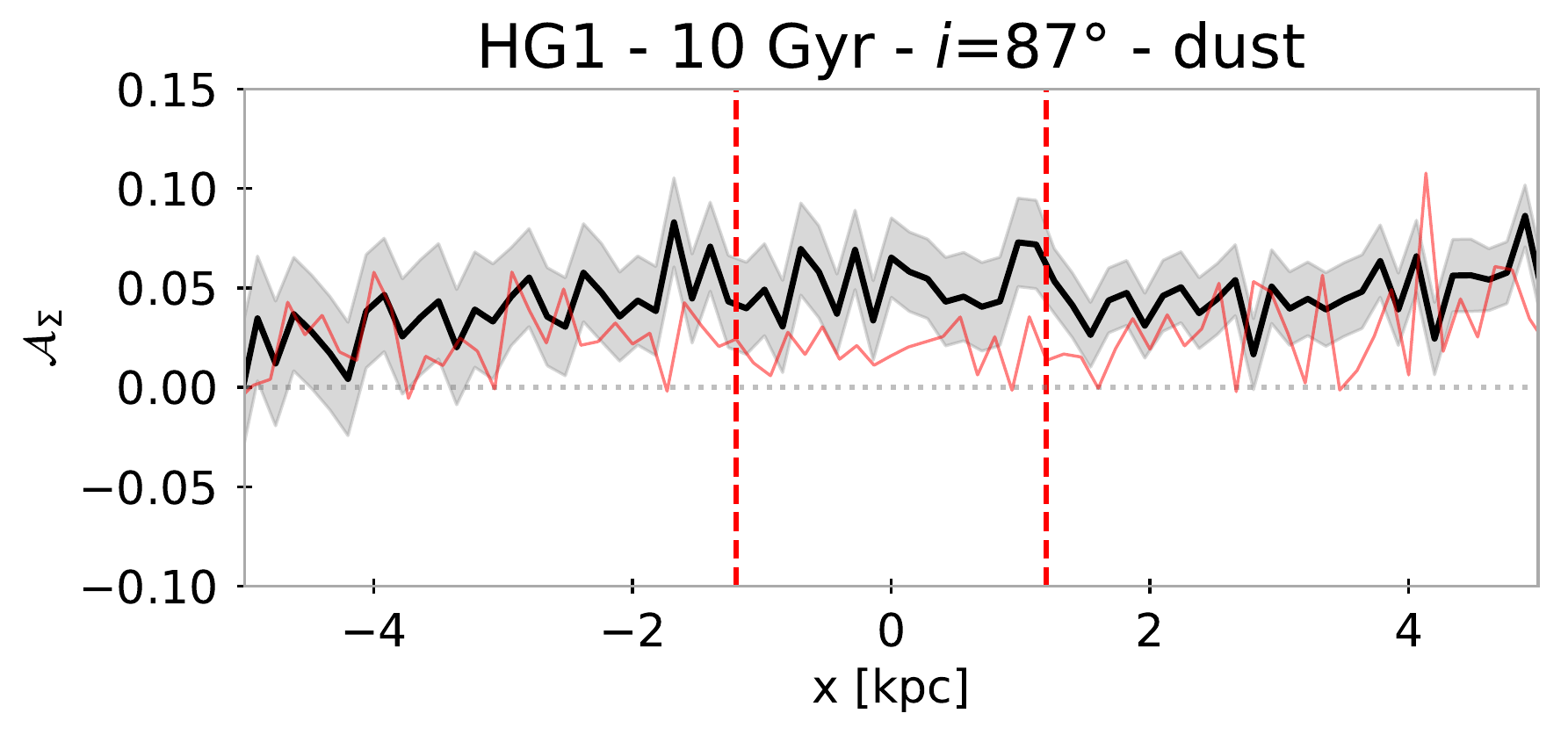}
    \includegraphics[scale=0.43]{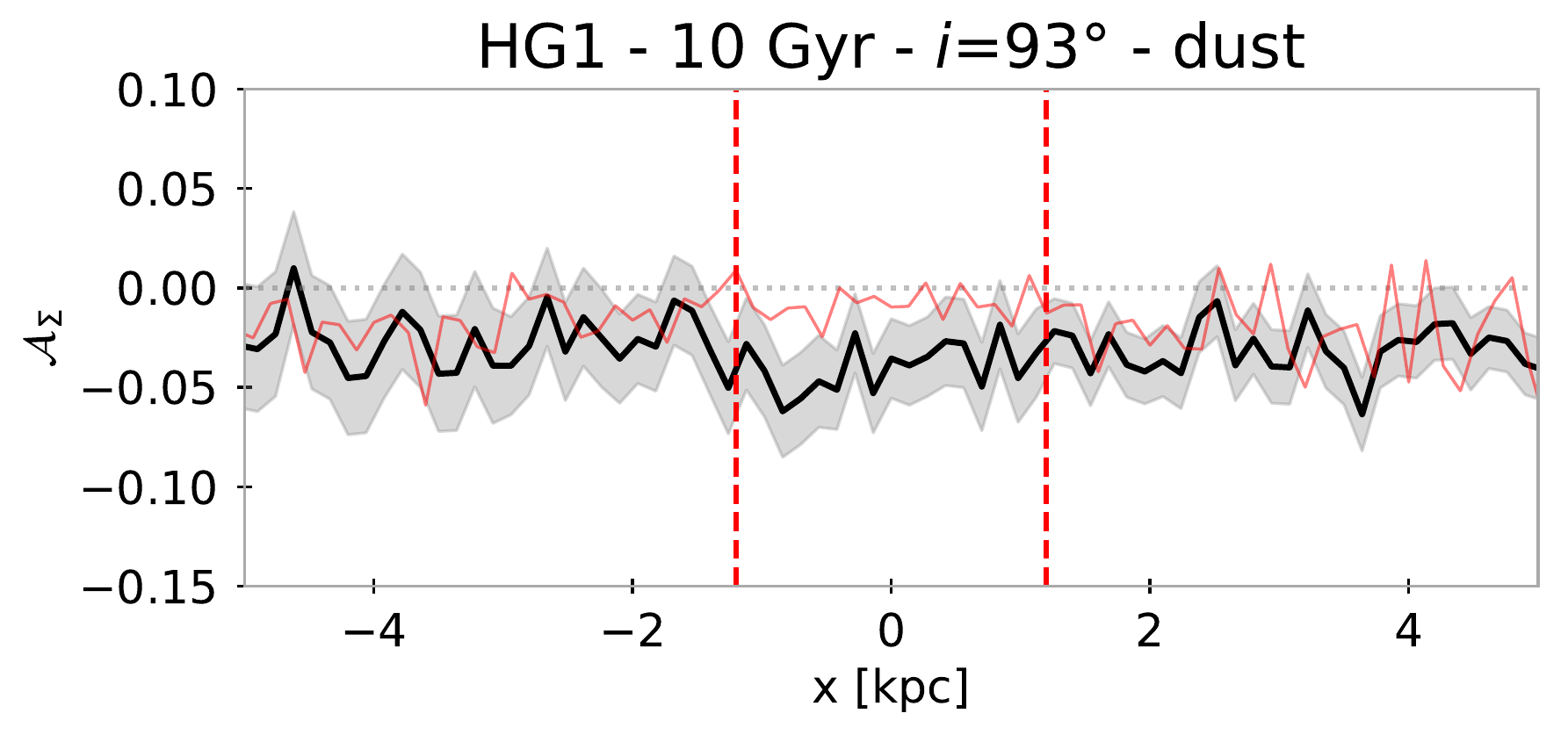}
    \includegraphics[scale=0.43]{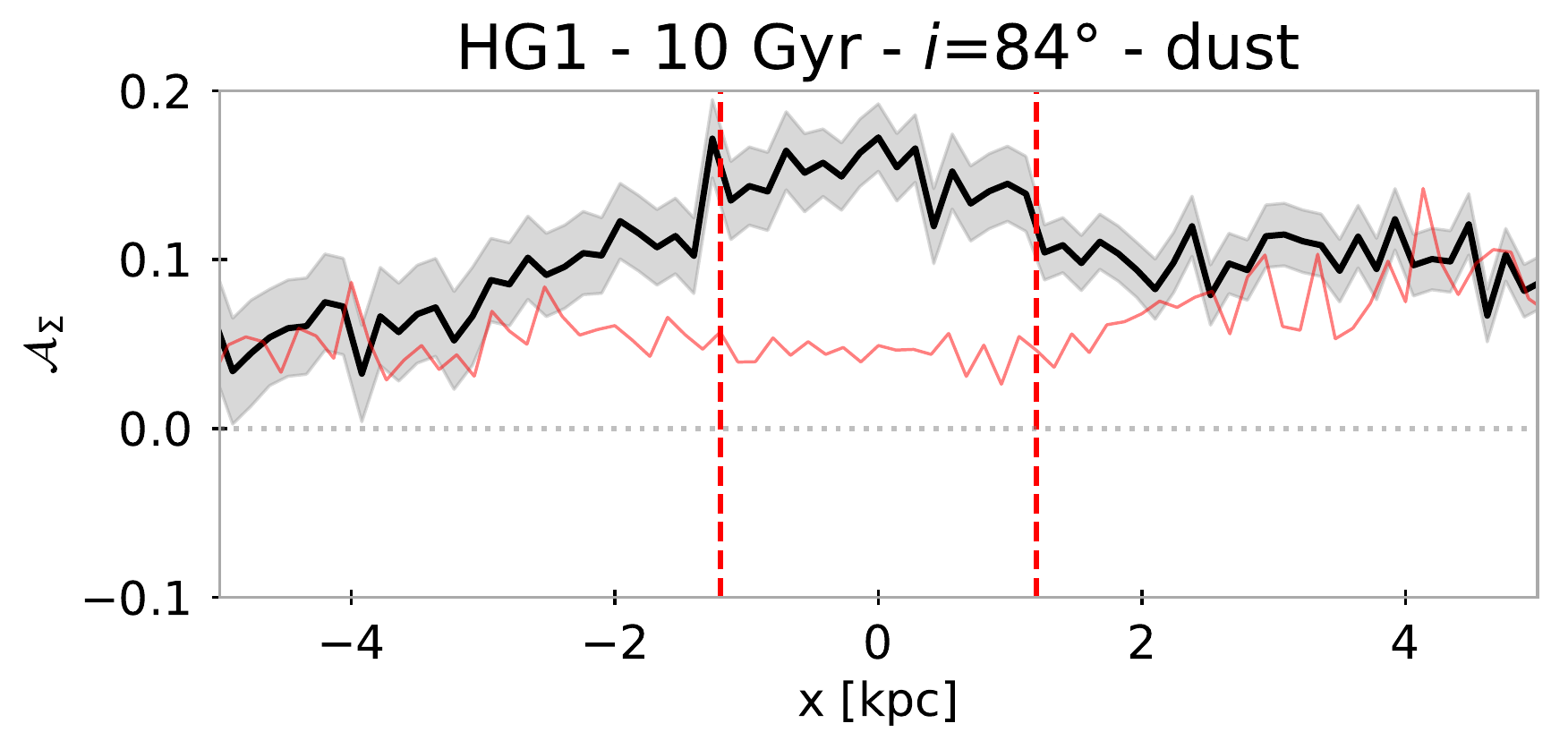}
    \includegraphics[scale=0.43]{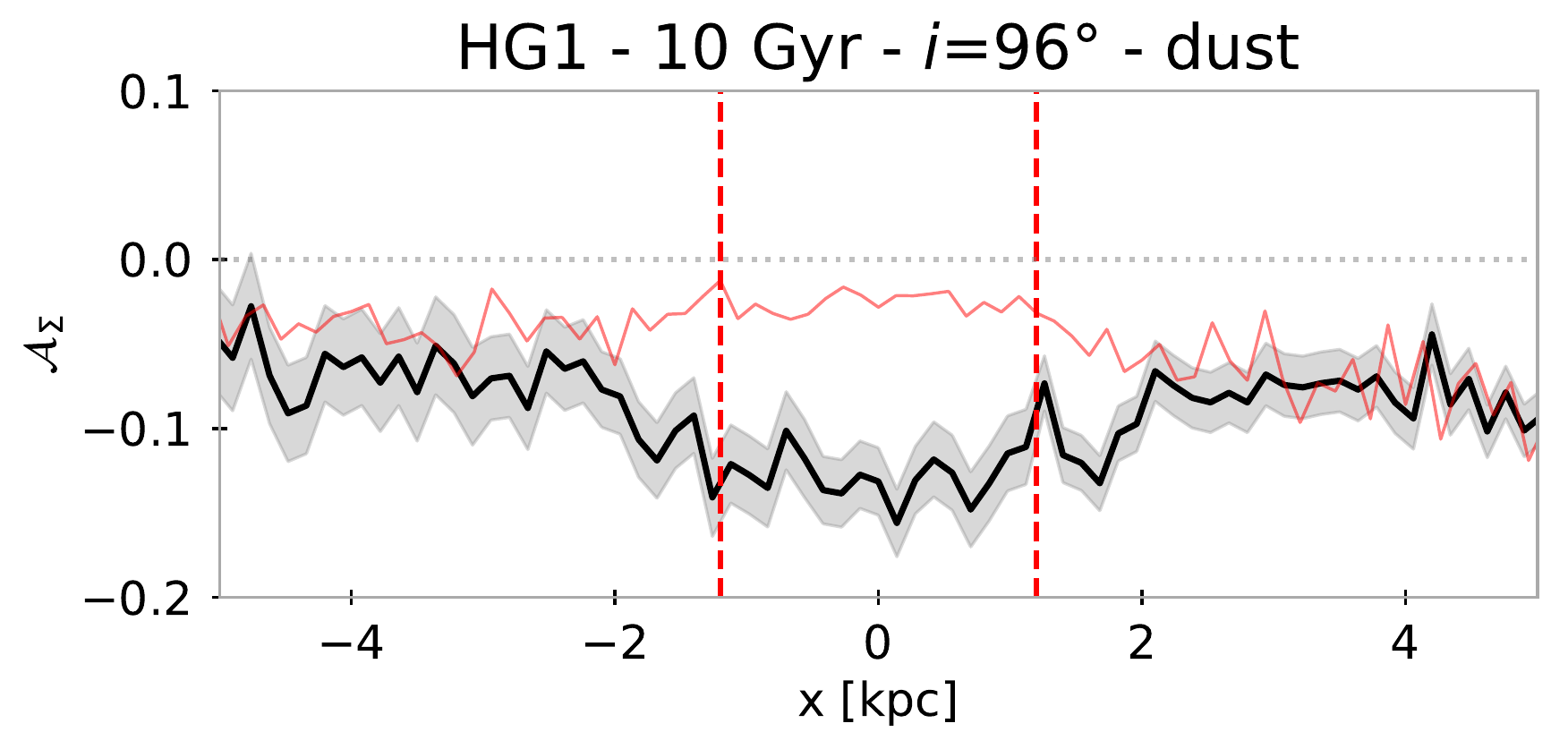}\\
    \includegraphics[scale=0.43]{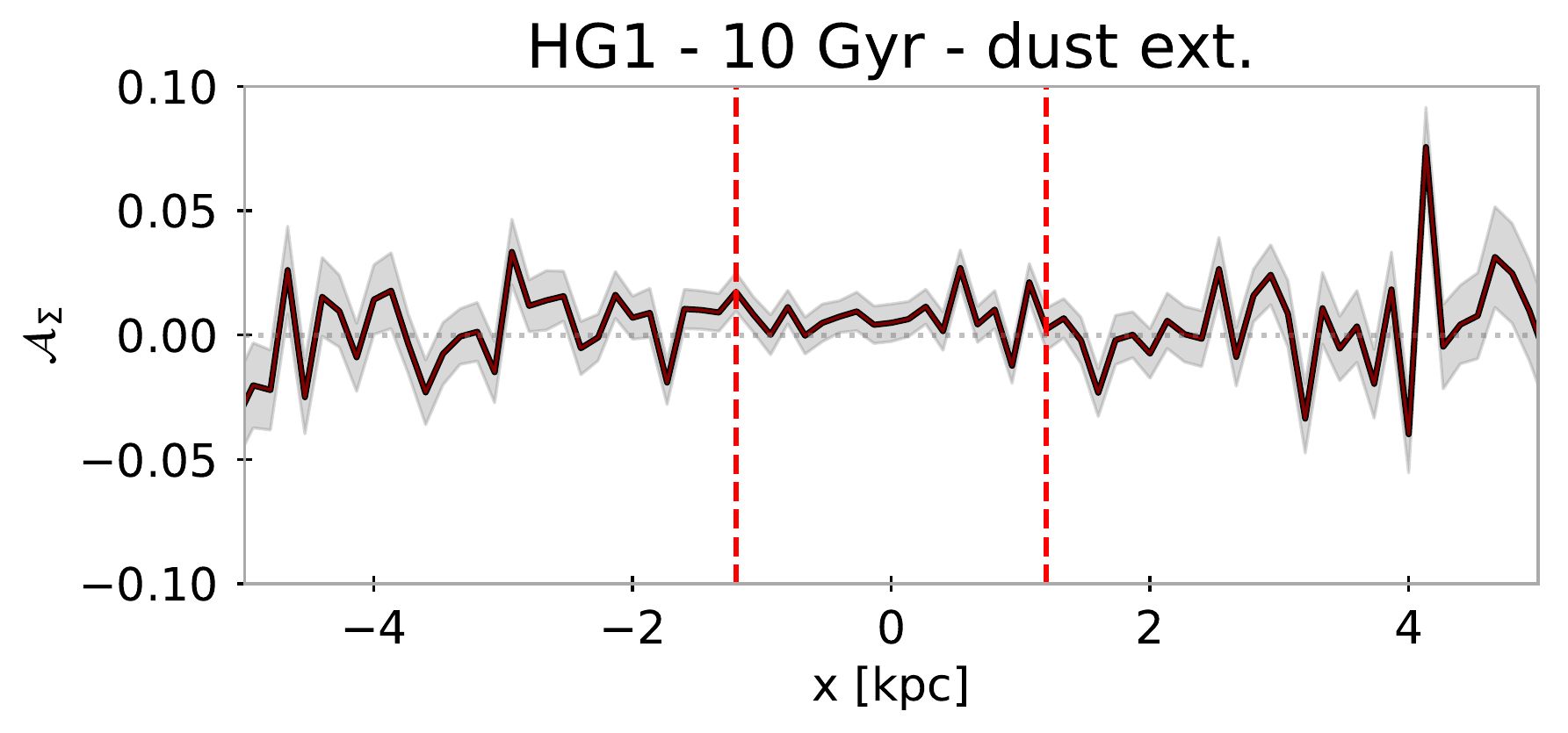}
    \caption{Mid-plane asymmetry profiles as in Fig.~\ref{fig:sim_dia} lower panel, but for model the HG1 (at 10 Gyr) after taking into account the effect of the dust (disc with $h_{\rm R,disc}=1$~kpc and $h_{\rm z,disc} = 0.1$~kpc + dust lanes), with different inclinations with respect to the side-on view ($i=90\degr$, first row; $\Delta i\pm3\degr$, second row row; $\Delta i\pm6\degr$, third row) and with a radially very extended dust disc with $h_{\rm R,disc}=1.5\times$ initial galaxy disc scalelength ($i=90\degr$, fourth row). Superimposed in red are shown the corresponding profiles obtained without taking into account the effect of the dust.}
    \label{fig:dust_pa_hg1}
\end{figure*}

\clearpage
\subsection*{Model D8}

Model D8 is a pure $N$-body one, presented in \cite{Anderson2022}. It suffered from a first and weak buckling event at 1.5 Gyr ($A_{\rm buck}\sim 0.015$ kpc), after which it develops a B/P bulge, and a second buckling event at 4 Gyr ($A_{\rm buck}\sim 0.04$ kpc), after which it develops strong asymmetries. After the second buckling event the bar grows constantly (from $A_{\rm bar}\sim 0.2$ to $\sim0.3$) till the end of the simulation (10 Gyr). In the following plots we describe the asymmetries visible at 5 Gyr (1 Gyr after the second buckling event). The B/P bulge has a semi-major axis of 3.5 kpc and a semi-minor one of 2.5 kpc.

\begin{figure}
    \centering
    \includegraphics[scale=0.5]{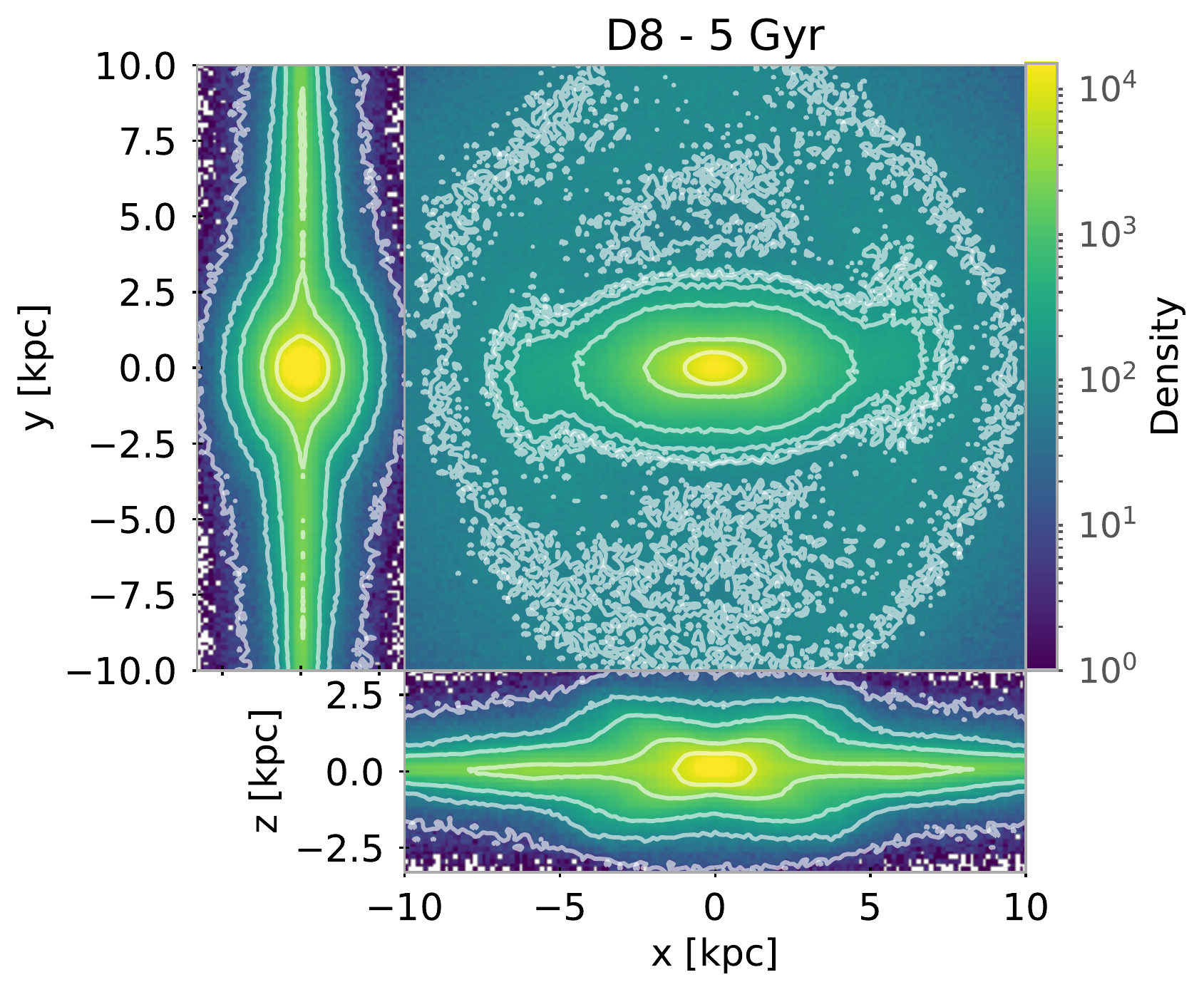}
    \caption{Same as Fig.~\ref{fig:sim_3d_view}, but for model D8.}
    \label{fig:d8_views}
\end{figure}

\begin{figure}
    \centering
    \includegraphics[scale=0.5]{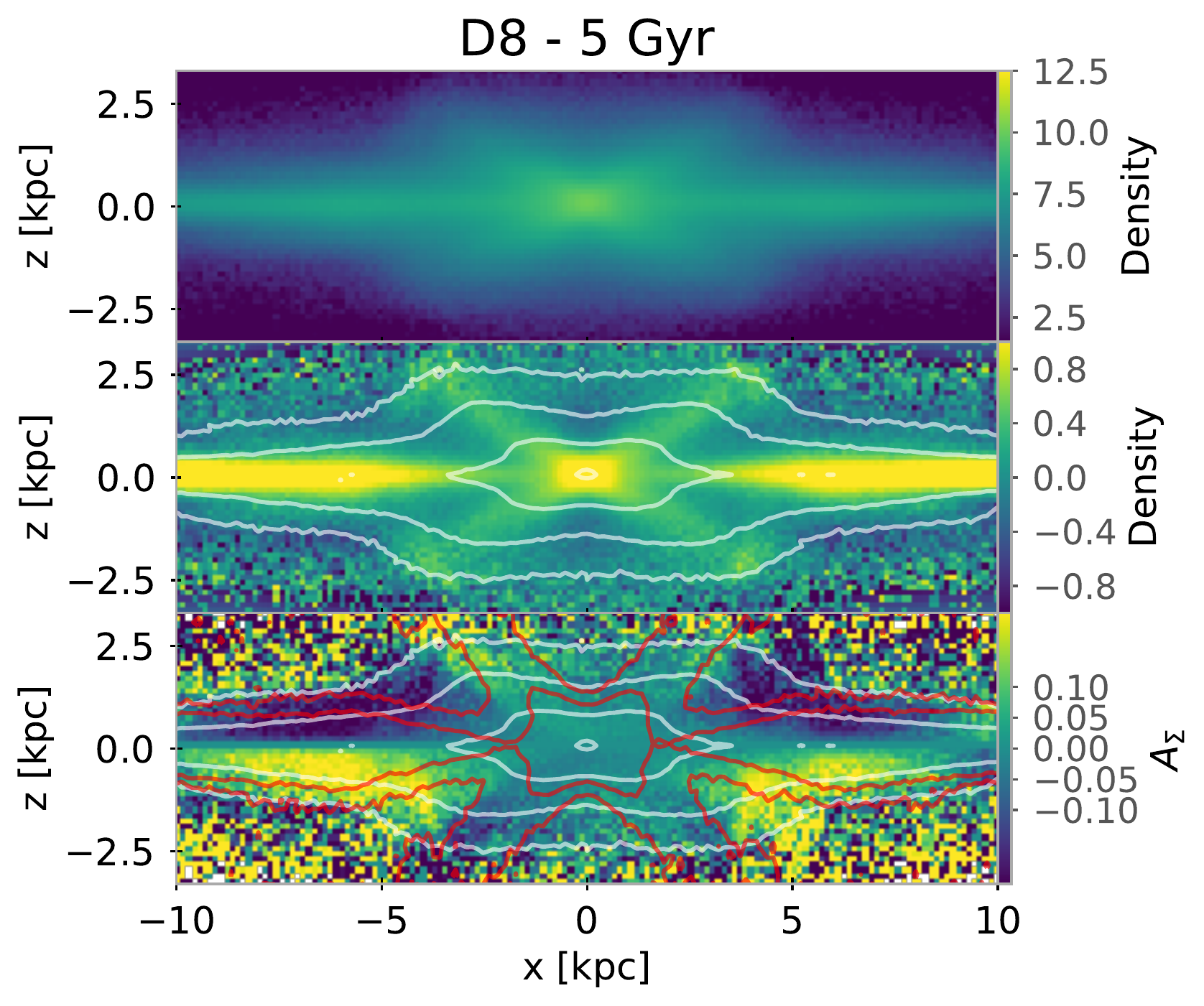}
    \caption{Same as Fig.~\ref{fig:dia}, but for model D8.}
    \label{fig:d8_side}
\end{figure}

\begin{figure}
    \centering
    \includegraphics[scale=0.4]{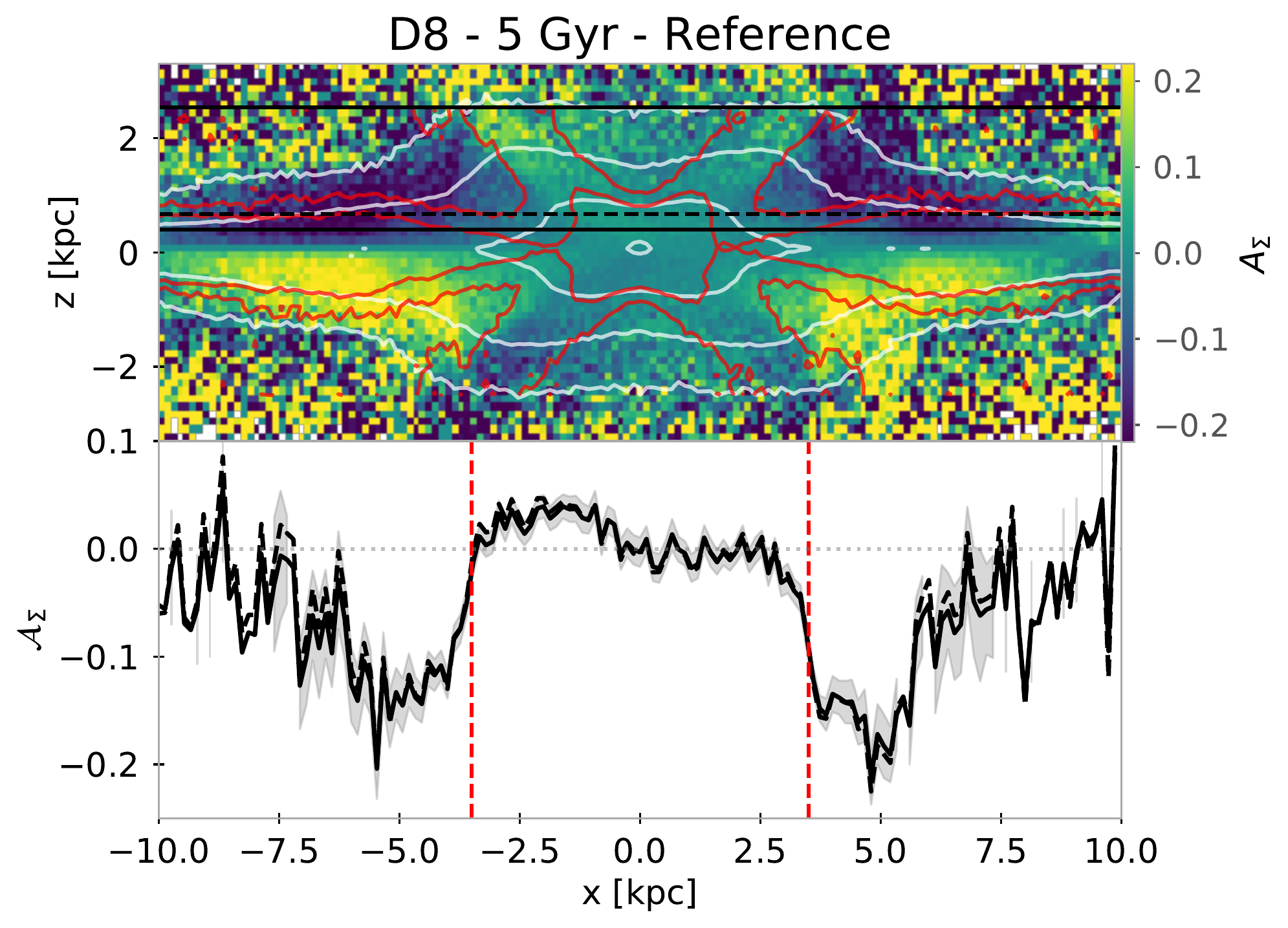}
    \caption{Same as Fig.~\ref{fig:sim_dia}, but for model D8 (at 5 Gyr) with $0.4< z<2.5$ kpc (solid line) and $0.7< z<2.5$ kpc (dashed line).}
    \label{fig:d8_diagn}
\end{figure}

\begin{figure}
    \centering
    \includegraphics[scale=0.5]{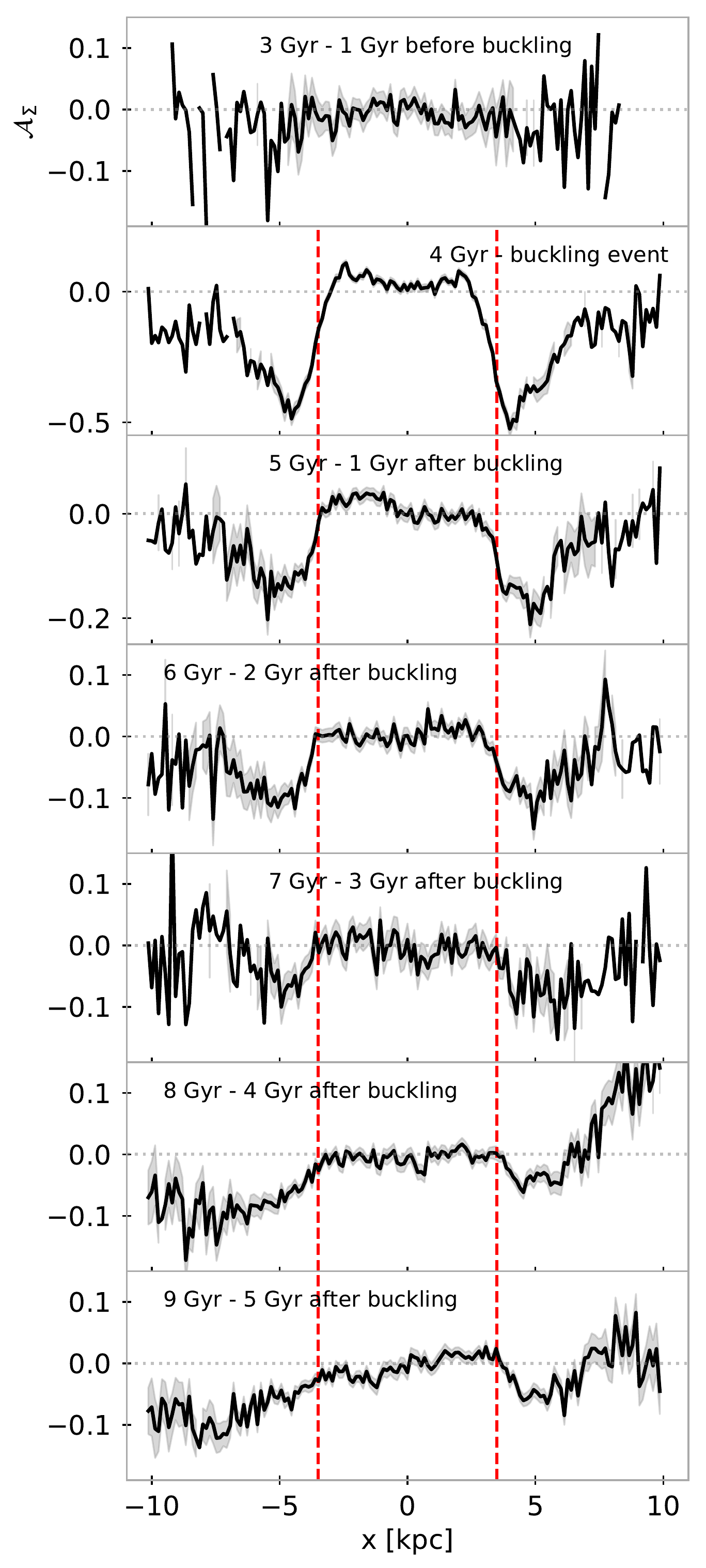}
    \caption{Same as Fig.~\ref{fig:evol} but for model D8.}
    \label{fig:d8_evol}
\end{figure}

\begin{figure*}
    \centering
    \includegraphics[scale=0.4]{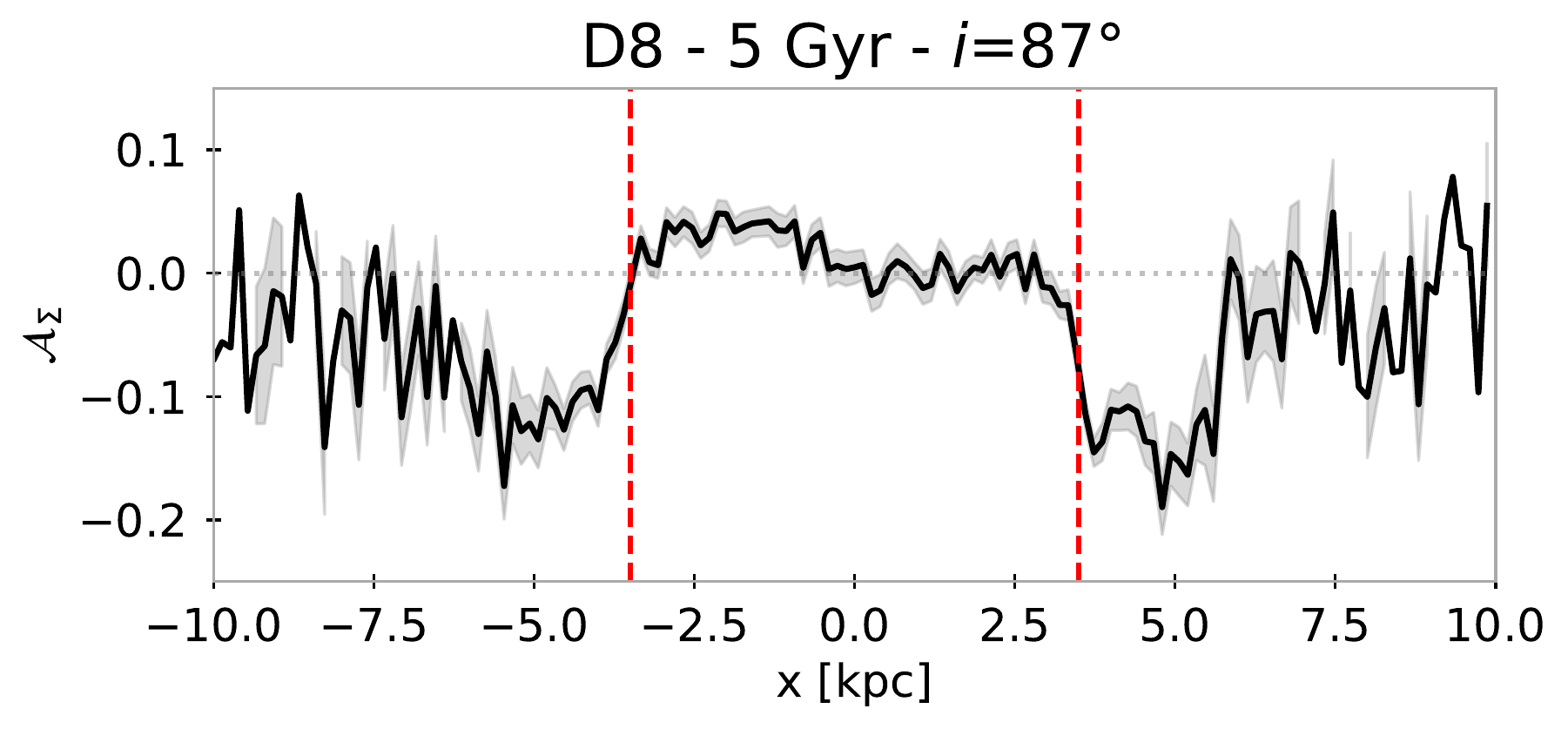}
    \includegraphics[scale=0.4]{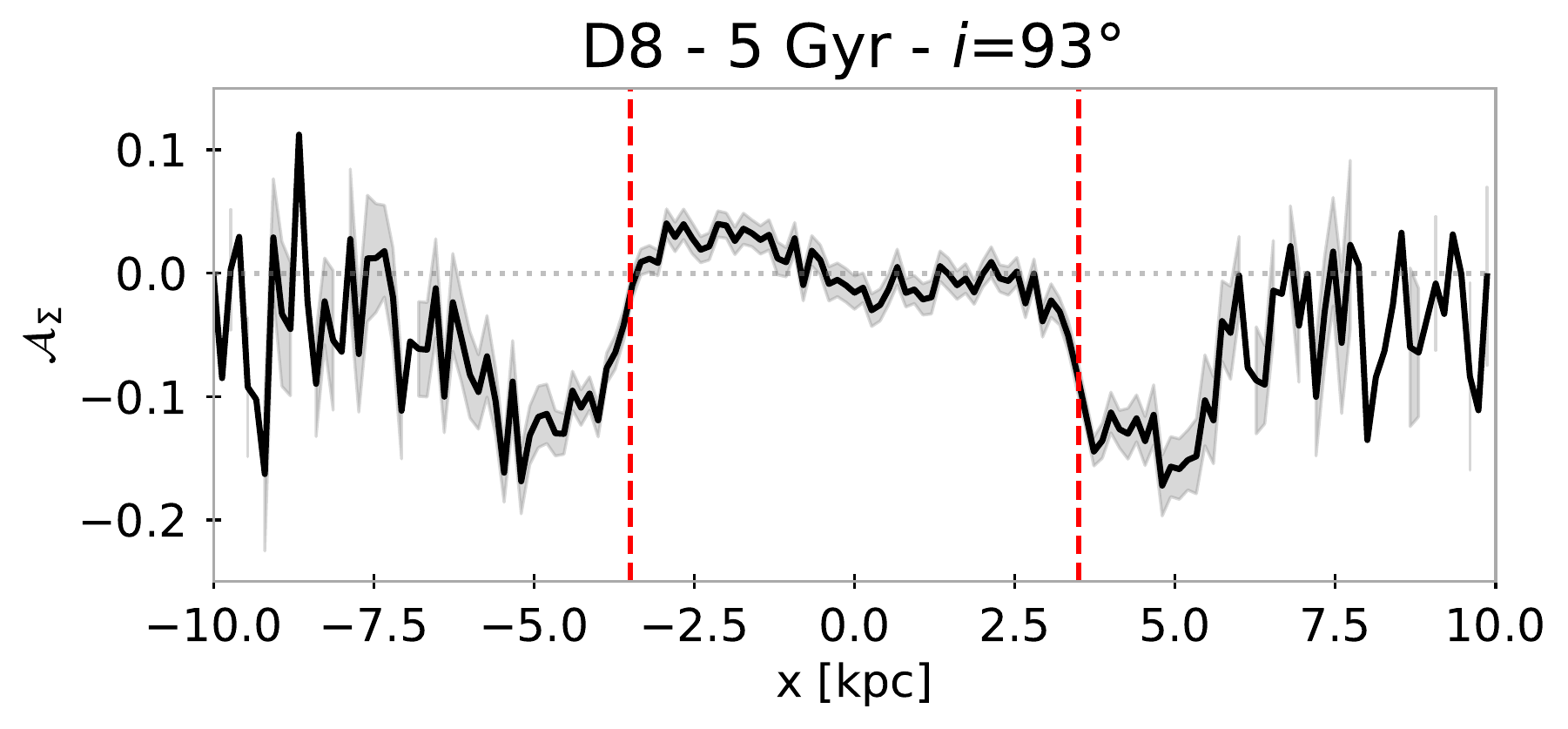}
    \includegraphics[scale=0.4]{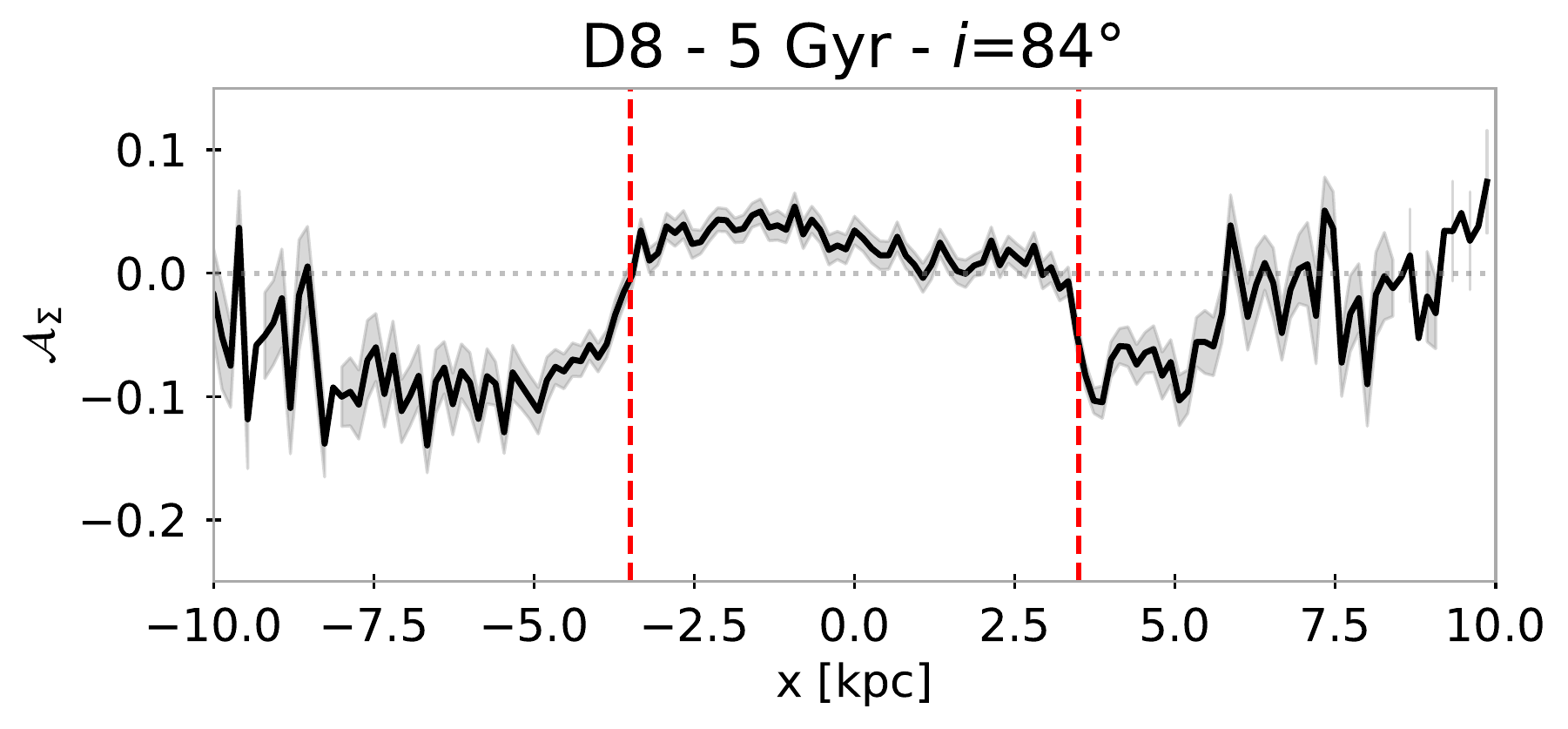}
    \includegraphics[scale=0.4]{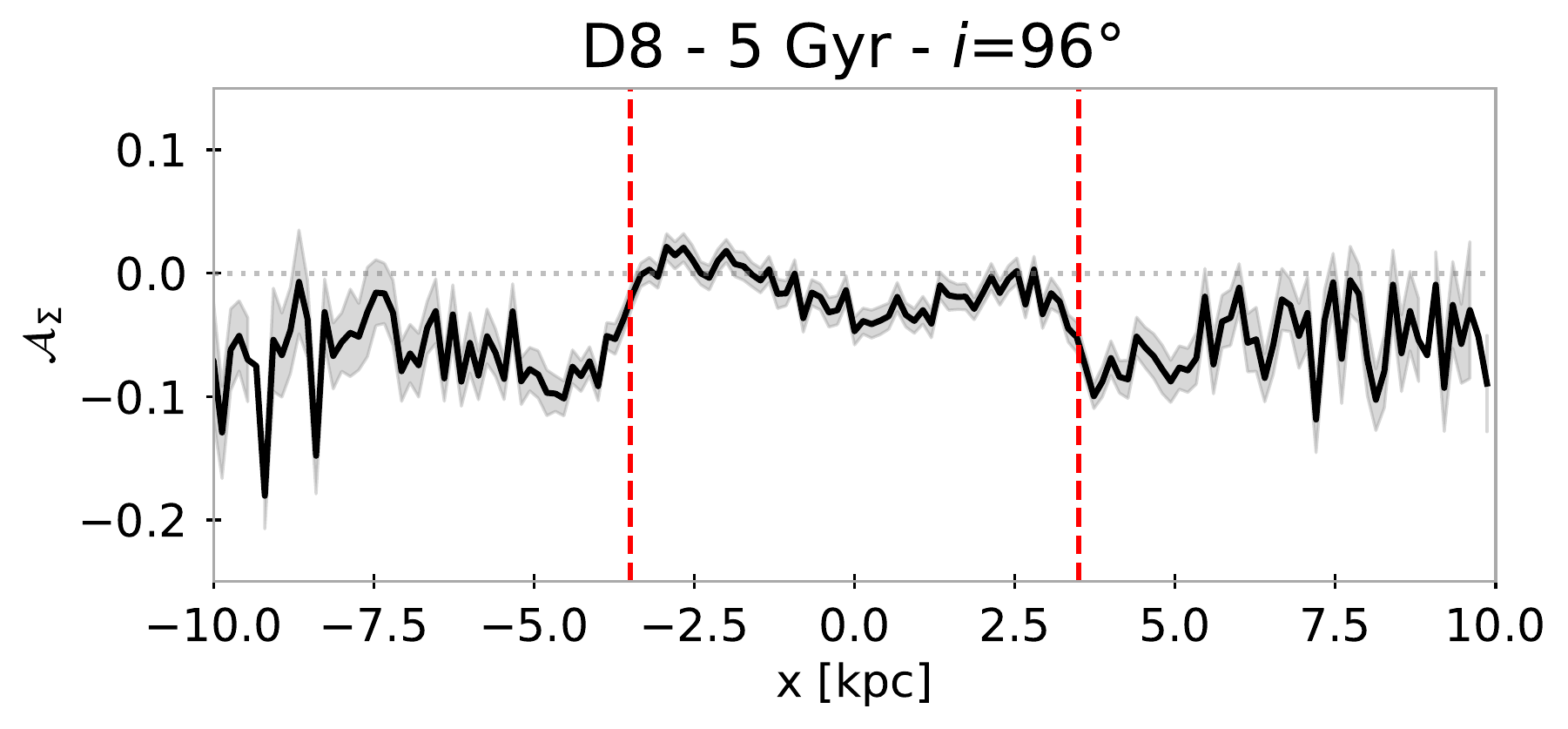}
    \includegraphics[scale=0.4]{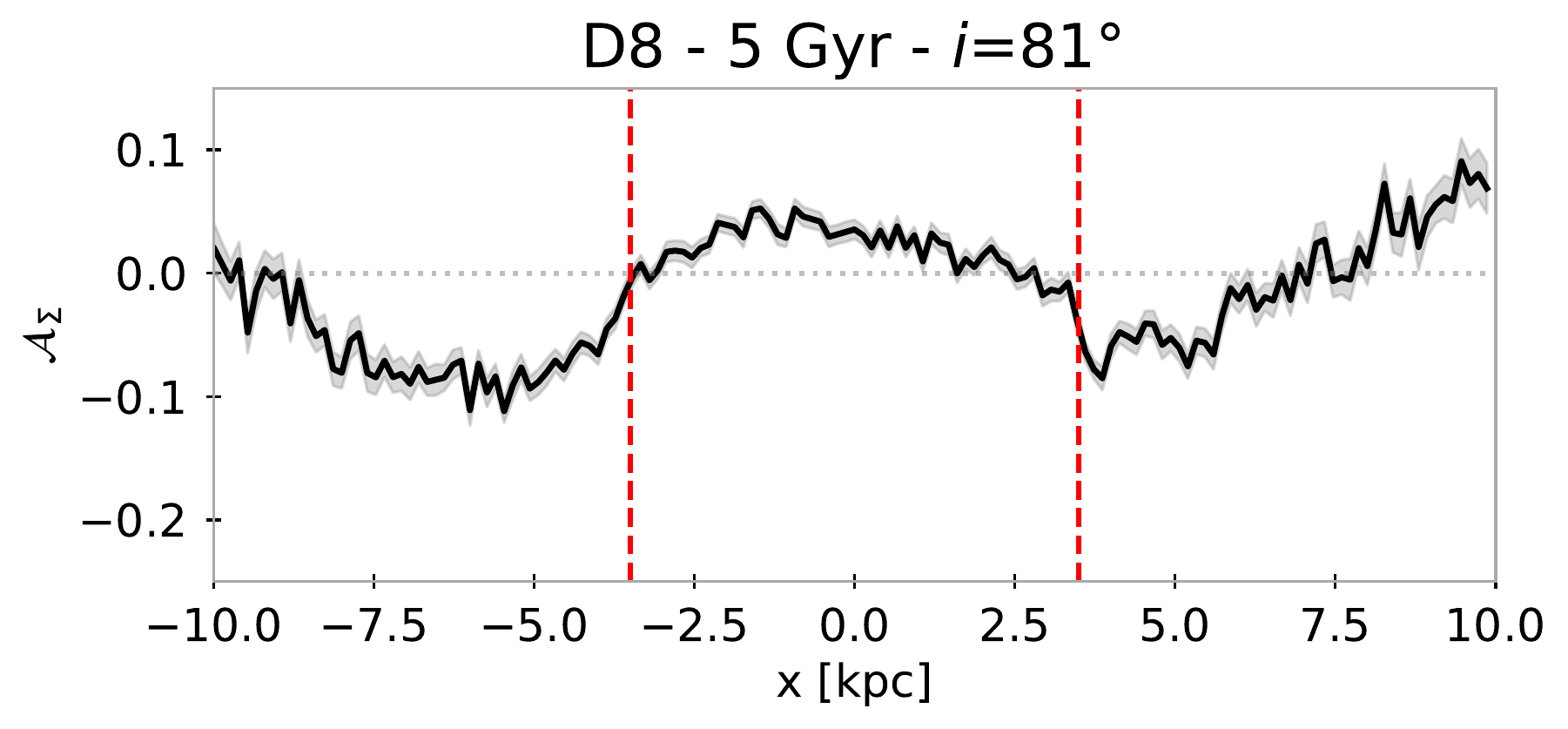}
    \includegraphics[scale=0.4]{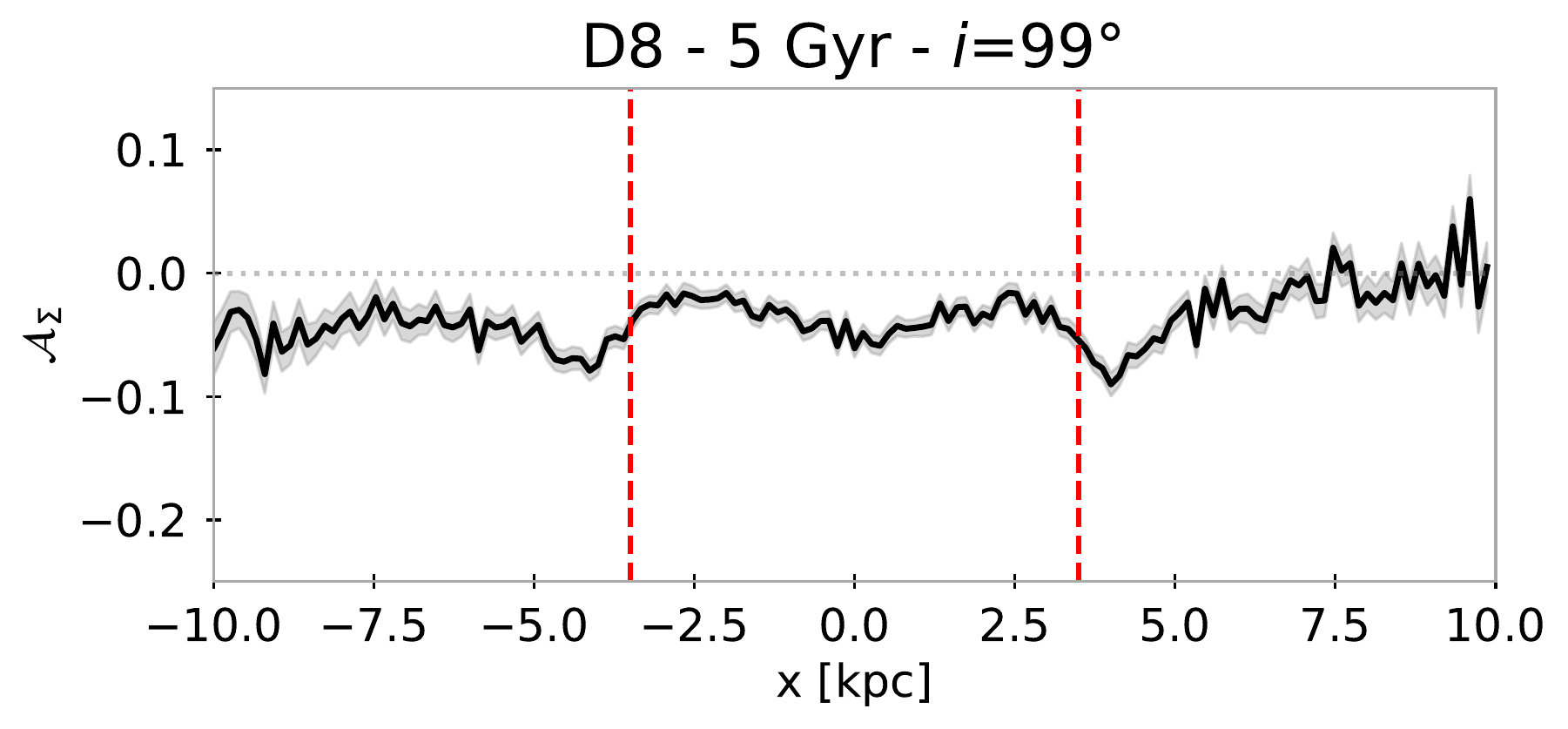}
    \vspace{-0.2cm}
    \caption{Same as Fig~\ref{fig:sim_d5_inc}, but for model D8.}
    \label{fig:d8_inc}
\end{figure*}

\begin{figure}
    \centering
    \includegraphics[scale=0.4]{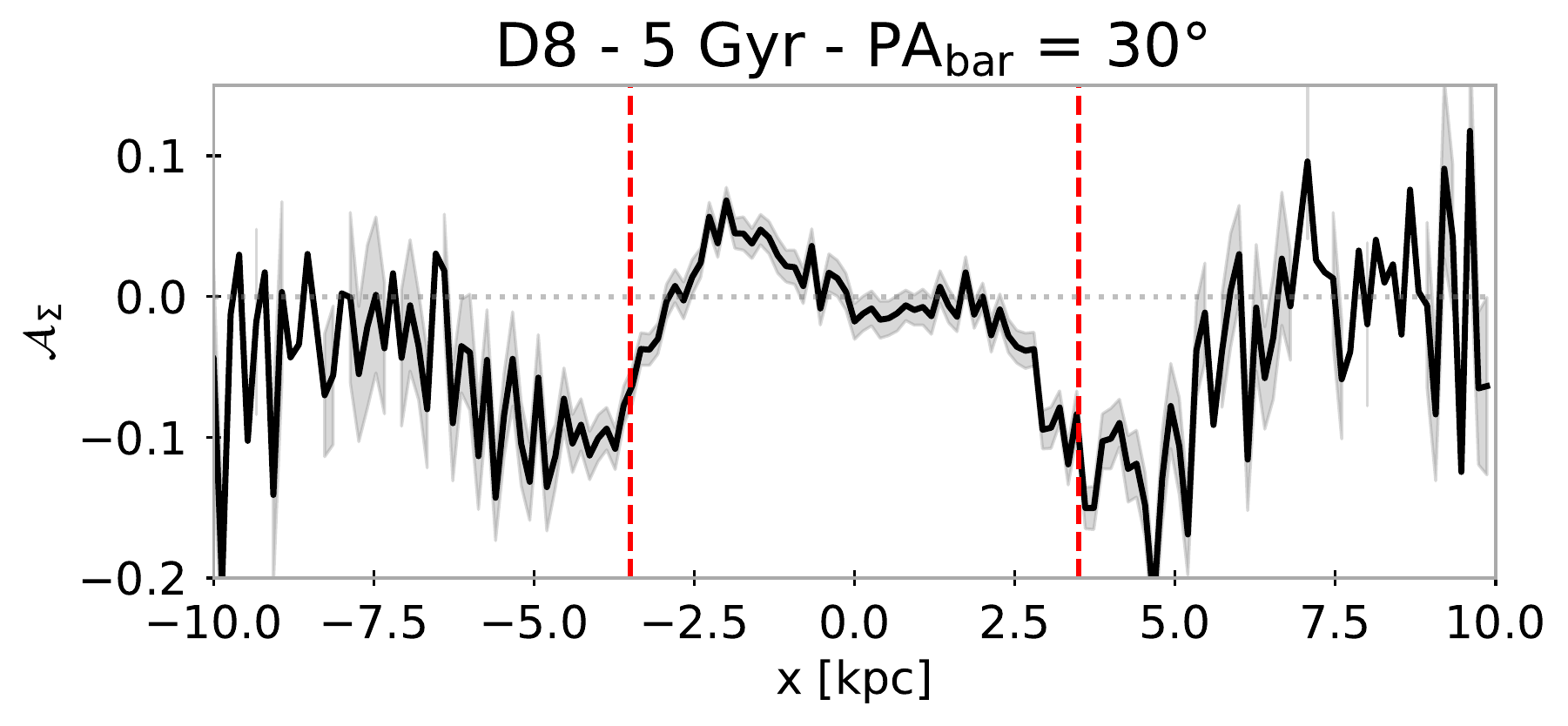}
    \includegraphics[scale=0.4]{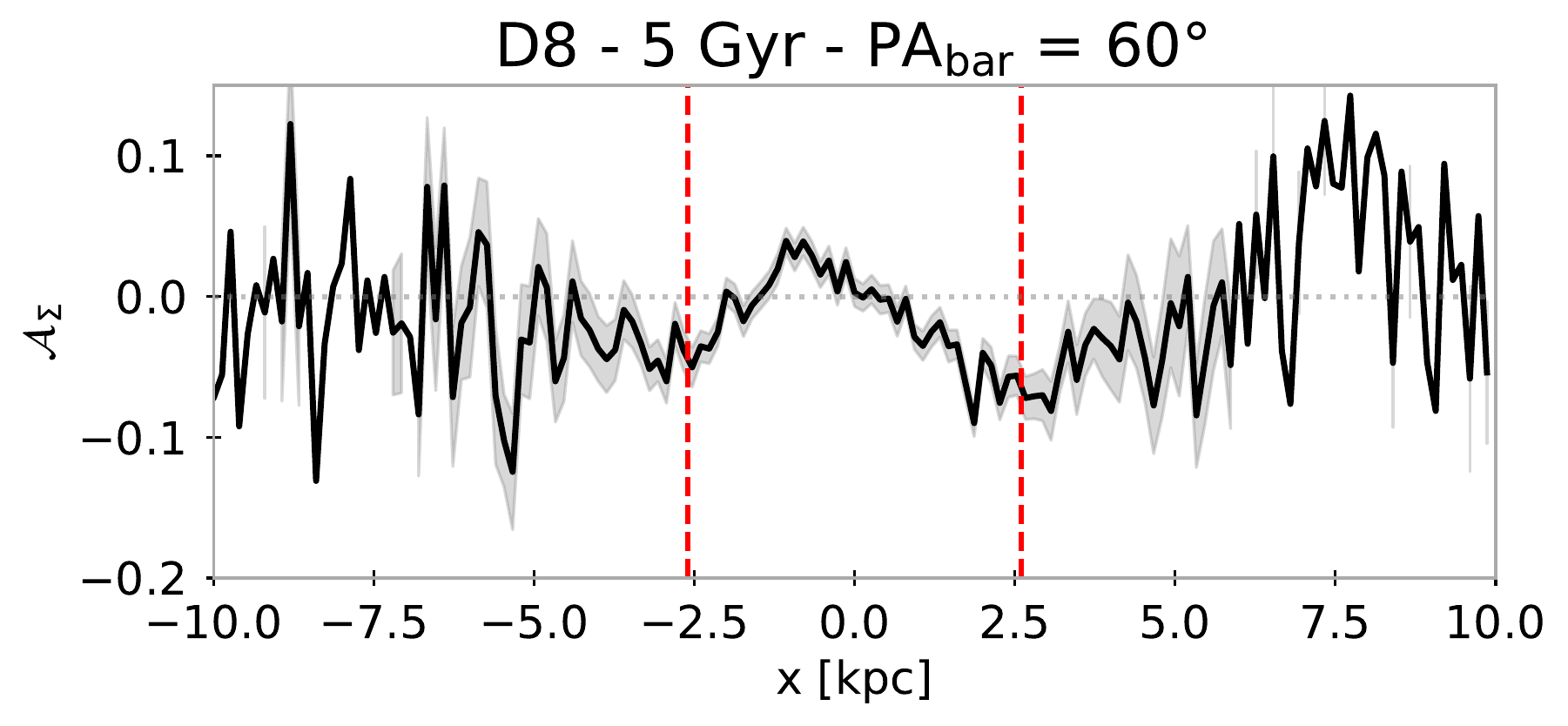}
    \includegraphics[scale=0.4]{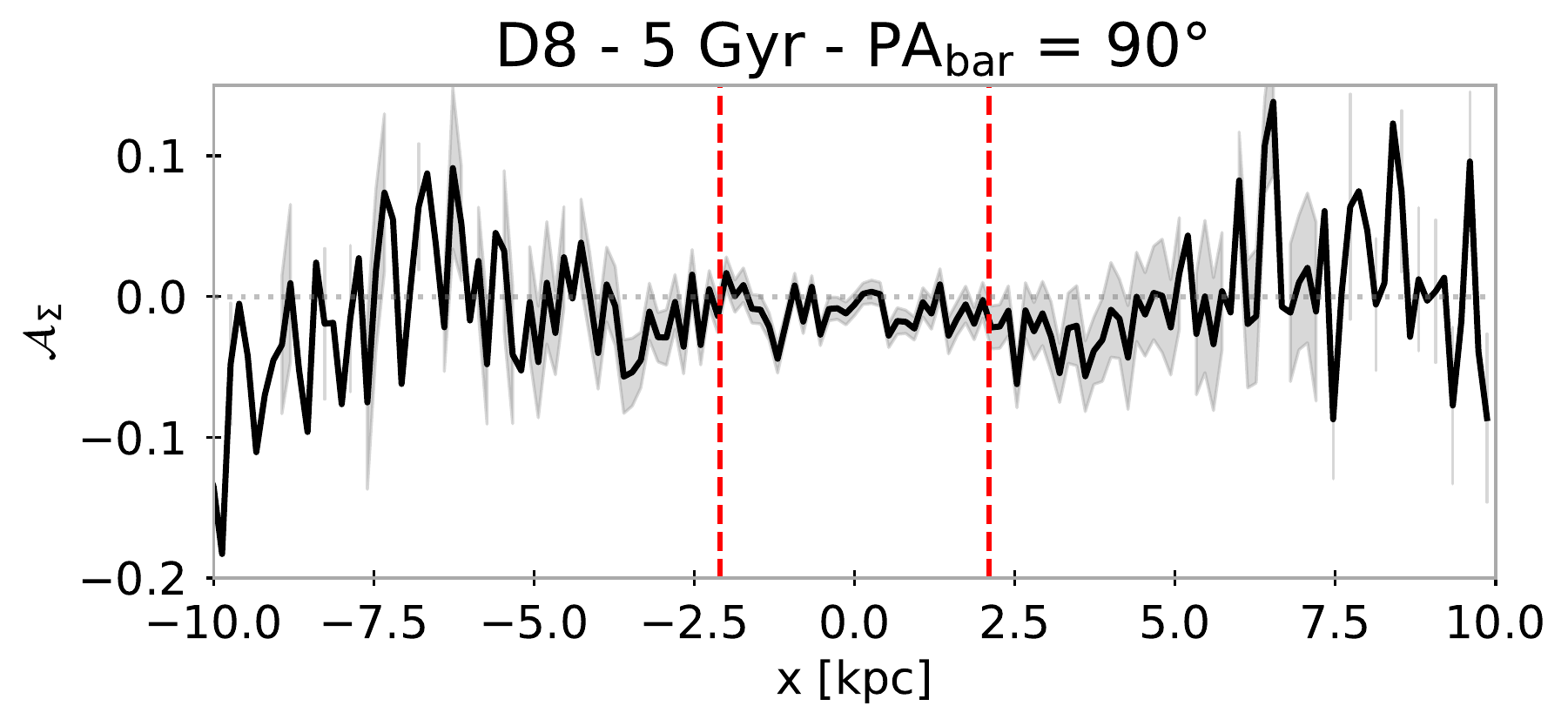}
    \caption{Same as Fig~\ref{fig:sim_bar}, but for model D8.}
    \label{fig:d8_bar}
\end{figure}

\begin{figure}
    \centering
    \includegraphics[scale=0.4]{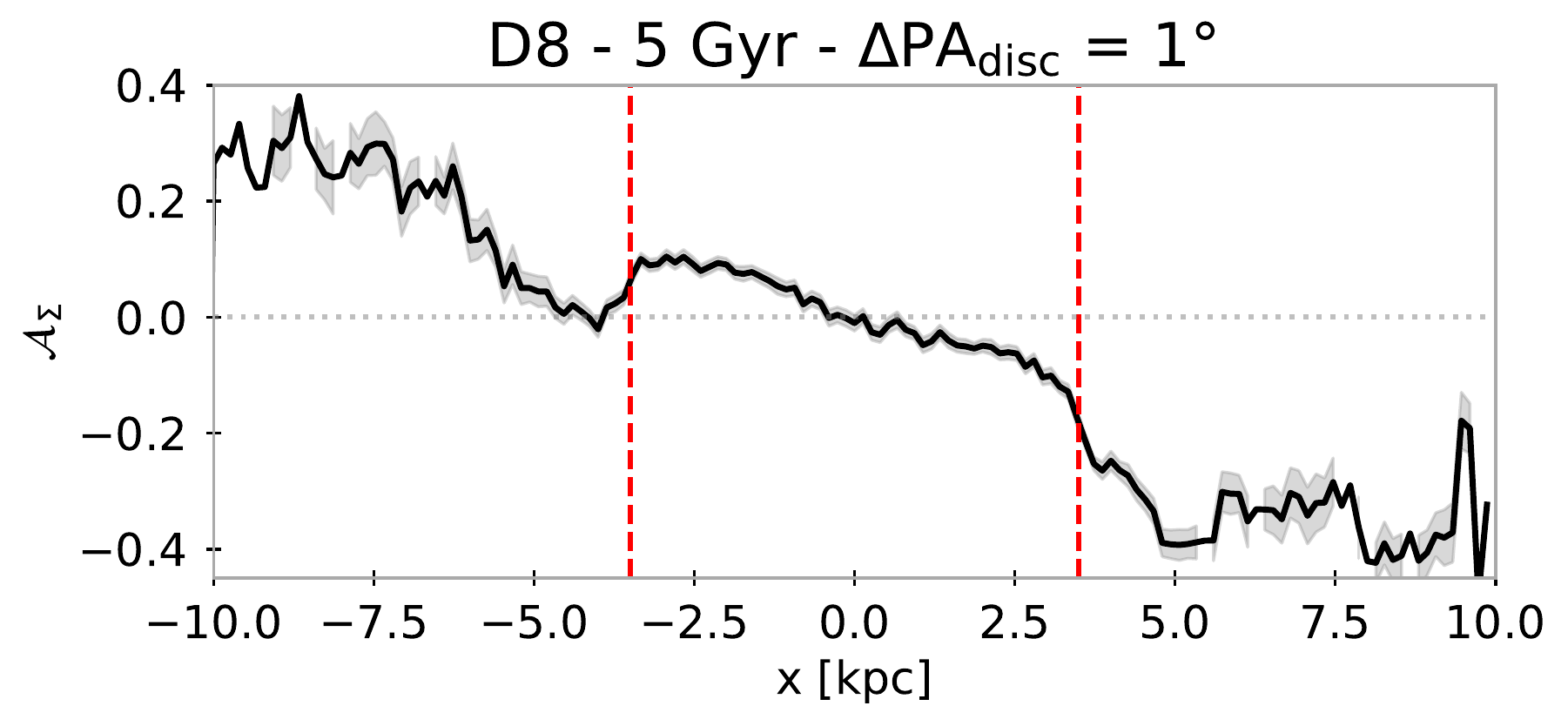}
    \includegraphics[scale=0.4]{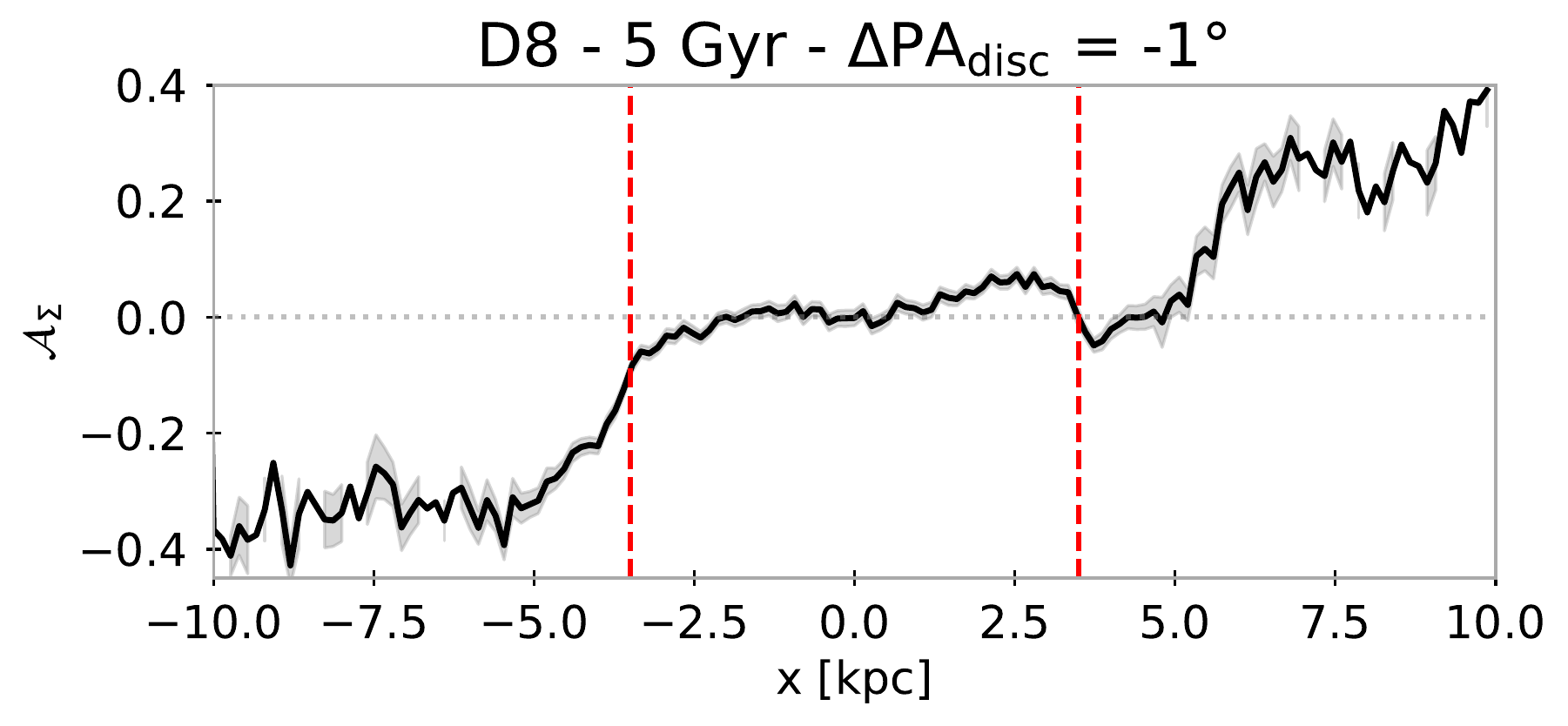}
    \caption{Same as Fig~\ref{fig:sim_pa}, but for model D8.}
    \label{fig:d8_pa}
\end{figure}

\begin{figure}
    \centering
    \includegraphics[scale=0.4]{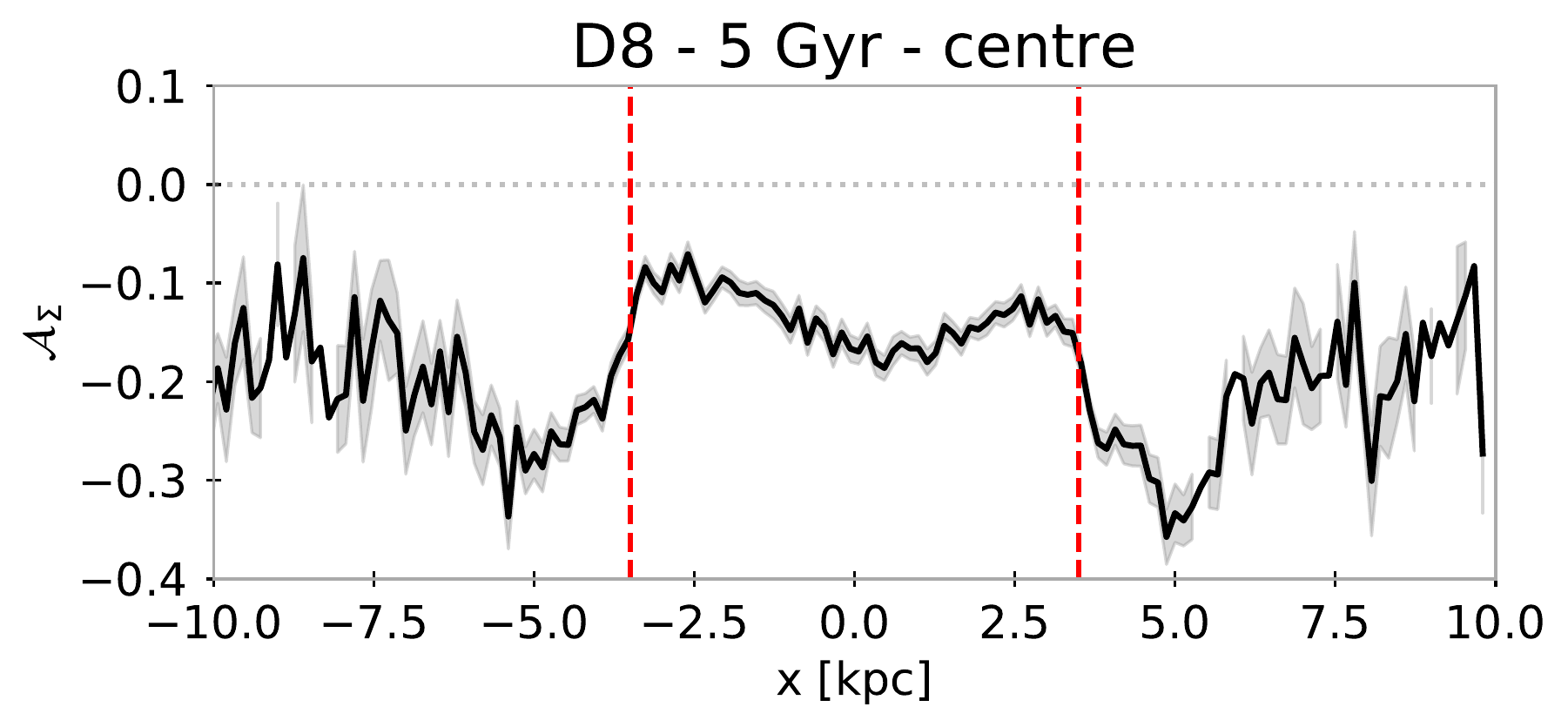}
    \caption{Same as Fig~\ref{fig:centre}, but for model D8.}
    \label{fig:d8_centre}
\end{figure}

\clearpage
\subsection*{Model T1}

Model T1 is a pure $N$-body one, presented in \cite{Anderson2022}. It suffered from a mild buckling event at 1.5 Gyr ($A_{\rm buck}\sim 0.02$), after which it develops a B/P bulge while the bar grows constantly (from $A_{\rm bar}\sim 0.1$ to $\sim0.25$) till the end of the simulation (10 Gyr). The model develops strong asymmetries just after the buckling event (at 2 Gyr) but they quickly disappear (they are no longer visible at 5 Gyr). In the following plots we describe the asymmetries visible at 2 Gyr and in comparison the corresponding plots at 4 Gyr. The B/P bulge has a semi-major axis of 2.5 kpc and a semi-minor one of 2.1 kpc.

\begin{figure}
    \centering
    \includegraphics[scale=0.5]{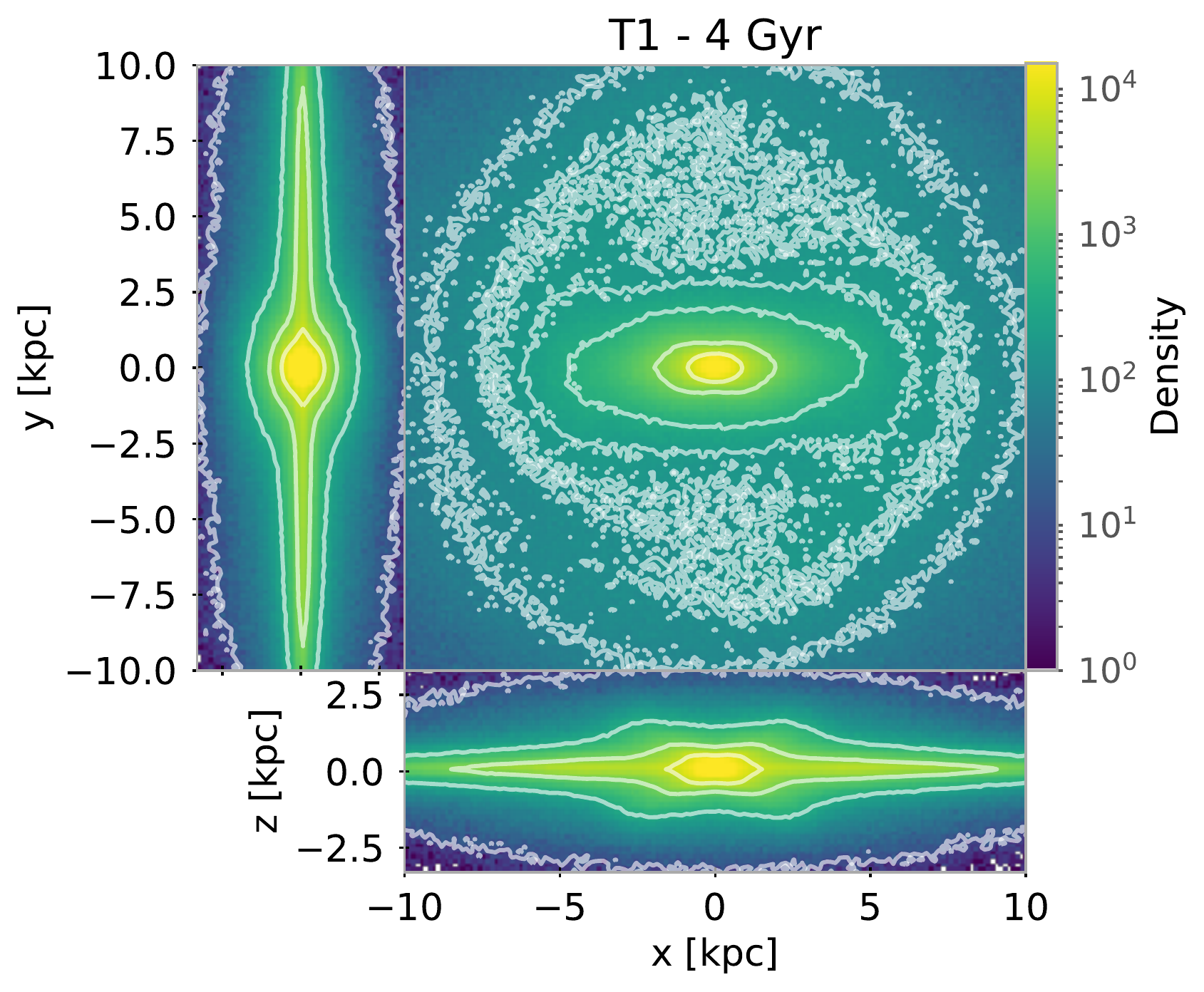}
    \caption{Same as Fig.~\ref{fig:sim_3d_view}, but for model T1 at 4 Gyr, i.e. 1.5 Gyr after the mild buckling event.}
\end{figure}

\begin{figure*}
    \centering
    \includegraphics[scale=0.5]{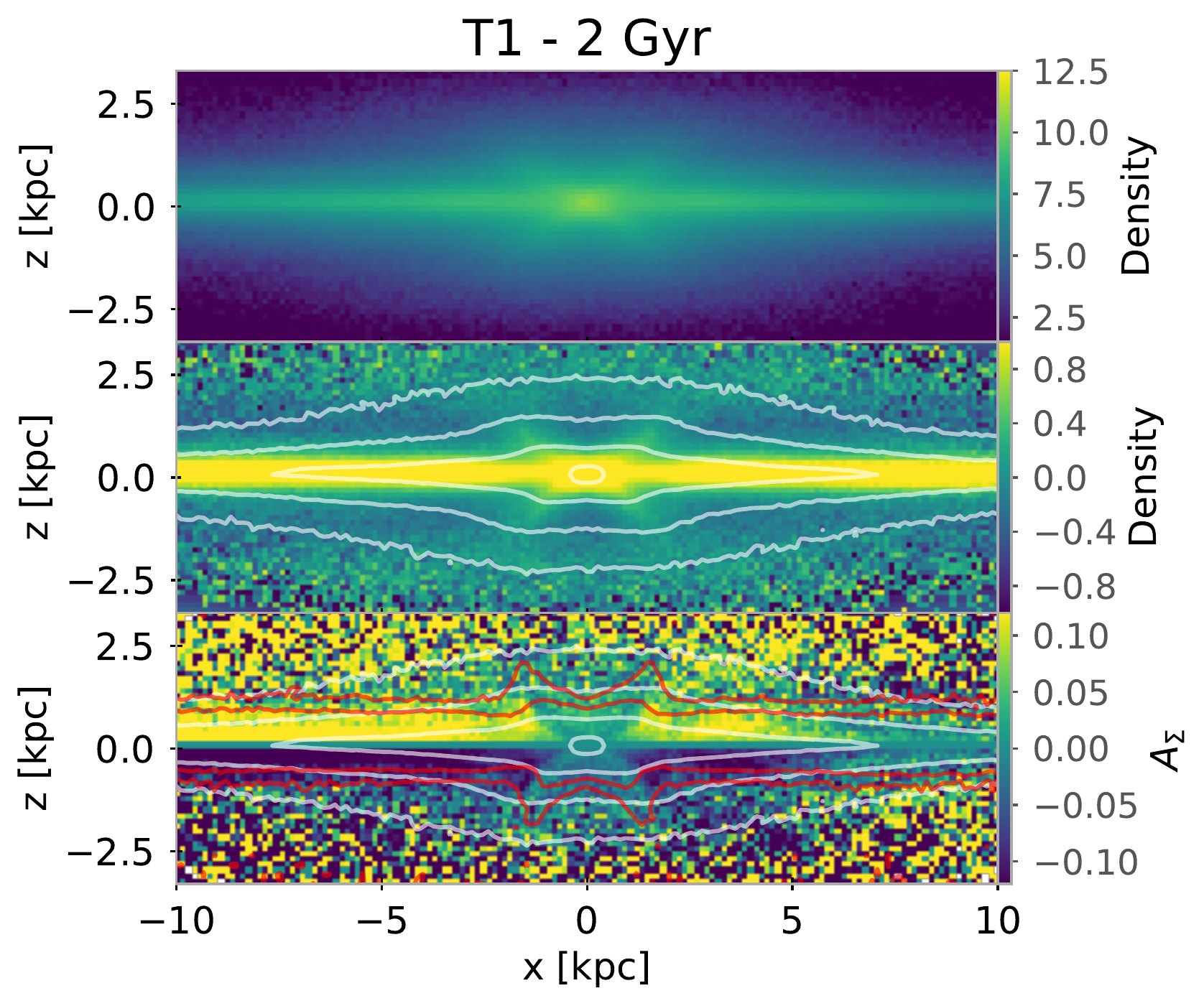}
    \includegraphics[scale=0.5]{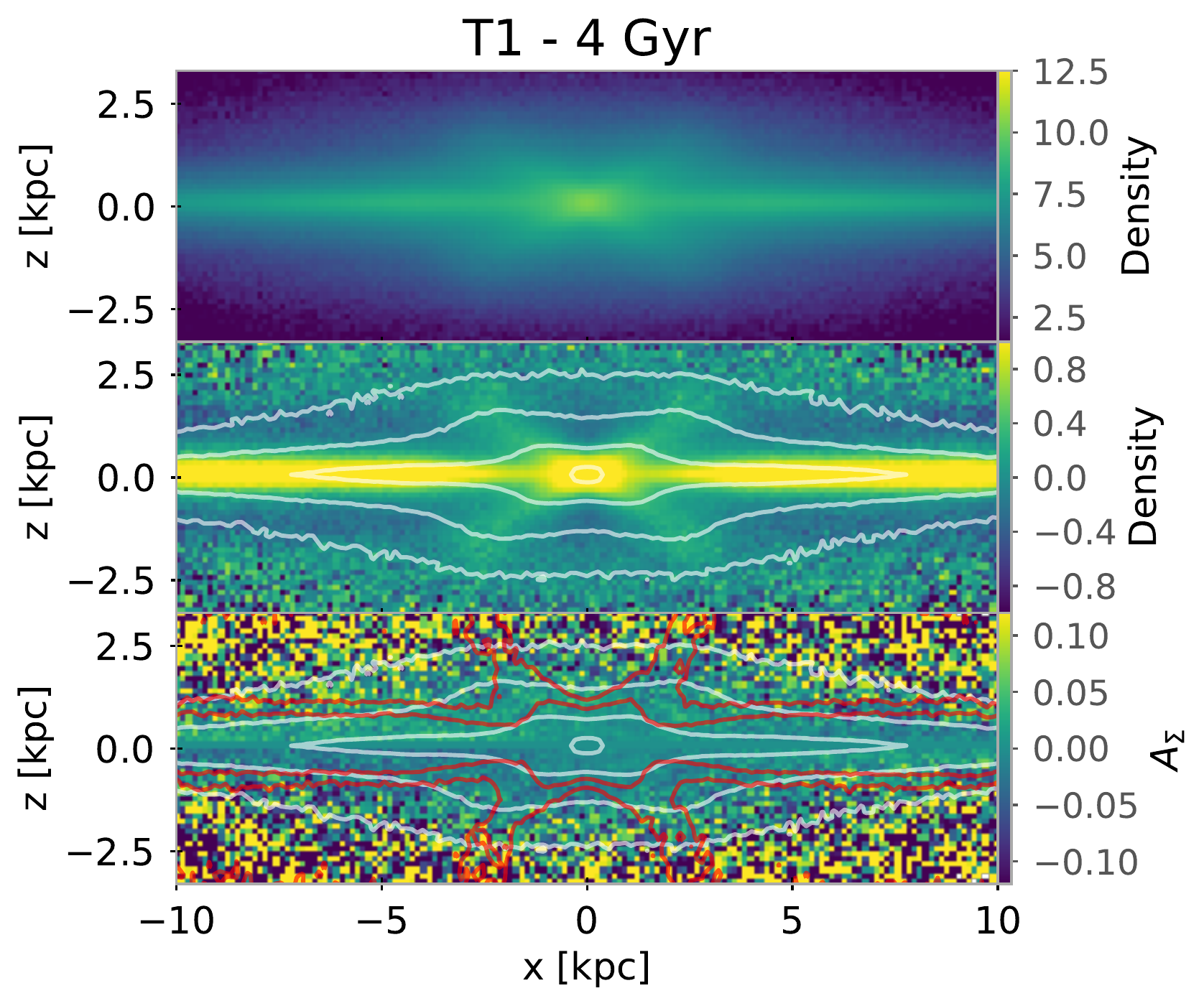}
    \caption{Same as Fig.~\ref{fig:dia}, but for model T1 at 2 Gyr (left-hand column) and 4 Gyr (right-hand column).}
\end{figure*}

\begin{figure*}
    \centering
    \includegraphics[scale=0.4]{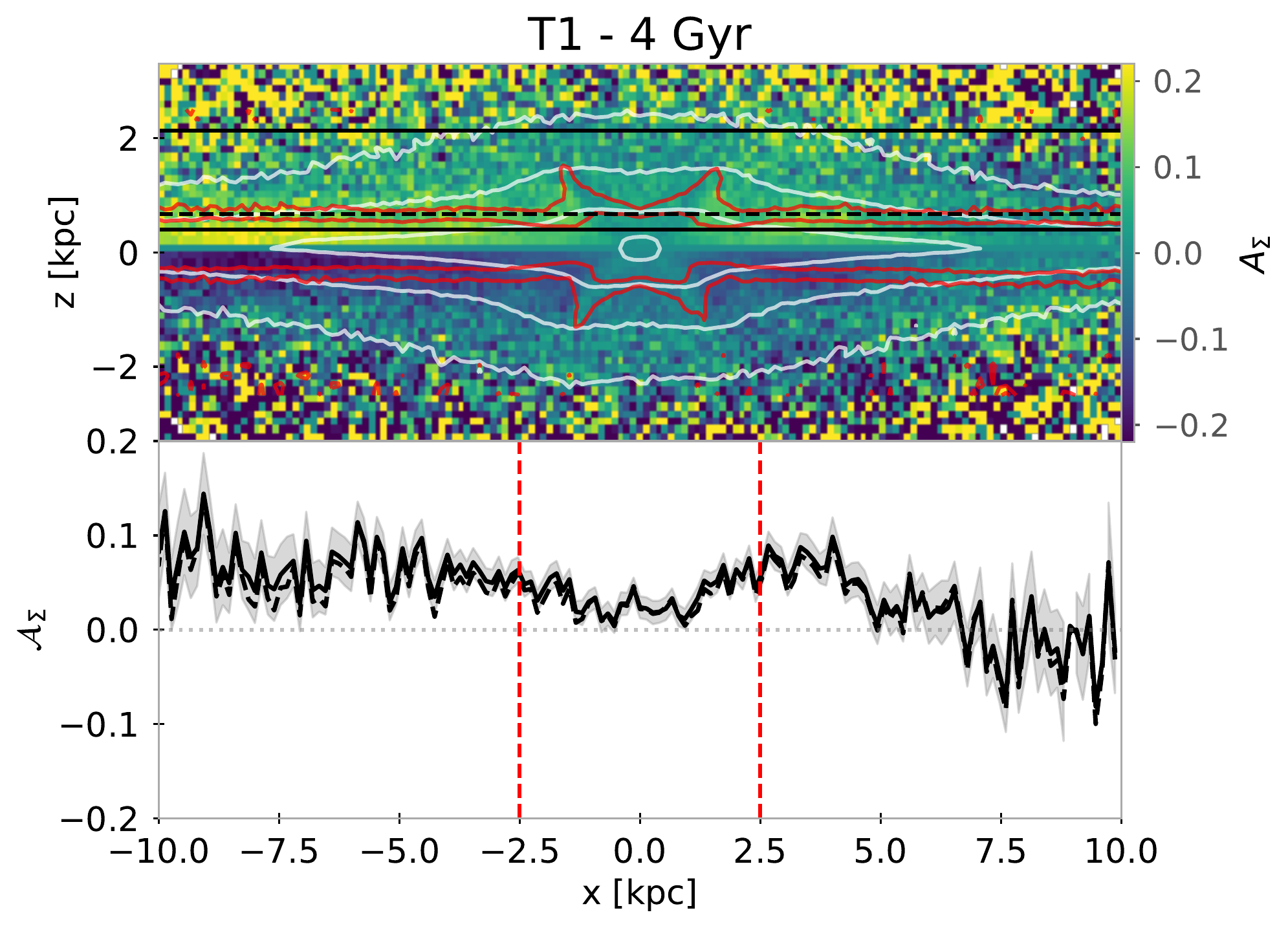}
    \includegraphics[scale=0.4]{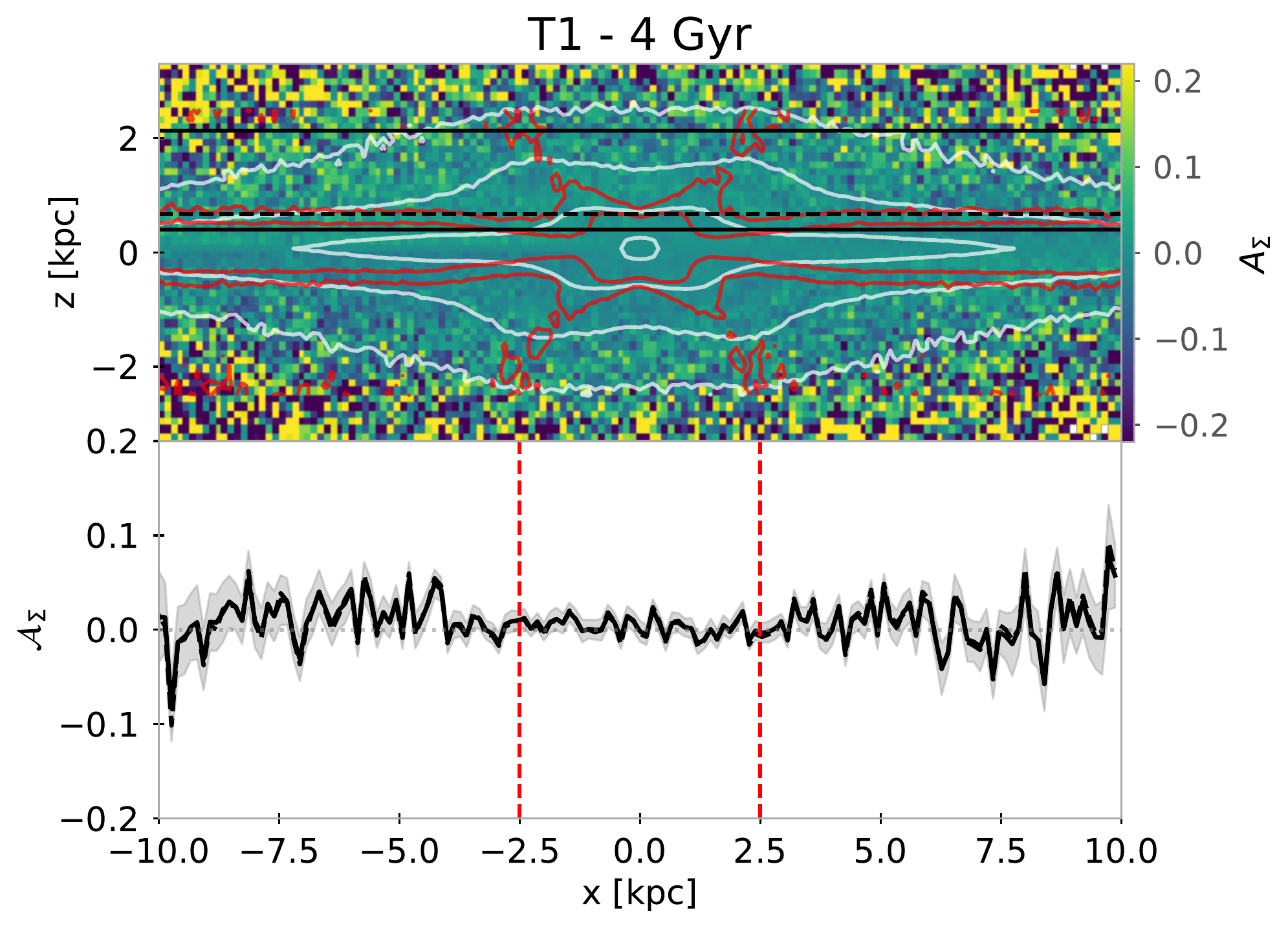}
    \caption{Same as Fig.~\ref{fig:sim_dia}, but for model T1 at 2 Gyr (left-hand column) and 4 Gyr (right-hand column) with $0.3<z<2.1$ kpc.}
\end{figure*}

\clearpage
\subsection*{Model SD1}

Model SD1 is a pure $N$-body one, presented in \citealt{Anderson2022}. It develops a B/P bulge without suffering a strong buckling event ($A_{\rm buck}$ remains below 0.008 kpc). The bar grows constantly reaching $A_{\rm bar}\sim0.25$ at the end of the simulation (10 Gyr). No asymmetries are visible during the formation and evolution of the B/P bulge. Here we present the results at 6 Gyr: the disc hosts asymmetric spiral arms, while the outermost region of the disc is warped: this explains the asymmetry appearing at the end of the evolution. The B/P bulge has a semi-major axis of 3.3 kpc and a semi-minor one of 2.6 kpc. 

\begin{figure}
    \centering
    \includegraphics[scale=0.5]{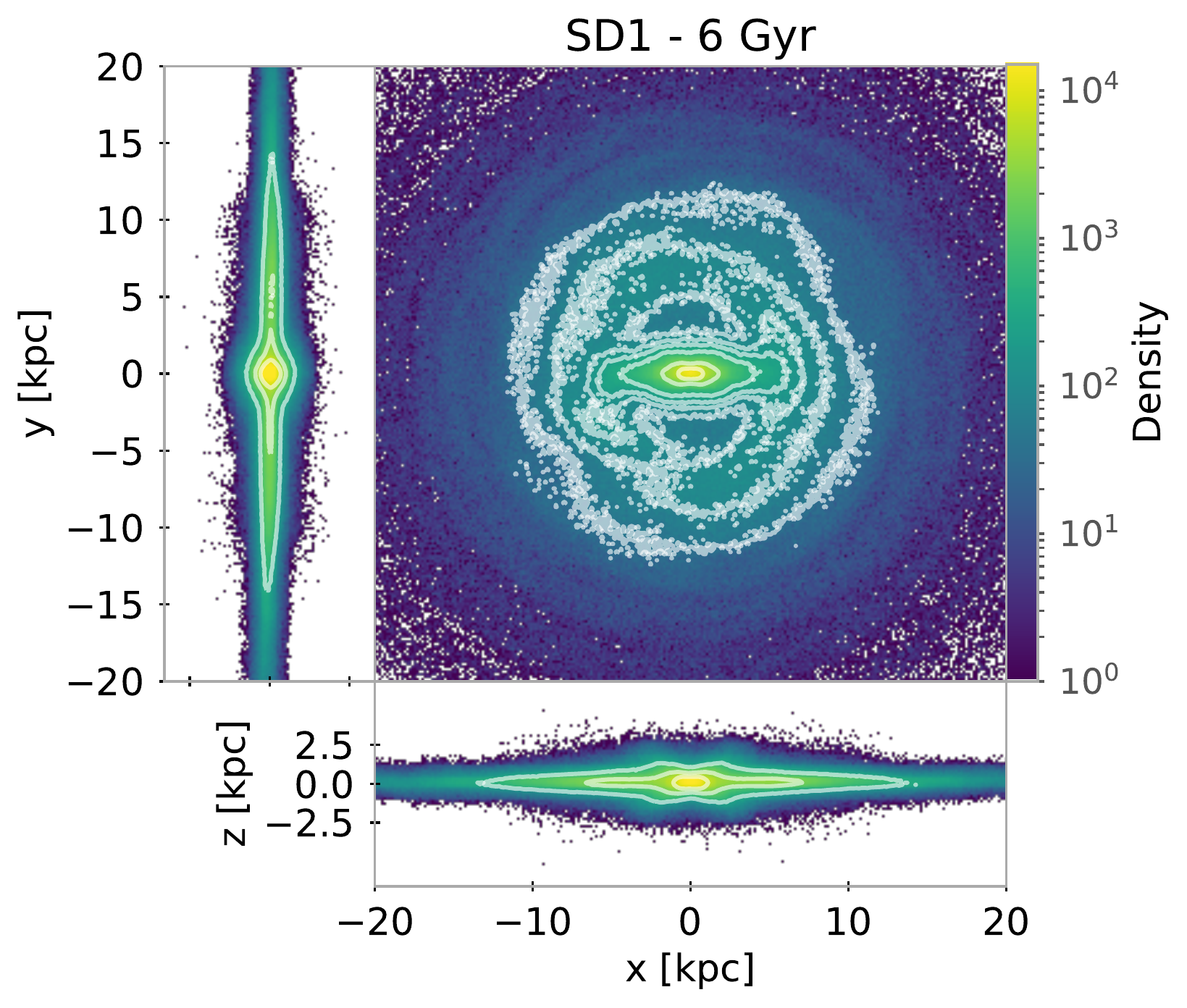}
    \caption{Same as Fig.~\ref{fig:sim_3d_view}, but for model SD1.}
\end{figure}

\begin{figure}
    \centering
    \includegraphics[scale=0.5]{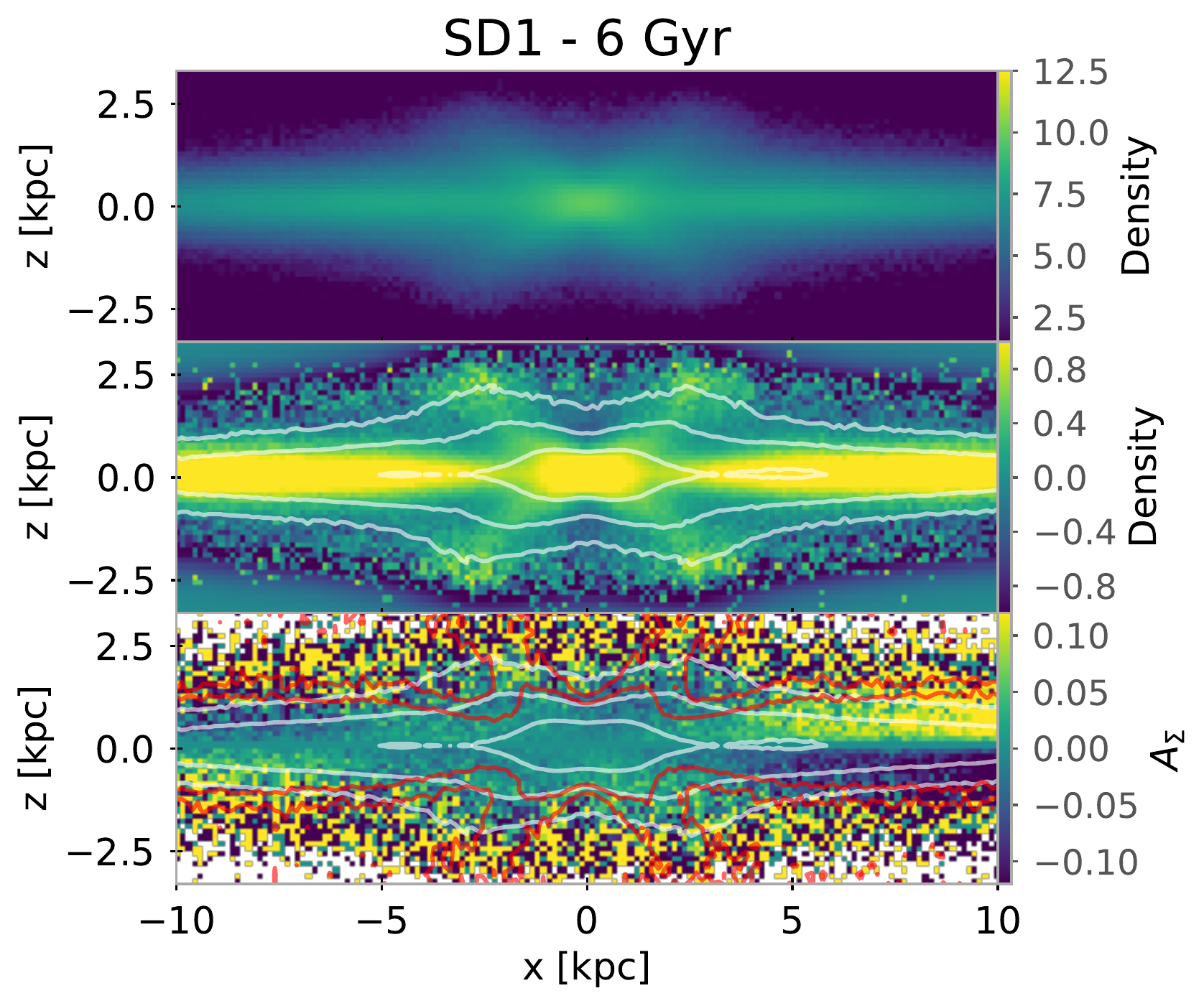}
    \caption{Same as Fig.~\ref{fig:dia}, but for model SD1.}
\end{figure}

\begin{figure}
    \centering
    \includegraphics[scale=0.4]{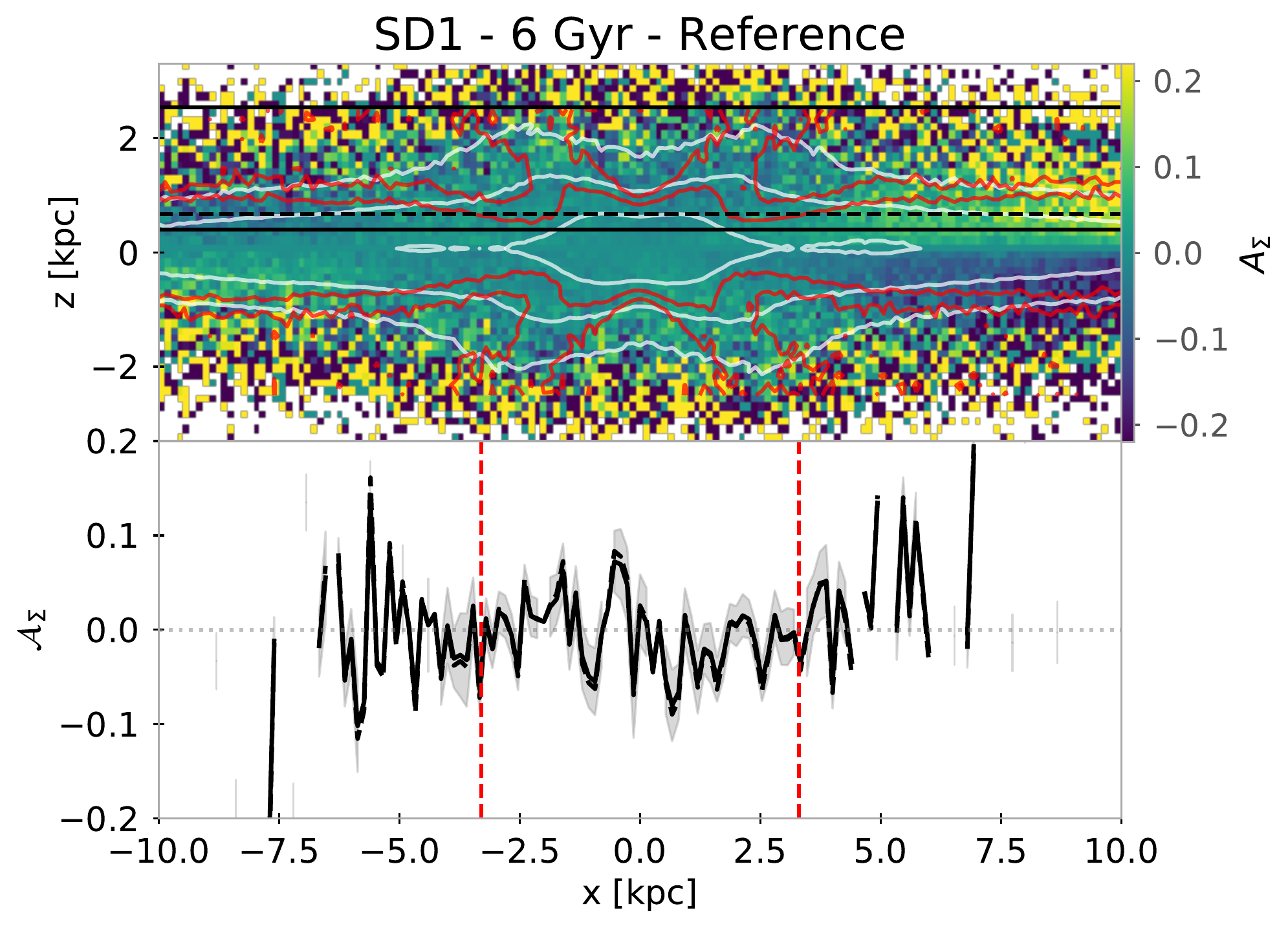}
    \caption{Same as Fig.~\ref{fig:sim_dia}, but for model SD1 (at 6 Gyr) with $0.7<z<2.6$~kpc (solid line) and $0.4< z<2.6$~kpc (dashed line).}
\end{figure}

\begin{figure*}
    \centering
    \includegraphics[scale=0.4]{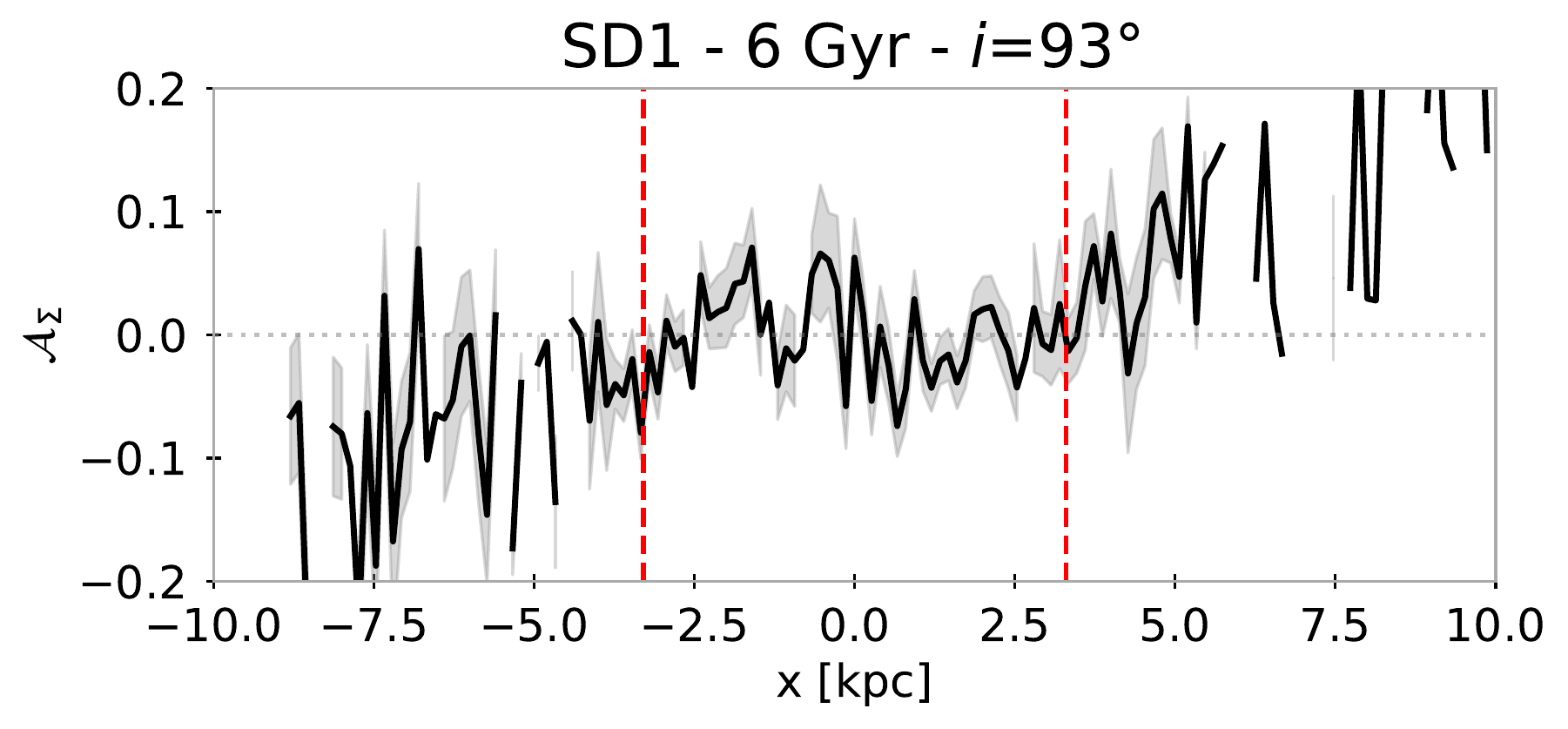}
    \includegraphics[scale=0.4]{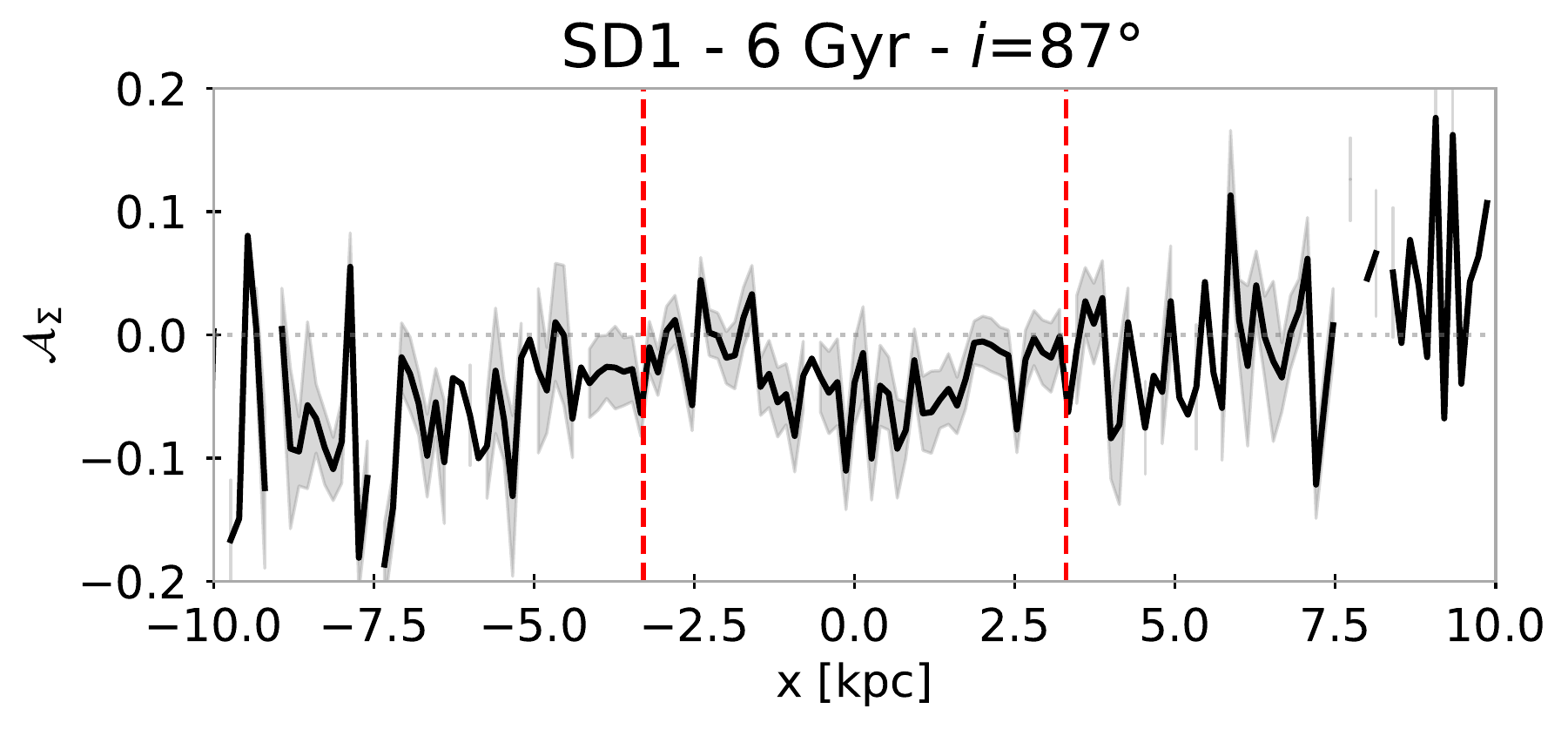}
    \includegraphics[scale=0.4]{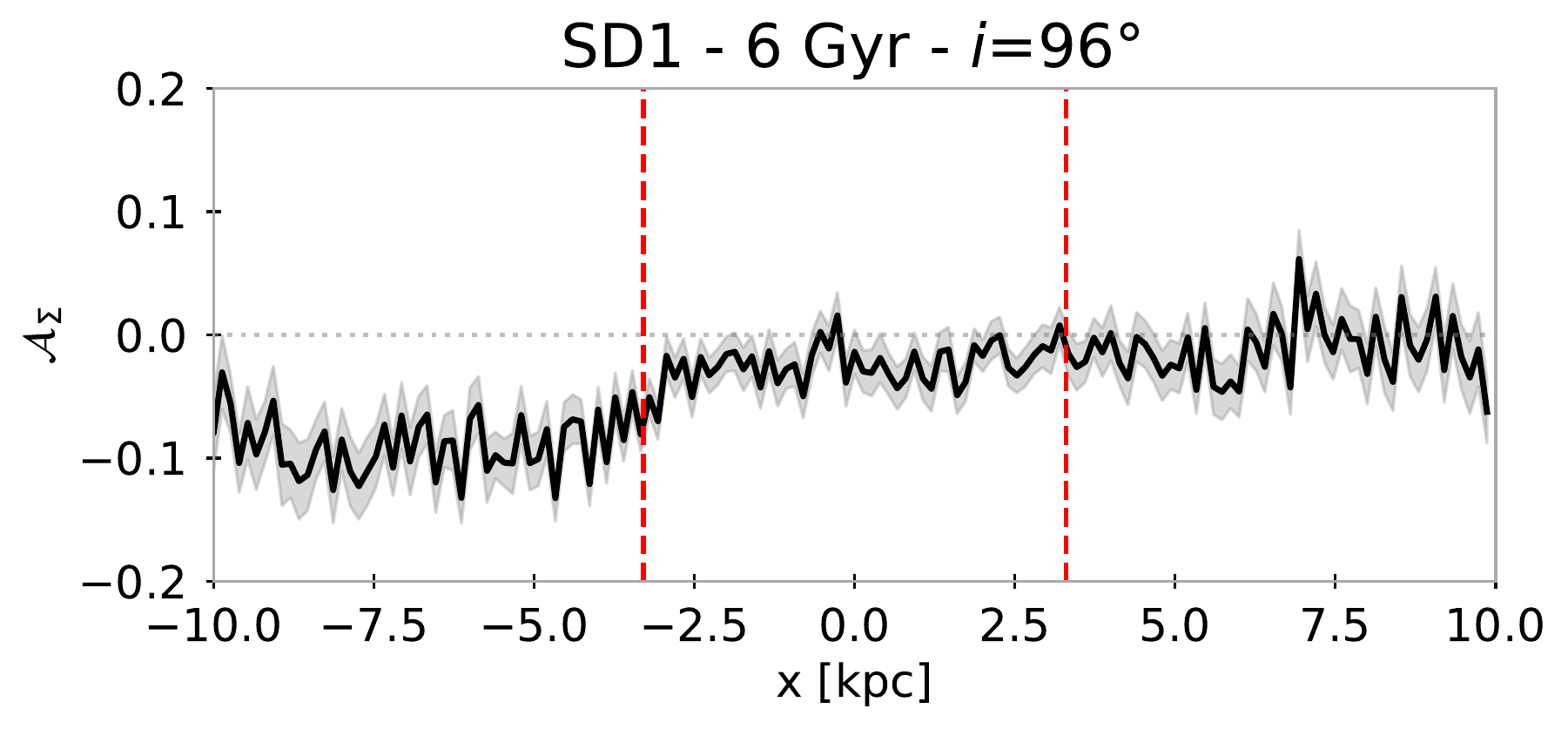}
    \includegraphics[scale=0.4]{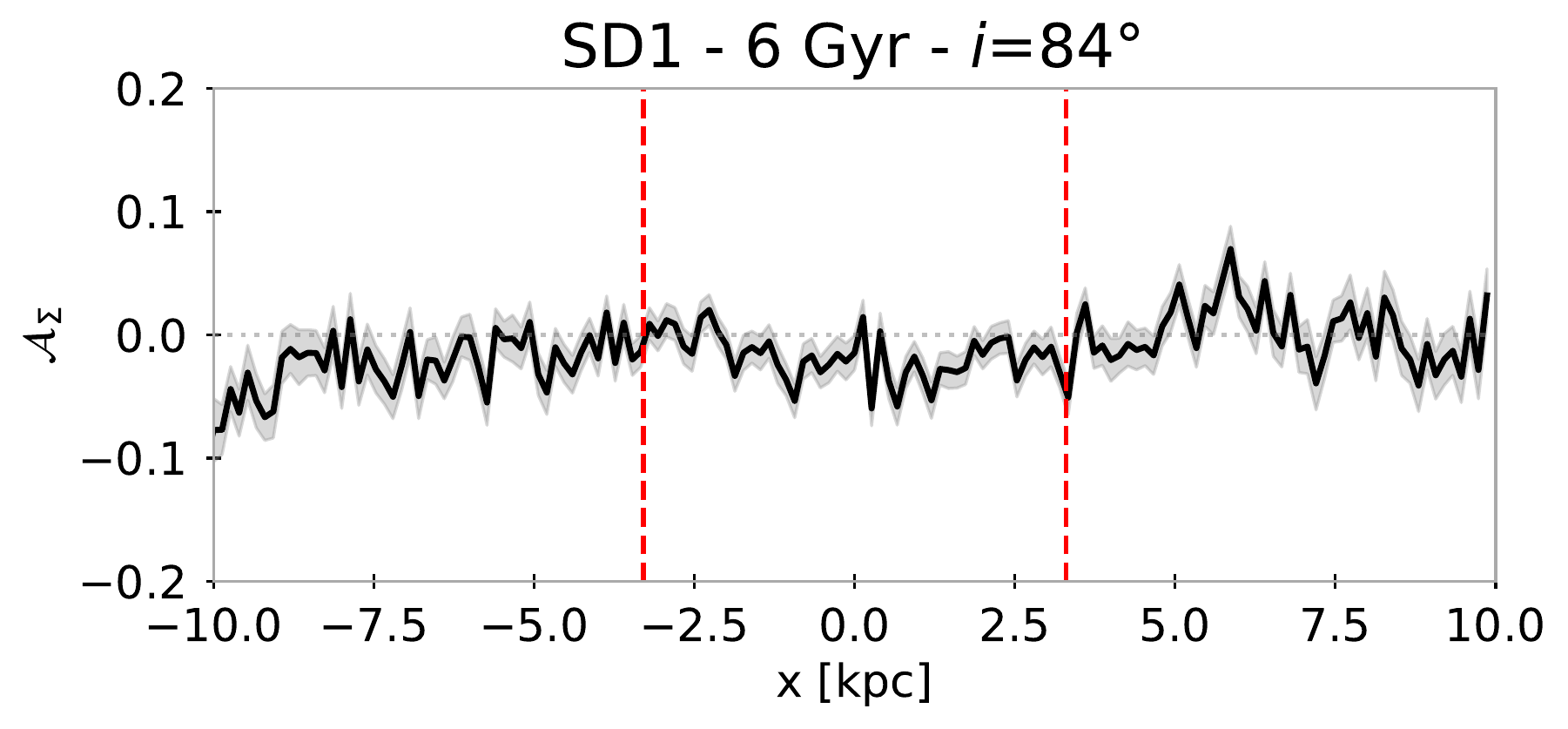}
    \includegraphics[scale=0.4]{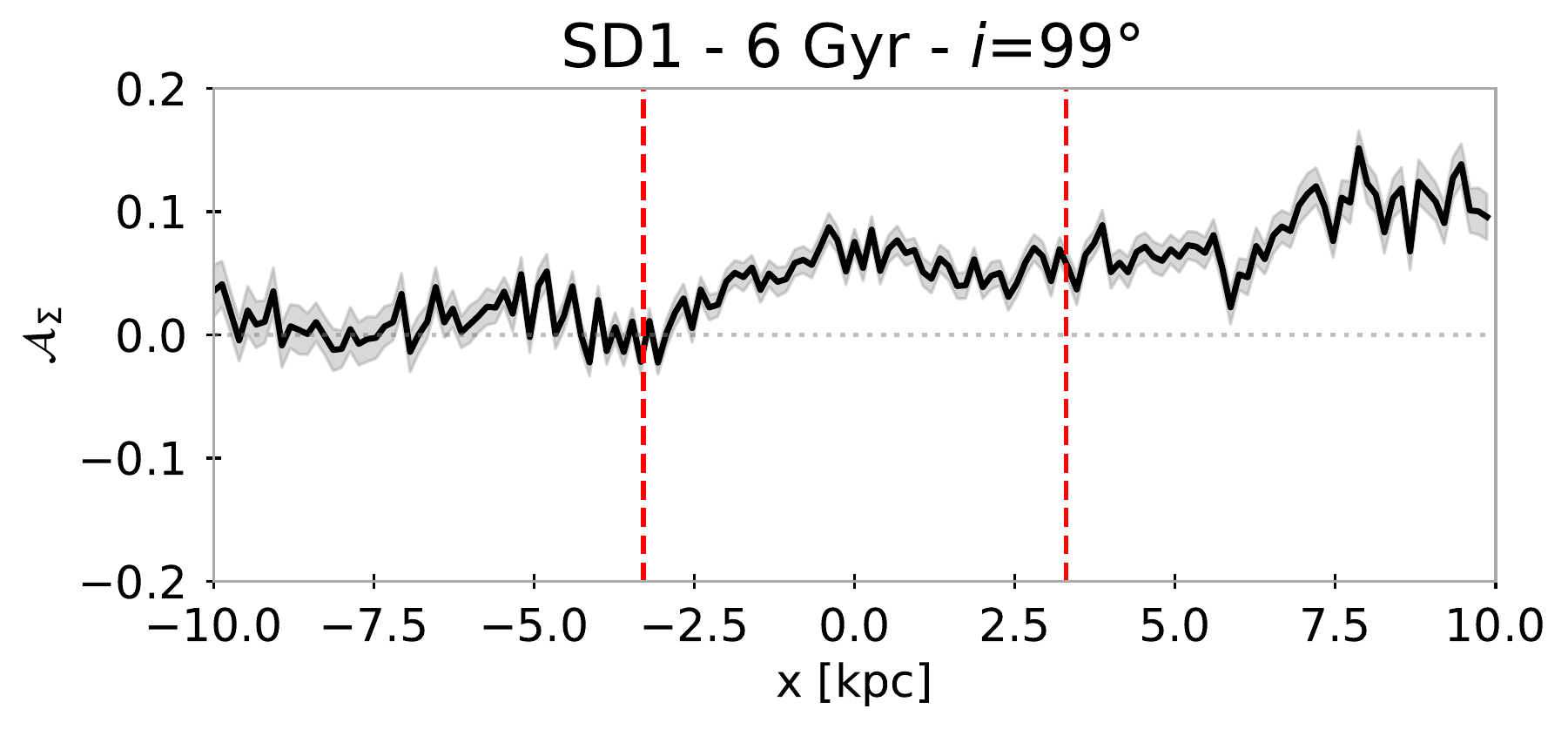}
    \includegraphics[scale=0.4]{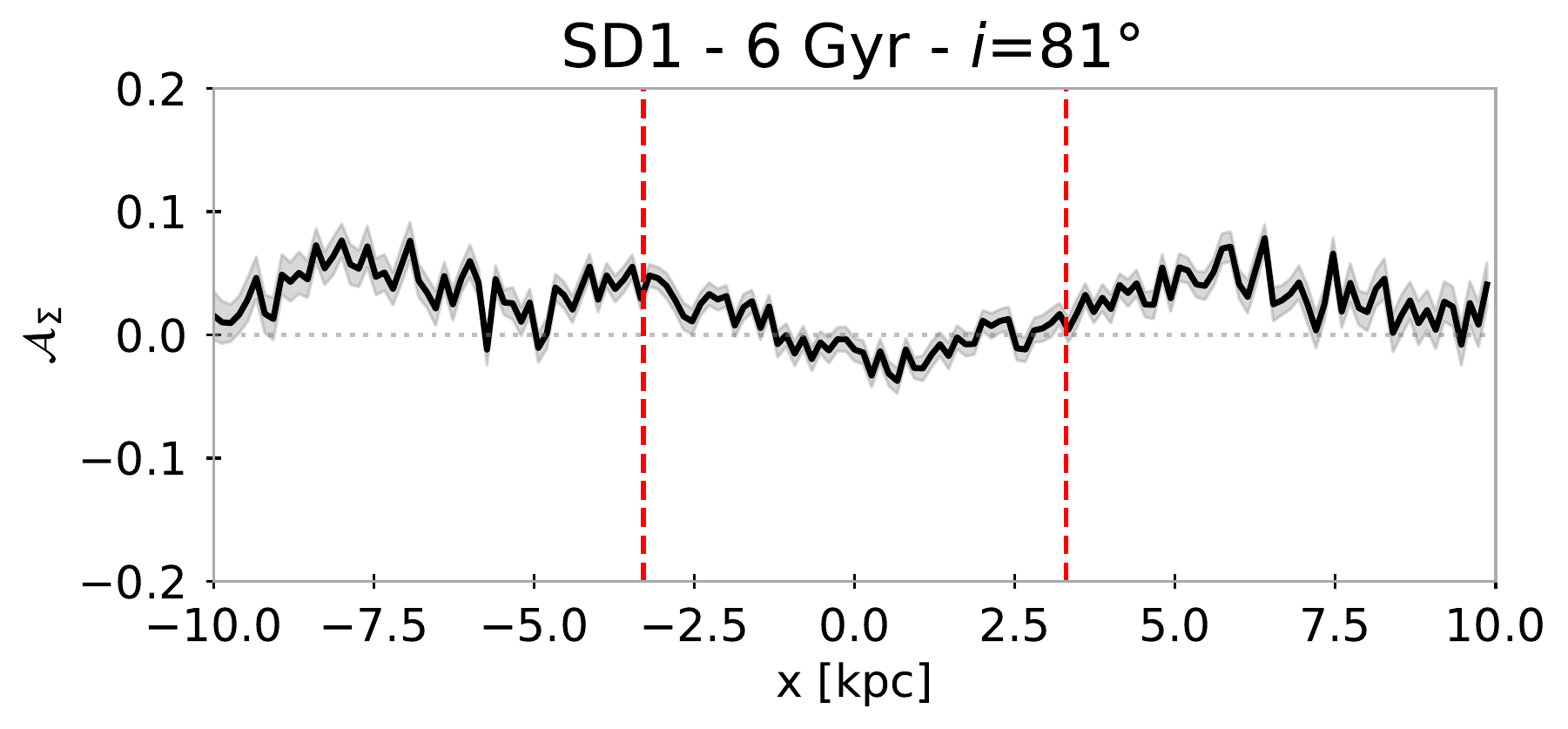}
    \vspace{-0.2cm}
    \caption{Same as Fig~\ref{fig:sim_d5_inc}, but for model SD1.}
\end{figure*}

\begin{figure}
    \centering
    \includegraphics[scale=0.4]{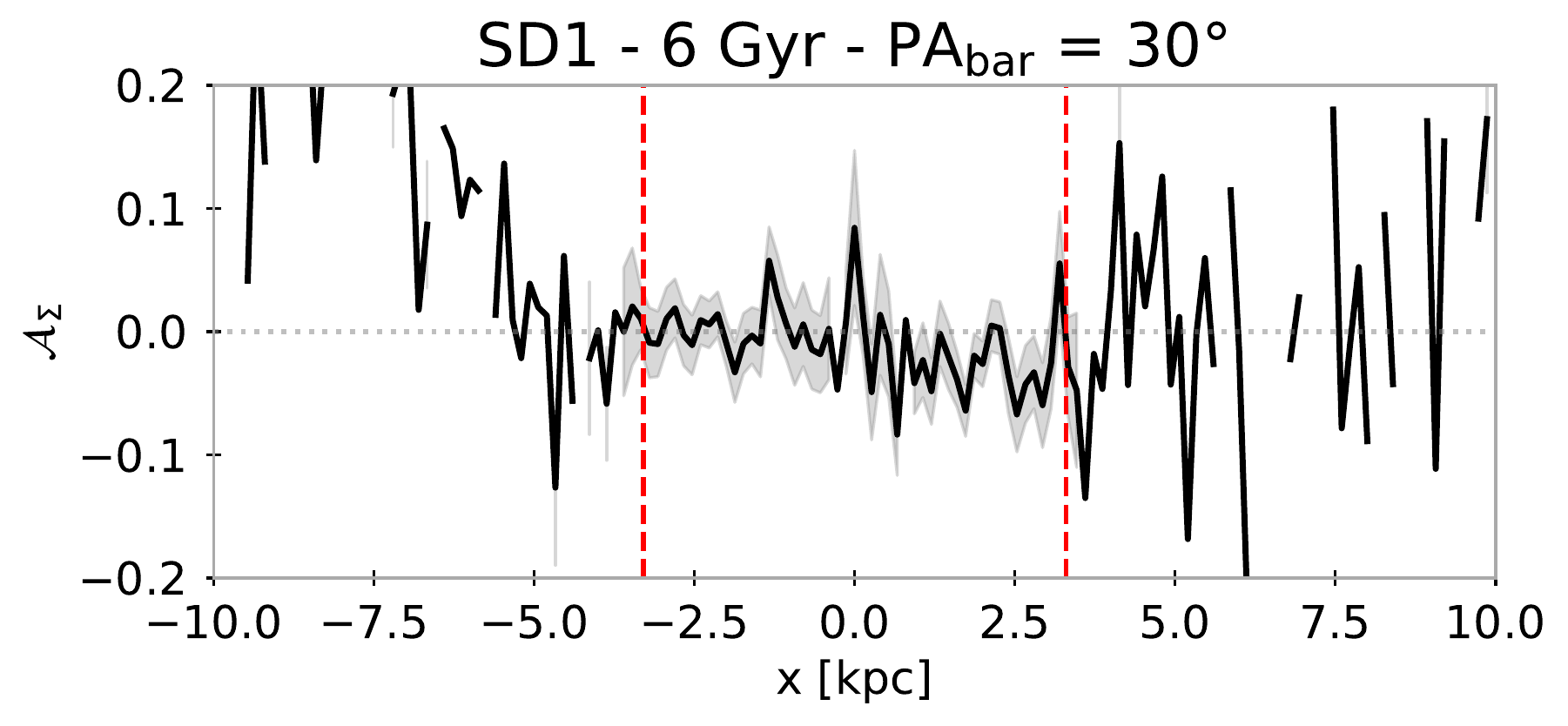}
    \includegraphics[scale=0.4]{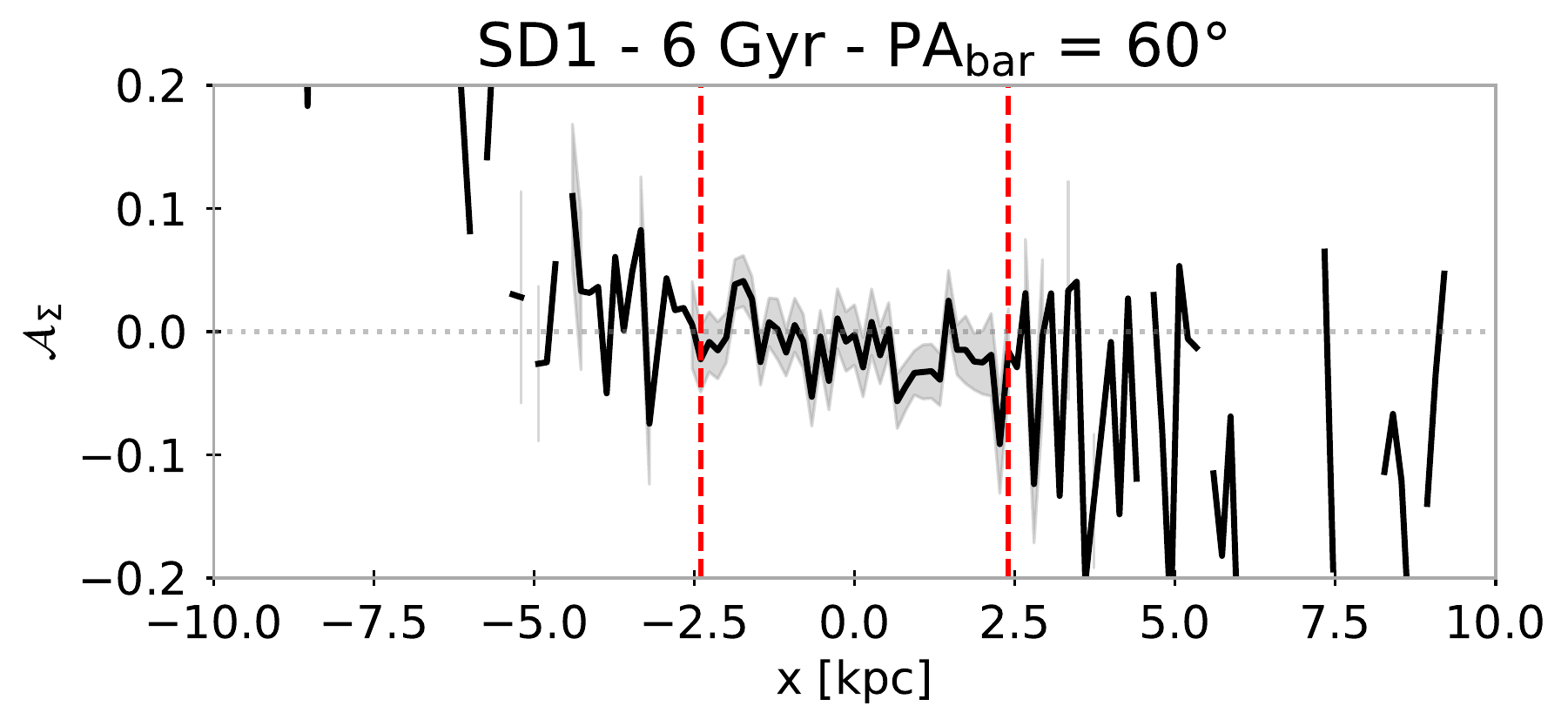}
    \includegraphics[scale=0.4]{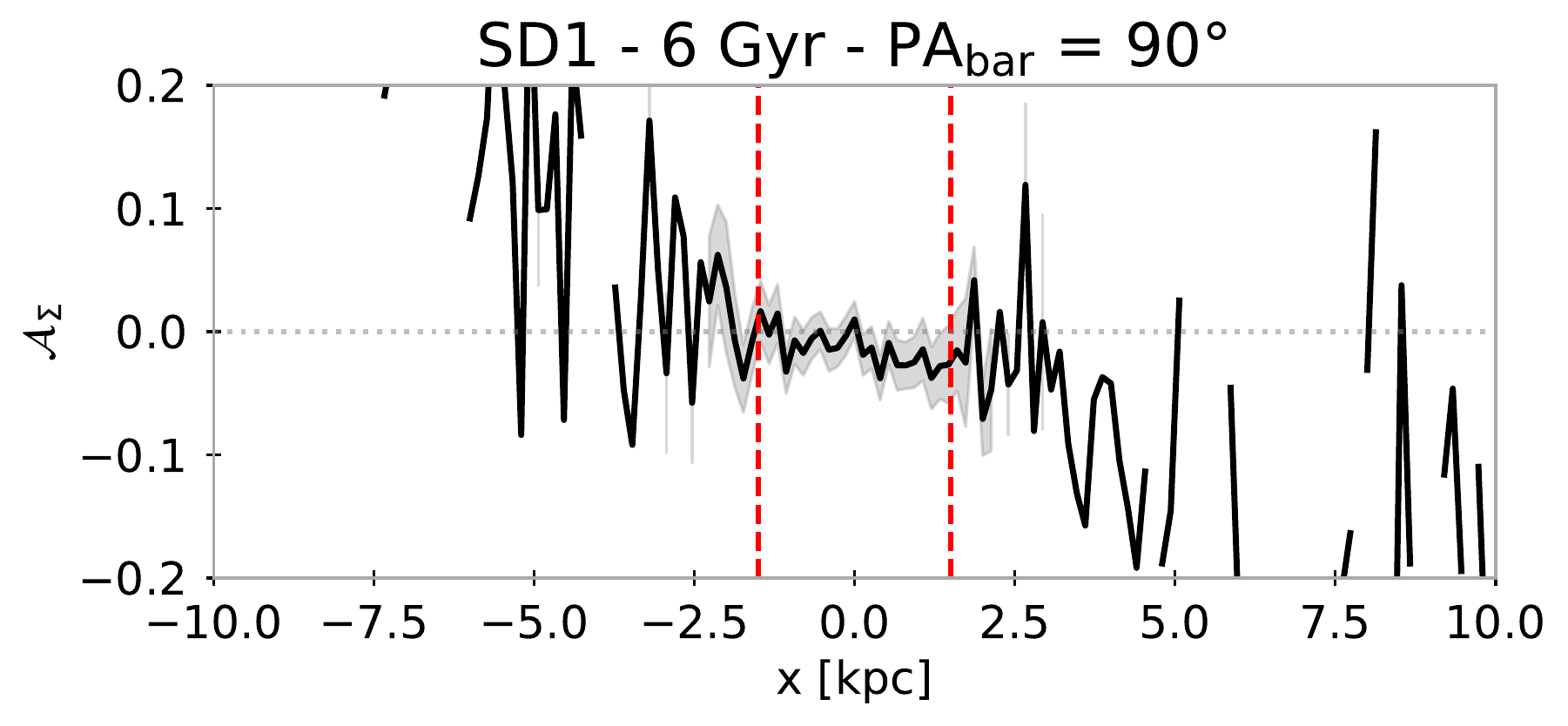}
    \caption{Same as Fig~\ref{fig:sim_bar}, but for model SD1.}
\end{figure}

\begin{figure}
    \centering
    \includegraphics[scale=0.4]{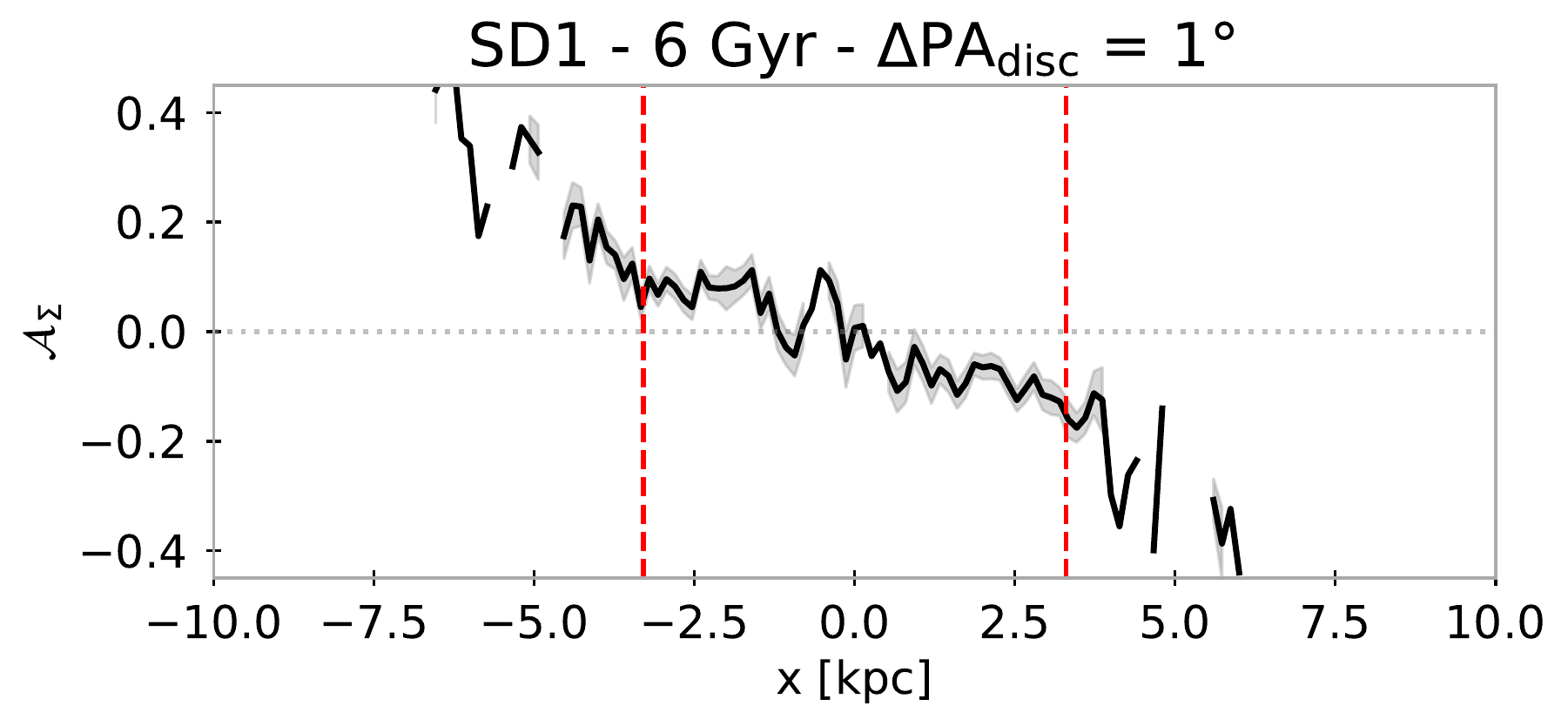}
    \includegraphics[scale=0.4]{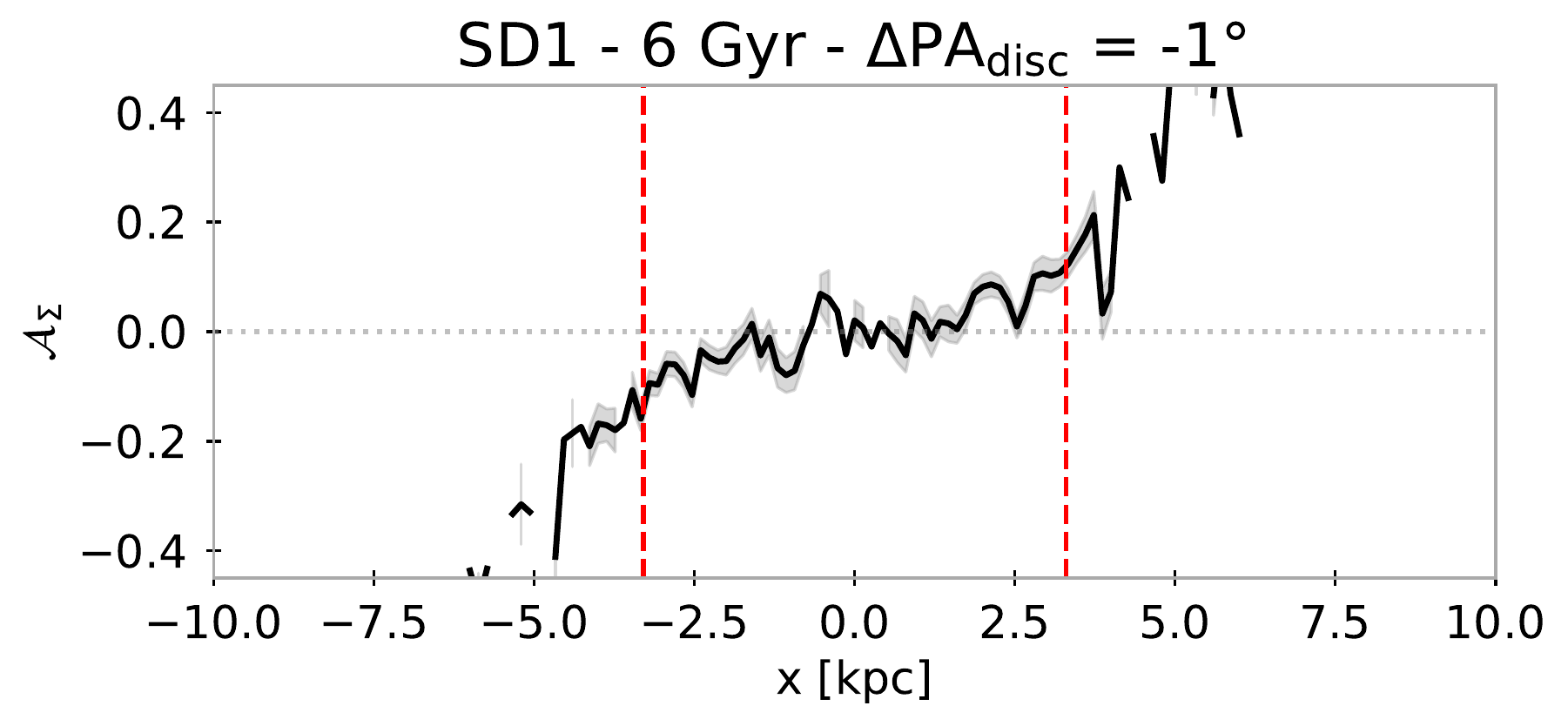}
    \caption{Same as Fig~\ref{fig:sim_pa}, but for model SD1.}
\end{figure}

\begin{figure}
    \centering
    \includegraphics[scale=0.4]{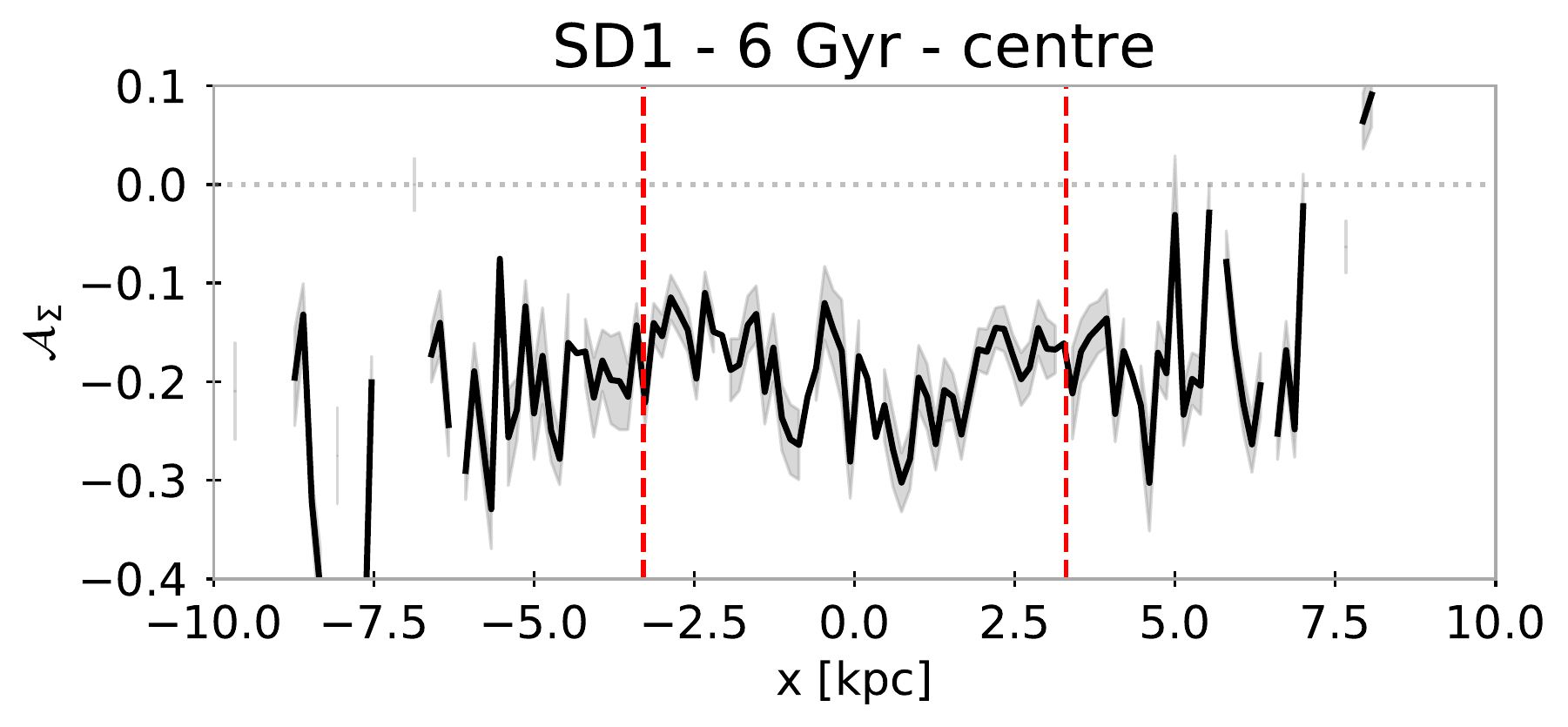}
    \caption{Same as Fig~\ref{fig:centre}, but for model SD1.}
\end{figure}

\clearpage
\section{Mid-plane asymmetry diagnostics for the discarded galaxies}
\label{appendix:b}

Four of the eight edge-on galaxies we selected from the S$^4$G catalog were discarded from the parent sample because of difficulties in analysing them. Here we present the mid-plane asymmetry diagnostics for these four galaxies. Table~\ref{tab:galaxies_poperties} gives the reason why each of these galaxies was discarded. 

\begin{figure*}
    \centering
    \includegraphics[scale=0.4]{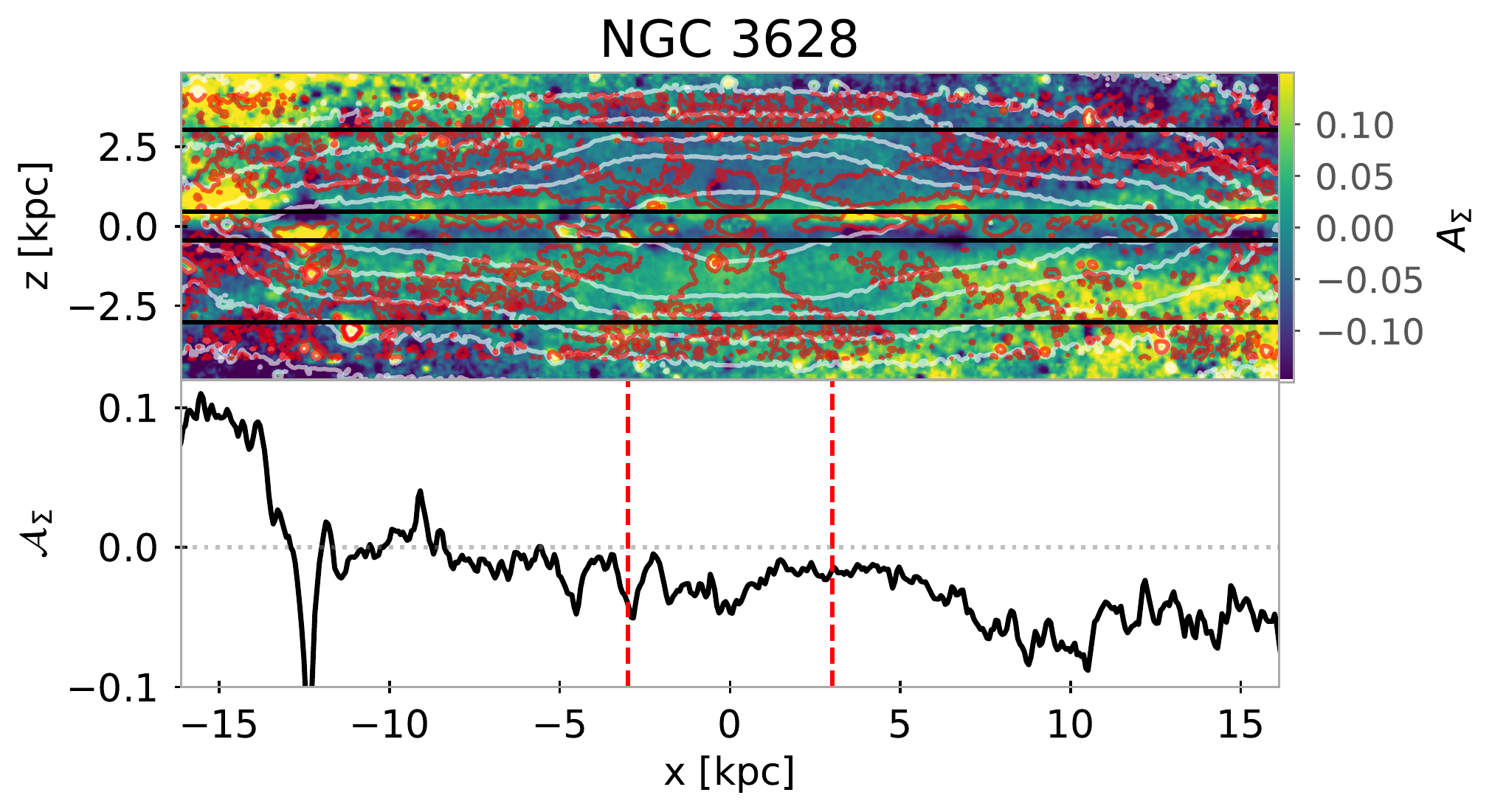}
    \includegraphics[scale=0.4]{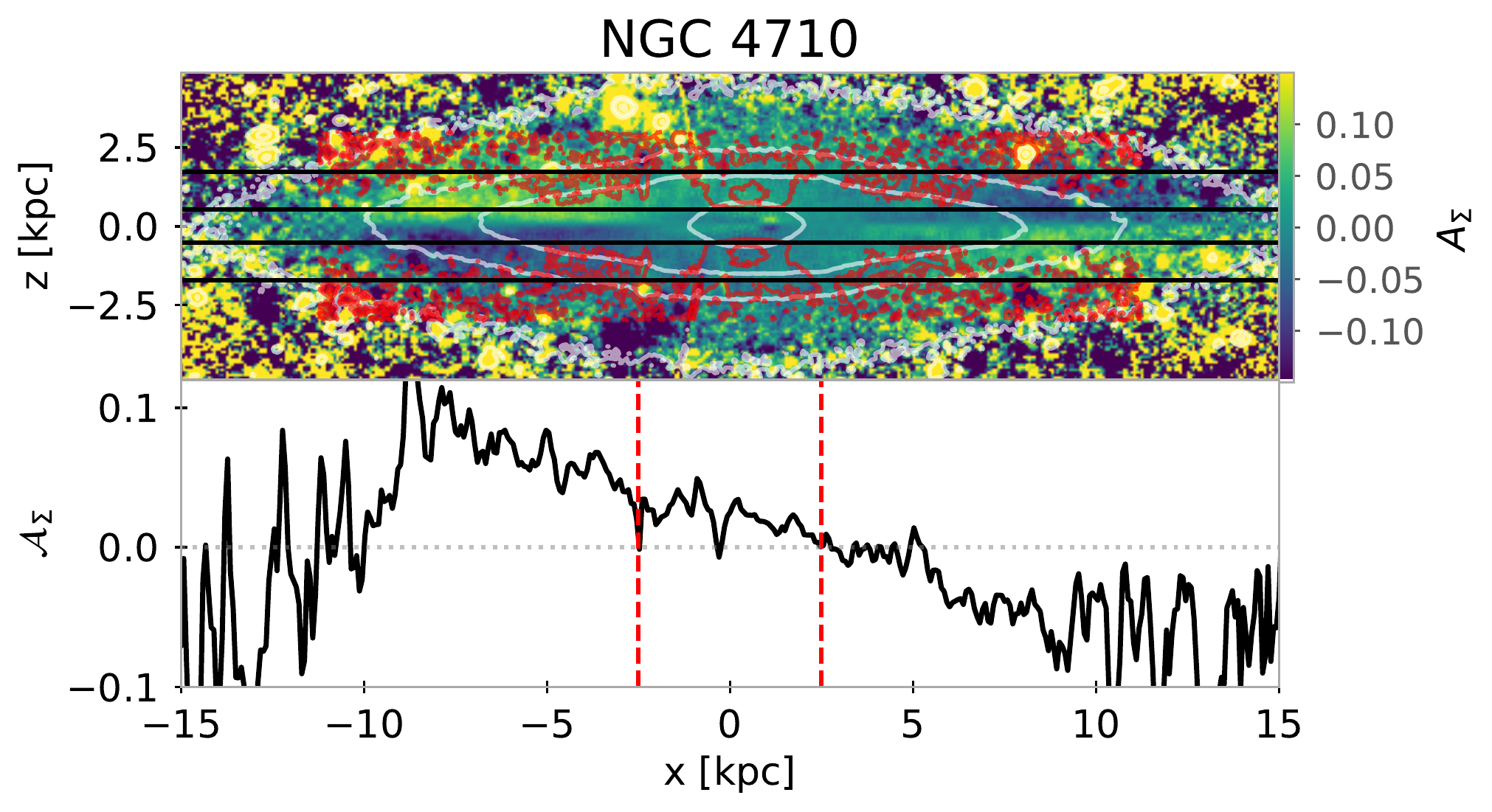}
    \includegraphics[scale=0.4]{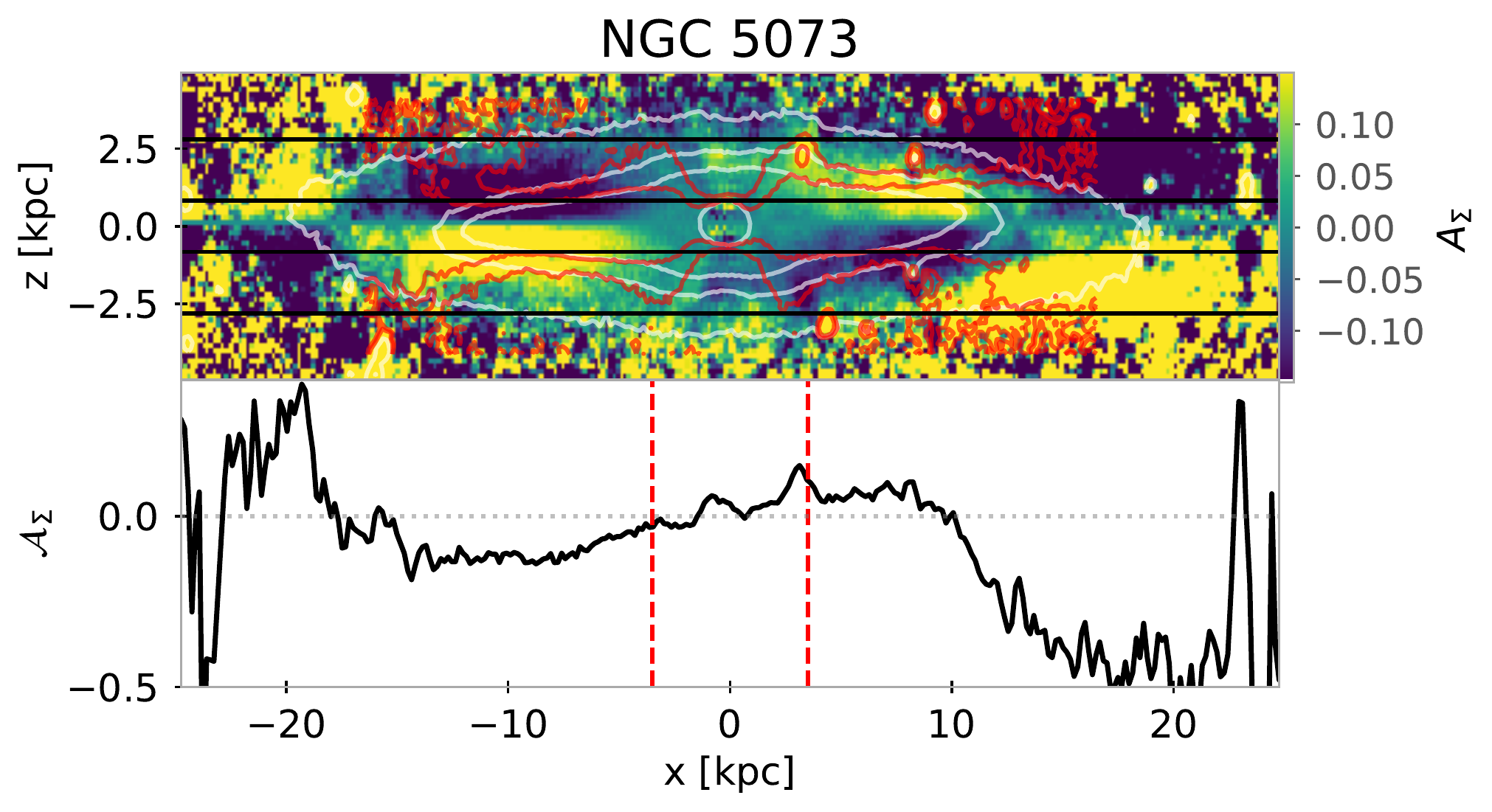}
    \includegraphics[scale=0.4]{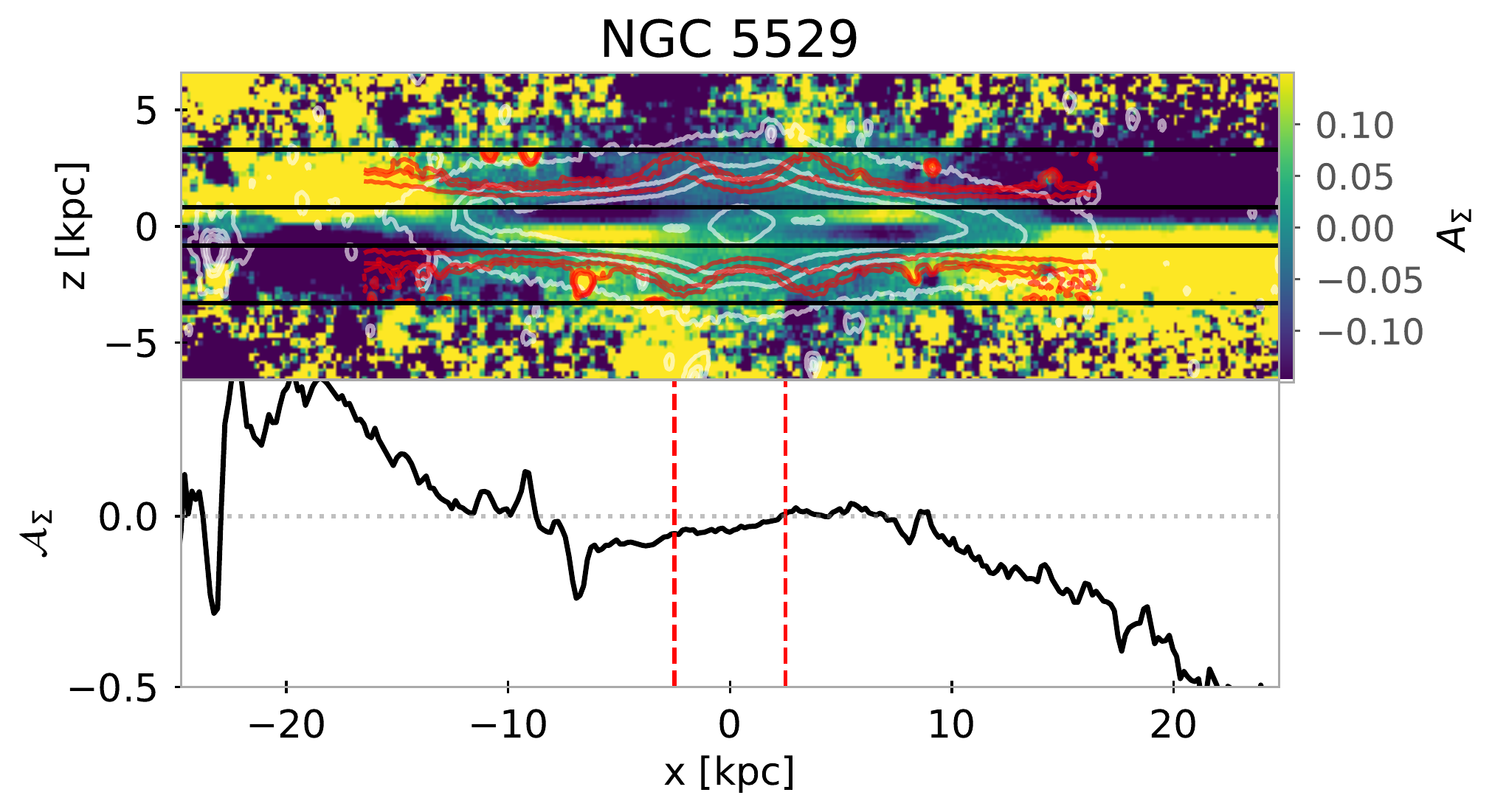}
    \caption{Same as Fig.~\ref{fig:AMP_real_galaxies}, but for the galaxies excluded from the final sample. Top-left panel: NGC~3628 with $0.5<z<3.0$ kpc. Top-right panel: NGC~4710 with $0.5<z<2.7$ kpc. Bottom-left panel: NGC~5073 with $0.8<z<2.8$ kpc. Bottom-right panel: NGC~5529 with $0.8<z<3.3$ kpc.}
    \label{fig:N3628}
\end{figure*}

\clearpage
\section{Mid-plane asymmetry diagnostics for the stellar mass maps of the galaxies}
\label{appendix:c}

Mid-plane asymmetry diagnostics for two of our final sample of galaxies obtained using the stellar mass maps of the galaxies. No stellar image is available for NGC~5170, while the image of ESO~443-042 is not suitable for this analysis.

\begin{figure*}
    \centering
    \includegraphics[scale=0.4]{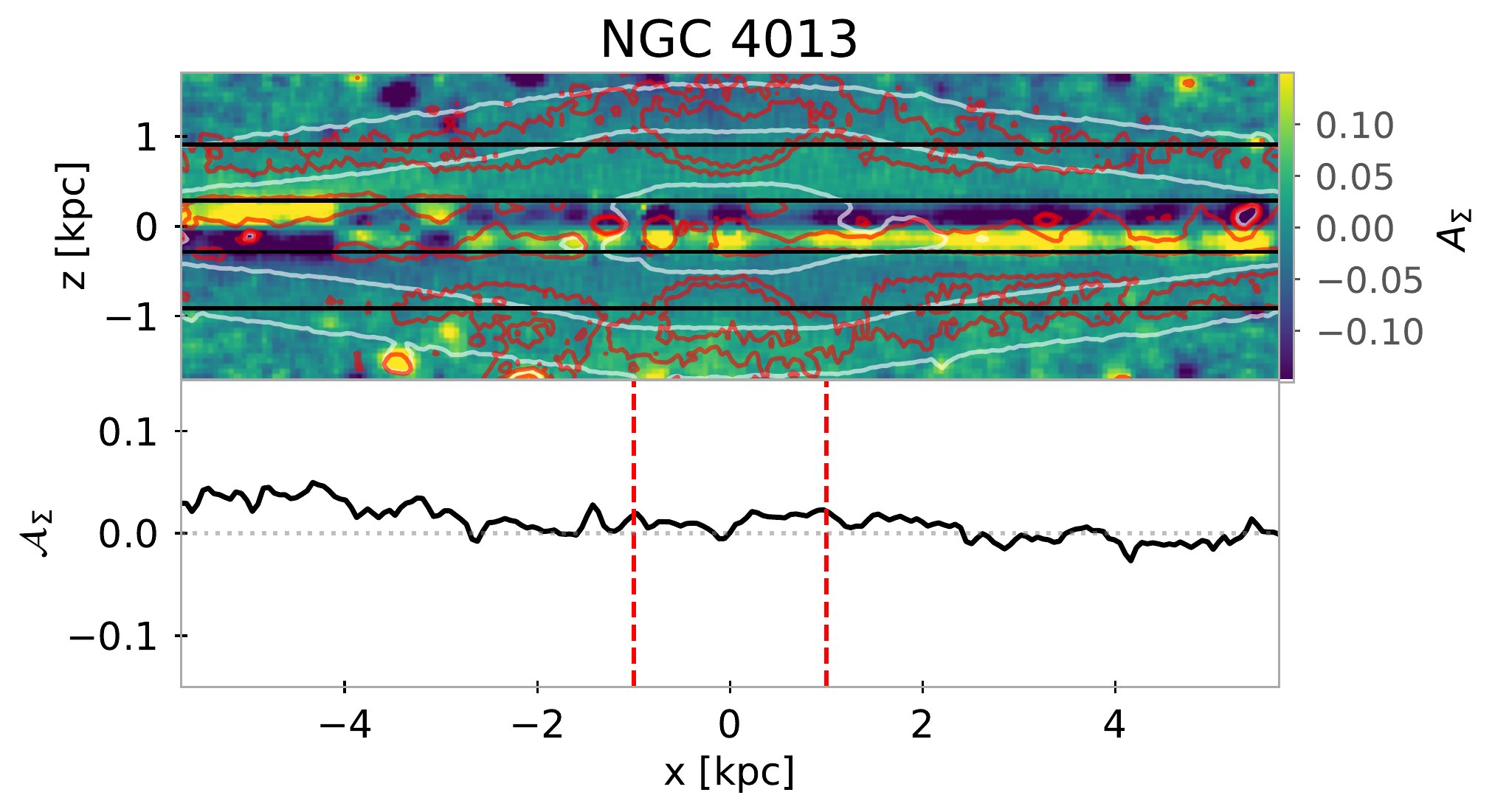}
    \includegraphics[scale=0.4]{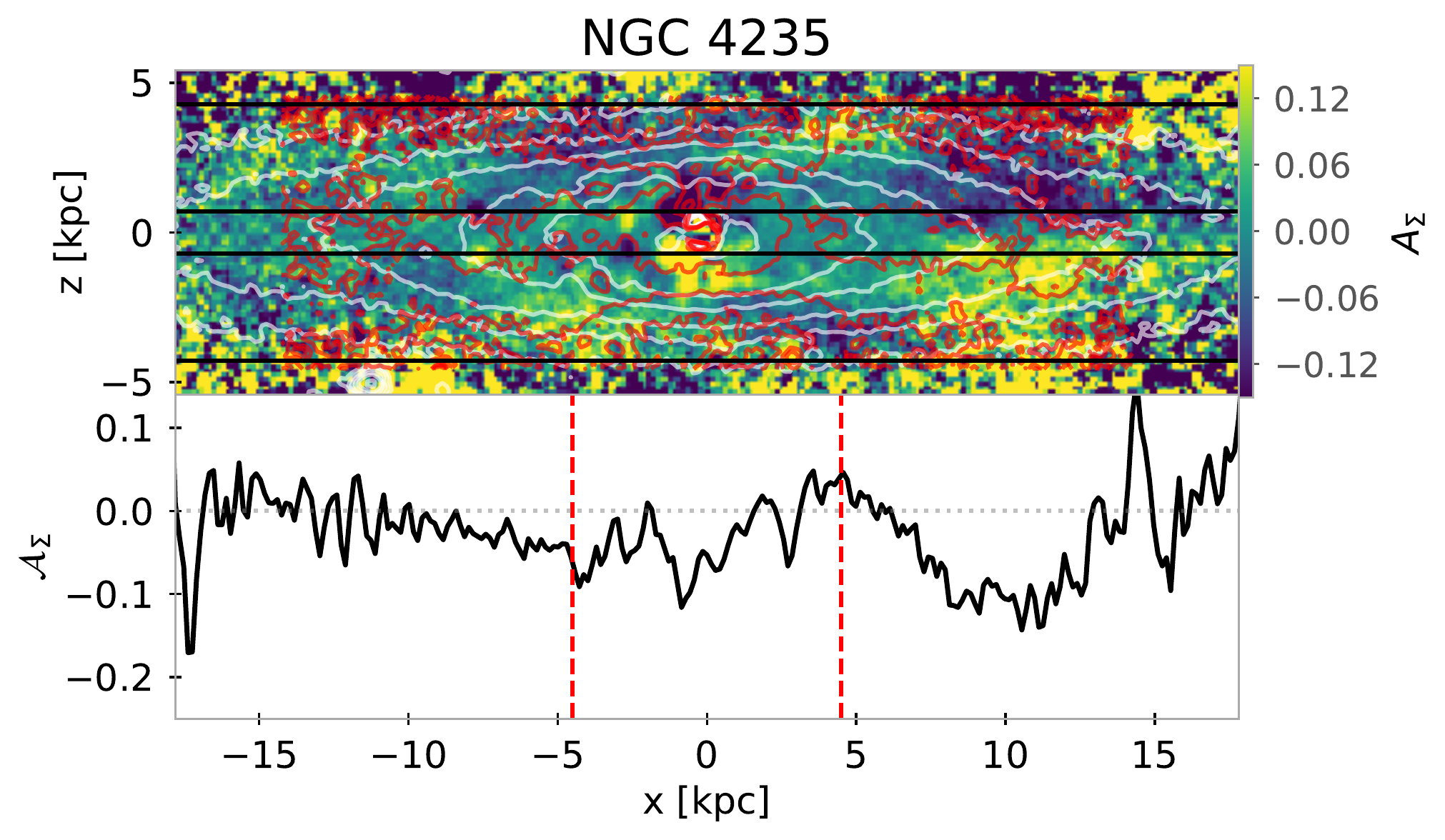}
    \caption{Same as Fig.~\ref{fig:AMP_real_galaxies}, but using the stellar mass map. Left panel: NGC~4013 with $0.3<z<1.0$ kpc. Right panel: NGC~4235 with $0.7<z<4.2$ kpc.}
    \label{fig:N3628m}
\end{figure*}

\end{appendix}

\bsp	
\label{lastpage}
\end{document}